\documentclass{aa}  

\usepackage{graphicx}
\usepackage{txfonts}
\usepackage{amsmath}	
\usepackage{amssymb}	
\usepackage{subcaption}
\usepackage{comment}
\usepackage{float}
\usepackage{enumerate}
\usepackage[export]{adjustbox}
\usepackage{xcolor}
\usepackage{lscape}

\newcommand{\kms}{$\mbox{km s}^{-1}~$}  
\newcommand{\ergs}{$\mbox{erg s}^{-1}~$}    
\newcommand{\lsun}{$L_{\sun}$}              
\newcommand{\lo}{L$_{\mbox{\rm [OIII]}}$}       
\newcommand{\lbol}{L$_{\rm bol}$}               
\newcommand{\hb}{$\mbox{H}\beta$~}          
\newcommand{\msun}{$\mbox{M}_{\sun}$}              
\newcommand{\msunyr}{$\mbox{M}_{\sun} \mbox{~yr}^{-1} $} 
\newcommand{\qfd}{\emph{QSOFEED }}
\newcommand{\tysp}{$t_{ysp}$}

\usepackage[colorlinks=true,linkcolor=bluea,allcolors=blue]{hyperref}

\begin{document}

   \title{\emph{QSOFEED}: The relationship between star formation and AGN Feedback}

   \author{P. S. Bessiere
          \inst{1,2}
          \and
          C. Ramos Almeida\inst{1,2}
          \and
          L. R. Holden\inst{3}
          \and
          C. N. Tadhunter\inst{3}
          \and
          G. Canalizo\inst{4}
          }

   \institute{Instituto de Astrofísica de Canarias, Calle Vía Láctea, s/n, E-38205, La Laguna, Tenerife, Spain\\
              \email{patricia.bessiere.astro@gmail.com}
         \and
             Departamento de Astrofísica, Universidad de La Laguna, E-38206, La Laguna, Tenerife, Spain
        \and
        Department of Physics \& Astronomy, University of Sheffield, S6 3TG Sheffield, UK
        \and
        Department of Physics and Astronomy, University of California, Riverside, 900 University Ave., Riverside CA 92521, USA
             }

   \date{Received November 20, 2023; }

 
  \abstract
   {Large-scale cosmological simulations suggest that feedback from active galactic nuclei (AGN) plays a crucial role in galaxy evolution. {\color {black} More specifically,} outflows are one of the mechanisms by which the accretion energy of the AGN is transferred to the interstellar medium (ISM), heating and driving out gas and impacting star formation (SF).}
   {The purpose of this study is to directly test this hypothesis utilising SDSS spectra of a well-defined sample of 48 low redshift (z<0.14) type 2 quasars (QSO2s).}
   {Exploiting these data, we characterised the kinematics of the warm ionised gas by performing a non-parametric analysis of the [OIII]$\lambda 5007$ emission line, as well as constrain the properties of the young stellar populations (YSP; \tysp$< 100 \mbox{~Myr}$) of their host galaxies through spectral synthesis modelling.}
   {These analyses revealed that 85\% of the QSO2s display velocity dispersions in the warm ionised gas phase larger than that of the stellar component of their host galaxies, indicating the presence of AGN-driven outflows. We compared the gas kinematics with the intrinsic properties of the AGN and found that there is a positive correlation between gas velocity dispersion and 1.4 GHz radio luminosity but not with AGN bolometric luminosity or Eddington ratio. This suggests that, {\color {black} either the radio luminosity is the key factor in driving outflows or that the outflows themselves are shocking the ISM and producing synchrotron emission. We found} that {\color {black} 98\%} of the sample host YSPs to varying degrees, with star formation rates (SFR) $0 \le SFR \le 92$ \msunyr, averaged over 100 Myr. 
We compared the gas kinematics and outflow properties to the SFRs to establish possible correlations which may suggest that the presence of the outflowing gas is impacting SF and find that none exists, leading to the conclusion that, on the scales probed by the SDSS fibre (between 2 and 7 kpc diameters), AGN-driven outflows do not impact SF on the timescales probed in this study. {\color{black} However, we found a positive correlation between the light-weighted stellar ages of the QSO2s and the black hole mass, which might be indicating that successive AGN episodes lead to the suppression of SF over the course of galaxy evolution.}}
   {}


   \maketitle
%

\section{Introduction}
\label{sec:introduction}

The impact of Active Galactic Nuclei (AGN) on their host galaxies is currently a topic of hot debate. Large-scale cosmological simulations (e.g. \citealt{bower06,croton06,schaye15,weinberger17,dave19}) rely on some fraction of the prodigious energy generated by gas accreting onto the central supermassive black hole (SMBH) coupling with the host's interstellar medium (ISM), disrupting gas which otherwise would be processed into stars. This process, known as AGN feedback, is thus credited with suppressing star formation (SF) and, consequently, preventing the growth of over-massive galaxies. AGN feedback is also invoked to explain the observed correlation between SMBH mass ($M_{BH}$) and the stellar velocity dispersion of the galaxy bulge \citep{magorrian98,ferrarese00,gebhardt00,tremaine02}, as both will grow in step with each other.

In these simulations, AGN feedback is typically separated into two distinct modes, the radiative or quasar mode and the mechanical or jet mode (e.g., \citealt{sijacki07,fabian12}). Feedback is incorporated into simulations using several different prescriptions, with the radiative mode associated with high accretion states, where the AGN is accreting at $> 1$ per cent of its Eddington limit. In this case, radiation emitted from the accretion disc couples with the ISM and drives the gas \citep{hopkins10,crenshaw15,fischer17,meena23}. In the case of objects in lower accretion states, radio jets interact with the ISM, driving shocks and carving out large-scale cavities in the inter-galactic medium, thereby preventing the cooling of gas and halting star formation \citep{mcnamara00,birzan04,rafferty06}.

In recent years, observational evidence for both modes of feedback, in the form of outflowing molecular and ionised gas outflows \citep{harrison14, cicone14,fiore17, fluetsch19,smethurst21,ramos22,audibert23}, has demonstrated the ability of AGN to disrupt the gas in a galaxy. The most accessible phase is the warm ionised gas, where the kinematics can be measured via the strong [OIII] $\lambda 5007$ emission line (e.g. \citealt{villar11,harrison14,woo16}), leading to this being the most commonly investigated tracer. Studies such as these have unequivocally demonstrated that AGN-driven outflows are ubiquitous at high and low redshift.

However, although we can be confident that AGN drive gas outflows, the impact of those outflows on SF is less clear. Some studies of stellar populations in AGN host galaxies have found evidence for suppressed SF \citep{ho05,mullaney15,shimizu15,stemo20}, whilst other works have found that star formation rates (SFR) in AGN hosts lay on or above the galaxy main sequence \citep{mullaney12,stanley15,scholtz18,grimmett20,jarvis20,shangguan20,zhuang20,zhuang22}, suggesting that AGN do live in star-forming galaxies.

One of the fundamental challenges in understanding the impact (if any) that AGN-driven outflows have on SF is the vastly different timescales over which AGN feedback and other processes that govern SF operate. AGN phases last for a few Myr \citep{martini01,hopkins05}, although the imprint of these phases in the ISM might be visible on significantly longer timescales of $\sim100 \mbox{~ Myr}$ \citep{king15}. Analytic modelling \citep{king11} also suggests that energy-driven outflows can propagate to radii larger than 10 kpc on timescales that are consistent with the current episode of AGN activity (1 -- 100 Myr). Therefore, to further our understanding of the potential of AGN feedback in galaxy evolution and evaluate whether it is capable of directly impacting SF via the heating/removal of gas, we must consider SFRs over timescales consistent with with those of AGN-driven outflows \citep{ramos22}. In a recent work, \citet{bessiere22} used integral-field spectroscopic data of the QSO2 Mrk 34 to show that both positive and negative feedback can occur simultaneously in different parts of the same galaxy, depending on the amount of energy and turbulence that the outflows inject in the ISM. This illustrates the complexity of the outflow-ISM interplay and the need to consider the same timescales. In the case of Mrk 34, the dynamical time of the ionised outflow traced by [O III] is $\sim$1 Myr, and the age of the stellar populations considered in the analysis is $\sim$1-2 Myr. 

Finally, to truly understand the impact that the \emph{current} episode of AGN activity has on the distribution of gas, particularly in terms of their ability to impact SF, it is crucial to understand the characteristics of the gas in several different outflow phases in \emph{the same} galaxy \citep{cicone18}.  We must try to understand in which gas phase the most significant disruption occurs, on what physical scales and whether this is capable of disrupting/suppressing SF.

In this paper, we provide a detailed analysis of the stellar populations and gas kinematics within the \qfd sample, presenting a comprehensive investigation into the interplay between AGN-driven outflows and star formation. In Section \ref{sec:sample}, we summarise the characteristics of the \qfd sample, outline the data utilised in this investigation and describe our methods of data analysis for both the stellar population and emission line modelling. In Section \ref{sec:results} we present our findings and investigate the possibility of correlations between gas kinematic properties and SFRs for the whole sample. Finally, in Section \ref{sec:discussion}, we consider the results presented in this work in the context of galaxy evolution.

Throughout this work, we assume a cosmology with $H_0 = 70 \mbox{km s}^{-1} \mbox{ Mpc}^{-1}$, $\Omega_m = 0.3$ and $\Omega_\Lambda = 0.70$.

\section{Data and analysis}
\label{sec:data}
\subsection{\qfd~sample selection and characteristics}
\label{sec:sample}

The purpose of the Quasar Feedback (\href{http://research.iac.es/galeria/cra/qsofeed/}{\it{QSOFEED}}) project is to quantify the impact of AGN feedback on galaxies by conducting a multi-wavelength survey aimed at characterising the multi-phase outflow properties of a well-defined sample of nearby, obscured quasars (QSO2s; \citealt{ramos17,ramos19,ramos22,speranza22,speranza24}). These outflow characteristics will then be compared with the intrinsic properties of the host galaxies, such as recent star formation \citep{bessiere22} and nuclear molecular gas reservoirs \citep{ramos22,audibert23}. We focus on obscured quasars because the overwhelming direct quasar light is shielded from our view by intervening material which acts as a natural chronograph. Attempting to study the properties of the host galaxies of unobscured quasars (e.g., stellar populations) is extremely challenging because it is difficult to disentangle the galaxy and quasar light. The obscuration in QSO2s alleviates many of these issues, allowing for a detailed analysis of the host galaxy and the central regions where we expect AGN-driven outflows to have the most obvious direct impact. 

The \qfd~sample is selected from the catalogue of Sloan Digital Sky Survey (SDSS; \citealt{york00}) narrow-line AGN of \citet{reyes08} and comprises all objects that fulfil the criteria \lo $> 10^{8.5}$ \lsun~ (log \lo > 42 \ergs) and $z<0.14$. These criteria were applied to ensure that the objects selected for this study represent the most luminous nearby AGN, whilst still allowing for the detailed study of their host galaxies. Applying these criteria results in a sample of 48 QSO2s with bolometric luminosities $44.9 < \mbox{log}$ \lbol $< 46.0$ \ergs\, assuming the bolometric correction of \citet{lamastra09} and using the extinction corrected [OIII] luminosities of \citet{kong18}.

\begin{figure}
    \centering
    \includegraphics[width = \columnwidth]{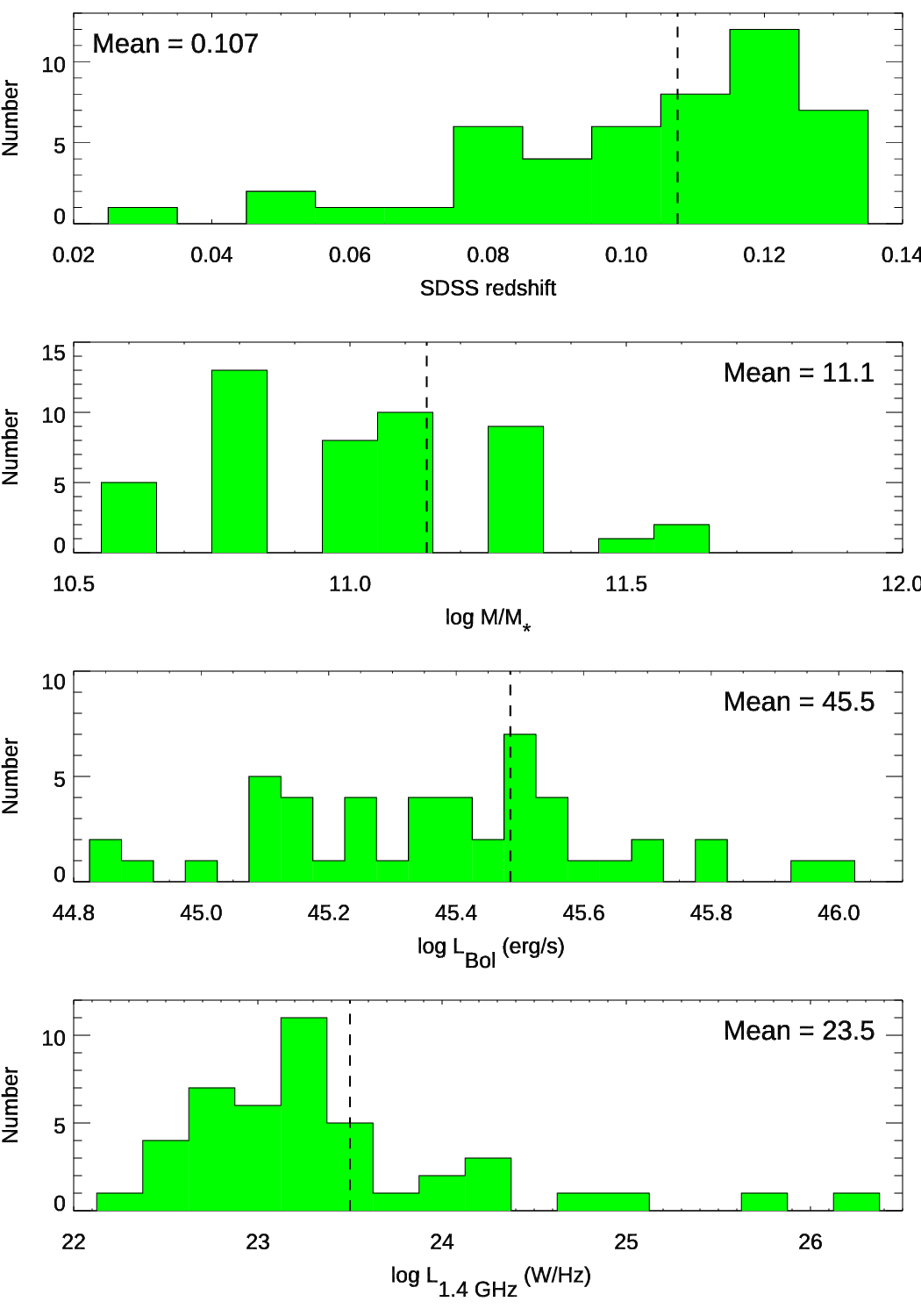}
    \caption{The distribution of the key properties of the \qfd~sample. From top to bottom, the panels show the distribution in redshift of the sample, in stellar masses of the host galaxies taken from \citet{pierce23}, in bolometric luminosities, and in 1.4 GHz radio luminosities. Note that only objects that have either FIRST or NVSS detections are included in the bottom panel and upper limits are given in Table \ref{tab:sample}. The mean value for each property is given in the individual panels.}
    \label{fig:histos}
\end{figure}

The \qfd host galaxies are all massive ($10.6 \le \log \mbox{M}_* \le 11.7$ \msun) and display a range of morphologies, including early types, spirals, and interacting systems. Visual inspection of deep optical imaging of the complete sample shows that at least $65^{+6}_{-7} \mbox{ per cent}$ are currently undergoing a merger event \citep{pierce23}. Figure \ref{fig:histos} shows the distribution of the redshift, mass, bolometric luminosity, and 1.4 GHz radio luminosity ($L_{1.4 GHz}$)\footnote{Calculated either from the Faint Images of the Radio Sky at Twenty-cm (FIRST; \citealt{becker95}) or NRAO VLA Sky Survey (NVSS; \citealt{condon98}) fluxes of the sample.}, and Table \ref{tab:sample} gives a classification of the main properties. Not all the objects were detected in the radio (44/48 detections), so the bottom panel of  Figure \ref{fig:histos} only includes objects that were detected by either FIRST or NVSS, whilst upper limits for $L_{1.4 GHz}$ are shown for these objects in Table \ref{tab:sample}.  The majority of the QSO2s have radio luminosities in the range [22.5,23.5] W~Hz$^{-1}$, with three objects qualify as radio-loud, having log L$_{1.4\text{GHz}}>$25 W~Hz$^{-1}$ (see the bottom panel of Fig. \ref{fig:histos}) and L$_{1.4\text{GHz}}$/L$_{\text{[O~III]}}$ ratios above the \citet{Xu99} division. The remaining 44 QSO2s are radio-quiet according to the two previous criteria.

The black hole masses and Eddington ratios reported in Table \ref{tab:sample} are from \citet{kong18} and were estimated using stellar velocity dispersions measured from SDSS spectra and the M$_{\rm BH}$-$\sigma_*$ relation. The black hole masses range from 10$^{6.8}$ to 10$^{8.8}M_{\sun}$ and the Eddington ratios (f$_{\rm Edd}$) between 0.01 and 4.57.
Thus, our QSO2s are near-Eddington to Eddington-limit obscured AGN in the local universe, unlike Seyfert 2 galaxies, which have typical Eddington ratios of f$_{\rm Edd}\sim$0.001--0.1. We refer the reader to \citet{ramos22} for more details on the {\it QSOFEED} sample.

To carry out this investigation into the stellar populations and gas kinematics in the \qfd~sample, we downloaded spectra for each object from the SDSS archive. The majority of the spectra (43) are derived from the Legacy survey \citep{abazajian09}, which used a 3 arcsec diameter fibre and covered an observed wavelength range of 3800 -- 9200 \AA\ ($\sim 3455 - 8360$ \AA\ at the average redshift of the sample, z=0.11) with a spectral resolution of R=1800--2200. The remainder of the objects (5) have been observed as part of the Baryon Oscillation Spectroscopic Survey (BOSS; \citealt{dawson13}), which used a 2 arcsec diameter fibre and covered an observed wavelength range 3600 -- 10000 \AA\ ($\sim 3270 -9090$ \AA\ at z=0.11) with a resolution 
of R=1300--2600. Where available, the BOSS spectra were used because the extended blue-ward coverage is useful when attempting to detect the presence of young stellar populations (YSPs) and these objects are marked with an asterisk in Table \ref{tab:sample}. The wide wavelength coverage means that key features for the modelling of stellar populations are covered, however, the signal-to-noise ratio of the data varies widely across the sample, ranging from $\sim 12 -40$. This can have a significant impact on our ability to constrain the proportion of flux allocated to the YSP in the model and consequently, the star formation rate.

\subsection{Stellar populations modelling}
\label{sec:stellar_mod}

Using the same stellar population modelling technique as in  \citet{bessiere22}, we made use of the  {\sc starlight} \citep{fernandes05} code in conjunction with the Binary Population and Spectral Synthesis (BPASS) synthetic stellar models \citep{eldridge17, stanway18}. Before carrying out the modelling, the spectra were first corrected for Galactic extinction using the E(B-V) values of \citet{schlafly11} and the \citet{cardelli89} extinction law and then shifted to the rest frame using the SDSS redshift values. The spectra were then resampled to 1 \AA\ pix$^{-1}$ as suggested in the {\sc starlight} user manual. 

We selected BPASS models with a broken power-law inital mass function (IMF) with $\alpha_1 = -1.30$ and $\alpha_2 = -2.35$ and an upper mass cut-off of 100\msun, where each template represents a simple stellar population formed in an instantaneous burst of $10^6$\msun. Three metallicities were included in the modelling, allowing for sub-solar, solar and super-solar stellar populations ($Z = 0.002,0.02,0.04$) and 27 ages for each metallicity ranging from 1 Myr to 12.5 Gyr, sampled to reflect the rapidity of the evolution of the stellar spectra at young ages.  The base templates can then be combined in different proportions by the {\sc starlight} code to produce the final model, assuming that 1) all the templates will be subject to the same reddening or 2) selected templates can have additional reddening applied. Throughout this study, the extinction curve of  \citet{calzetti00} has been assumed.

To better fit the stellar populations, the nebular continuum and higher-order Balmer emission lines were first modelled and subtracted from the SDSS spectra using the same technique outlined in \citet{bessiere17}. The most prominent emission lines were masked in this first step. To briefly summarise, initially 50 independent fits to each spectrum were carried out allowing for all ages and metallicities outlined above. In this initial run, it was assumed that templates with ages $<7$ Myr could have additional reddening applied compared to those of older ages. The fit with the lowest $\chi^2/N$ was then selected as the best fit and subtracted from the data, leaving a pure emission line spectrum. The [OIII]$\lambda\lambda 5007,4959$ and \hb lines were then fit simultaneously, using up to 5 Gaussian components, applying the standard technique of tying the velocities and widths of the \hb Gaussian components to those of the [OIII] components. In this way, it is possible to better constrain the fit to the \hb line when multiple Gaussian components are required to produce an acceptable fit. Assuming Case B recombination \citep{osterbrock06}, the total flux of the \hb line (the sum of Gaussian components) is then used to construct the nebular model, which consists of the nebular continuum and the higher-order Balmer emission lines (e.g., H9 (3835 \AA), H10 (3798 \AA), H11 (3771 \AA), and H12 (3750 \AA)). The resulting nebular model is then subtracted from the SDSS spectrum,  producing the nebular subtracted spectrum used for the remainder of the stellar modelling process. A detailed discussion of the impact of the nebular subtraction on the results of the stellar population modelling is presented in Section \ref{stellar}.

The next step in the modelling process is determining the best combination of metallicities to fit the nebular subtracted data. To facilitate this, three age bins for the stellar populations were defined, and stellar population templates from each bin were included in all the following modelling step. The defined age bins are:

\begin{itemize}
      \item Young stellar population (YSP; \tysp $ < 100 \mbox{ Myr}$)
    \item Intermediate stellar population (ISP; $100 \mbox{ Myr} < t_{isp} < 2 \mbox{ Gyr}$)
    \item Old stellar populations (OSP; $t_{osp} > 2 \mbox{ Gyr}$)  
\end{itemize}

A grid of 26 combinations of age bins and metallicities was defined by assigning one of the above metallicities to each age bin in turn. For each combination two scenarios were considered, one in which all the templates were subject to the same reddening and one in which models with ages $<$7 Myr were allowed to have additional reddening, resulting in a total of 54 combinations tested for each spectrum. Each combination was run through {\sc starlight} 20 times and the combination that produced the lowest mean $\chi^2/N$ was selected as the correct stellar model. 

To estimate the errors on the flux allocated to each of the stellar age bins outlined above, random Gaussian noise, consistent with the SDSS error spectrum, was added to each spectrum 100 times and run again using the same combination of metallicities as found in the previous step. The errors in the proportion of the flux in each age bin are then considered to be the standard deviation of the results of these 100 runs.

Examples of the {\sc starlight} fits to the nebular subtracted data are shown in Figure \ref{fig:sl_exemp}, where the left panel shows the fit to J0818+36 in which 7 per cent of the flux in the normalising bin is associated with the YSP, whilst the right panel shows the fit for J1548-01, which has a large contribution from the YSP (88 per cent). In both cases, the bottom panel shows the residuals of the fit, with the grey-shaded areas showing the regions masked out during the fit. The pink-shaded areas show regions given double weighting because they are strongly associated with stellar absorption features not otherwise infilled by emission lines. The green crosses mark pixels flagged in the original SDSS spectrum and masked out of the fit. 

The results of the {\sc starlight} modelling also allow us to calculate star formation rates (SFR) within the physical area covered by the SDSS fibre {(between 2 and 7 kpc diameters depending on the redshift). We sum the total mass associated with each input stellar template with ages $< 100 \mbox{~Myr}$ and divide by $10^8 \mbox{~ years}$, thereby averaging the SFR over the timescale of interest in this work.

\begin{figure*}
    \centering
    \begin{subfigure}{\columnwidth}
    \includegraphics[width = 1\columnwidth]{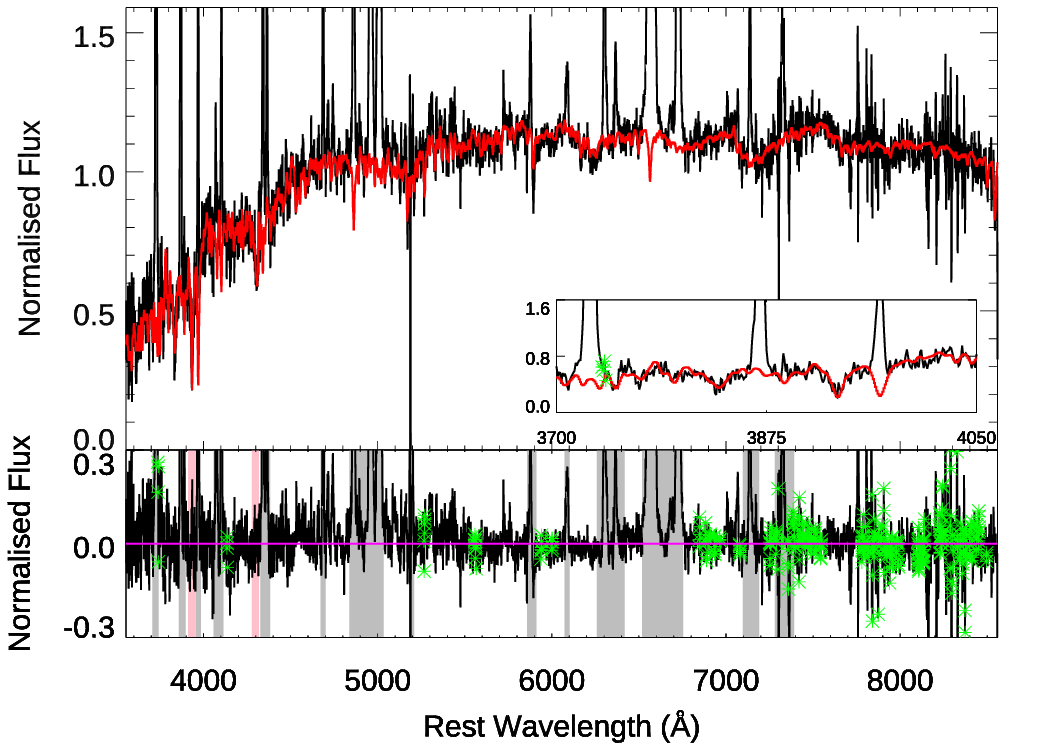}
	\end{subfigure}
    \begin{subfigure}{\columnwidth}
    \includegraphics[width = 1\columnwidth]{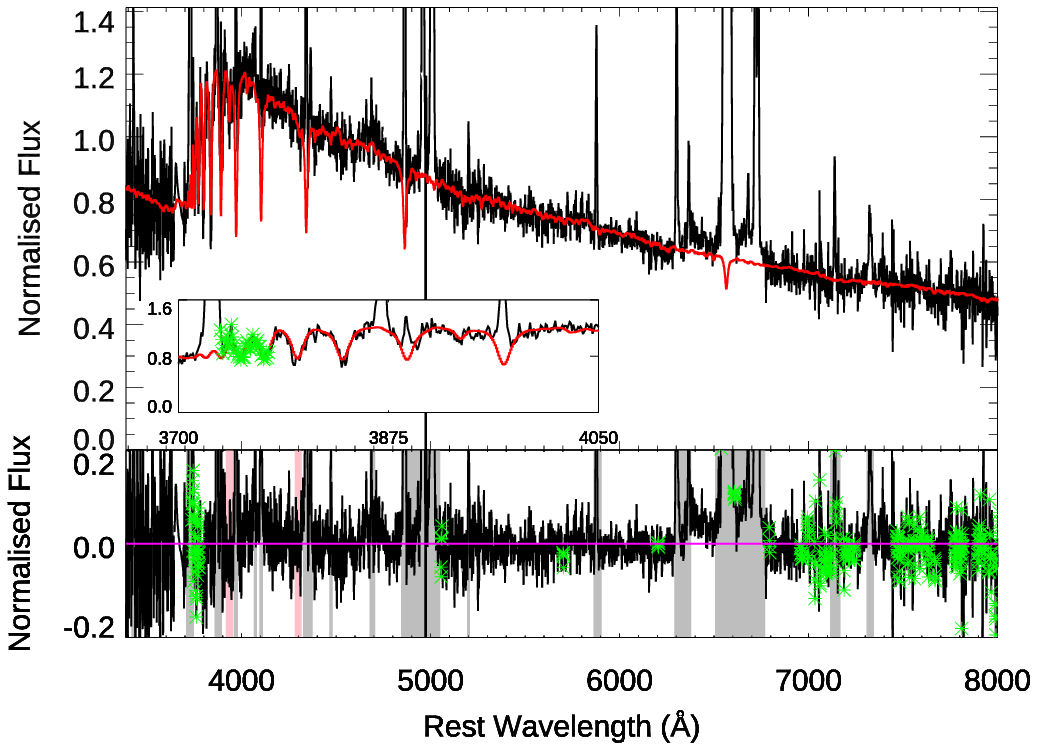}
	\end{subfigure}
    \caption{Two examples of the results of the {\sc starlight} fitting. The left panel shows the fit to J0818+36, which has a low fraction of YSP (7 per cent), where the top panel shows the data in black and the overall fit in red. The bottom panel shows the residuals of the fit with the residuals in black, the regions masked out during the fit shaded in grey and those that were double-weighted shaded in pink. Pixels flagged in the SDSS spectrum are marked by green crosses and the magenta line marks zero. The inset in the top panel shows the fit in the region of the higher-order Balmer absorption features associated with young stars. The right panel shows the same but for J1548-01, which has one of the highest fractions of YSP (88 per cent).}
    \label{fig:sl_exemp}
\end{figure*}

Although all the objects in the \qfd~sample are obscured, making it unnecessary to introduce a direct AGN component into the stellar modelling, it is not possible to rule out potential contamination of the stellar spectrum by direct AGN light being scattered into our line of sight by the intervening obscuring material. As this scattering is wavelength dependent ($F_\lambda \propto \lambda^\alpha$), it would be expected to contaminate shorter (bluer) wavelengths more strongly, thus having the potential to skew the results of the stellar population modelling to younger ages. 

\citet{bessiere17} carried out a similar study to that presented here based on a sample of 20 QSO2s at slightly higher redshift ($0.3 \leq z \leq 0.41 $) within the same AGN luminosity range as those of the \qfd~sample. In that work, the potential impact of introducing a power-law (PL) component into the model to account for this scattered component was tested by carrying out the population modelling using the same stellar templates both with and without a PL. In common with the \qfd~sample, it was not possible to impose constraints on either the proportion of the flux contributed by the PL or $\alpha$ as no spectropolarimetric information was available. Therefore, the PL flux was allowed to vary between $0 \leq \mbox{PL\%} \leq 100$ and $\alpha$ between $-15 \leq \alpha \leq 15$. The results of that work clearly demonstrated that the inclusion of an unconstrained PL component results in the modelling becoming highly degenerate between YSP age, YSP reddening,  PL flux, and $\alpha$ and thus no meaningful constraints could be placed on the YSP from the modelling alone. For this reason, we do not attempt to include a power-law component into the stellar population modelling.

Although we are unable to place any constraints on a potential scattered AGN component, we are able to make a basic assessment of whether not including a power-law component is a valid assumption. As described above, one of the key steps in performing the stellar population modelling is to produce an emission line spectrum by subtracting the stellar model from the data. Scattered light from the obscured AGN will carry with it the imprint of the broad line region (BLR) and, therefore, we should expect to see some evidence of broad \hb in the emission line spectrum if a scattered component is contributing strongly to the total observed flux. However, in the individual objects, we do not find evidence of any broad \hb component in the emission line spectrum and \hb is well fit using the same kinematic components as those used to fit the [OIII]$\lambda\lambda4959,5007$ emission lines.

However, we would expect traces of a broad \hb component to be faint, but to become more significant with increasing AGN luminosity. Therefore, as a further check on the potential significance of a scattered component to the total light spectrum, we construct stacks of the \hb and [OIII] emission lines, consisting of all objects with {\color {black} observed} $\log L_{[OIII]} > 8.9$ (8 objects) excluding J1517+33, which has a double peaked [OIII] line profile. These comprise the most luminous objects in the sample, where any contamination in the form of broad \hb, associated with the BLR, would be most evident. 

The stacks for each emission line were generated independently by first normalising the individual spectra to the peak flux of the \hb or [OIII]$\lambda 5007$ line and then taking the mean value of flux at each of the pixels (excluding flagged pixels). We then fit an increasing number of Gaussian components to the [OIII] composite, applying the same constraints as outlined in Section \ref{sec:emline}, until a reasonable fit was achieved. The resulting model was then applied to fit the \hb composite, constraining the widths and velocities of each of the Gaussian components to be the same as those obtained from the [OIII] fit, with only the amplitudes of each of the components left as free parameters. The results of this procedure are shown in Figure \ref{fig:scattered}, where the top panel shows the [OIII] composite in black, the total fit in red and the individual components are also shown in different colours. The bottom panel shows the resulting fit to \hb and the individual components which are constrained to have the same width and velocity as those shown above. The residual of the fits are shown at the bottom of each panel and clearly demonstrate that, even when considering only the most luminous objects in the \qfd~sample, the \hb line is well fit using the same components as [OIII] and there is no evidence that a broader component associated with light scattered from the BLR is required to adequately fit the \hb line. 

As a further check on our assumption that scattered AGN light does not significantly contaminate our results, we test the relationship between $L_{[OIII]}$ and the monochromatic continuum luminosity in the bluest common wavelength bin {\color {black} among the QSO2s} (3684 -- 3704 \AA). If a scattered component significantly contributes to the total measured light,  increasing $L_{[OIII]}$ should be accompanied by increasing near-UV continuum flux because both are powered by the AGN.  Therefore, we calculate the Pearson correlation coefficient between the monochromatic luminosity (calculated as the average over the bin) of the continuum and $L_{[OIII]}$ and find $r=0.437$, suggesting that no correlation exists between the two. Taking into account the results of the tests presented above, we do not consider the role that scattered light may play in the results of our stellar population analysis further.

\begin{figure}
    \centering
    \begin{subfigure}{\columnwidth}
    \includegraphics[width = \columnwidth]{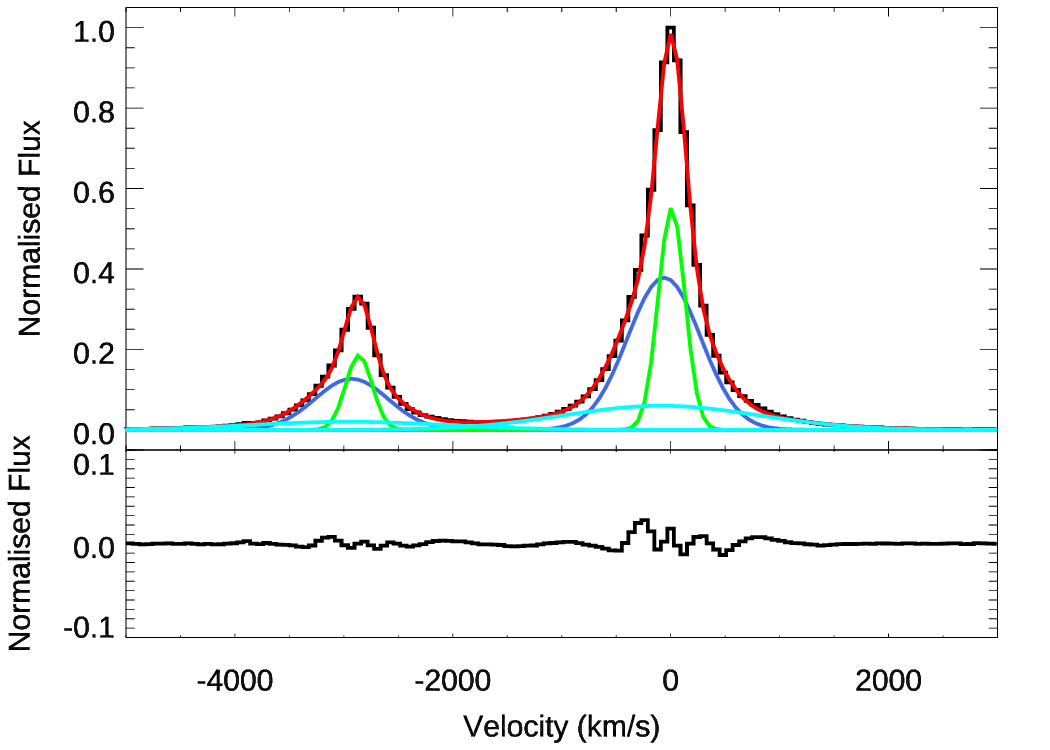}
	\end{subfigure}
    
    \begin{subfigure}{\columnwidth}
	\includegraphics[width = \columnwidth]{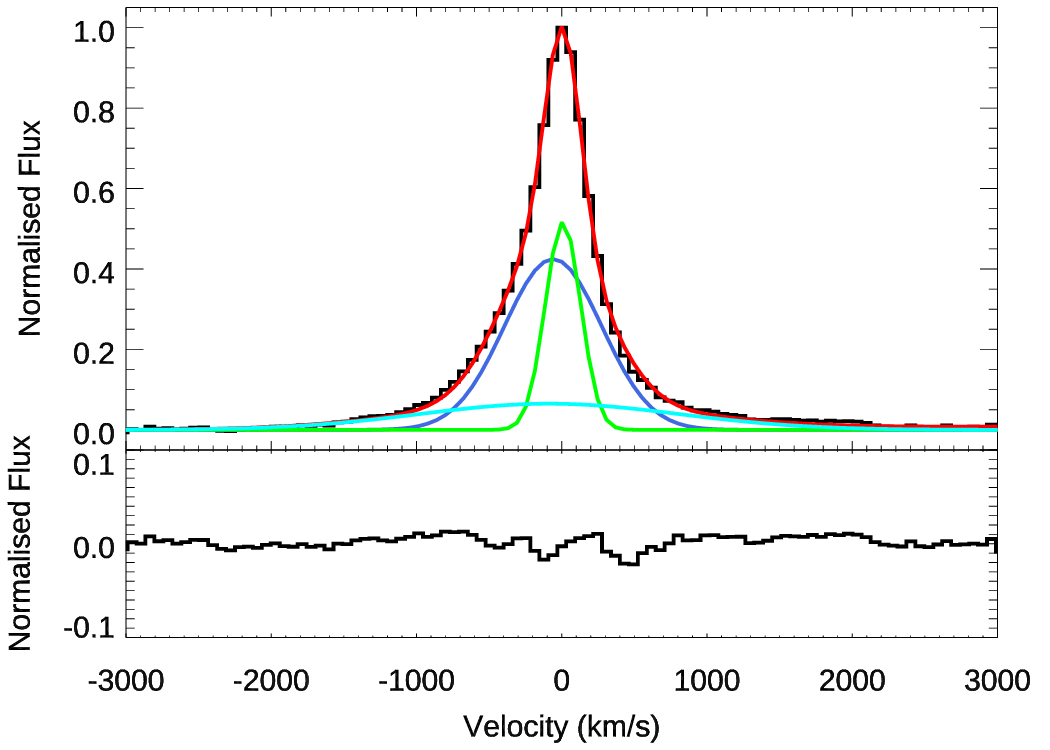}
	\end{subfigure}
    \caption{The [OIII] (top) and \hb (bottom) emission line stacks (black) and total fits (red). The individual Gaussian components which comprise the total fit are shown in different colours, with the same components shown in the same colours in both fits. The individual components of the \hb fit are constrained to have the same width and velocity as those obtained from the [OIII] fit, thereby demonstrating that no additional broad component is required to account for light scattered from the BLR. The residuals are shown underneath each fit.}
    \label{fig:scattered}
\end{figure}

The main results of the stellar population modelling are given in Table \ref{tab:starlight}, which shows whether the Balmer absorption lines were detectable in the unsubtracted data, the proportion of the flux allocated to each age bin, and the SFR calculated from the {\sc starlight} results. $\Delta V_{SL}$  measures the velocity shift between the data and stellar templates and $\sigma_{SL}$ is the velocity dispersion of the stellar populations. The reddening (A$_{\rm V}$) and additional reddening (YA$_{\rm V}$) applied to populations with ages less than 7 Myr, as well as the metallicities of the templates used in each age bin are also listed.

\subsection{Emission line fitting}
\label{sec:emline}

 To measure the kinematics of the warm ionised gas, we adopt a non-parametric approach \citep{harrison14,zakamska14} to measuring the properties of the $\mbox{[OIII]}\lambda 5007$ emission line. Initially we define the values V05,V10,V90, and V95, which measure the velocities at the 5, 10, 90 and 95 per cent points of the normalised cumulative function of the emission line flux. V05 and V95 are considered to be the maximum outflow velocities,  W80 is a measure of the width of the line containing 80\% of the total flux and is equivalent to 1.088$\times$FWHM = $2.563\times\sigma$ (W80 = V90 -- V10), whilst $\Delta\mbox{V}$ measures the velocity offset of the broad wings of the emission line from systemic ($\Delta \mbox{V} = (\mbox{V}05 + \mbox{V}95)/2$); see \citealt{zakamska14} for a detailed explanation) . We also measure the asymmetry of the line, which is a measure of the difference of the flux in the blue and red wings of the line as $A = |V_{90}-V_{50}|-|V_{10}-V_{50}|$.  

We find a wide range of [OIII] emission line profiles in the \qfd~sample, which, in some cases, require several Gaussian components to match the profile adequately. Therefore, we first used the {\sc idl} non-linear least-squares fitting routine {\color{black}{\sc mpfit} \citep{markwardt09}} to simultaneously fit the  [OIII]$\lambda\lambda4959,5007$ lines, adopting the standard technique of fixing the lines to share the same kinematic components and fixing the flux ratio to a value of 3. Due to the analysis method adopted here, we ascribe no physical meaning to the individual components, and are solely concerned with the goodness of the overall fit (as in \citealt{harrison14}). After subtracting the continuum, we begin by fitting two Gaussian components to each line, adding components (up to a maximum of six in total) until the improvement in reduced $\chi^2 < 10\%$.  Each time an additional component was included, the initial input values were randomly generated (within reasonable constraints) and the fit was run $nol \times 30$ times, where $nol$ is the number of Gaussian components. The fit with the lowest reduced $\chi^2$ from the $nol \times 30$ runs was the retained model for that number of Gaussian components, and the associated reduced $\chi^2$ was used to determine whether the additional component was required.

The resulting $\mbox{[OIII]}\lambda 5007$ line model, comprising the sum of all Gaussian components, was then used as the basis for the non-parametric measurements outlined above. Figure \ref{fig:emline} shows examples of how the method described was carried out, showing two of the diverse objects in the sample. The left column shows the results of the line fitting using {\sc mpfit} where the data is shown in black and the total fit is shown in red. The colours of the individual Gaussian components are the same for each line. The right panels show examples of how the non-parametric values were derived, with the model now in black and the grey shaded area representing W80.

In order to estimate the $1\sigma$ errors on these non-parametric values, noise was added to the spectrum that was consistent with the SDSS error spectrum and the fit was performed again and the non-parametric values recalculated. This process was repeated 100 times for each of the objects.

\begin{figure}
    \centering
    \begin{minipage}{\columnwidth}
        \centering
        \begin{subfigure}{0.5\columnwidth}
            \includegraphics[width=\linewidth]{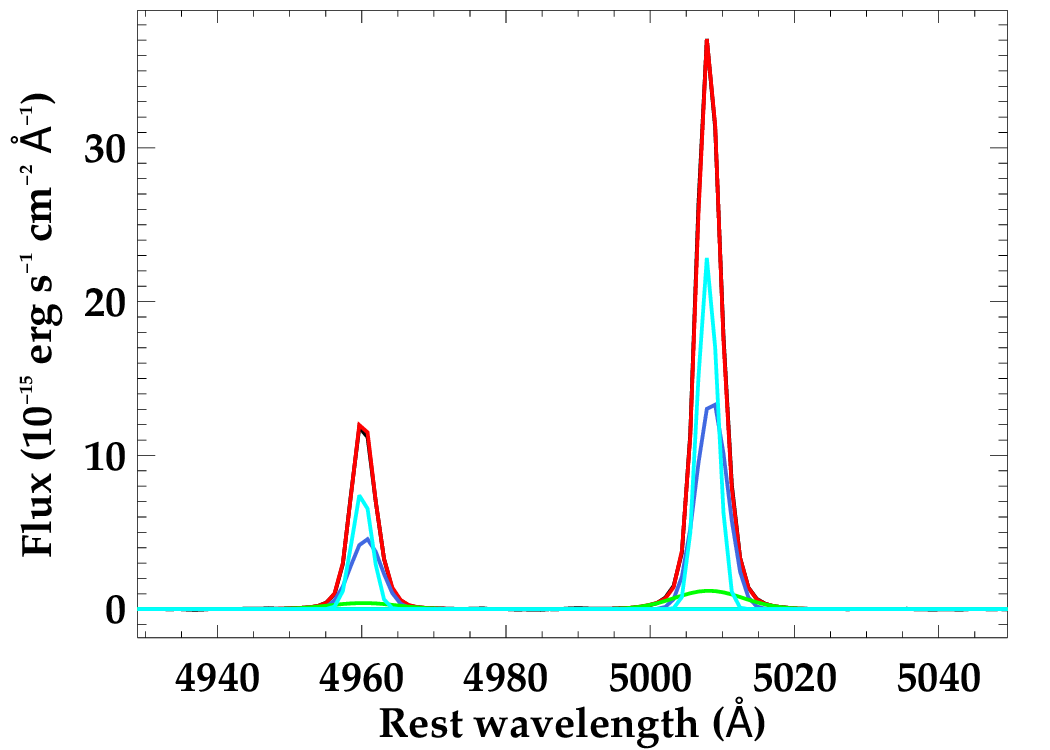}            
        \end{subfigure}\hfill
        \begin{subfigure}{0.5\columnwidth}
            \includegraphics[width=\linewidth]{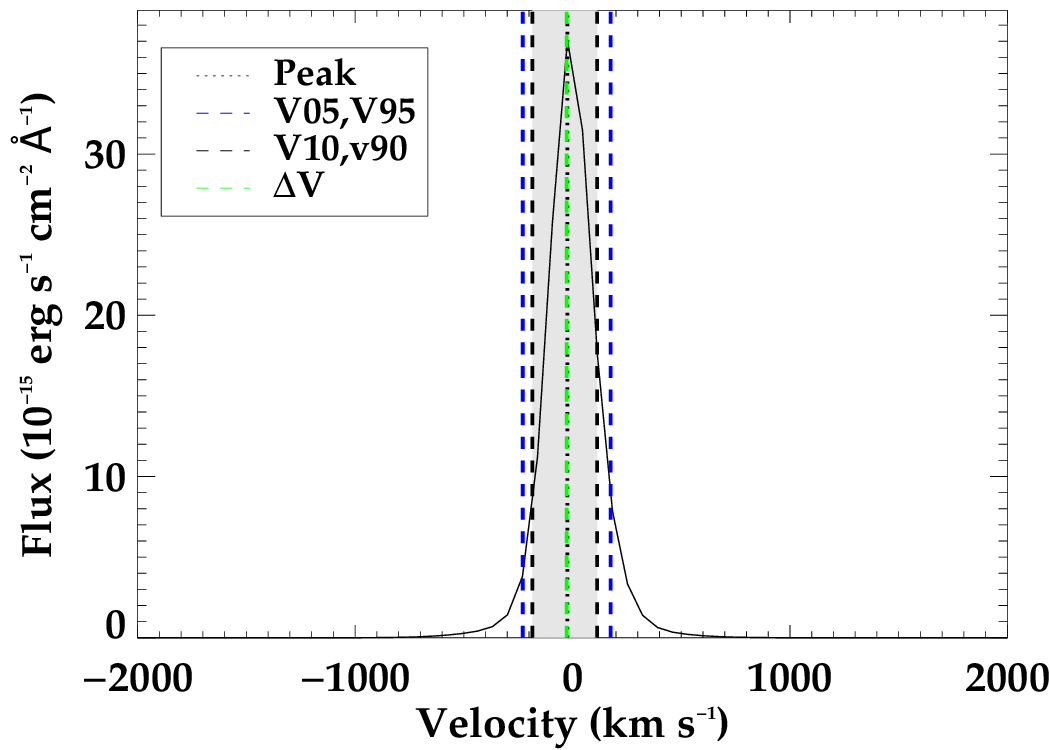}            
        \end{subfigure}
        
        \begin{subfigure}{0.5\columnwidth}
            \includegraphics[width=\linewidth]{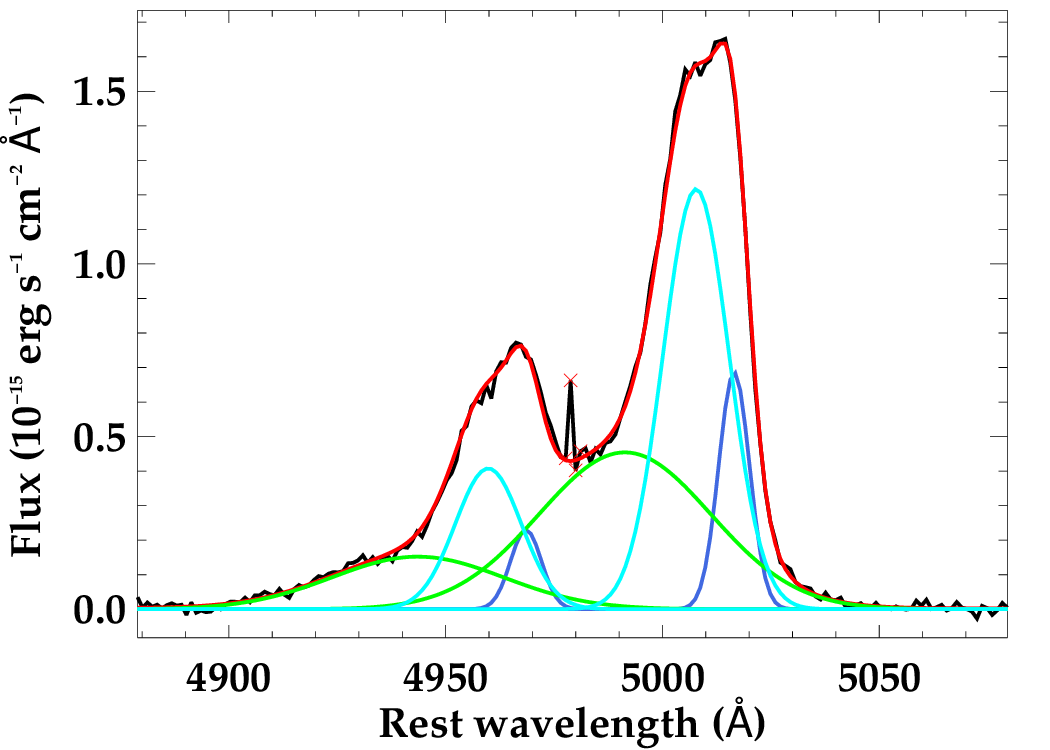}            
        \end{subfigure}\hfill
        \begin{subfigure}{0.5\columnwidth}
            \includegraphics[width=\linewidth]{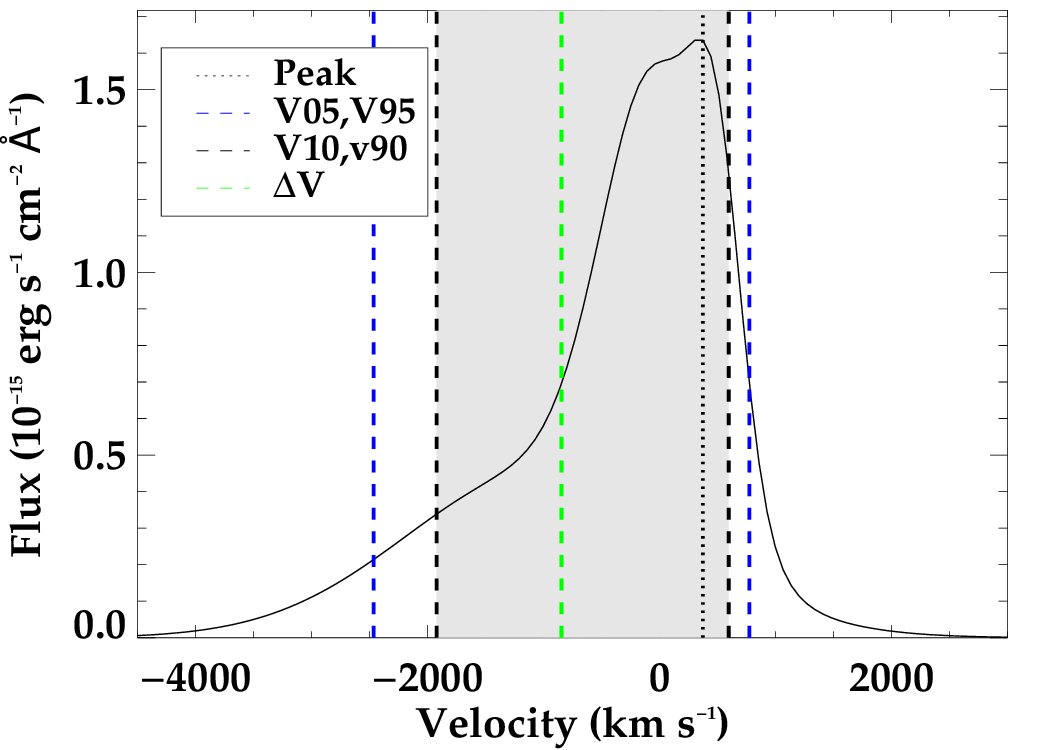}            
        \end{subfigure}
    \end{minipage}
    \caption{Examples of the emission line fitting technique used. The left panels show examples of the Gaussian fits obtained using {\sc mpfit} to J1300+54 and J1347+12, which have the narrowest and broadest line profiles in the sample. The black line shows the data and and the red line shows the sum of the Gaussian components which are denoted in shades of blue and green. The red crosses show pixels that were masked out of the fit. The right panels show the corresponding non-parametric values derived from the emission line models.}
    \label{fig:emline}
\end{figure}

\subsection{Outflow rates}
\label{sec:outflow_rates}

Using the results of the non-parametric fitting outlined in Section \ref{sec:emline} and the results presented in Table \ref{tab:results-sfr-gas}, it is also possible to estimate outflow masses, {\color {black} mass rates, and kinetic} energies in the warm ionised gas phase. To make these estimations, it was assumed that all gas at absolute velocities greater than V05 and V95 is outflowing.

To estimate the mass outflow rate,  the electron densities were derived separately for each object using the transauroral line technique  \citep{holt11,santoro18,davies20,holden23,holden23a}, where possible, which is based on the ratio of the fluxes of the transauroral lines (TR([OII]) = [OIII](3726 + 3729)/[OIII](7319 + 7331); TR([SII]) = [SII](4068 + 4076)/[SII](6717 + 6731)). In six cases, it was not possible to use this technique because the required lines were not well detected in the SDSS data, so the [SII](6717/6731) doublet flux ratio was used instead.

Fluxes were measured from the pure emission line spectrum (i.e., data - stellar model) after fitting a first-order polynomial to remove any remaining continuum. In most cases, a single Gaussian profile was fit to each line, however,  some cases required an additional Gaussian to account for the entire line profile. This was most often the case when measuring the fluxes of the [SII]6717,6731 doublet, which sometimes required an additional broader component. In cases where it was possible to measure the fluxes of the transauroral lines, the measured fluxes were used to calculate the line ratios which were then compared to those expected from photoionisation models, as first described by \citet{holt11}. The photoionisation models were created with the CLOUDY code \citep{ferland17}, and assume a solar-metallicity, radiation-bounded, single-slab, plane-parallel cloud. The central ionising continuum is assumed to have a spectral index of $\alpha = 1.5$, and the ionisation parameter is $\log U = -2.3$\footnote{Note that while this choice of parameters is an assumption, it only affects the derived densities by 0.1-0.3 orders of magnitude (see Appendix B in \citealt{santoro20}). {\color {black} However,} if the gas is ionised by shocks rather than photoionised as we assume, the derived densities will vary by a factor of 2 \citep{holden23a}.}. The density of the cloud was varied between $ 2<\log(n_e[cm^{-3}])<5$ and the TR ratios were reddened from the model using the \citet{cardelli89} extinction law. The resulting ratios from the model were plotted and the measured ratios from the SDSS spectra were overplotted, with the electron densities being derived from this grid {\color {black} (see Figure 5 in \citealt{speranza22} for an example).}

In a further two cases (J1218+47 and J1316+44), it was not possible to measure the electron density using either technique because the SDSS flags cover the transauroral lines and [SII]6717,6731 doublet, therefore we adopted the mean of the measured values of electron density ($n_e = 4607 \mbox{ cm}^{-3}$).  The values of electron density derived from the SDSS spectra are shown in Table \ref{tab:results-sfr-gas} where electron densities measured using the [SII](6717/6731) doublet ratio are {\color {black} given between parenthesis}. Where we have used the transauroral technique to measure $n_e$, we find values in the range $210 < n_e < 38020 \mbox{ cm}^{-3}$ with a mean value of $4230 \mbox{ cm}^{-3}$ and using the [SII] ratio we find a range $275 < n_e < 550 \mbox{ cm}^{-3}$, with a mean value of $455 \mbox{ cm}^{-3}$. Where we have been able to measure $n_e$ via both methods, we have found that, on average, the electron density derived using the transauroral technique is 7 times larger than when using the [SII] ratio. 

Because we are primarily interested in the high velocity outflowing gas associated with the wings of the line, it would be preferable to measure the emission line ratios used to calculate electron density (for both techniques adopted here) from the wings of the relevant emission lines. However, in practise, this is not feasible as the emission lines can be weak and difficult to deblend. Therefore, the electron densities used in these outflow calculations are derived from the ratios of the total line flux as previously described. However, it should be noted that outflowing gas is denser than non-outflowing gas \citep{villar99,speranza22,holden23}, meaning that we are likely to be somewhat underestimating electron densities.

After correcting the [OIII] emission line model for internal reddening using the E(B-V) values calculated from the Balmer decrements measured when constructing the nebular emission model, we then calculate the total mass of the outflow by integrating the total flux in the blue and red wings of the [OIII]$\lambda$5007 emission line model below (above) V05 (v95). 
After converting this value to a luminosity using the values for luminosity distance given in Table \ref{tab:sample}, we use Equation \ref{Eq1} to calculate the outflow mass \citep{fiore17}.

\begin{equation}
    M_{\rm [OIII]} = 4 \times 10^7 M_{\odot} \left(\frac{C}{10^{\rm O/H}}\right)   \left(\frac{L_{\rm [OIII]}}{10^{44}}\right)   \left(\frac{10^{3}}{\langle n_e \rangle}\right)
\label{Eq1}
\end{equation}

To then estimate the total ionised gas mass of the outflows, we assume that it is three times the [OIII] outflow mass (M$_{of}$ = 3$\times$M$_{\rm [OIII]}$; \citealt{fiore17}). The velocity of the outflowing gas ($v_{50})$ is estimated as the 50 percentile of the cumulative function of the flux between the continuum and V05 or V95 in the blue and red wings of the line independently, {\color {black} following \citet{2023arXiv230910572H}} and \citet{speranza24}. The kinetic energy of the outflow is then:

\begin{equation}
    E_{kin} = \frac{1}{2}M_{of}v^2_{50}
    \label{eq:ofek}
\end{equation}

The estimate of mass outflow rates and the kinetic energy they carry is dependent on the physical size of the outflow (Equations \ref{eq:ofr} and \ref{eq:ofrek}). The use of SDSS spectroscopy in this study precludes the ability to measure the sizes of the outflow regions. However, using a combination of HST narrow-band imaging and STIS long-slit spectroscopy, \citet{fischer18} measured the sizes of the outflow regions in a sample of 12 QSO2s that also form part of the \qfd~sample. They found deprojected outflow sizes ranging from 0.15 -- 1.89 kpc, with a mean and median of 0.62 and 0.57 kpc respectively, although in 10/12 cases, the outflow radii are less than 1 kpc. Therefore, when estimating the mass outflow rates in the \qfd~sample, here we assume an outflow radius of 0.62 kpc, so the mass outflow rate is given by:

\begin{equation}
    \dot{M}_{of}= 3\times v_{50} \left(\frac{M_{of}}{R_{of}}\right)
    \label{eq:ofr}
    \end{equation}
    
And the kinetic power is:    
    
\begin{equation}
    \dot{E}_{of} = \frac{1}{2} \times \dot{M}_{of} \times v^2_{50}
    \label{eq:ofrek}
\end{equation}

The values for the blue and red wings of the lines are calculated independently and then summed to obtain the total values which are given in Table \ref{tab:results-sfr-gas}. Figure \ref{fig:outflow_calc} demonstrates how this process was carried out, where the flux in the regions shaded in blue and red were integrated to determine the luminosities as input for Equation \ref{Eq1}. The solid blue and red lines mark the $v_{50,blue}$ and $v_{50,red}$ velocity values that are used as the outflow velocities in Equations \ref{eq:ofek}, \ref{eq:ofr}, and \ref{eq:ofrek}. 

\begin{figure}
    \centering
    \includegraphics[width = \columnwidth]{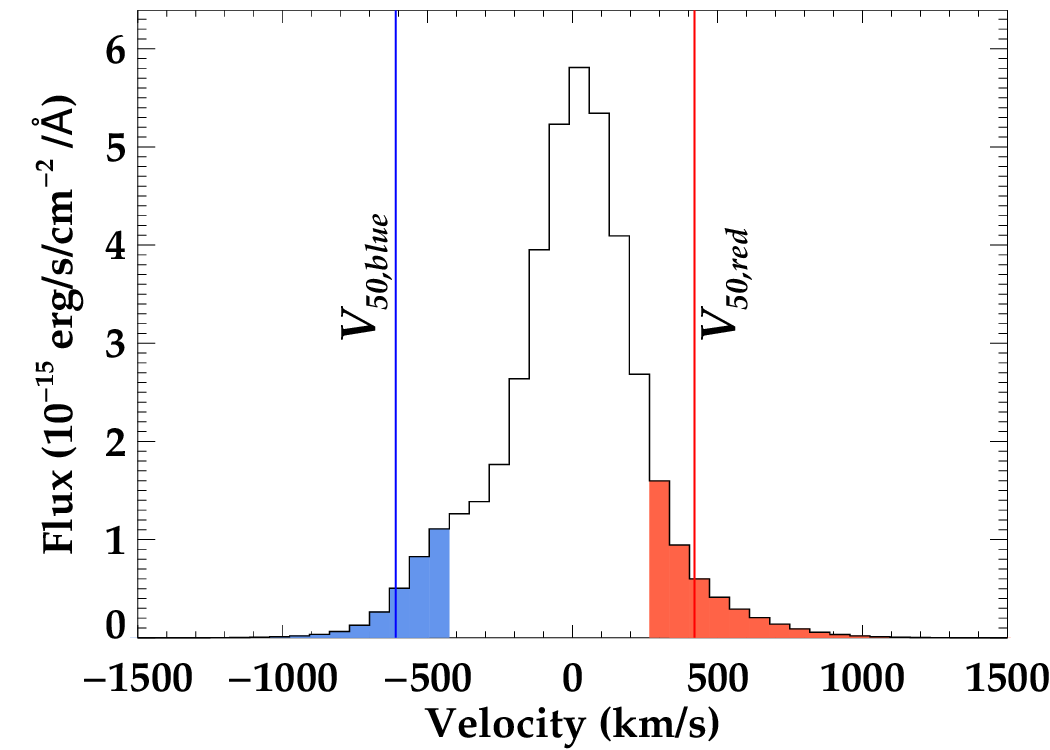}
    \caption{An example of how {\color {black} outflow fluxes and velocities} were calculated. The black line shows the model of the [OIII]$\lambda 5007$ emission line and the blue and red shaded regions show the velocities at which gas is considered to be included in the outflow. The solid blue and red lines represent $v_{50,blue}$ and $v_{50,red}$ which are the outflow velocities.}
    \label{fig:outflow_calc}
\end{figure}

\section{Results}
\label{sec:results}

The aim of this work is to not only characterise the stellar populations and warm ionised gas kinematics of the full \qfd~sample, but to also compare the results of these two strands of investigation  to clarify whether evidence that the presence of the AGN is directly impacting star formation can be established. Due to the likely timescales associated with AGN feedback, we are primarily concerned with the YSP because this is the population that will potentially be affected by the presence of AGN-driven outflows.

\subsection{Stellar populations}
\label{stellar}

The results of the stellar population modelling are given in Table \ref{tab:starlight}, which shows the percentage of the total flux allocated to each age bin (before reddening), at the normalising wavelength, as well as the extinction (A$_{\rm V}$) and additional extinction (YA$_{\rm V}$) applied by {\sc starlight} to the populations with ages $< 7 ~\mbox{Myr}$ (where applicable). In total, we find that 47/48 (98 per cent) of the objects in the \qfd~sample require the inclusion of a YSP younger than 100 Myr in order to adequately model their SDSS spectrum. If we further discount J0052-01, which has a YSP flux consistent with 0 per cent within the errors, we find that 46/48 (96 per cent) of the objects then require the inclusion of a YSP. Figure \ref{fig:sfr-ms} shows the proportion of the total flux associated with the YSP, which varies between 0 and 100 per cent, against the current stellar mass within the SDSS aperture, whilst the right panel shows the SFR within the SDSS aperture, which ranges between 0 and 92 \msunyr, against the stellar mass contained within the aperture.

\begin{figure}
    \includegraphics[width =\columnwidth]{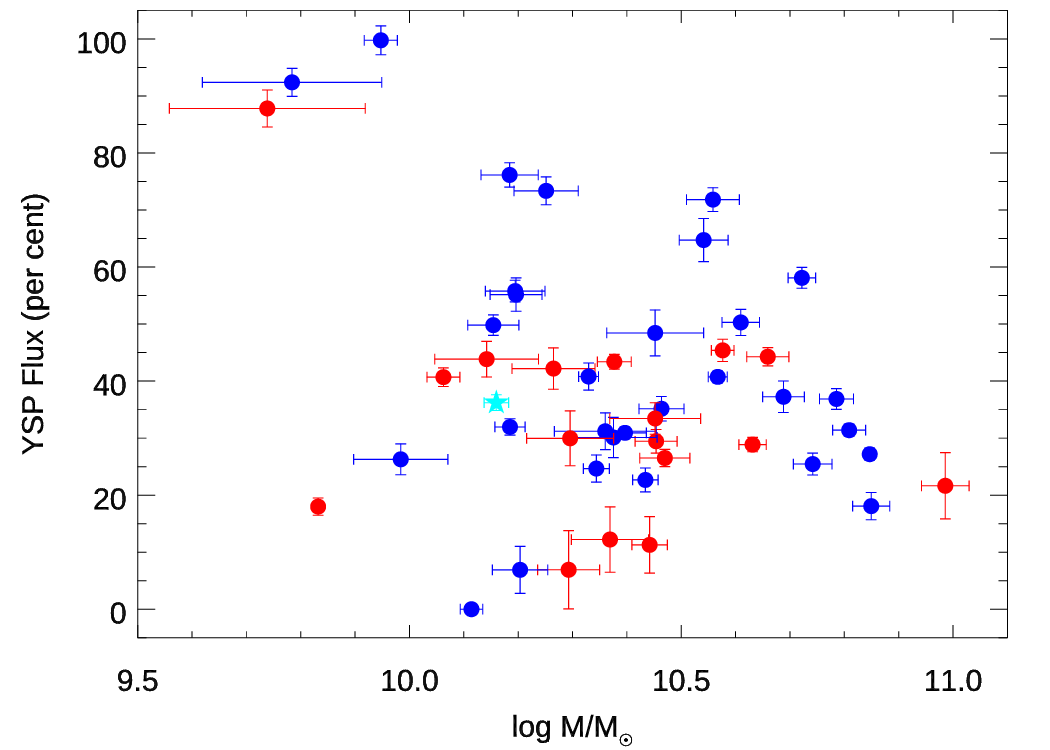}
    \caption{The proportion of the flux that is attributed to the YSP by the {\sc starlight} modelling against the current stellar mass within the SDSS aperture. Red symbols represent undisturbed host galaxies whilst blue represent merging systems. J1034+60  (Mrk\,34) is represented by a cyan star (see Section \ref{sec:SF_outflows}) for details.}
    \label{fig:sfr-ms}
\end{figure}

A full morphological classification of the \qfd~sample, based on deep optical imaging, has been presented in \cite{pierce23} and in order to understand whether a host galaxies' interaction status is a determining factor in the stellar and/or gas kinematic properties found in this work, we have differentiated between these groups. Throughout this work, unless otherwise stated, blue symbols denote QSO2 host galaxies that have been classified as morphologically disrupted (i.e., evidence of galaxy mergers), whilst red symbols denote the galaxies that have been classified as morphologically undisturbed.

SFRs for 7 of the objects were also measured in \citet{ramos22} using rest frame infrared {\color {black} (IR) luminosities and in one case (J0232-08), we find a significantly higher SFR (33 \msunyr) than that measured from the IR (3 \msunyr).} 
A potential reason for this difference is that this object may be a case where a scattered AGN component is a significant factor and we have not accounted for it in our modelling. In the remainder of the cases, our results are either consistent with or less than the SFR that was found in \citet{ramos22}. When considering this, we must bear in mind that, when calculating the SFR from IR luminosities, the flux of the entire galaxy is taken into account, whereas the SFRs calculated from the SDSS spectra only include the SF encompased by the fibre. It could be that, in these cases, the SF is distributed throughout the galaxy rather than in the central region on which the fibre is trained.

The objects in the \qfd~sample were not selected based on any property of their host galaxies, only on AGN luminosity and redshift. This results in a sample which includes a range of host galaxy morphologies, as well as mergers and non-mergers. These are properties that play an important role in star formation. Both simulations and observations show that, under certain conditions, galaxy mergers can trigger rapid star formation. As {\color {black} at least} 65 per cent of the \qfd~sample show morphological disruption consistent with merger activity, it is instructive to consider whether mergers have a more significant impact on the SFR of a QSO2s host galaxy than any properties associated with the AGN. To test this, after excluding the 3 objects in which the SFR is unconstrained, we split the sample into undisturbed (15) and merging (30) galaxies and find that for undisturbed QSO2s the mean SFR is $18.5 \pm 12.4$ \msunyr~(median of 16.5 \msunyr), whilst for the disturbed QSO2s we find a mean of $23.5 \pm 21.0$\msunyr~(median of 20.8 \msunyr). A two-sided K-S test returns $D=0.200$ with a significance $P=0.771$, so we do not find a significant difference between the SFRs of merging and non-merging galaxies in this sample.

\begin{figure}
    \centering
    \includegraphics[width = 1.1\columnwidth]{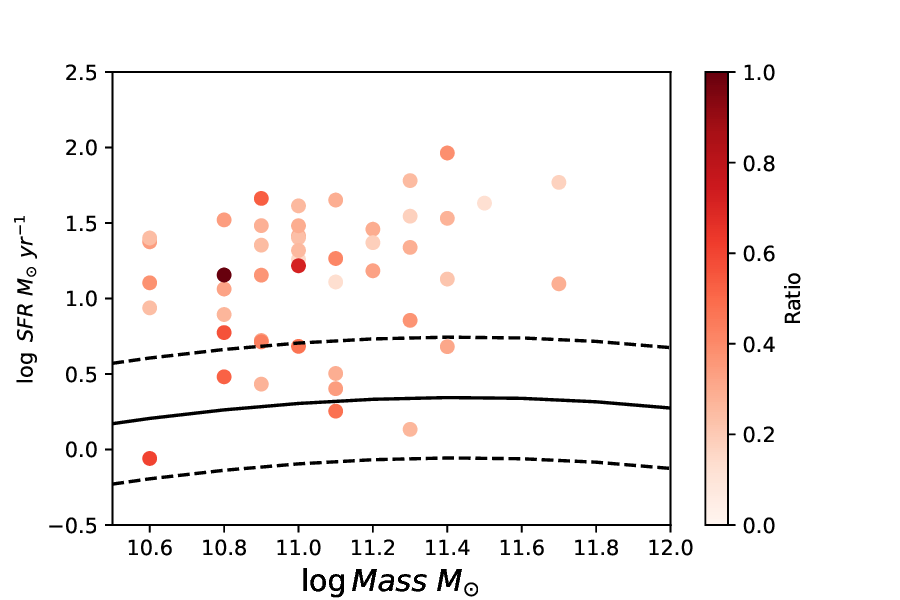}
    \caption{The SFR measured from the {\sc starlight} modelling of the SDSS spectra against the total stellar mass of the host galaxies \citep{pierce23}. The colour scale shows the ratio of the physical size on the galaxy covered by the SDSS fibre against the r-band Petrosian diameter. The black solid line shows the SF main sequence (MS) defined in \citet{saintonge16} and the dashed black lines the range of SFR that is considered to cover the MS {\color {black} ($\pm$0.4 dex)}. }
    \label{fig:main_seq}
\end{figure}

The \qfd~galaxies vary in physical size, while the range of redshifts and the use of spectra from both the Legacy (3 arcsec fibre) and BOSS (2 arcsec fibre) archives means that the physical scale covered by the SDSS fibre varies from object to object. The SDSS r-band Petrosian radius \citep{petrosian76} measures the radius at which the local surface brightness is equal to 0.2 times the mean surface brightness within that radius, so if we consider this to be a reasonable measure of the size of the galaxy, the physical diameters of the hosts are in the range 6.3 -- 56 kpc with a mean of 18 kpc, whilst the physical scales covered by the SDSS fibres are 2.2 -- 7.4 kpc with a mean of 5.6 kpc. If we make the assumption that all the star formation occurring within the galaxy is captured by the SDSS fibre (although this may be a lower limit), we can compare the values obtained to the main sequence of galaxy formation to gain an insight into whether the outflows driven by the QSO2s affect SF in their host galaxies. If this lower limit puts the galaxies on or above the expected main sequence values, then this is an indication that SF is not being suppressed. However, for galaxies that lay below the relation, we cannot definitively conclude that SF is being suppressed as it may instead be the case that all the SF that is occurring has not been captured. Figure \ref{fig:main_seq} shows the SFR against the total mass of the host galaxies \citep{pierce23}, where the colour scale shows the ratio of the physical scale covered by the fibre and the Petrosian radius. The main sequence defined in \citet{saintonge16} is shown as the solid black line with the dashed black lines showing the range of main sequence values ($\pm 0.4$ dex). As can be seen, the majority of the objects in the sample (47/48) lie on or above the main sequence except for one object in which no star formation activity is detected within the previous 100 Myr. These findings suggest that the \qfd~host galaxies are still actively forming stars and in some cases, at rates usually associated with  {\color {black} luminous or ultra-luminous galaxies (U)LIRGs \citep{ramos22,2022A&A...668A..45L}}.

In addition to considering the absolute proportion of flux allocated {\color {black} to} the YSP by the {\sc starlight} modelling, Table \ref{tab:starlight} shows whether it is possible to detect the presence of the higher-order Balmer absorption features in the spectrum \emph{before} nebular subtraction for each object. This feature is indicative of the presence of YSPs and therefore, it is instructive to consider what proportion of the objects have detectable Balmer absorption lines as an independent check. We find that 26/48 (54 per cent) of the sample clearly show the presence of Balmer absorption lines in their spectra \emph{before} nebular subtraction, which increases to 42/48 (88 per cent) after the nebular subtraction, consistent with the proportion determined by the spectral synthesis modelling. 

To understand why this significant increase in detection occurs after nebular subtraction, the left panel of Figure \ref{fig:nebsub} shows an example of the nebular subtraction process for an object in which the Balmer absorption lines were not detected in the original data but became apparent when the nebular subtraction process was carried out. The black line shows the original data, the blue line is the model of the nebular continuum and higher-order Balmer emission lines, and the red line shows the nebular subtracted data. This example illustrates the impact of the removal of the emission line infilling of the underlying stellar absorption features and also the reduction in the total continuum flux below the Balmer edge (3646 \AA). These two effects have differing results on the final model because the increase in the strength of the stellar Balmer absorption features (if this occurs) will result in the final stellar population model being more weighted towards younger stellar ages, whilst the reduction in the near-UV flux has the opposing effect. The right panel of Figure \ref{fig:nebsub} shows the impact that the nebular subtraction procedure has on the proportion of the flux in the normalising bin that is allocated to the YSP by {\sc starlight} both before and after the nebular subtraction. The black line shows the one-to-one relation and demonstrates the importance of considering nebular contamination of the stellar component when attempting to characterise the stellar populations of QSO2s host galaxies. Failure to do so can result in significant over or under-estimations in flux associated with the YSP.

\begin{figure*}
    \centering
    \begin{subfigure}{\columnwidth}
    \includegraphics[width = \columnwidth]{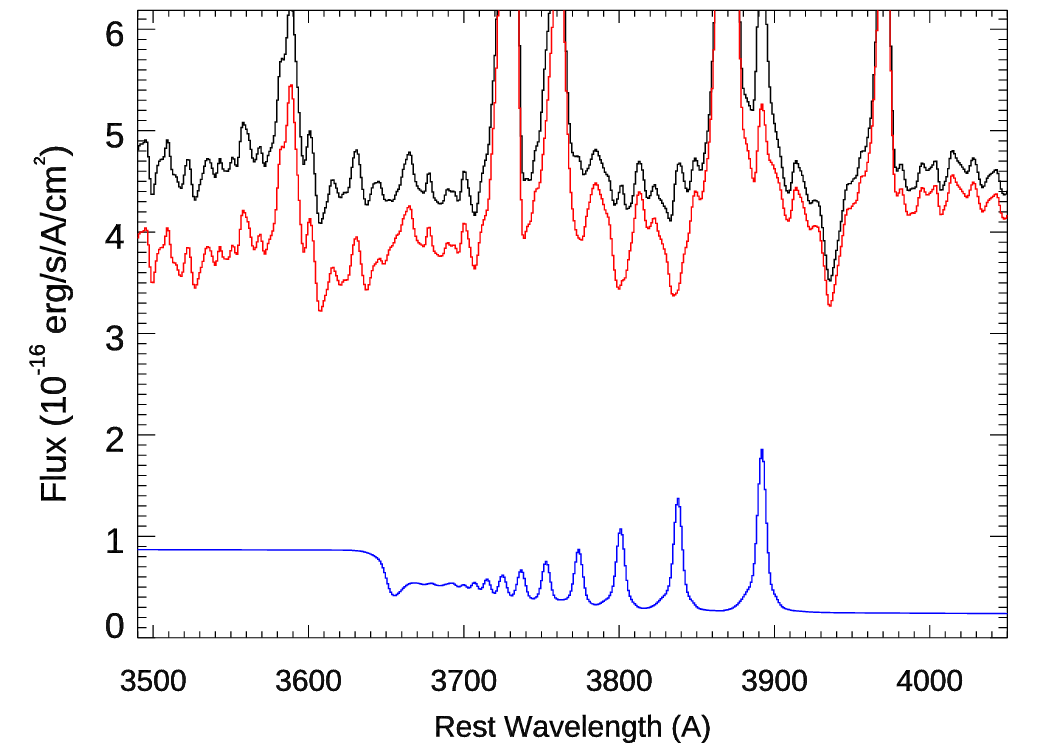}
	\end{subfigure}
    \begin{subfigure}{\columnwidth}
    \includegraphics[width = \columnwidth]{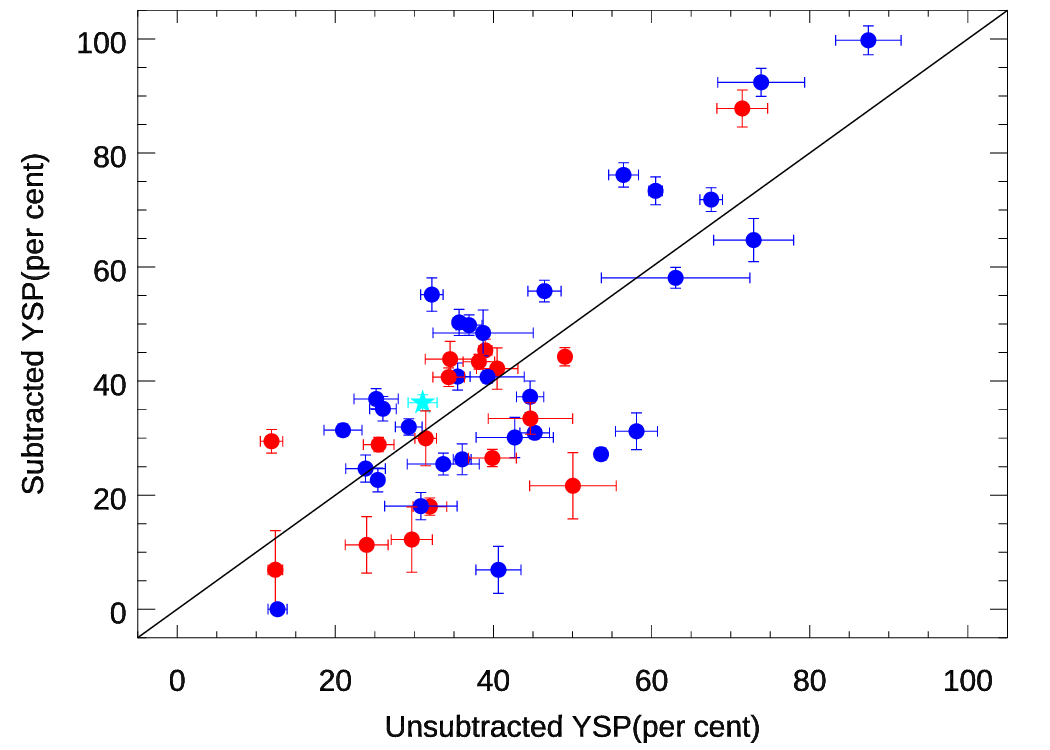}
    \end{subfigure}
    \caption{Left: An example of the impact of the nebular subtraction process on the SDSS spectrum of J1316+44. The black and red lines show the original and nebular subtracted data, while the blue line shows the subtracted nebular model. Right: A comparison of the percentage of the total flux allocated to the YSP in the unsubtracted and nebular subtracted data. {\color {black} The black line shows the one-to-one relation between the two values. Red and blue points represent objects classified as morphologically undisturbed and disturbed, respectively.}} 
    \label{fig:nebsub}
\end{figure*}

\begin{table}
\centering
	
     \caption{Summary of the main results. Column 1 gives the parameter and corresponding units whilst columns 2, 3, 4, and 5 give the minimum, maximum, mean, and median calculated for that parameter. The values for individual objects can be found in Tables \ref{tab:starlight} and \ref{tab:results-sfr-gas} in Appendix \ref{tabs}.}
    \label{tab:summary}
	\begin{tabular}{c|l|c|c|c|c}
\hline											
\multicolumn{2}{c}{Variable}	&	Min	&	Max 	&	Mean 	&	Median	\\											
 \hline
YSP	&(\%) &	0	&	100	&	39	&	36	\\
SFR	&(\msunyr) &	0	&	92	&	21	&	16	\\
W80	&(\kms) &	298	&	2519	&	804	&	717	\\
|$\Delta V$| &(\kms)	&	2	&	845	&	99	&	61		\\
$n_{e[TR]}$	&($\mbox{cm}^{-3}$) &	210	&	38020	&	4230	&	2579	\\
$n_{e[S[II]]}$	&($\mbox{cm}^{-3}$) &	275	&	550	&	455	&	495		\\
$\dot{M}_{of}$	&(\msunyr) &	0.04	&	1.9	&	0.5	&	0.4	\\
$\log \dot{E}_{of}$	&(\ergs) &	39.3	&	41.8	&	40.7	&	40.6		\\
 	\hline 
    \hline
	\end{tabular}
\end{table}

\subsection{Gas kinematics}
\label{gas_kin}

As previously stated, we find a wide variety of [OIII] line profiles in the \qfd~sample and therefore, a wide range of gas kinematic properties. Table \ref{tab:results-sfr-gas} shows the results of the non-parametric analysis of these line profiles. There is a wide variety in the widths of the [OIII] emission lines, ranging from $300 < \mbox{W80} < 2500 $ \kms, with a mean value of 804 \kms (median 717 \kms).

Figure \ref{fig:w80_slvd} shows a comparison of W80 and the velocity dispersion of the stellar component of the host galaxy ($W80_{SL} = 2.563\sigma_{SL}$) as measured from the {\sc starlight} modelling of the SDSS spectra. The black line shows the one-to-one relation between the two properties and demonstrates that, in the majority of cases, the [OIII] dispersion is higher than would be expected purely from the dispersion of the galaxy. We find that in 41/48 (85\%) of cases, the dispersion of the [OIII] line lays above the one-to-one relation, suggesting that the ionised gas in these galaxies is outflowing, driven either by the AGN, star formation or a combination of both. For comparison, if we require W80>600 \kms~for the gas to be considered outflowing (as in \citealt{harrison16,kakkad20,Scholtz20}) the percentage of QSO2s with ionised outflows would be 67\% (see Table \ref{tab:results-sfr-gas}).

When we consider the W80 values, where W80 is larger than $W80_{SL}$,  in the merger and non-merger groups separately, we find that in the merging galaxies (27), the mean W80 is $\approx$837 \kms (median 763 \kms) whilst in QSO2s classified as undisturbed (14), the mean W80 is $\approx 909$ \kms (median 853 \kms). A Kolmogorov-Smirnov (K-S) test to determine whether the W80 values for the disturbed and undisturbed groups of galaxies are drawn from the same underlying distribution returns p=0.57, so we conclude that there is no significant difference between W80 in merging and non-merging galaxies, suggesting that mergers are not the dominant cause of the disrupted gas kinematics observed.

\begin{figure}
    \includegraphics[width = \columnwidth]{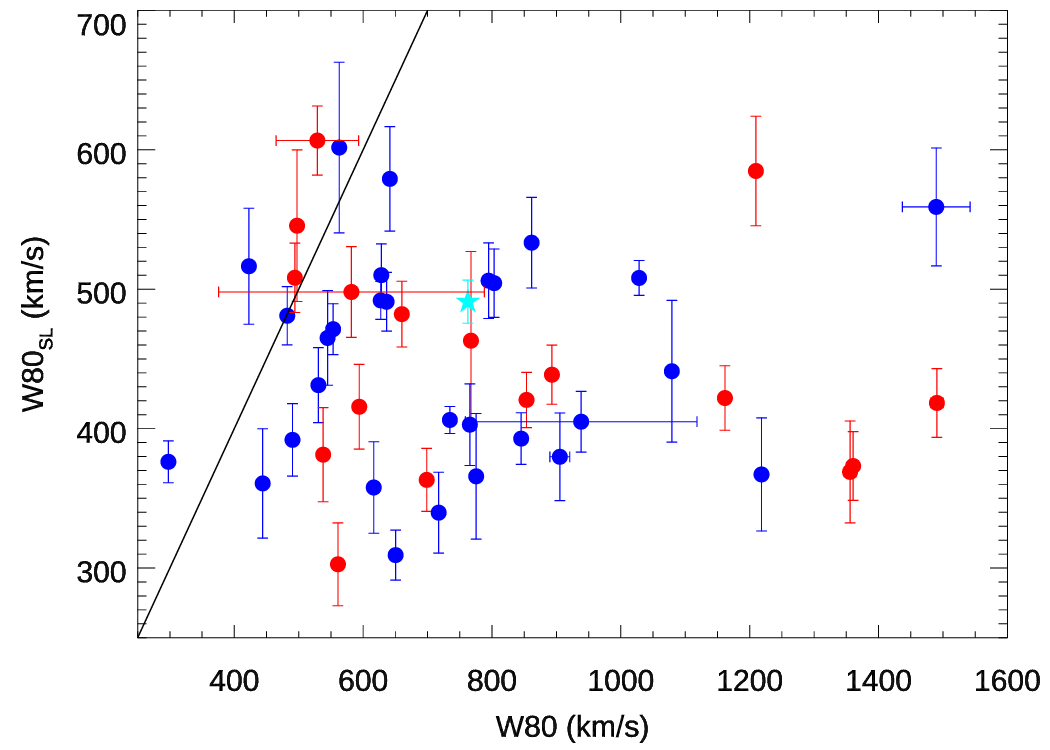}
    \caption{The results of the non-parametric [OIII] line measurements compared to the velocity dispersion of the stellar component of the host galaxies measured by {\sc starlight}. The black line shows the one-to-one relation between the two values. Note that for presentation purposes, J1347+12 has been omitted from this plot due to it's high value of W80 ($\approx 2520$ \kms).}
    \label{fig:w80_slvd}
\end{figure}

\begin{figure*}
    {\par\includegraphics[width = \columnwidth]{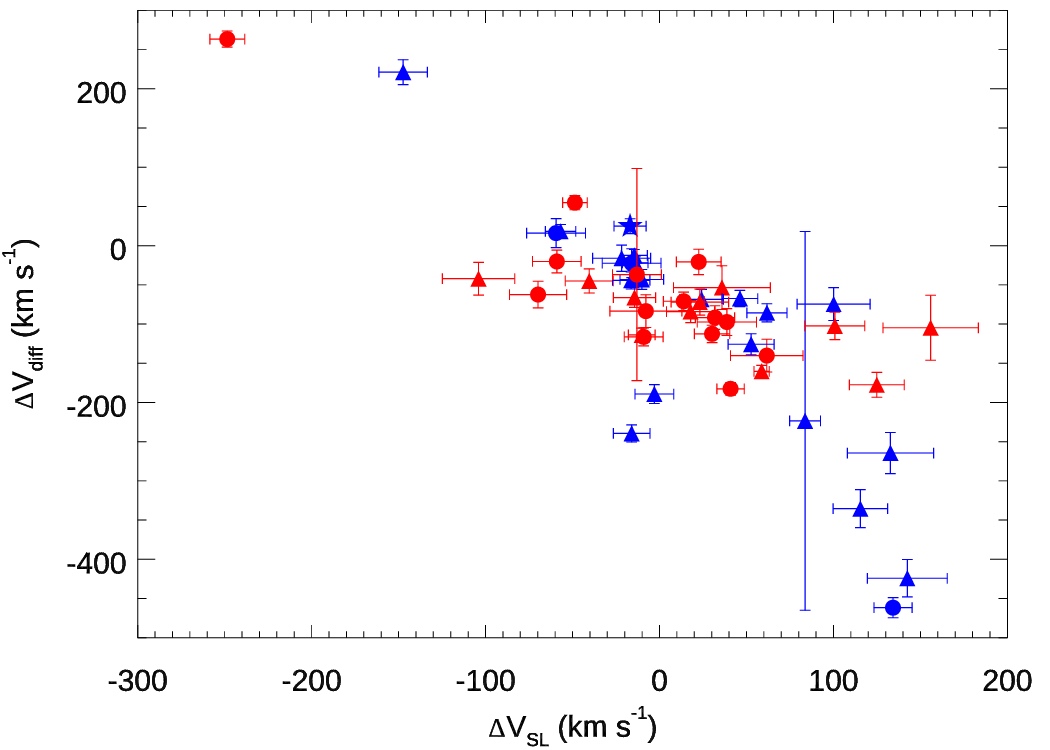}
    \includegraphics[width = \columnwidth]{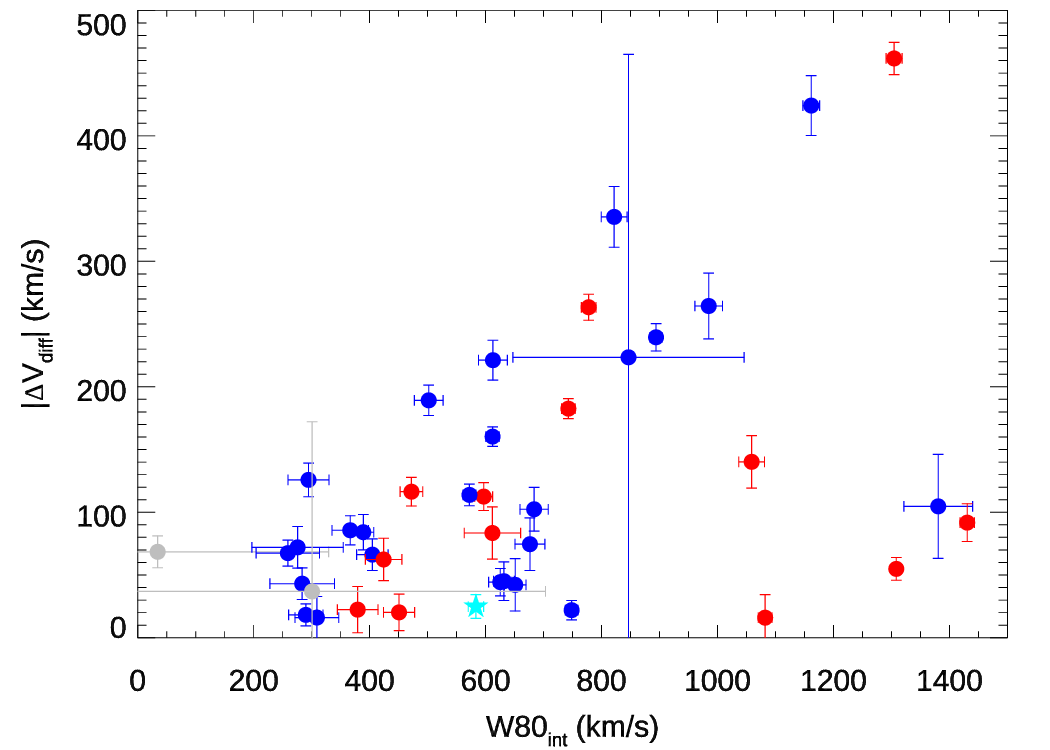}\par}
    \caption{Left: The difference in the velocity offset of the [OIII] emission line and that of the host galaxy. Right: The deconvolved $W80_{int}$ values against $|\Delta V_{diff}|$, where $W80_{int}=\sqrt{W80^2 - W80_{SL}^2}$ and $|\Delta V_{diff}|=\Delta V_{[OIII]}$ - $\Delta V_{SL}$. {\color {black} Grey symbols indicate objects in which no outflows were detected according to our criterium.}}
    \label{fig:decon_vdif}
\end{figure*}

The \qfd~host galaxies cover the range of stellar masses $10.6 < \log (M/M_{\sun}) < 11.7$ and more massive galaxies will have a higher intrinsic stellar velocity dispersion, therefore higher values of W80 do not necessarily indicate broader widths in relation to those that should be expected from that of the underlying host galaxy. To investigate this, we define the term $W80_{int} = \sqrt{W80^2 - W80_{SL}^2}$, thereby deconvolving the galaxy and [OIII] velocity dispersions. When considering the 41 galaxies where $\mbox{W80} > W80_{SL}$, we find the values of $\mbox{W80}_{int}$ in the range 260 - 2460 \kms\ with a mean 714 \kms\ and a median of 613 \kms, demonstrating that, in the majority of the \qfd~sample, the velocity dispersion of the gas is significantly higher than that of the stellar component of the host galaxy, suggesting the presence of powerful AGN-driven outflows. 

The SDSS pipeline uses various stellar and emission {\color {black} line} features to compute redshifts, which works well when the emission lines are narrow and symmetric. However, in cases where emission lines are asymmetric, this can lead to errors in the derivation of the redshifts. When carrying out the stellar population modelling, the emission lines are masked and the code relies solely on the stellar absorption features to determine how much to shift the synthetic templates to match the data, making the {\sc starlight} velocity shifts ($\Delta V_{SL}$; Table \ref{tab:starlight}) a more accurate estimate of the systemic velocity of the host galaxy. Therefore, we define the property $\Delta V_{diff} = \Delta V_{[OIII]}$ - $\Delta V_{SL}$  which is a measure of the velocity shift of the wings of the emission lines $\Delta V_{[OIII]}$ relative to $\Delta V_{SL}$ as a measure of whether the velocity of the ionised gas is shifted relative to the host galaxy, which we consider to be indicative of outflowing gas. {\color {black} 
Four of the objects (J0805+28, J1015+00, J1100+08, and J2154+11) have [OIII] emission line velocity shifts consistent, within the errors, with that of the underlying host galaxy ($\Delta V_{diff}\sim$0) and the shift is likely to be smaller when $\Delta V_{SL}$ is close to 0, as expected for the reasons outlined above (see left panel of \ref{fig:decon_vdif})}. However, it is interesting to note that these four objects all have values of W80 than are higher than those expected purely from the dispersion of the host galaxy {\color {black} (see right panel of Figure \ref{fig:decon_vdif})}. In particular, J1100+08 has $\Delta V_{diff} = 16 \pm 18$ \kms\ whilst the $W80 = 1082 \pm 7$ \kms\, which is significantly higher than the $W80_{SL} = 422$ \kms\ found from the {\sc starlight} modelling. It is possible that the reason that the [OIII] emission line is so broad whilst still being at the systemic velocity of the galaxy because the outflow that it is tracing is spherically symmetric.

Taking the above into account, we therefore only consider the condition that $W80 > W80_{SL}$ as determining whether it is considered that an outflow has been detected, meaning that 85\% of the QSO2s host outflows. However, the nature of the SDSS data means that we are unable to place constraints on the sizes of these outflows.

\subsection{Correlations}
\label{sec:correlations}

The key question the \qfd~project aims to investigate is \emph{how do luminous AGN impact their host galaxies?}, requiring us to consider two distinct aspects. Firstly, we must investigate whether QSO2s can drive significant outflows and, secondly, whether those outflows directly impact star formation. Therefore, in this section, we consider whether there are any correlations between the kinematics of the warm ionised gas and the intrinsic properties of the QSO2s (e.g., bolometric luminosity and Eddington ratio). We then compare SFRs with the kinematics of the warm ionised gas to investigate whether any correlations exist between these two properties. 

\subsubsection{AGN properties and gas kinematics}

To determine whether the gas kinematics are dependent on the AGN properties, we calculate the Pearson correlation coefficients ($r$) between intrinsic AGN characteristics ($L_{bol}$, $L_{1.4GHz}$, $M_{BH}$, $\lambda/\lambda_{Edd}$) and the ionised gas kinematics derived in this study ($W80_{int}$, $\Delta V_{diff}$, $\dot{E}_{kin}$, {\color {black} and $\dot{M}_{of}$)} for {\color {black} the 41 QSO2s} in which outflows were detected. The resulting correlation coefficients for each combination of parameters tested are given in Table \ref{tab:corr_gas}. 

The results presented in Table \ref{tab:corr_gas} demonstrate that no strong correlations ($r > 0.7$) exist between the gas kinematics and AGN properties. The strongest correlation found is between $L_{1.4GHz}$ and $W80_{int}$, which is shown in Figure \ref{fig:corr}, where $r = 0.58$, 
{\color {black} followed by that between $L_{1.4GHz}$ and $\dot{E}_{kin}$, where $r = 0.48$. This may indicate that radio luminosity is the most significant factor in driving outflows \citep{mullaney13} or alternatively, that the outflows might be inducing shocks in the ISM that produce radio emission \citep{2023arXiv230615047F}.} The majority of the \qfd~sample are radio-quiet, however, three objects are classified as radio-loud (J0939+35, J1137+61, and J1347+12) and, of these three, only J1347+12 has an outflow. Whilst J1347+12 is a compact steep spectrum source (CSS), the other two are FRII sources, with large-scale radio jets extending beyond the physical scale of the galaxy.

\begin{figure}
\centering
\includegraphics[width= \columnwidth]{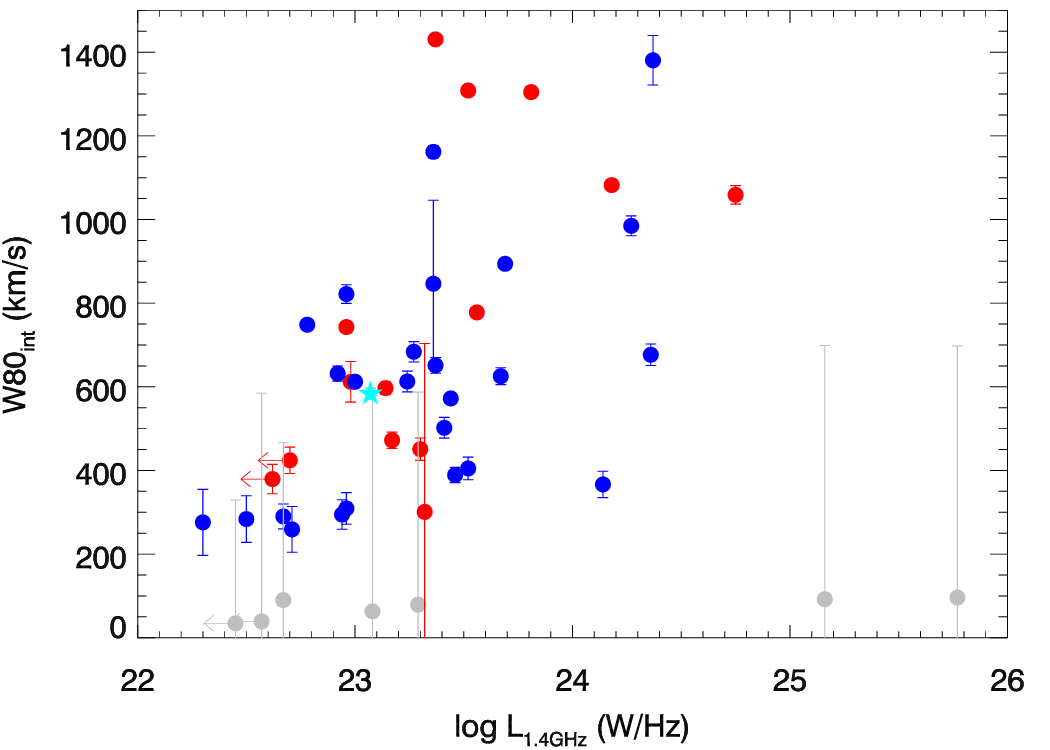}
    \caption{Derived $W80_{int}$ against observed $L_{1.4GHz}$, demonstrating the positive correlation between radio luminosity and disrupted gas kinematics. Grey symbols indicate objects in which no outflow was detected {\color {black} according to our criterium}.}
    \label{fig:corr}
\end{figure}

Potentially, the radio output of a QSO2/host galaxy can be powered by accretion onto the BH, star formation or a combination of the two, creating a possible source of confusion when attempting to interpret the weak correlation between $L_{1.4GHz}$ and  $W80_{int}$. To overcome this, we utilise the relation between SFR and $L_{1.4GHz}$ derived by \citet{davies17}, using a sample of non-AGN galaxies, to investigate this potential correlation in greater depth. Using the SFRs presented in Table \ref{tab:starlight}, we calculate the 1.4 GHz radio luminosity expected from star formation alone ($L_{1.4GHz,SF}$), allowing us to calculate the 1.4 GHz luminosity attributable solely to the AGN ($L_{1.4GHz,AGN}$), so that $\log L_{1.4GHz,AGN} = \log(L_{1.4GHz,obs} - L_{1.4GHz,SF})$, where $L_{1.4GHz}$ is the observed luminosity (see Table \ref{tab:sample}). We use these values to calculate $r$ between $\log L_{1.4GHz,AGN}$ and $W80_{int}$  (for objects with outflows) and find  $r = 0.665$. Although still not considered a strong correlation, the fact that it is strengthened when only considering the radio luminosity of the AGN demonstrates that this is likely to be the dominant factor. To further emphasise this point, we consider the correlation between $L_{1.4GHz,SF}$ and $W80_{int}$, which yields $r = 0.170$, demonstrating that it is not the radio luminosity generated by SF that is driving these outflows.

\begin{table}

    \centering
        \caption{The Pearson correlation coefficients between the various outflow properties determine from our non-parametric analysis and stellar population modelling of the SDSS spectra and the intrinsic properties of the AGN. These coefficients only include objects in which outflows are detected.} 
        \label{tab:corr_gas}
    \begin{tabular}{c|c|c|c|c}
    \hline
	&	$W80_{int}$	&	$\Delta V_{diff}$	&	$\dot{E}_{kin}$ & $\dot{M}_{of}$\\
	\hline
$L_{bol}$	&	0.279	&	-0.228	&	0.508 & 0.112\\
$L_{1.4GHz}$	& 0.576	&	0.239	&	0.543 & 0.176\\
$M_{BH}$	&	0.196	&	0.144	&	0.535 & 0.353\\
$\lambda/\lambda_{Edd}$	&	0.154	&	0.04	&	-0.40 & -0.260\\
\hline
\hline

    \end{tabular}   
\end{table}

{\color{black} In a recent work, \citet{2023arXiv230910572H} searched for correlations between AGN, host galaxy, and outflow properties in a sample of 19 QSO2s of the same luminosity as the \qfd~sample and at redshift z=0.3-0.41. Based on the Spearman's rank correlation coefficients ($\rho$), they found moderate correlations between the bolometric luminosity of the AGN and the outflow mass and outflow rate, both with $\rho$=0.52. This implies that more luminous AGN have more massive ionised outflows and higher outflow mass rates. On the other hand, \citet{2023arXiv230910572H} did not find any correlation between the radio luminosity and the outflow properties.}

\subsubsection{Gas kinematics and star formation rates}

Having compared the properties of the ionised gas with the properties of the AGN itself, we must now consider whether the detected outflows have an impact on the {\color {black} SFR}. If AGN-driven outflows are responsible for quenching star formation, then it may be reasonable to assume that the more powerful the outflow, the more likely it is to directly impact the star formation. {\color {black} In Section \ref{gas_kin},} we have measured the kinematics of the ionised gas and derived outflow rates and kinetic energies. Here we compare the results obtained for the outflows with the SFR. 

\begin{table}

    \centering
        \caption{The Pearson correlation coefficients when testing the correlations between star formation rate and various properties of the outflows.}
        \label{tab:corr_sfr}
    \begin{tabular}{c|c|c|c|c}
    \hline
	&	$W80_{int}$	&	$\Delta V_{diff}$	&	$\dot{E}_{kin}$ & $\dot{M}_{of}$\\
	\hline
SFR	&	0.223	&	0.347	&	0.074 & -0.110\\
SFR10 & 0.256   &   0.369   &   0.101 & -0.068\\
SFR50 & 0.222   &   0.345   &   0.101 & -0.049\\
\hline
\hline
    \end{tabular}   
\end{table}

Table \ref{tab:corr_sfr} shows the Pearson correlation coefficients between SFR, $W80_{int}$, $|V_{diff}|$ and $\dot{E}_{of}$, where only objects with detected outflows are considered. These results clearly demonstrate that no correlations are found between the key properties of the ionised gas and SFR in QSO2 host galaxies. These findings strongly suggest that the AGN-driven outflows do not directly impact SF on global scales, even at the highest outflow velocities and energies detected in the \qfd~sample.

However, thus far, we have considered YSPs to be populations with ages $<100 \mbox{~Myr}$. We must consider the possibility that this timescale is too long in comparison to the predicted lifetime of an AGN {\color {black} and the gas outflow that it drives, which may have the effect of washing out evidence of the direct impact of outflows on SF. Indeed, the measurements of the dynamical timescales of the ionised and cold molecular outflows reported in the literature for QSO2s when outflow radii have been constrained from observations are of $\sim$1-20 Myr \citep{bessiere22,ramos22,speranza24}.} Therefore, we also calculate SFRs on timescales of 10 and 50 Myr and calculate the Pearson correlation coefficients against the same outflow properties. These values are also given in Table \ref{tab:corr_sfr} as SFR10 and SFR50 and show that considering timescales that may be more in line with the lifetime of AGN-driven outflows does not lead to any significant change in the level of correlations between SFRs and outflow properties.

\section{Discussion}
\label{sec:discussion}

In Section \ref{sec:results}, we derived the SFR, gas kinematics, and outflow properties of the \qfd~sample and tested for correlations between these properties. Our objective was to investigate whether luminous quasars drive outflows and whether, in turn, these outflows directly impact SF. {\color {black} Here} we discuss our findings, comparing them to the results of previous studies and evaluating them in the context of galaxy evolution.

\subsection{AGN and outflow properties}
\label{sec:agn_outflows}

Due to the perceived importance of AGN feedback in galaxy evolution, there have been a wealth of studies aimed at quantifying the properties of AGN-driven outflows, particularly in the warm ionised gas phase {\color {black} (see \citealt{harrison18} for a review)}. Although these works adopt different approaches to characterising outflow kinematics, they unanimously conclude that {\color {black} gas outflows are a common phenomena in AGN, and practically ubiquitous at high bolometric luminosities}. In common with these works, we also find that outflows in the warm ionised gas phase are a common occurrence in QSO2, with 85\% of the sample having [OIII] line profiles that cannot be explained solely by the underlying gravitational potential of the galaxy. 

The analysis of [OIII] kinematics presented here has enabled us to calculate mass outflow rates and outflow kinetic powers. However, although we find clear evidence for fast outflows in the majority of the \qfd~objects, this does not translate to high mass outflow rates or kinetic powers (see Tables \ref{tab:summary} and \ref{tab:results-sfr-gas}). One key reason for the discrepancy between the outflow rates presented here and those found in various previous studies is the electron densities used to calculate outflow rates. Previous studies commonly made use of the [SII](6717/6731) ratio to determine electron densities, however, this technique saturates at $n_e > 10^3 \mbox{ cm}^{-3}$, often leading to low {\color {black} ($\sim$200-500 cm$^{-3}$)} densities being estimated or assumed. In this work, where possible, we have measured electron densities using the transauroral technique, which is sensitive to higher electron densities than the [SII] ratio. Densities measured using this technique are, in general, an order of magnitude larger than those assumed by previous studies. Equation \ref{Eq1} in Section \ref{sec:outflow_rates} demonstrates the significant impact that this assumption has on the calculated total outflow mass, as the outflow mass decreases with increasing electron density.

Another factor in the discrepancies between the mass outflow rates detected in this study and the significantly higher mass outflow rates detected in other works might be how we determine which gas is considered to be outflowing. Here, we have made conservative estimates of the gas that is included in the outflow by assuming that only gas that has velocities below V05 and above V95 can be considered to be outflowing.  For comparison, in Table \ref{tab:test_vd} we report the outflow mass rates and kinetic energy rates calculated by assuming that all gas with velocities greater than 1.5 times the FWHM of the stellar features derived from the {\sc starlight} modelling is outflowing. In principle this approach is less conservative than the one that we considered, but the results are similar. We also checked that the correlation coefficients do not vary by using this outflow definition.

Finally, something that might contribute to explaining the relatively low values of the outflow mass rates and kinetic energies is having} characterised the bulk velocity of that gas as the velocity at the 50 per cent cumulative function between the continuum and V05(V95). These assumptions result in lower estimates of the total mass in the outflow, which is calculated from the total luminosity of [OIII] in the wings, and lower outflow velocities {\color {black} than those calculated using parametric methods and maximum velocities (v$_{\rm of}$=v$_s$+2$\sigma$ or v$_{\rm of}$=v$_s$+FWHM/2) leading to lower mass outflow rates (see \citealt{2023arXiv230910572H} for a comparison of the different outflow properties using parametric and non-parametric methods).

A number of works claim positive correlations between outflow kinematics and intrinsic AGN properties, such as luminosity and Eddington ratio, with more powerful AGN driving more powerful outflows \citep{cicone14,fiore17,davies20,2023arXiv230910572H}. In contrast, when we test for correlations between {\color {black} these intrinsic parameters in the \qfd~sample and the various outflow properties here considered}, we find no such correlations. Potentially, our contrasting result can be explained by the differences in the AGN luminosity ranges covered by the samples used in these studies. The \qfd~sample was selected based on redshift and $L_{[OIII]}$, ensuring it contains only the most luminous AGN in the nearby Universe. The consequence of our selection criteria is that $L_{[OIII]}$ covers one order of magnitude ($42.1\le \log L_{[OIII]} \le 42.9$ \ergs), whilst works that claim such correlations tend to cover a broader range of luminosities. For example, the sample analysed by \citet{jin23} covers the luminosity range $41.5\le \log L_{[OIII]} \le 44.0$ \ergs, thus extending to significantly {\color {black} lower} luminosities than \qfd, and {\color {black} \citet{davies20} compiled mass outflow rates for AGN covering 5 orders of magnitude in L$_\text{bol}$.} Therefore, we must consider the possibility that the luminosity range probed in this investigation is too small to detect any such correlation.

Although not a selection criterion for \qfd, the sample probes a significantly broader range of radio luminosities ($22.3 \le \log L_{1.4GHz} \le 26.2 \mbox{W Hz}^{-1}$ for detected sources) than AGN luminosities, giving us increased scope to investigate the dependence of outflow strength on radio power. Previous work has shown that the width of the [OIII] emission lines correlates with radio-luminosity \citep{mullaney13,zakamska14}. \citet{mullaney13} used stacked SDSS spectra of {\color {black} thousands} of AGN to measure emission line profiles as a function of several variables, including $L_{1.4GHz}$, and found that the width of the [OIII] emission line increased as a function of radio-luminosity at intermediate radio luminosities {\color {black} (up to log $L_{1.4GHz}\sim$25 W~Hz$^{\rm -1}$), declining again at higher radio luminosities.} Those findings are consistent with the results presented here, where we see a moderate positive correlation between $L_{1.4GHz}$ and $W80_{int}$ at intermediate radio luminosities, which strengthens when radio emission associated with star formation is accounted for. Interestingly, the two objects which harbour large-scale radio jets (log $L_{1.4GHz}>$25 W~Hz$^{\rm -1}$) have $W80_{int}$ values which are consistent with those of the stellar velocity dispersion. 

We can understand this correlation in the context of simulations which investigate the interaction between low-power radio jets and the host's ISM \citep{mukherjee16,talbot22}. Such simulations indicate that low-power jets will remain confined within the ISM longer than their high-power counterparts, which will rapidly punch through the intervening material. The increased time that the low-power jet remains trapped means it will be more disruptive, impacting a larger volume of the host's ISM, injecting more turbulence and leading to the broader [OIII] line profiles observed.

However, not all studies comparing the outflow kinematics with radio luminosity find a similar correlation. \citet{ayubinia23} compare the [OIII] profile of a large sample of AGN against their radio properties and, in contrast to the results presented here, claim that accounting for the underlying galaxy gravitational potential weakens the positive correlation between the two and instead, it is the {\color {black} Eddington ratio} that correlates most strongly with [OIII] velocity dispersion. However, even though we account for the underlying gravitational potential, we still find a moderate correlation between the $W80_{int}$ and $L_{1.4GHz}$. {\color {black} This correlation, together with the lack of it} between [OIII] kinematics and other properties of the \qfd~sample, such as Eddington ratio, $M_{BH}$, and $L_{bol}$ suggests that small-scale (kpc), low-power radio jets may play an important role in driving outflows. Indeed, this finding is supported by observational studies in both the ionised \citep{jarvis19,cazzoli22,speranza22} and molecular \citep{morganti05,garcia19,audibert23} gas phases. {\color {black} However, as mentioned before, it is also possible that not all the kiloparsec-scale radio structures that we see in radio-quiet AGN are jets, but synchrotron radiation produced when the multi-phase outflows interact with the ISM \citep{zakamska14,2023arXiv230615047F}. This would also explain the spatial coincidence between the extended radio structures and the gas outflows and the correlation.}

\subsection{Star formation and outflow properties}
\label{sec:SF_outflows}

As we have shown, although there is clear evidence that AGN have the potential to drive fast ionised outflows, it is not clear that these outflows directly suppress SF. We have shown that 98 per cent (92 per cent excluding those with unconstrained SFRs) of the galaxies contain stellar populations with ages $< 100 \mbox{ Myr}$, consistent with the results of previous studies of YSPs in  QSO2 \citep{bessiere14,bessiere17,woo20,jin23}. \citet{bessiere17} used apertures of a fixed physical scale (8 kpc) in their study of the stellar populations of 21 QSO2s at $0.3 < z < 0.41$ and found that to adequately fit the data, 71 per cent required the inclusion of a YSP $< 100 \mbox{~Myr}$. The ability to control the physical size of the region probed allowed the authors to make direct comparisons between objects, however, when utilising SDSS spectra to measure stellar populations and gas kinematics, we have no control over the physical scale of the galaxy covered by the fibre. In addition, all spatial information is lost, since the fibre is collapsed to a 1d spectrum. When considering the relationship between AGN-driven outflows and SF, it is most likely that this will be traced within the central few kiloparsecs rather than on galaxy-wide scales as {\color {black} observational evidence from spatially resolved studies  
demonstrates \citep{bessiere22,ramos22}.}

Previously, we have carried out a spatially resolved study of both the ionised gas kinematics and stellar populations of the nearby (z = 0.050) QSO2 Mrk 34 \citep{bessiere22}, which forms part of the \qfd~sample (J1034+60). In that work, we demonstrated a clear spatial correlation between the 
edges of the approaching side of the {\color {black} kiloparsec-scale ionised outflow (t$_{\rm dyn}\sim$1 Myr) and an enhancement in the fraction of the stellar light attributable to the YSP (ages of $\sim$1-2 Myr as constrained from the modelling). On the other hand, at the edges of receding side of the outflow, where the turbulence and the energy injection are higher, we do not find the same enhancement of the YSP flux.} Here, we have analysed the available SDSS spectrum of Mrk 34 in the same manner as the remainder of the \qfd~sample as presented above. In Figures \ref{fig:sfr-ms}, \ref{fig:nebsub}, \ref{fig:w80_slvd}, \ref{fig:decon_vdif}, and \ref{fig:corr}, Mrk 34 is plotted as a cyan star and clearly shows that, when using the integrated fibre spectrum to characterise the outflows and YSP, it does not appear to be in any way different from the other objects in the sample. It is only through the use of spatially resolved information that we can detect the positive {\color {black} and/or negative impact} that the AGN-driven outflow is having on star formation in the host galaxy. This suggests that it is imperative to increase the number of spatially resolved studies {\color {black} as the one presented in \citet{bessiere22}}, measuring both gas kinematics and stellar populations, to expand our understanding of if, how and under which circumstances AGN-driven feedback directly impacts SF. The advent of the {\it James Webb} Space Telescope (JWST) now makes this type of studies possible at high redshift (e.g.; \citealt{d'eugenio23}).

Another, yet related to the previous one, possible explanation for the lack of correlation between the outflow properties and SFR is that the effect of AGN on SF is a cumulative rather than an instantaneous process \citep{Scholtz20,scholtz21,lamperti21,baker23}. AGN are now thought to be episodic events that can happen throughout the life of a galaxy, flickering on time-scales as short as $10^5$ years \citep{hickox14}. Observationally, when investigating the impact of the AGN on the host galaxy, we are only able to probe the effects of the current episode of activity, only allowing for the possibility of characterising direct impacts on the host galaxy. However, if the effects are accumulated over several AGN episodes, it may be that SMBH mass is a more relevant indicator because it is through episodes of accretion, leading to AGN activity, that SMBH grows their mass \citep{martin21,2022MNRAS.512.1052P,bluck22}. If the series of AGN episodes has led to the removal and/or heating of the gas, leading to a suppression of SF over time, we would expect to find older populations associated with more massive BHs as the capacity to form new stars is reduced. {\color {black} Indeed, studying the star formation histories of nearby massive galaxies with accurate BH mass measurements, \citet{2018Natur.553..307M} showed that quenching of star formation takes place earlier and more efficiently in galaxies that host higher-mass central black holes. To test this hypothesis in the \qfd~sample,} we calculated the light-weighted ages of each of the QSO2s and found a moderate positive correlation ($r=0.572$) {\color {black} with the BH mass}, shown in Figure \ref{fig:bhm_sfr}. 
Although many other factors, such as galaxy morphology and merger history also play an important role in the constituent stellar population, this finding could be interpreted as evidence that successive AGN episodes lead to the suppression of SF over the course of galaxy evolution.

\begin{figure}
    \includegraphics[width = \columnwidth]{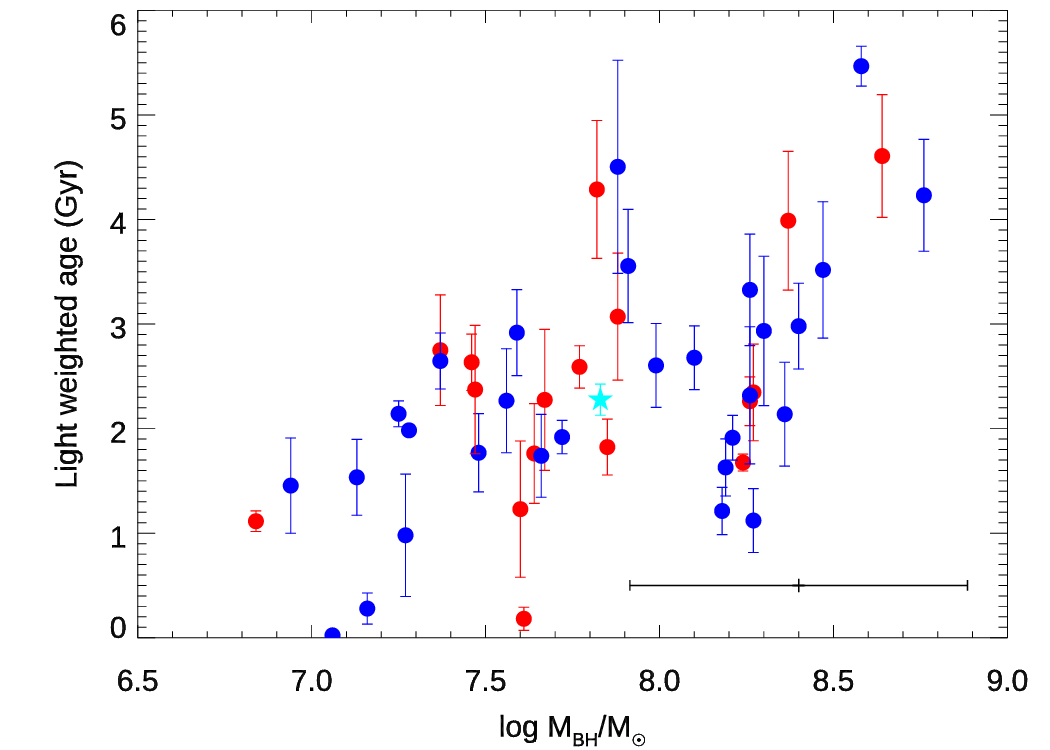}
    \caption{The light weighted ages of the stellar populations {\color {black} of the \qfd~sample as a function of black hole mass. A moderate positive correlation, with a Pearson coefficient of $r=0.572$ is found. The average error in the black hole mass measurements is shown by the black bar in the lower right corner, with an average error of $\pm 0.48$ dex.}} 
    \label{fig:bhm_sfr}
\end{figure}

\section{Conclusions}
In this investigation, we have derived stellar population ages and SFR, warm ionised gas kinematics, and outflow rates for a 
complete sample of 48 low redshift, obscured quasars. We have then compared the results of both strands of the investigation to gain insight into whether AGN-driven outflows directly impact SF in their host galaxies. The key findings of this study can be summarised thus:

\begin{itemize}
\item Our starlight modelling shows that 98\% of the \qfd~galaxies contain a stellar populations {\color {black} younger than 100 Myr, with percentages of light fraction ranging from 0 to 100\%.}

\item  98\% of the QSO2s lay on or above the main sequence of SF.

\item When taking into account stellar velocity dispersion, we find that 85\% of the QSO2s show evidence of {\color {black} ionised} outflows.

\item In the majority of cases, we use a technique involving the transauroral lines of [OII] and [SII] to determine electron densties and find that these are significantly higher than often assumed, with a mean value of  $4230 \mbox{ cm}^{-3}$.

\item Assuming an outflow radius of 0.62 kpc, we find mass outflow rates for the \qfd~sample ranging from 0.04 to 1.9 \msunyr, with a mean of 0.5 \msunyr.

\item We find no correlation between AGN luminosity, black hole mass or Eddington ratio and the kinematics of the warm ionised gas and/or outflow kinetic powers.
\item We do find a moderate positive correlation between $L_{1.4GHz}$ and $W80_{int}$, suggesting that radio luminosity is the most significant factor in driving {\color {black} ionised} outflows.

\item We do not find a correlation between {\color {black} the outflow properties} and SFR, suggesting that AGN-driven outflows do not directly suppress SF on the physical scales probed by this study.

\item {\color {black} We find a moderate positive correlation between the light-weighted age of the QSO2s' stellar populations and BH mass, which might be indicating that successive AGN episodes lead to the suppression of SF over the course of galaxy evolution.}
\end{itemize}

\begin{acknowledgements}
The authors would like to thank the anonymous referee for their insightful and constructive feedback, enabling them to enhance and refine the results presented here. PSB and CRA acknowledge the
projects “Feeding and feedback in active galaxies”, with reference PID2019-106027GB-C42, funded by MICINN-AEI/10.13039/501100011033, ``Quantifying the
impact of quasar feedback on galaxy evolution'', with reference
EUR2020-112266, funded by MICINN-AEI/10.13039/501100011033 and the European Union NextGenerationEU/PRTR, and ``Quasar feedback and molecular gas reservoirs'', with reference ProID2020010105, ACCISI/FEDER, UE from the Consejer\' ia de Econom\' ia, Conocimiento y Empleo del Gobierno de 
Canarias and the European Regional Development Fund (ERDF).

\end{acknowledgements}

%
   \bibliographystyle{aa} 
   \bibliography{ref_list} 
%
\begin{appendix}


\section{Stellar population fits}
\label{app:stellar}

Here we present examples of the stellar population modelling for each object, using the best fitting combination of metallicities presented in Table \ref{tab:starlight}. In each plot, the main panel shows the SDSS data in black with the results of the {\sc starlight} modelling overlayed in red. The inset in the main panel shows a zoom of the fit around the region of the higher-order Balmer absorption features characteristic of YSPs. The bottom panel shows the residuals of the fit, with zero marked with a magenta line. Regions masked out of the fit are highlighted in grey and regions given double weighting are shaded in pink. Pixels flagged in the SDSS spectra are marked with green crosses and were excluded from the fit.

\begin{figure*}
    \centering
    \begin{subfigure}{\columnwidth}
    \includegraphics[width = 0.95\columnwidth]{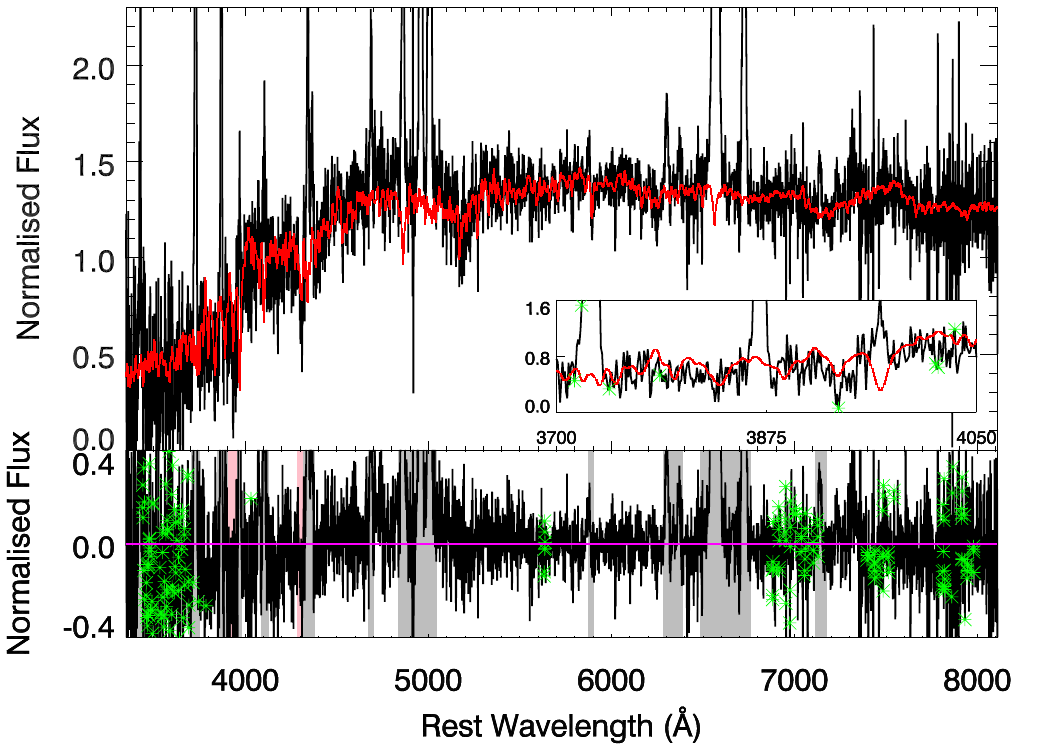}
	\end{subfigure}
    \begin{subfigure}{\columnwidth}
    \includegraphics[width = 0.95\columnwidth]{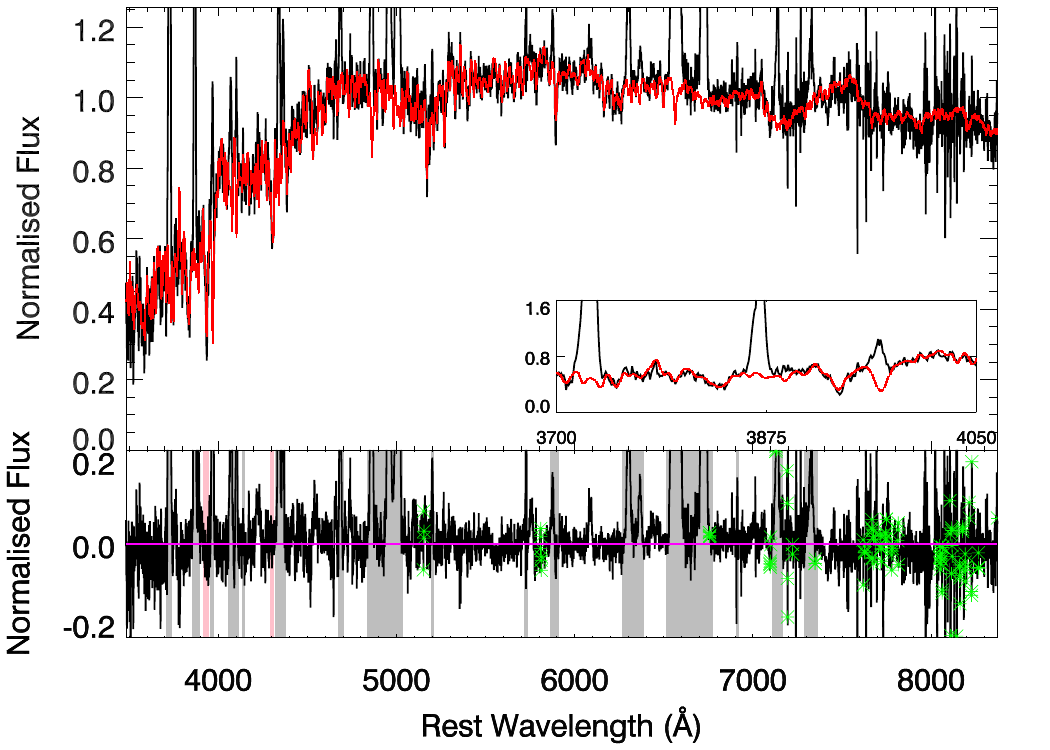}
	\end{subfigure}   
    
    \begin{subfigure}{\columnwidth}
    \includegraphics[width = 0.95\columnwidth]{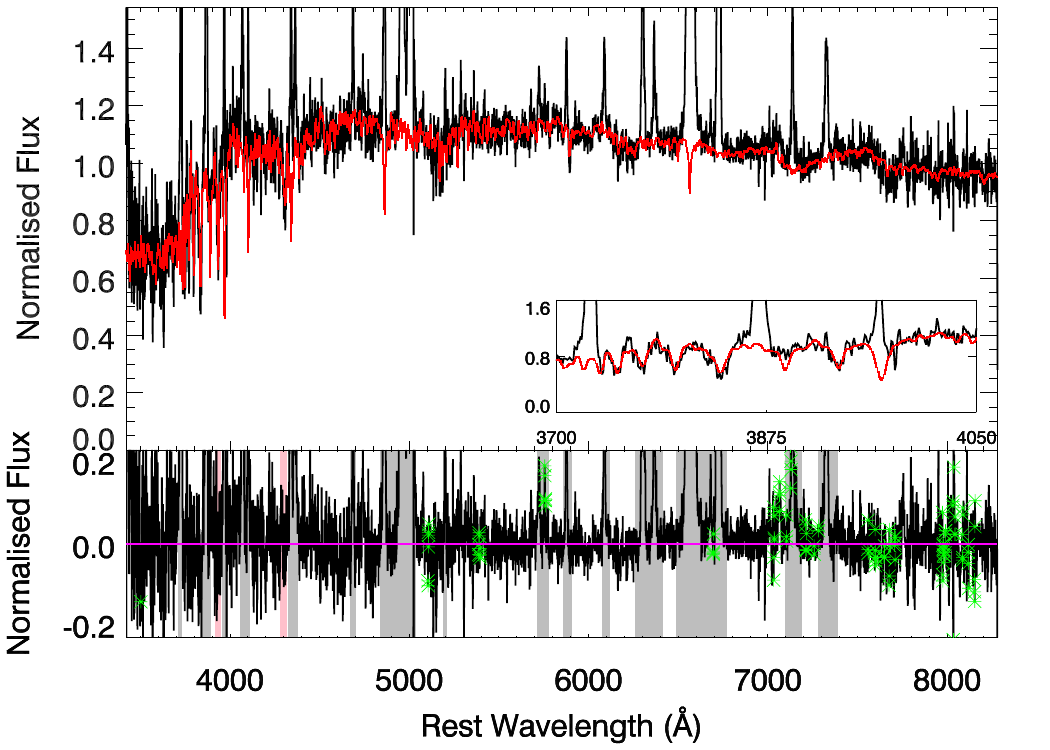}
	\end{subfigure}
    \begin{subfigure}{\columnwidth}
    \includegraphics[width = 0.95\columnwidth]{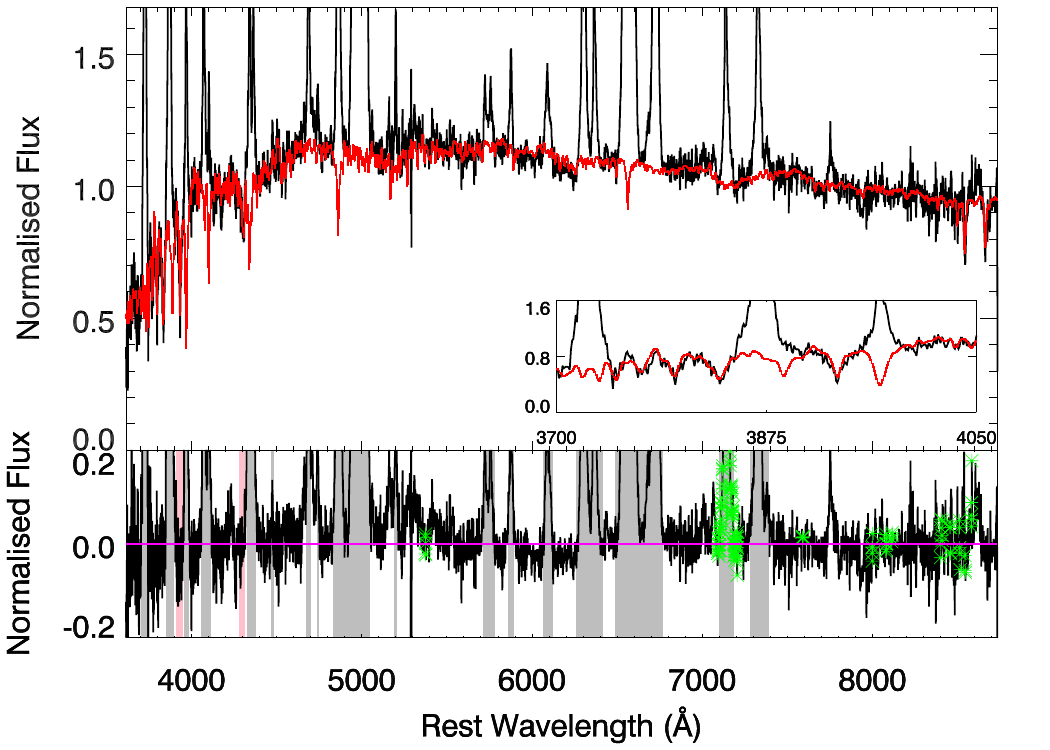}
	\end{subfigure}

    \begin{subfigure}{\columnwidth}
    \includegraphics[width = 0.95\columnwidth]{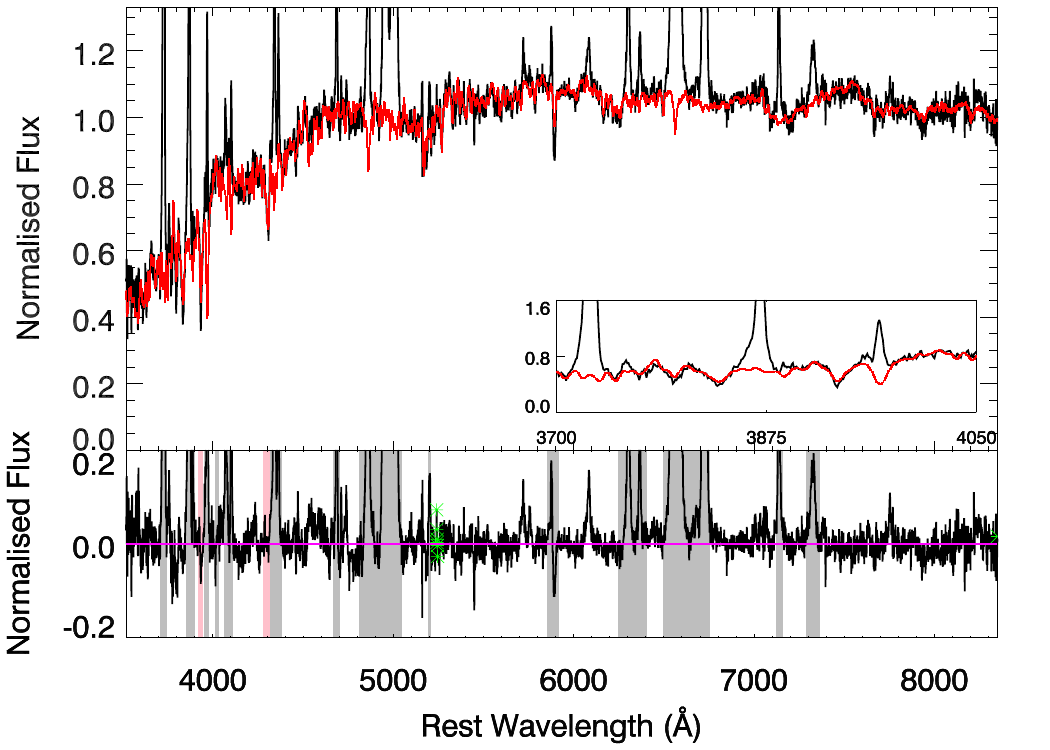}
	\end{subfigure}
    \begin{subfigure}{\columnwidth}
    \includegraphics[width = 0.95\columnwidth]{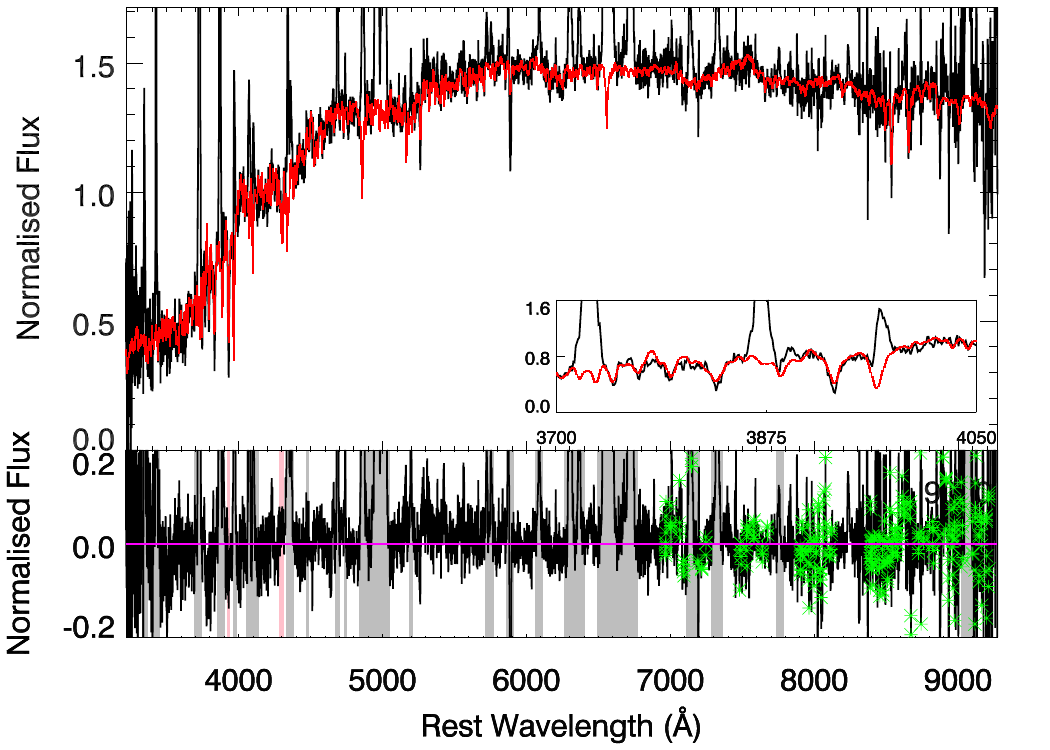}
	\end{subfigure}
    \caption{Stellar population fits of J0052-01 and J0232-08 (top row), J0731+39 and J0759+50 (middle row), and J0802+25 and J0802+46 (bottom row).}
    \label{fig:sl_pg1}
\end{figure*}

\clearpage

\begin{figure*}
    \centering
    \begin{subfigure}{\columnwidth}
    \includegraphics[width = 1\columnwidth]{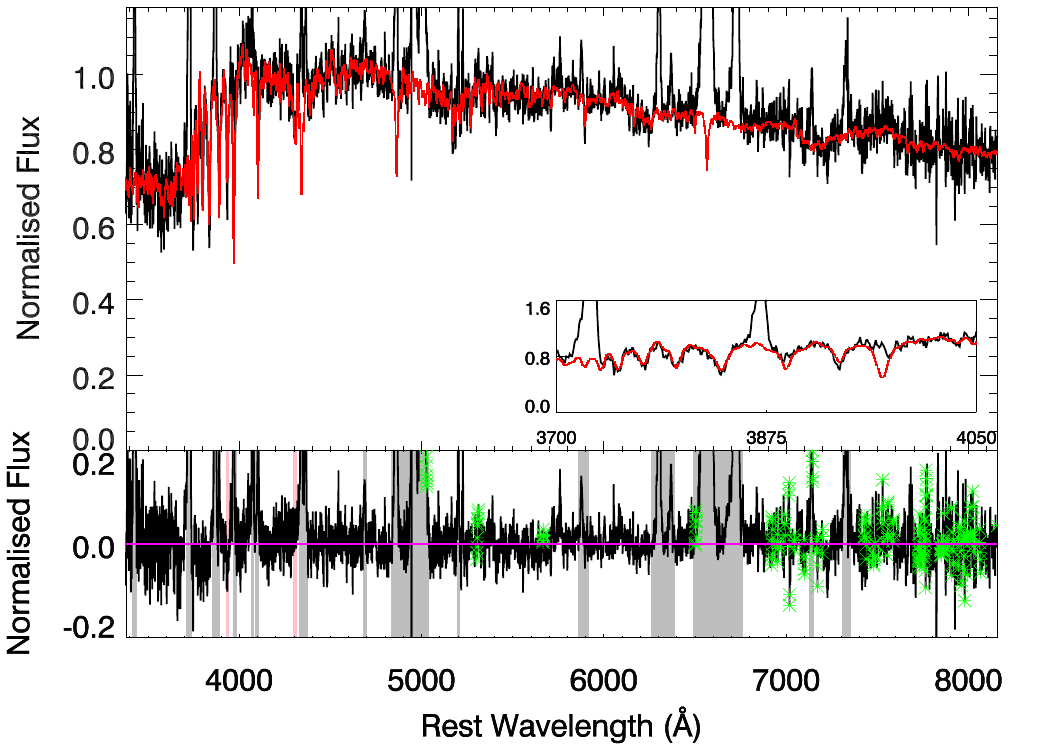}
	\end{subfigure}
    \begin{subfigure}{\columnwidth}
    \includegraphics[width = 1\columnwidth]{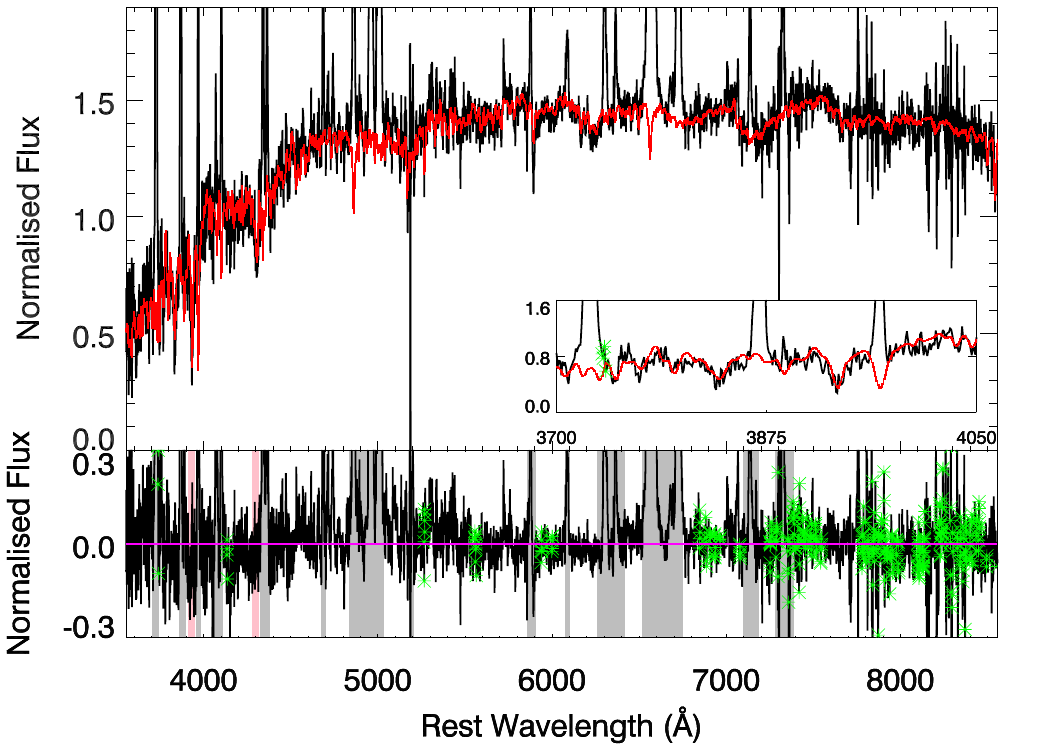}
	\end{subfigure}   
    
    \begin{subfigure}{\columnwidth}
    \includegraphics[width = 1\columnwidth]{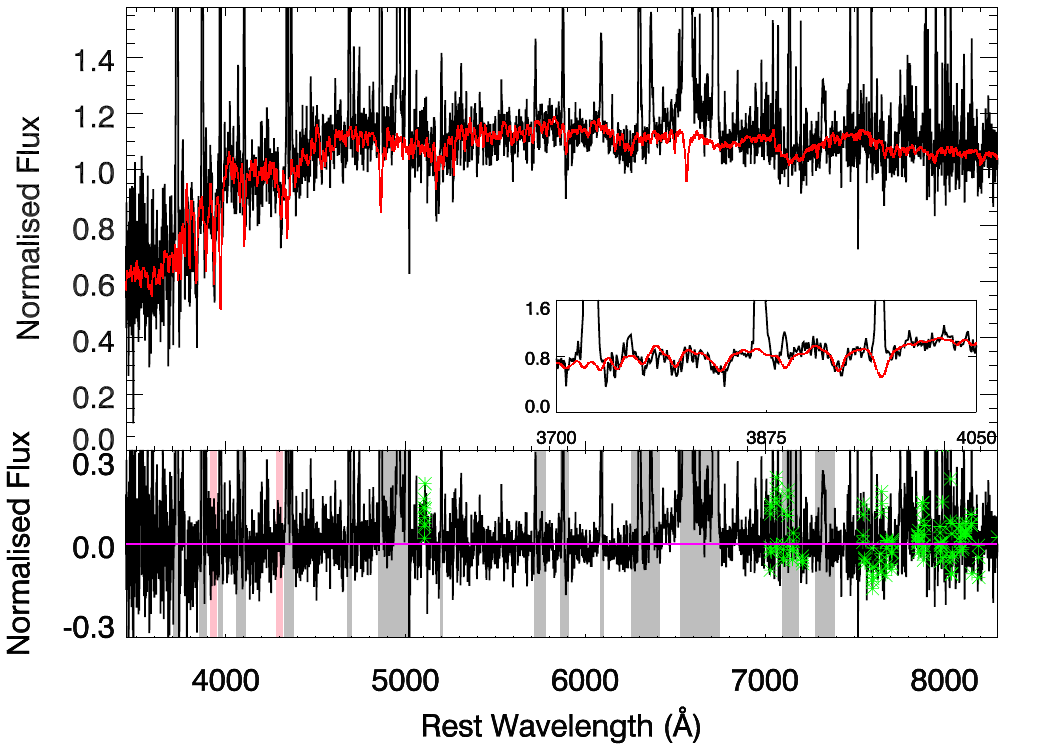}
	\end{subfigure}
    \begin{subfigure}{\columnwidth}
    \includegraphics[width = 1\columnwidth]{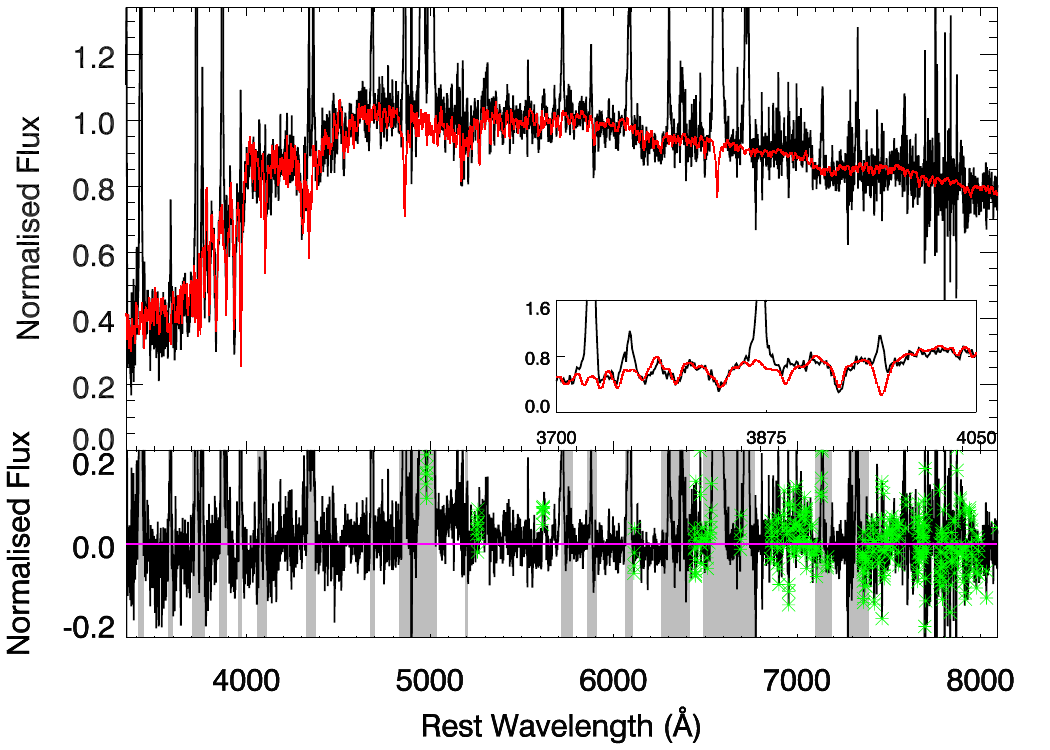}
	\end{subfigure}

    \begin{subfigure}{\columnwidth}
    \includegraphics[width = 1\columnwidth]{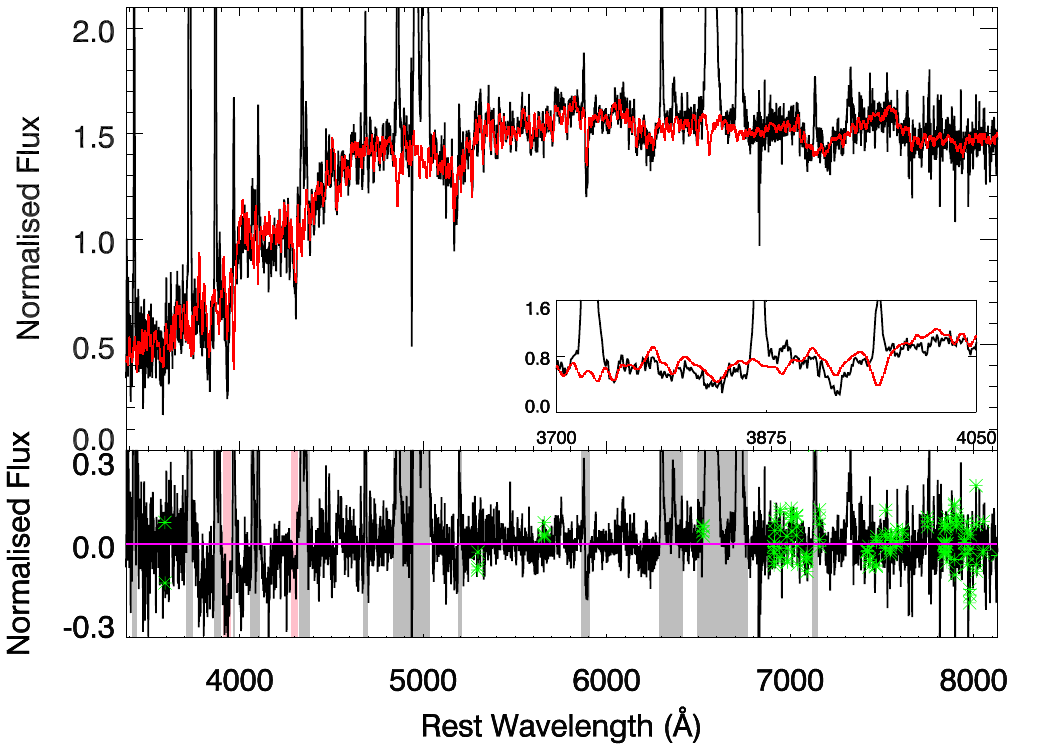}
	\end{subfigure}
    \begin{subfigure}{\columnwidth}
    \includegraphics[width = 1\columnwidth]{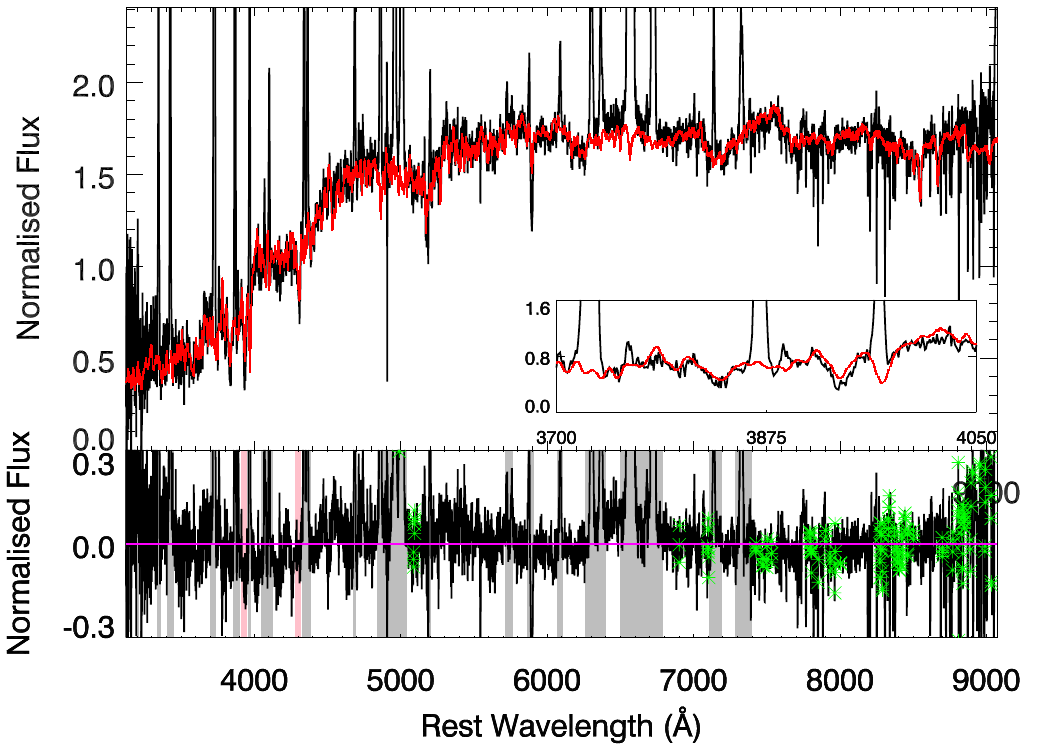}
	\end{subfigure}
    \caption{Stellar population fits of J0005+28 and J0818+36 (top row), J0841+01 and J0858+31 (middle row), and J0915+30 and J0939+35 (bottom row).}
    \label{fig:sl_pg2}
\end{figure*}

\clearpage

\begin{figure*}
    \centering
    \begin{subfigure}{\columnwidth}
    \includegraphics[width = 1\columnwidth]{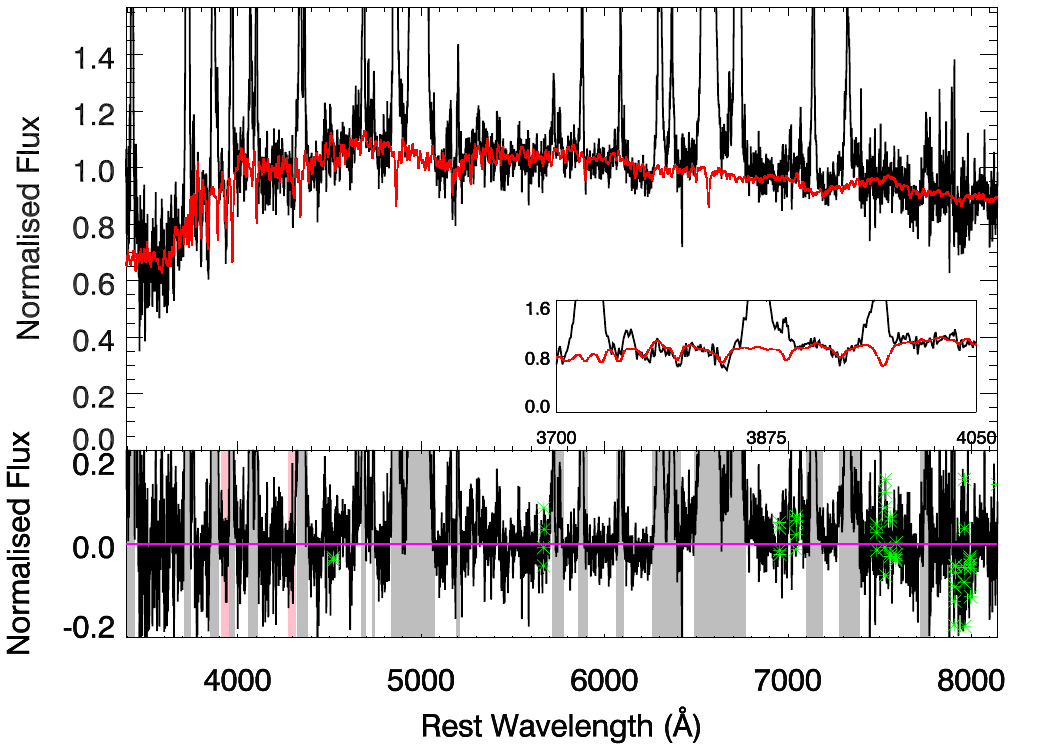}
	\end{subfigure}
    \begin{subfigure}{\columnwidth}
    \includegraphics[width = 1\columnwidth]{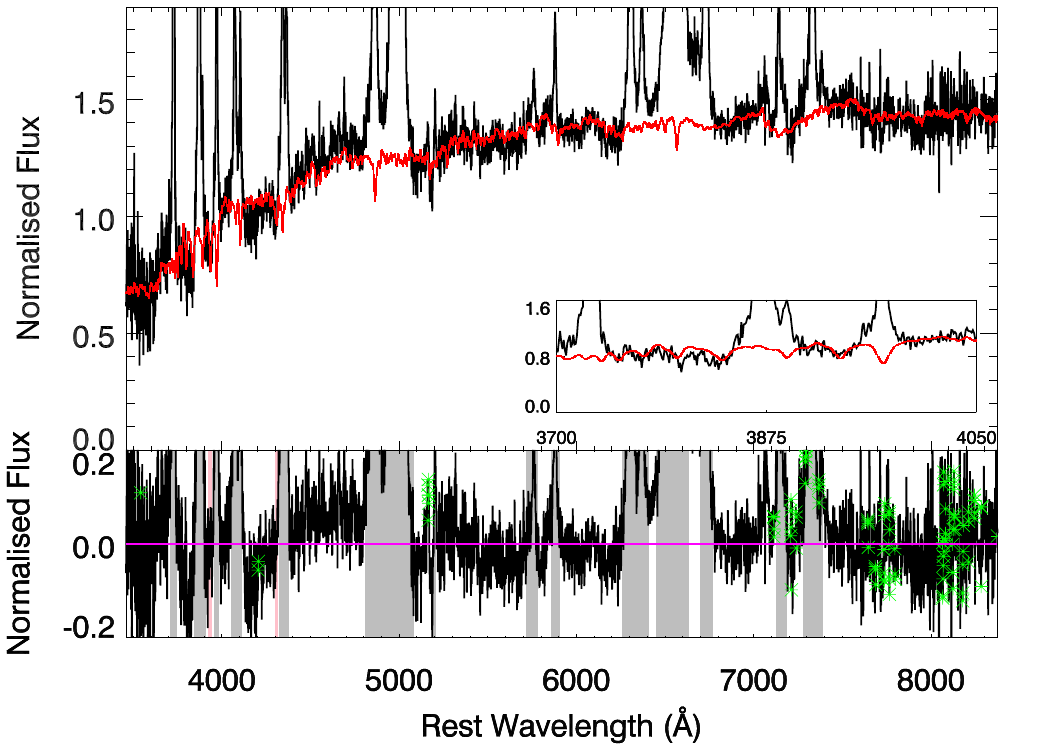}
	\end{subfigure}   
    
    \begin{subfigure}{\columnwidth}
    \includegraphics[width = 1\columnwidth]{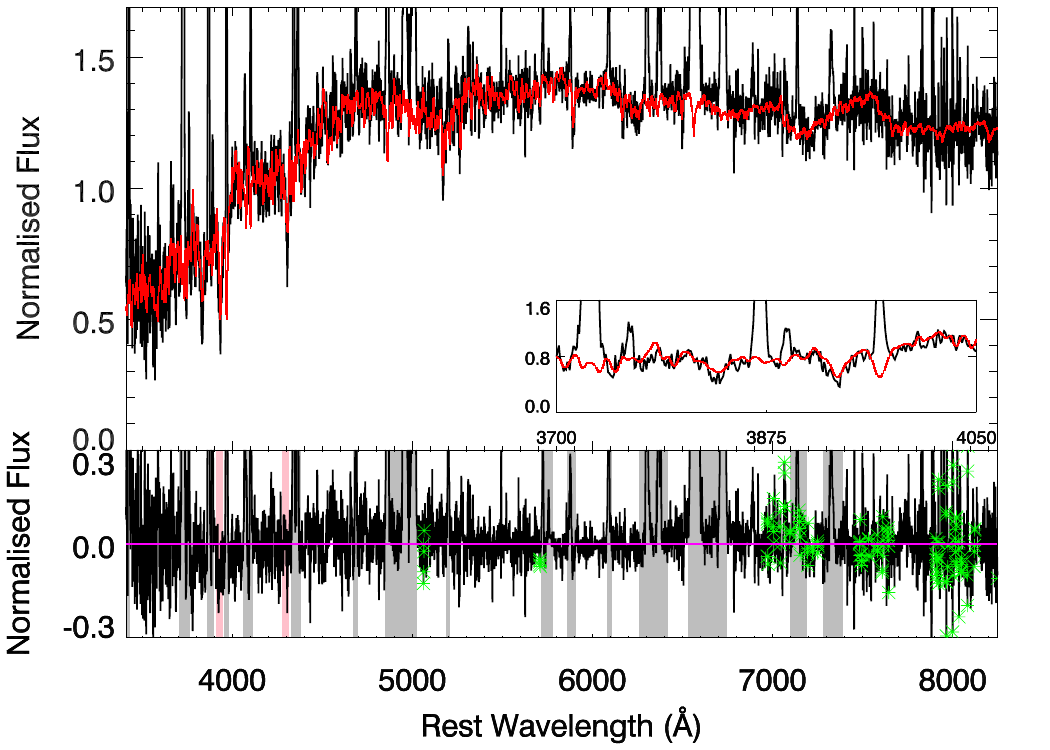}
	\end{subfigure}
    \begin{subfigure}{\columnwidth}
    \includegraphics[width = 1\columnwidth]{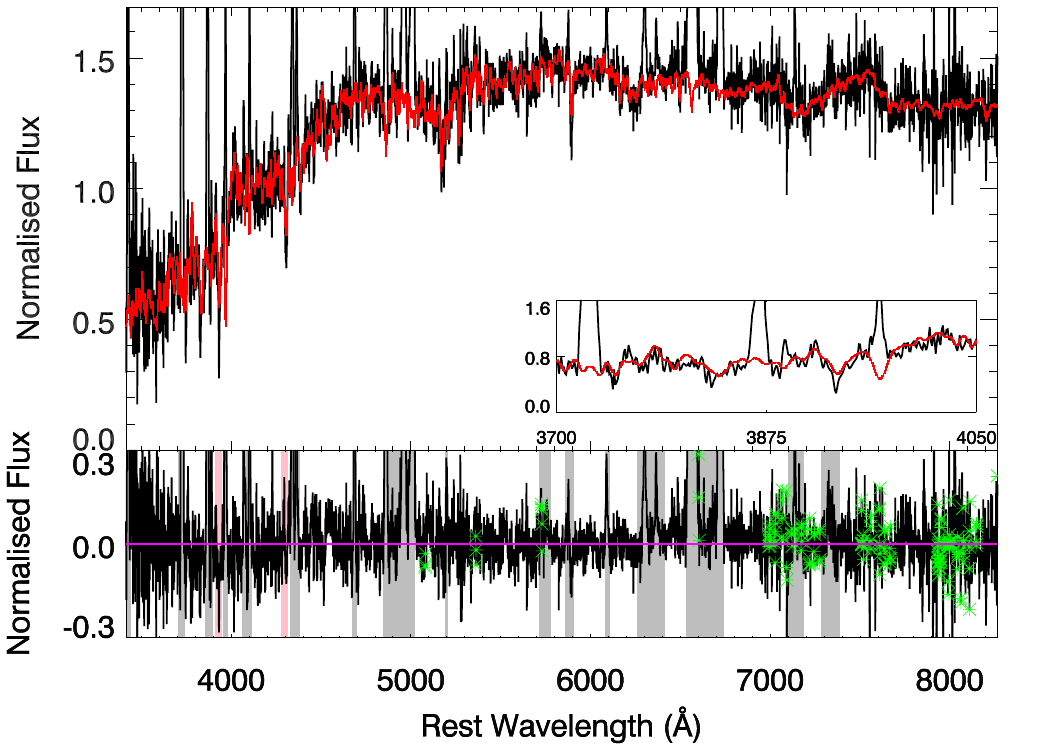}
	\end{subfigure}

    \begin{subfigure}{\columnwidth}
    \includegraphics[width = 1\columnwidth]{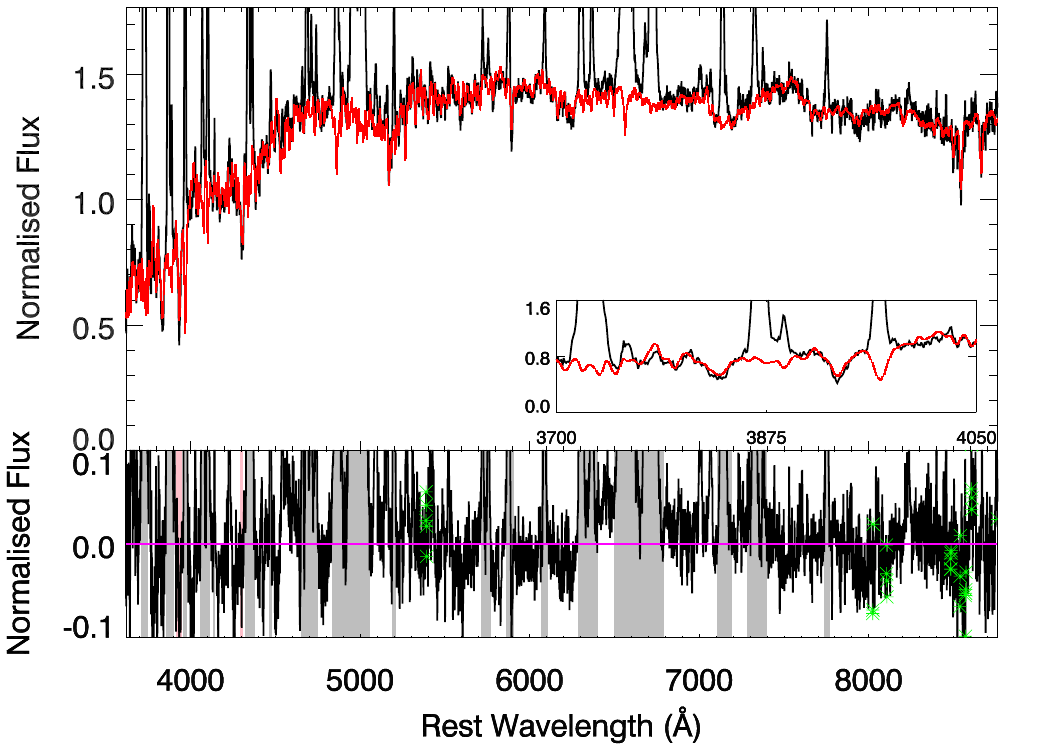}
	\end{subfigure}
    \begin{subfigure}{\columnwidth}
    \includegraphics[width = 1\columnwidth]{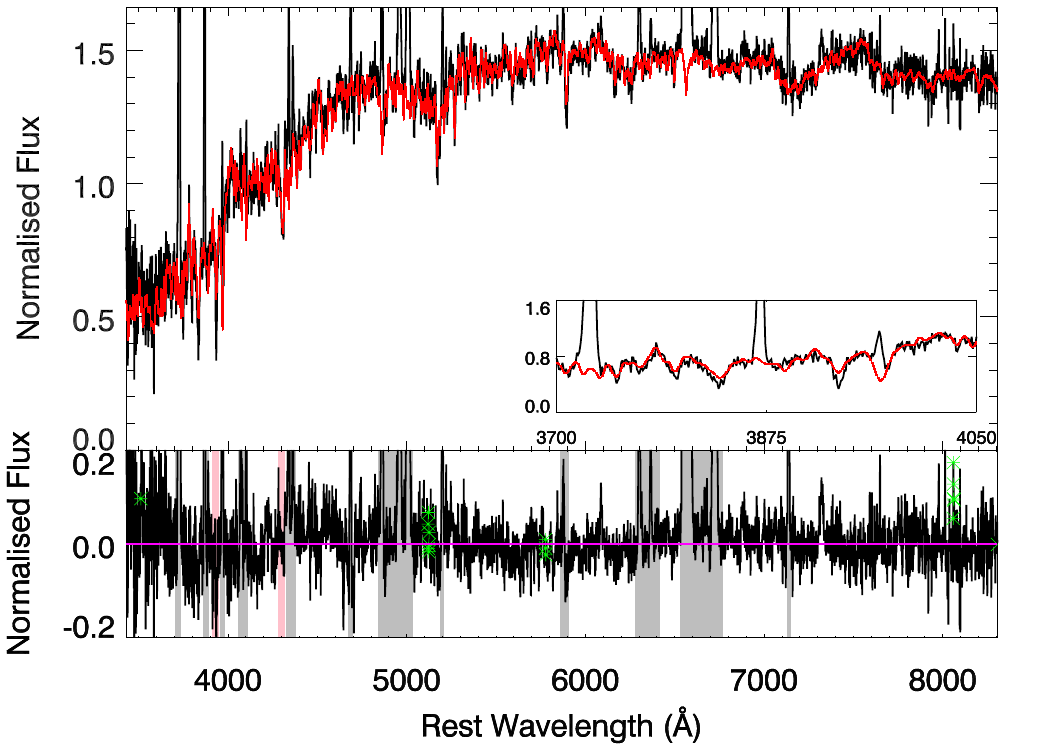}
	\end{subfigure}
    \caption{Stellar population fits of J0945+17 and J01010+06 (top row), J1015+00 and J1016+00 (middle row), and J1034+60 and J1036+01 (bottom row).}
    \label{fig:sl_pg3}
\end{figure*}

\clearpage

\begin{figure*}
    \centering
    \begin{subfigure}{\columnwidth}
    \includegraphics[width = 1\columnwidth]{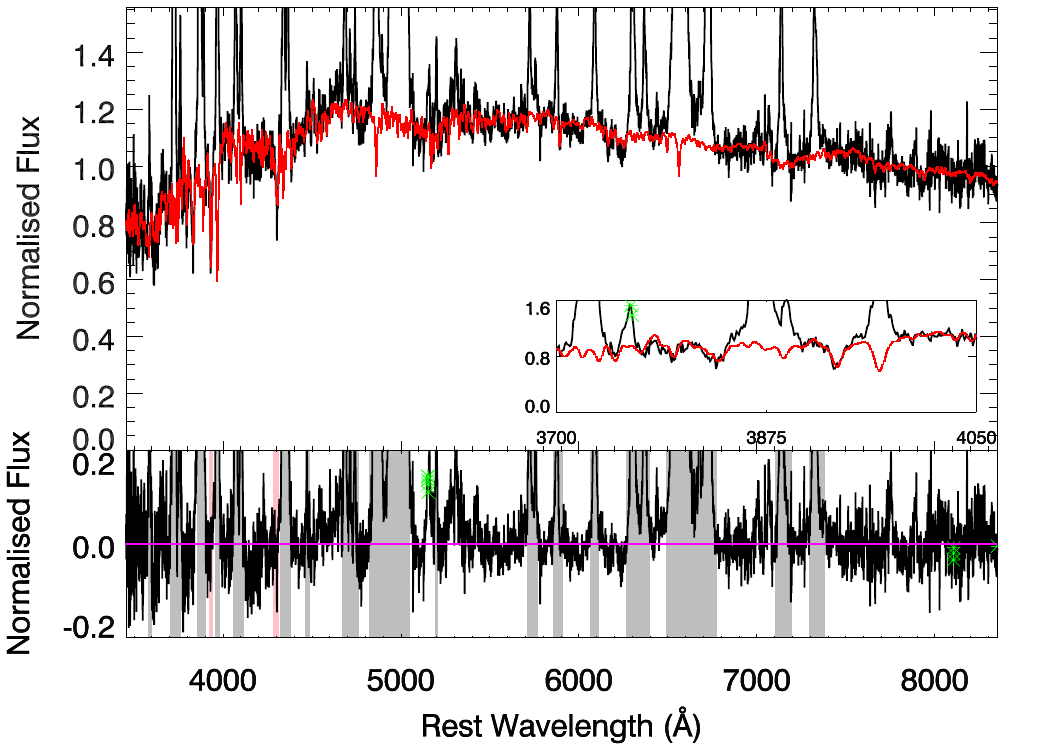}
	\end{subfigure}
    \begin{subfigure}{\columnwidth}
    \includegraphics[width = 1\columnwidth]{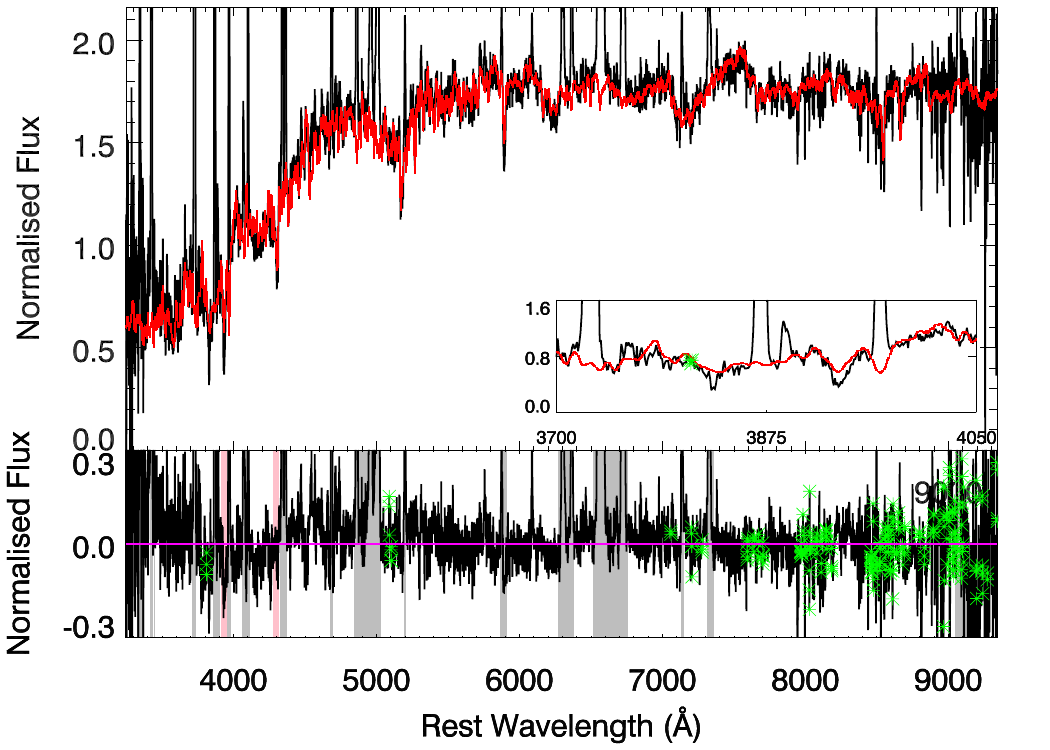}
	\end{subfigure}   
    
    \begin{subfigure}{\columnwidth}
    \includegraphics[width = 1\columnwidth]{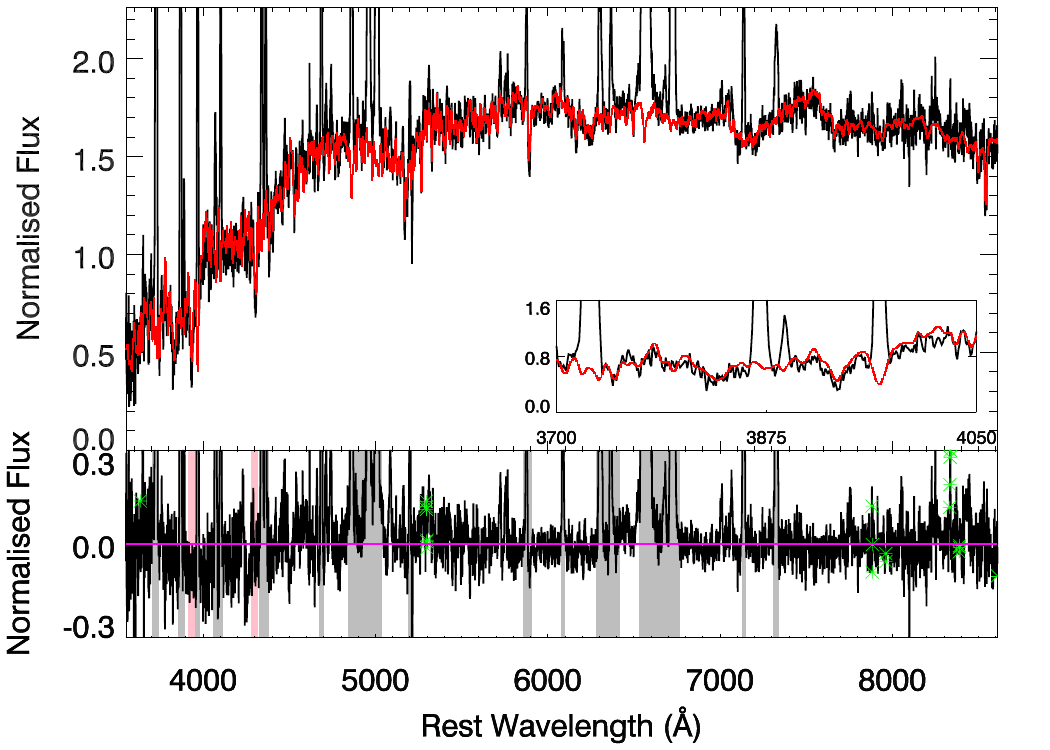}
	\end{subfigure}
    \begin{subfigure}{\columnwidth}
    \includegraphics[width = 1\columnwidth]{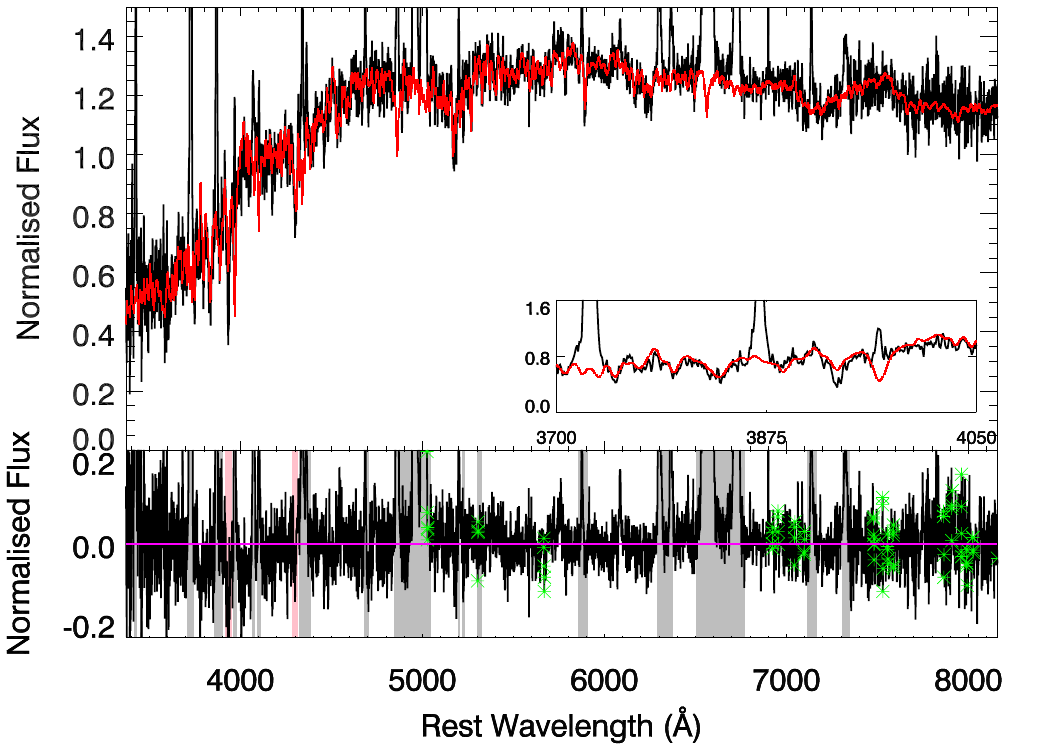}
	\end{subfigure}

    \begin{subfigure}{\columnwidth}
    \includegraphics[width = 1\columnwidth]{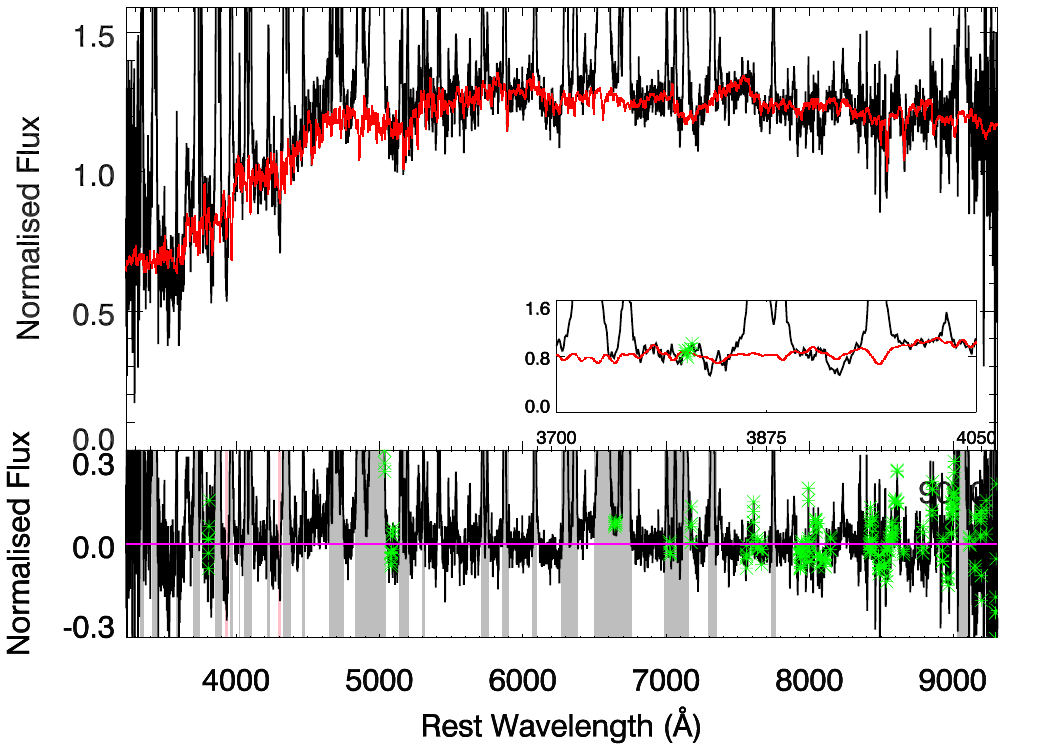}
	\end{subfigure}
    \begin{subfigure}{\columnwidth}
    \includegraphics[width = 1\columnwidth]{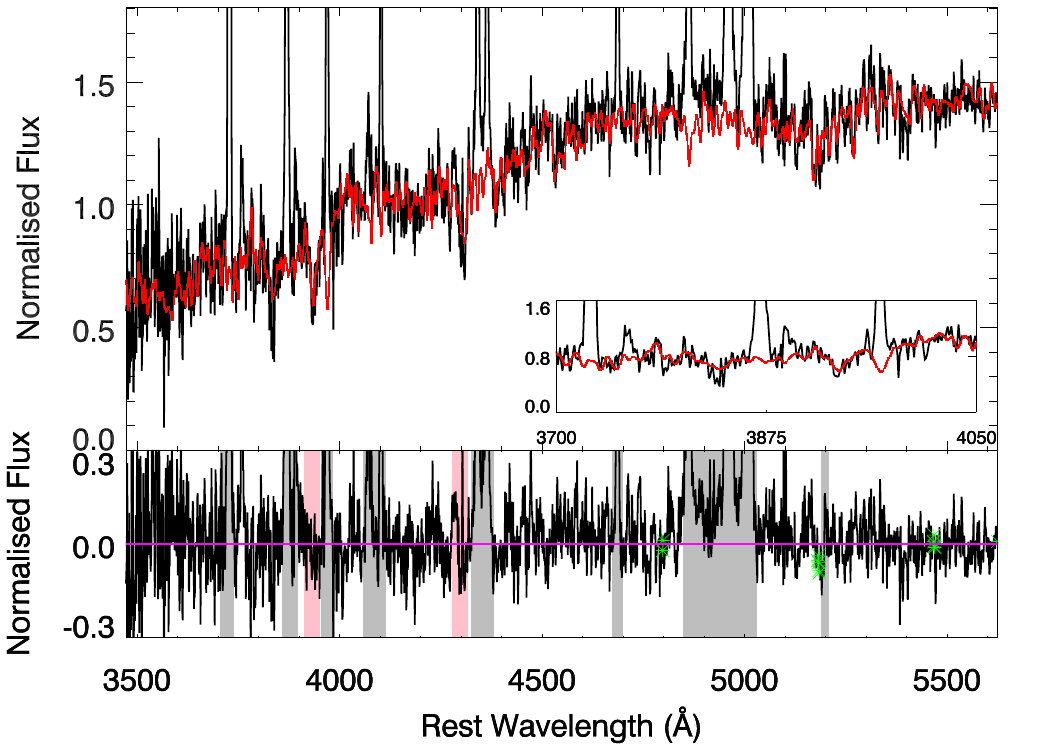}
	\end{subfigure}
    \caption{Stellar population fits of J1100+08 and J1137+61 (top row), J1152+10 and J1157+37 (middle row), and J1200+31 and J1218+47 (bottom row).}
    \label{fig:sl_pg4}
\end{figure*}

\clearpage

\begin{figure*}
    \centering
    \begin{subfigure}{\columnwidth}
    \includegraphics[width = 1\columnwidth]{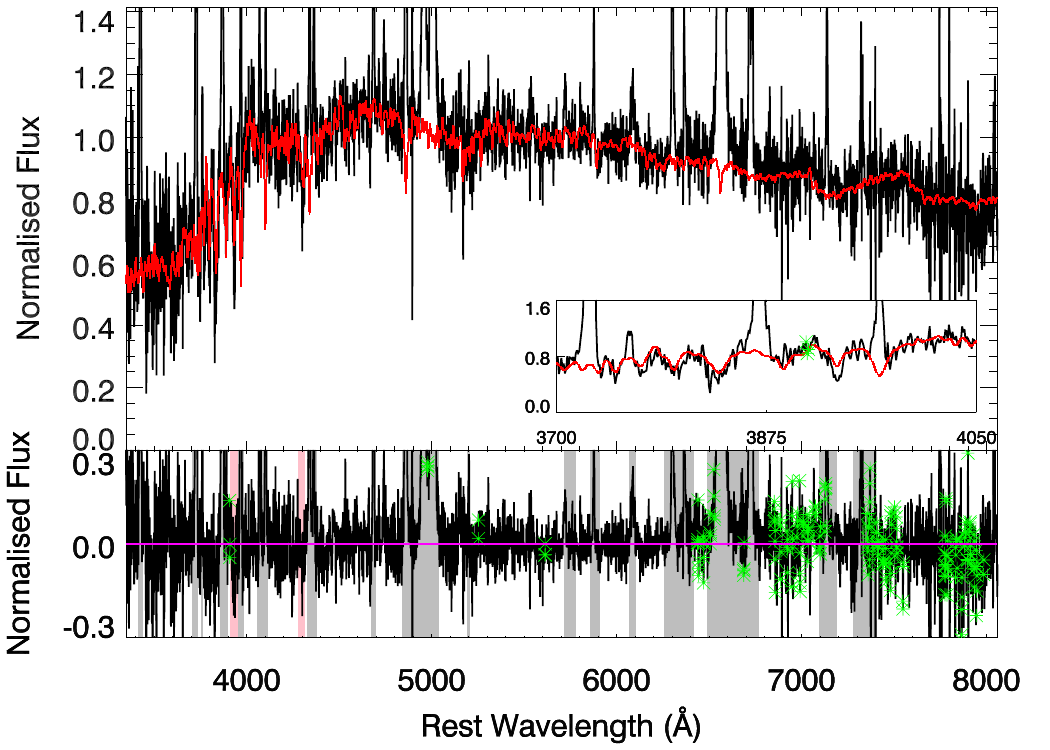}
	\end{subfigure}
    \begin{subfigure}{\columnwidth}
    \includegraphics[width = 1\columnwidth]{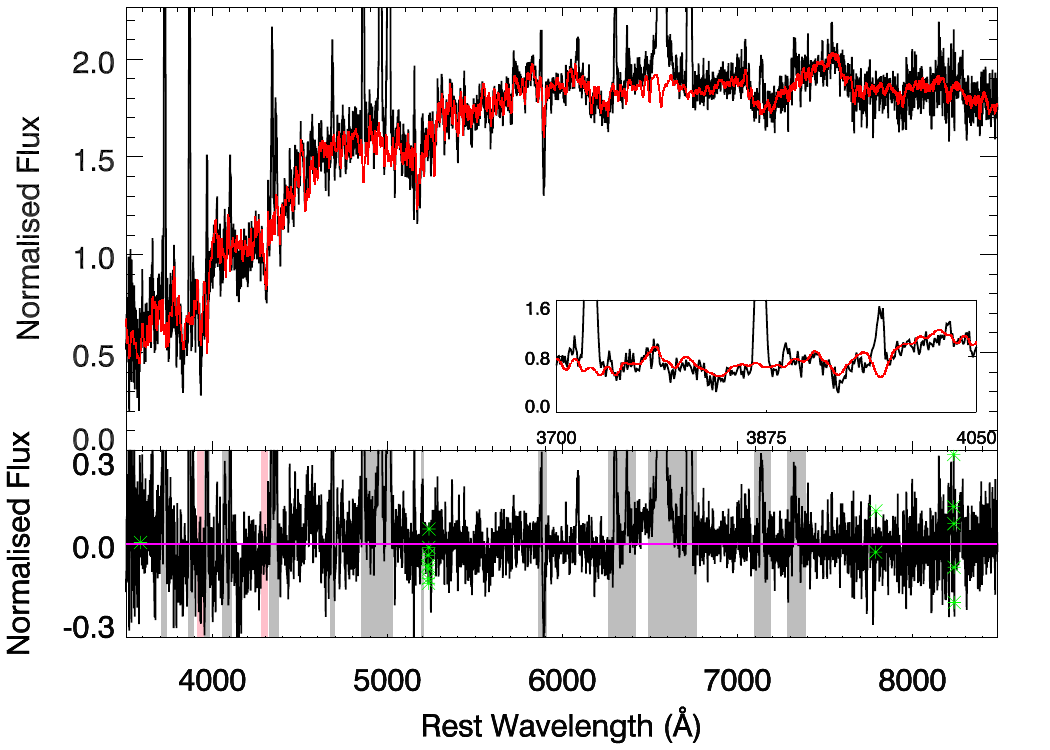}
	\end{subfigure}   
    
    \begin{subfigure}{\columnwidth}
    \includegraphics[width = 1\columnwidth]{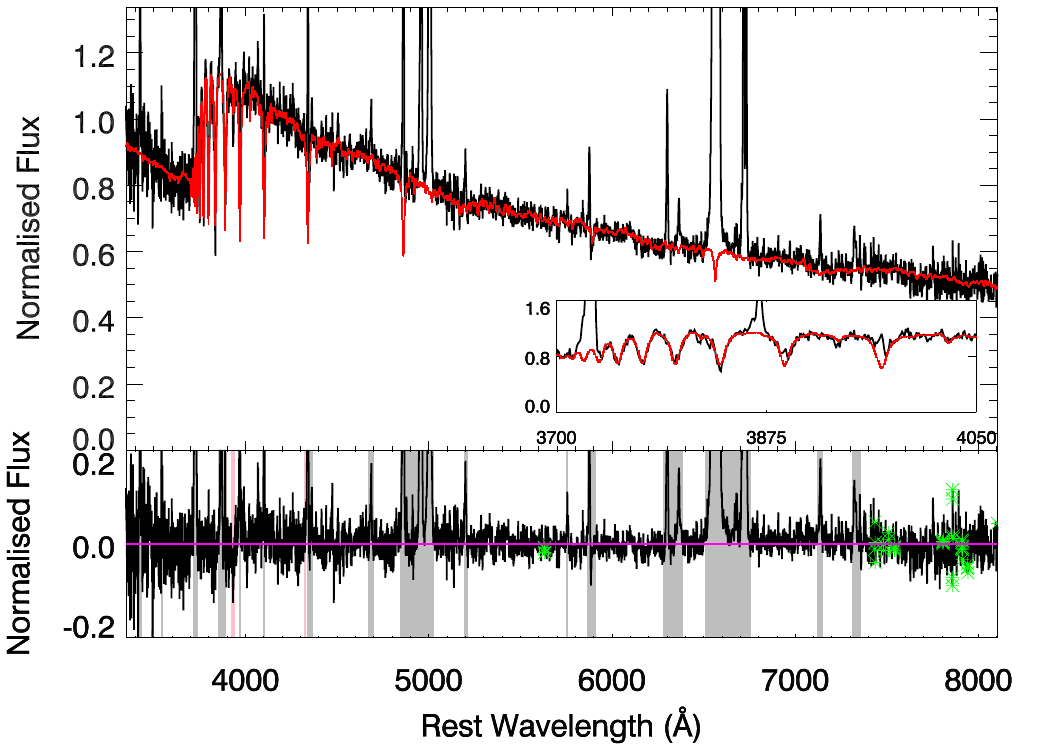}
	\end{subfigure}
    \begin{subfigure}{\columnwidth}
    \includegraphics[width = 1\columnwidth]{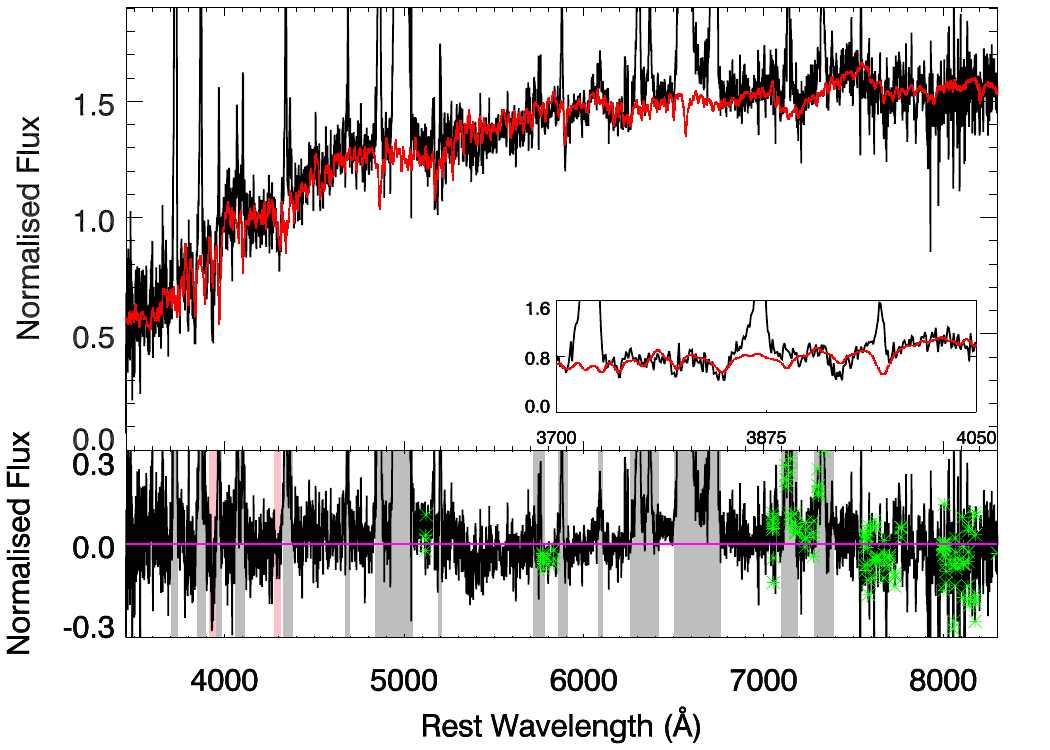}
	\end{subfigure}

    \begin{subfigure}{\columnwidth}
    \includegraphics[width = 1\columnwidth]{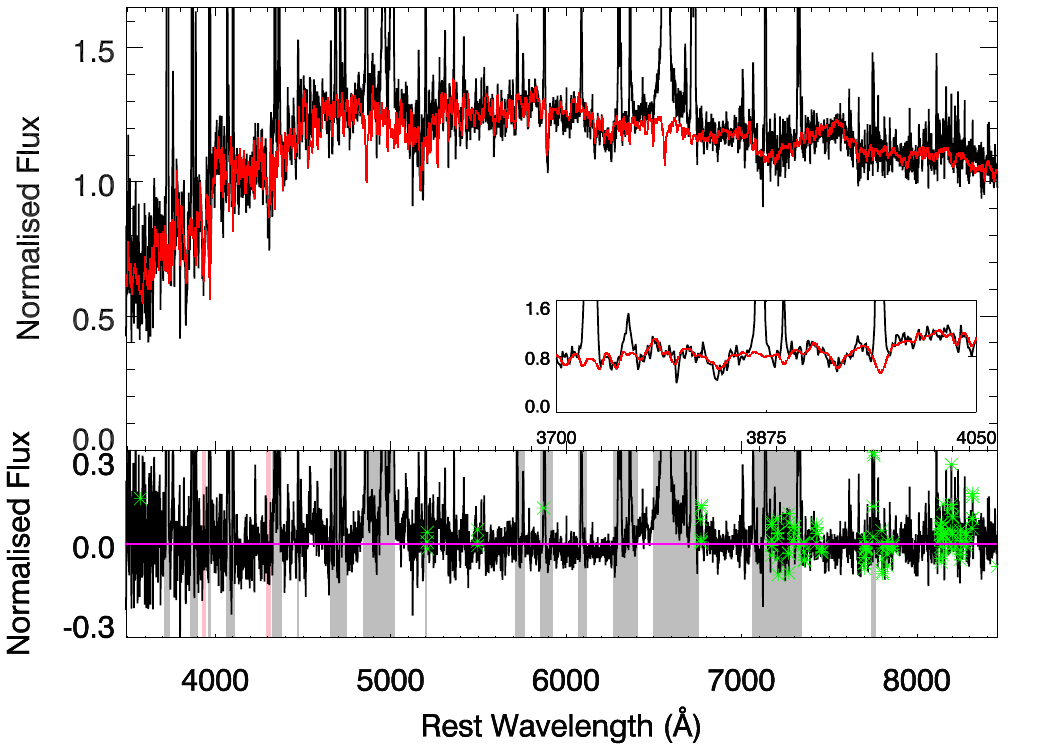}
	\end{subfigure}
    \begin{subfigure}{\columnwidth}
    \includegraphics[width = 1\columnwidth]{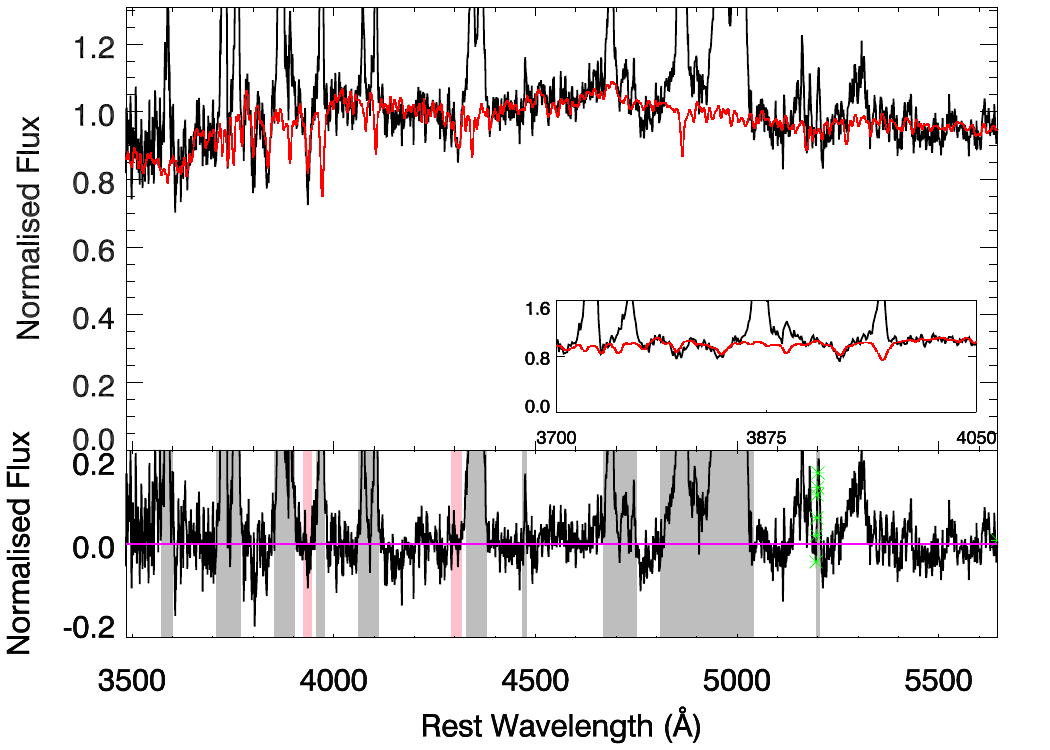}
	\end{subfigure}
    \caption{Stellar population fits of J1223+08 and J1238+09 (top row), J1241+61 and J1244+65 (middle row), and J1300+54 and J1316+44 (bottom row).}
    \label{fig:sl_pg5}
\end{figure*}

\clearpage

\begin{figure*}
    \centering
    \begin{subfigure}{\columnwidth}
    \includegraphics[width = 1\columnwidth]{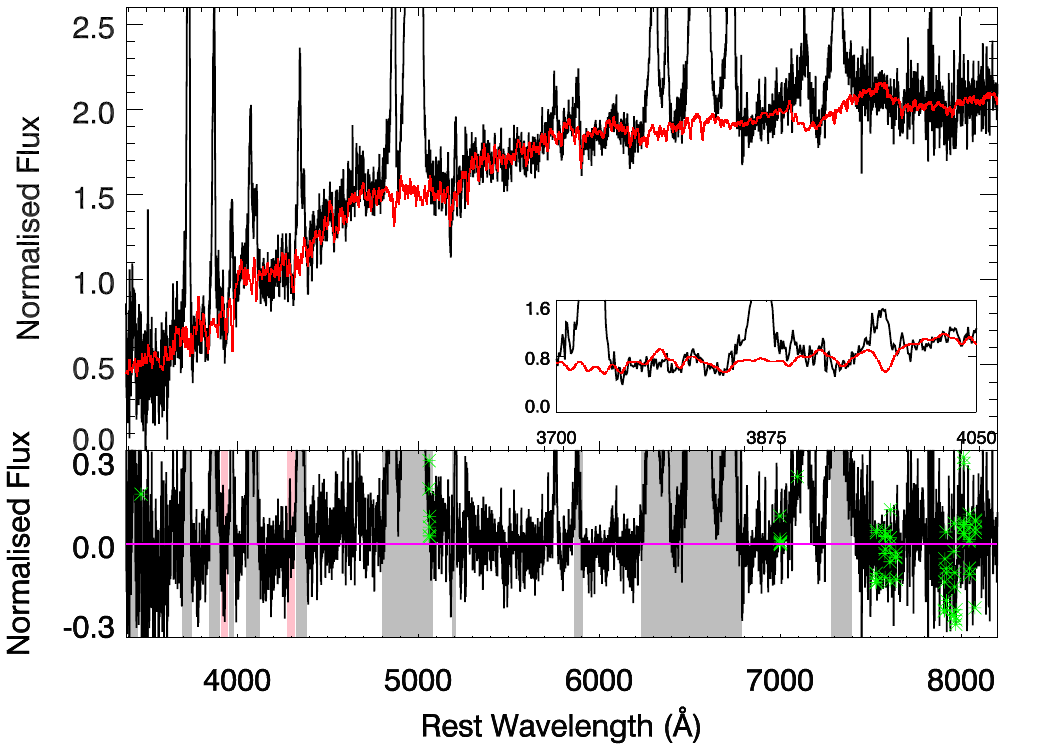}
	\end{subfigure}
    \begin{subfigure}{\columnwidth}
    \includegraphics[width = 1\columnwidth]{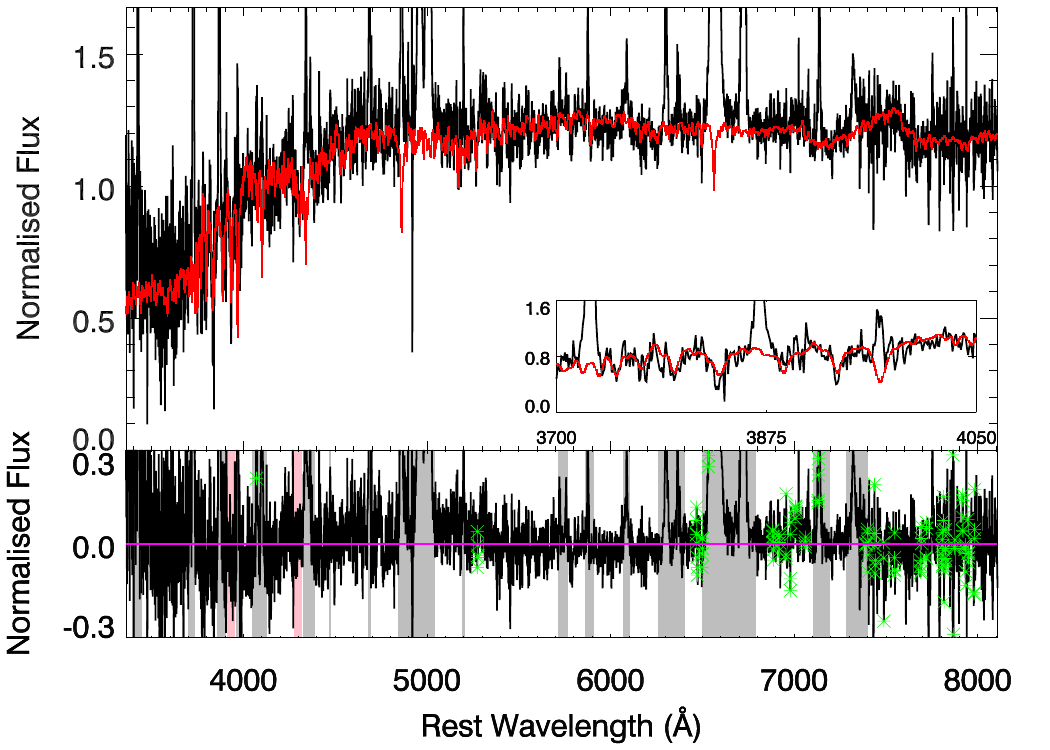}
	\end{subfigure}   
    
    \begin{subfigure}{\columnwidth}
    \includegraphics[width = 1\columnwidth]{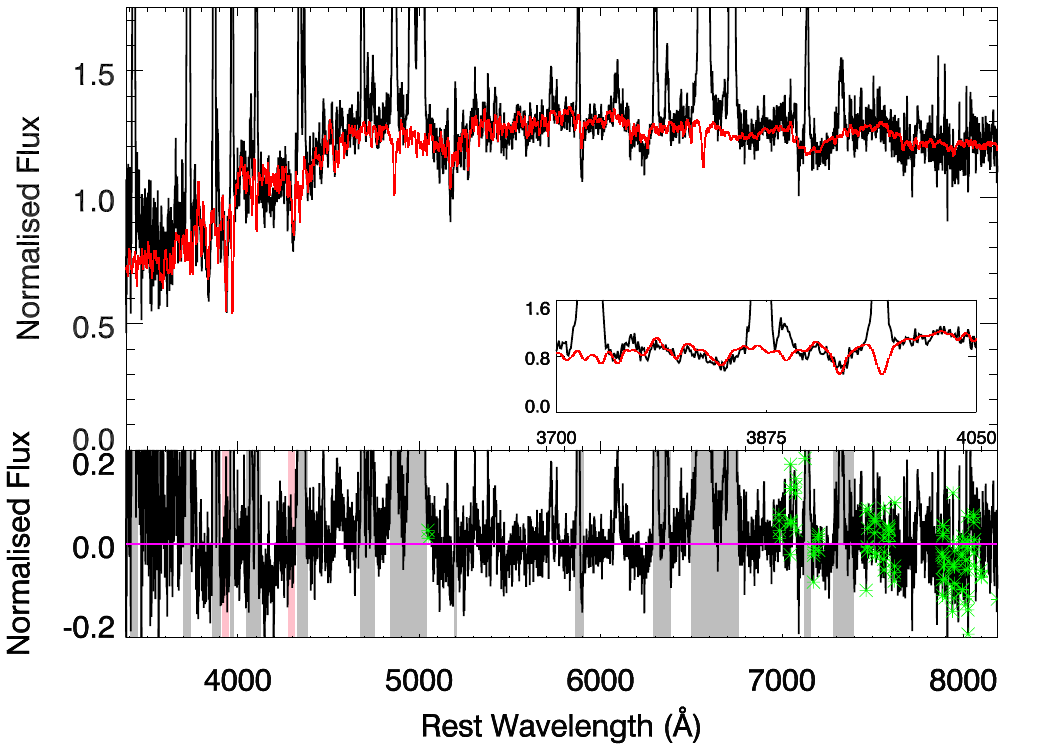}
	\end{subfigure}
    \begin{subfigure}{\columnwidth}
    \includegraphics[width = 1\columnwidth]{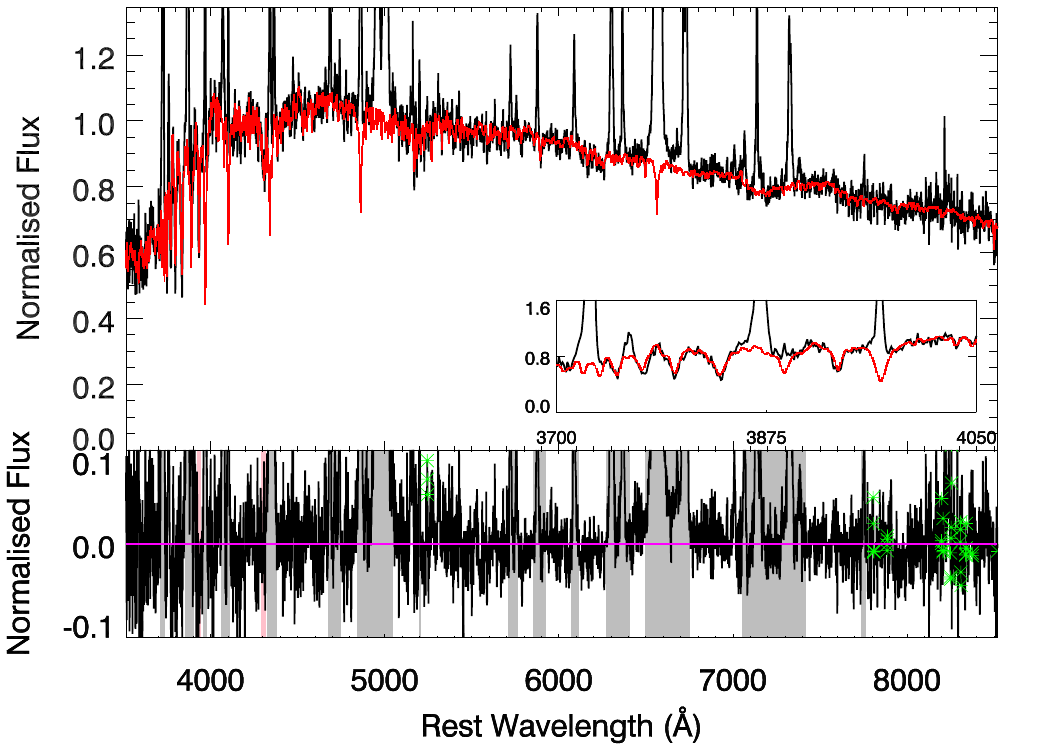}
	\end{subfigure}

    \begin{subfigure}{\columnwidth}
    \includegraphics[width = 1\columnwidth]{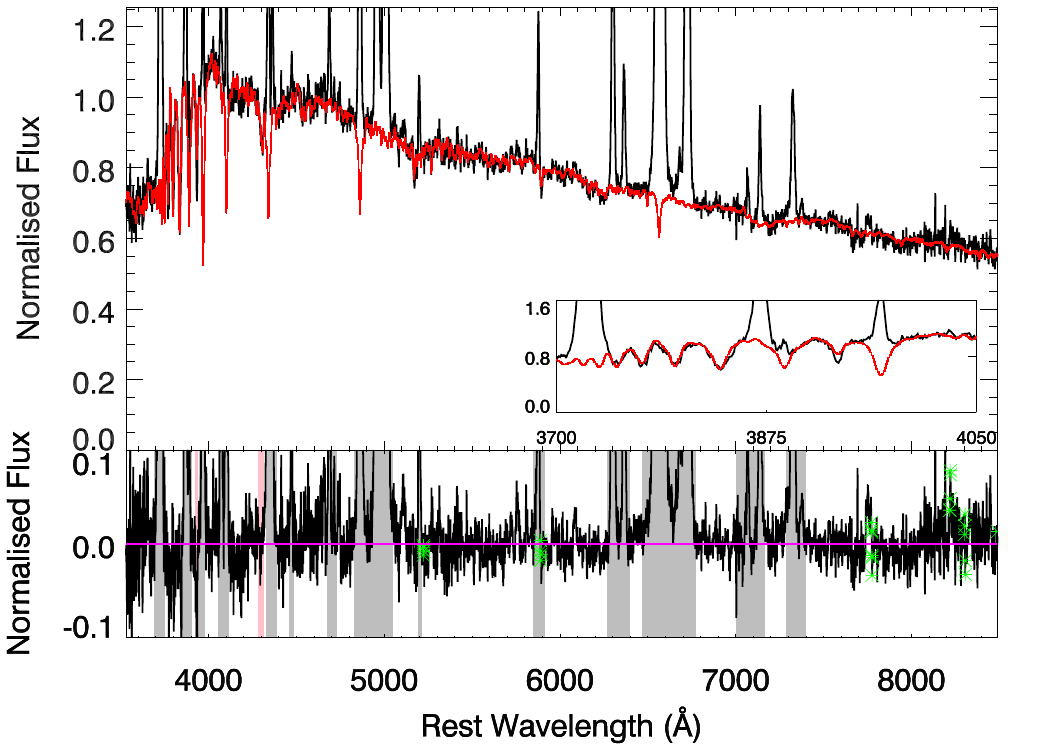}
	\end{subfigure}
    \begin{subfigure}{\columnwidth}
    \includegraphics[width = 1\columnwidth]{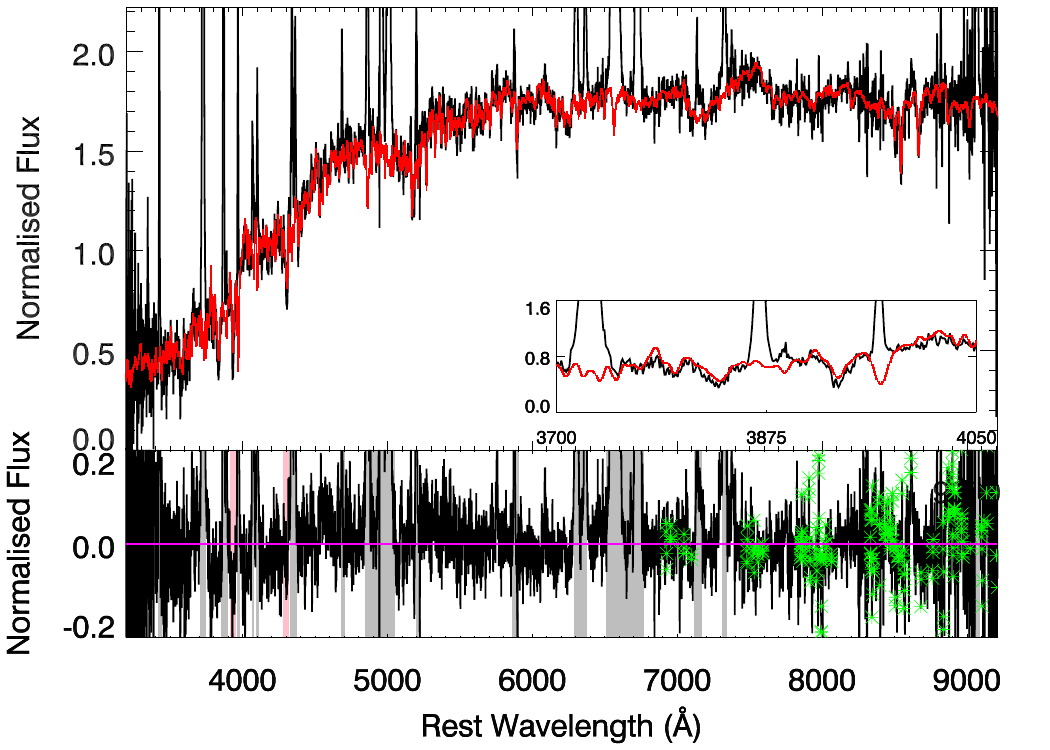}
	\end{subfigure}
    \caption{Stellar population fits of J1347+12 and J1405+40 (top row), J1356-02 and J1356+10 (middle row), and J1430+13 and J1436+13 (bottom row).}
    \label{fig:sl_pg6}
\end{figure*}

\clearpage

\begin{figure*}
    \centering
    \begin{subfigure}{\columnwidth}
    \includegraphics[width = 1\columnwidth]{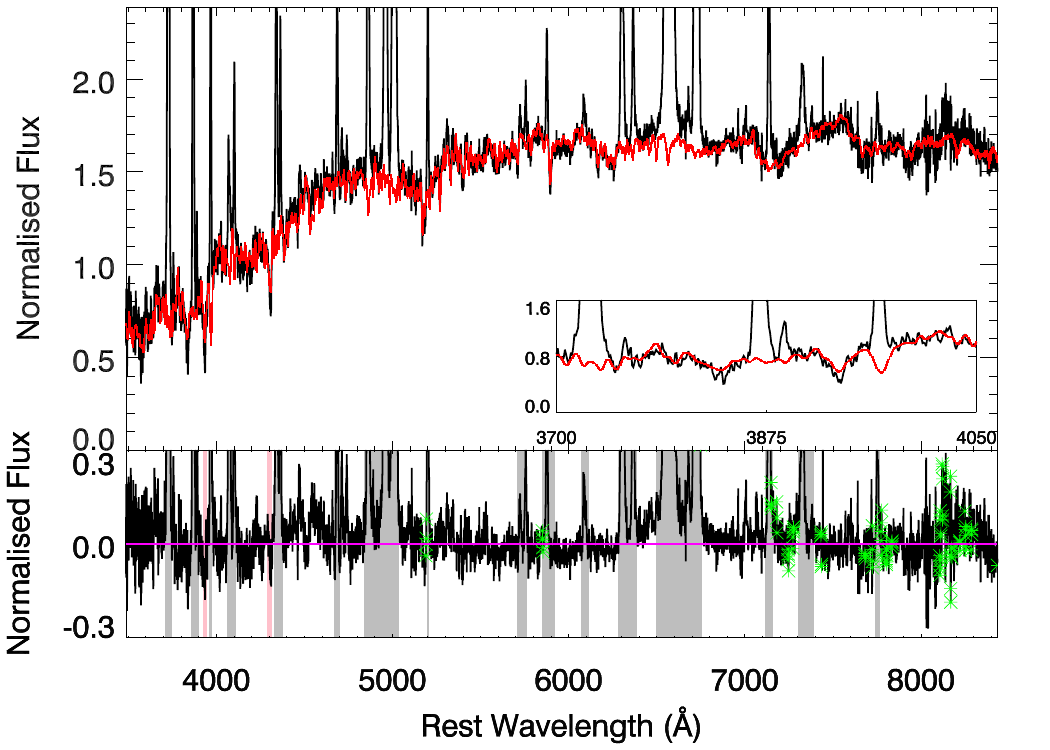}
	\end{subfigure}
    \begin{subfigure}{\columnwidth}
    \includegraphics[width = 1\columnwidth]{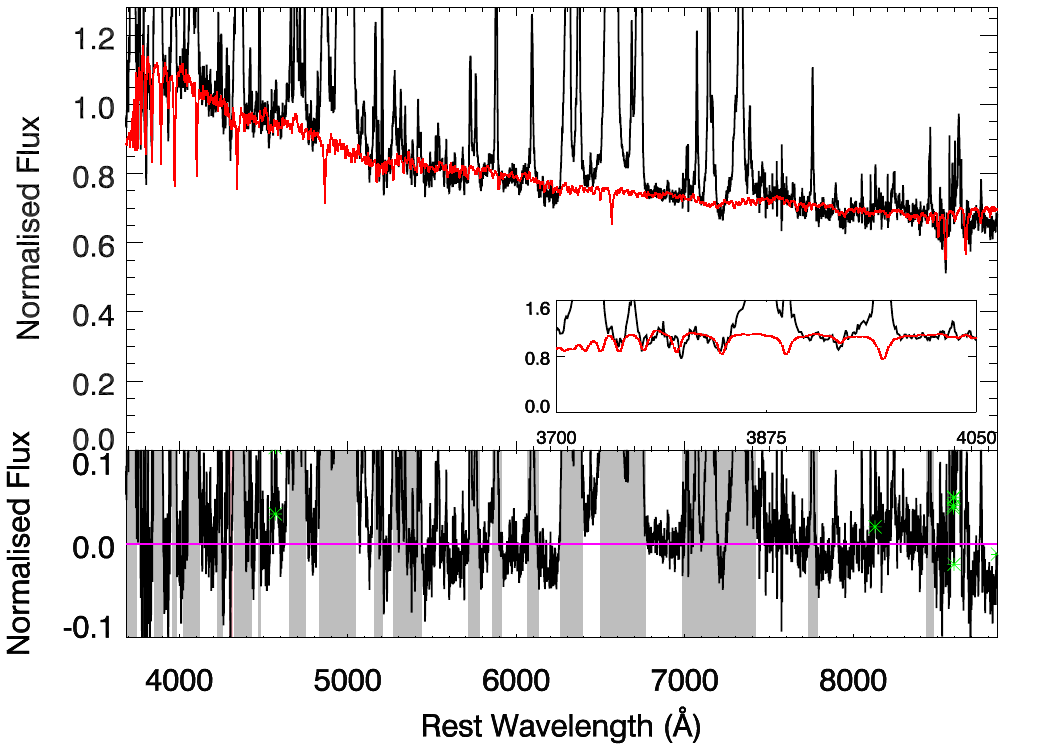}
	\end{subfigure}   
    
    \begin{subfigure}{\columnwidth}
    \includegraphics[width = 1\columnwidth]{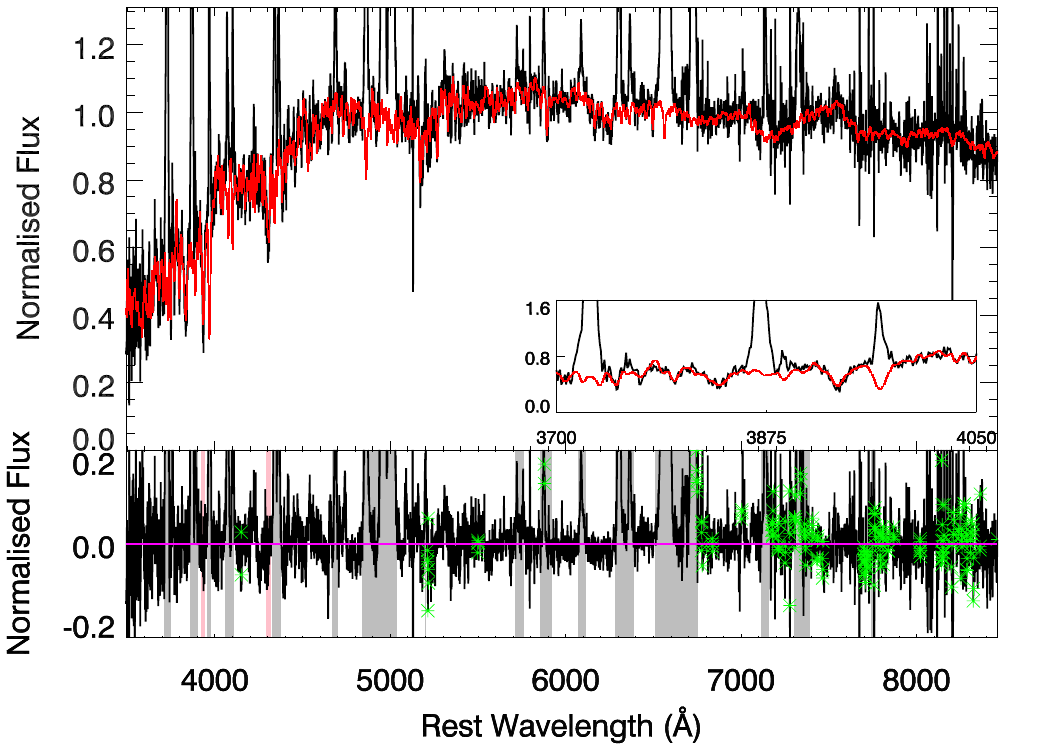}
	\end{subfigure}
    \begin{subfigure}{\columnwidth}
    \includegraphics[width = 1\columnwidth]{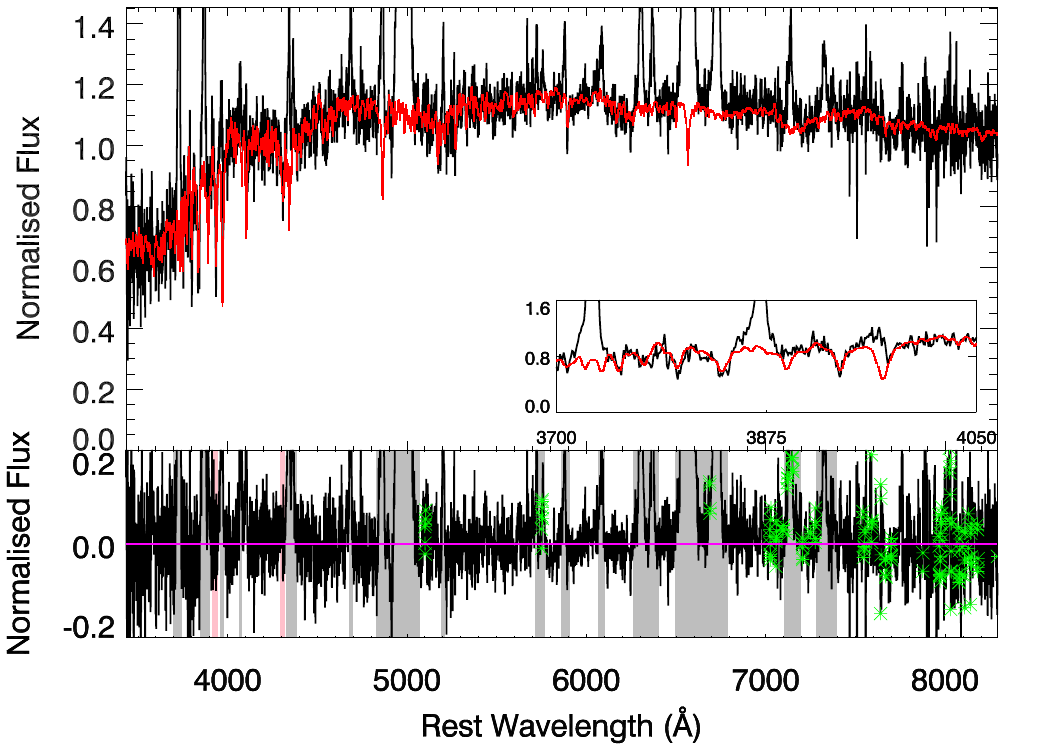}
	\end{subfigure}

    \begin{subfigure}{\columnwidth}
    \includegraphics[width = 1\columnwidth]{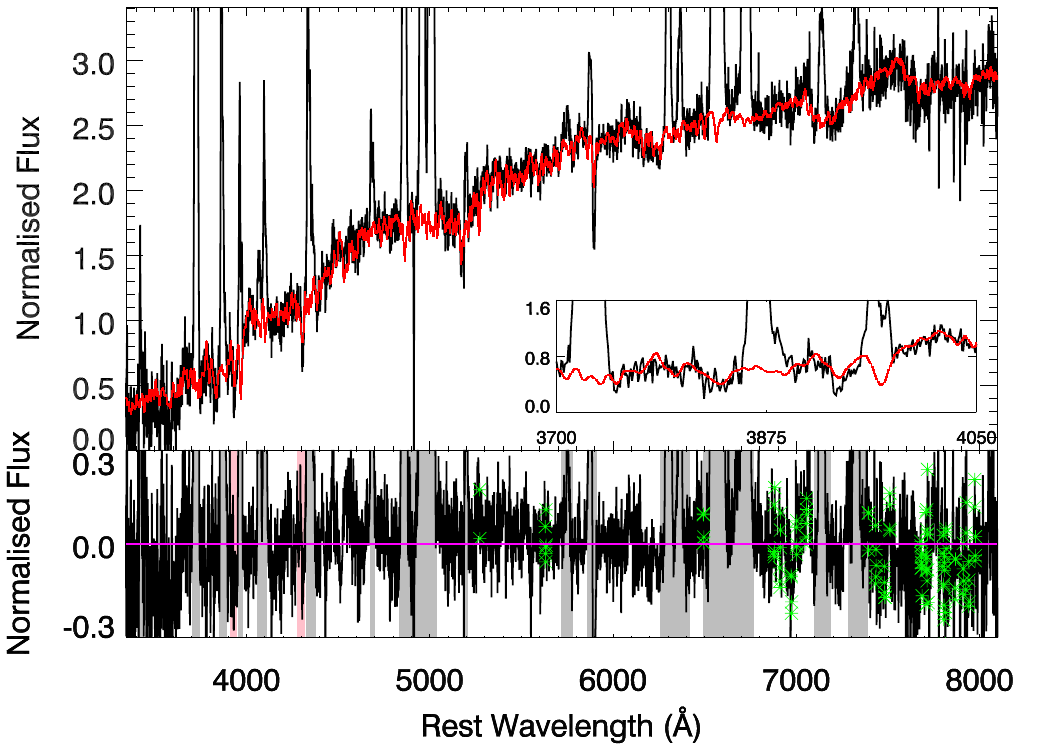}
	\end{subfigure}
    \begin{subfigure}{\columnwidth}
    \includegraphics[width = 1\columnwidth]{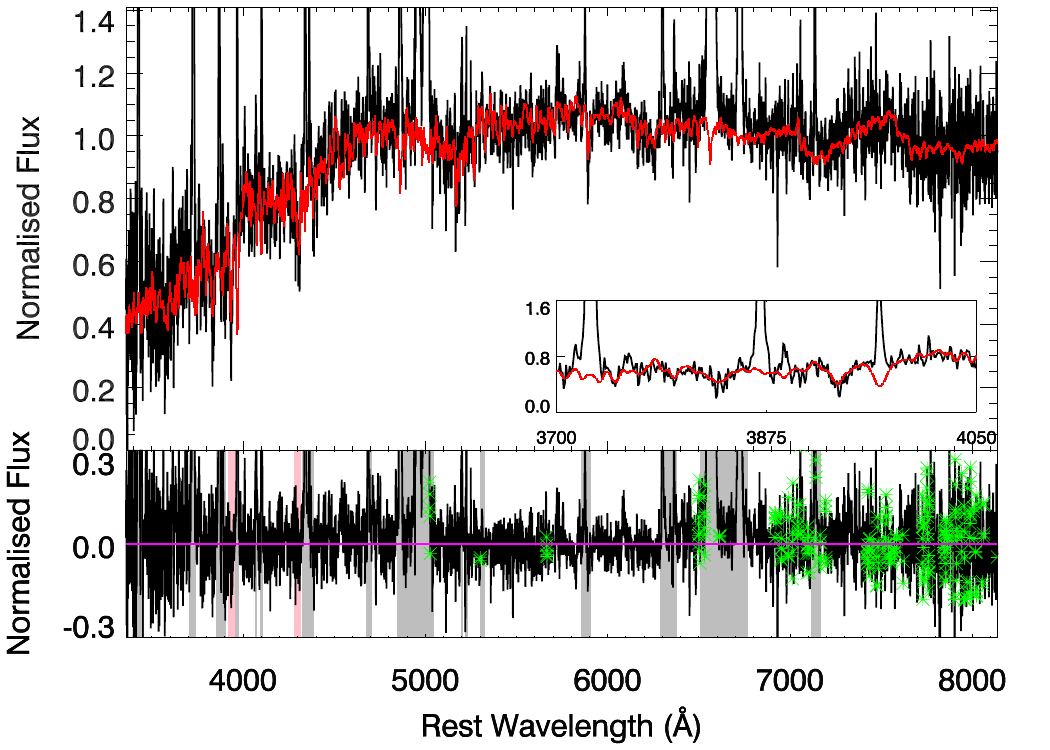}
	\end{subfigure}
    \caption{Stellar population fits of J1437+30 and J1440+53 (top row), J1455+32 and J1509+04 (middle row), and J1517+33 and J1533+35 (bottom row).}
    \label{fig:sl_pg7}
\end{figure*}

\clearpage

\begin{figure*}
    \centering
    \begin{subfigure}{\columnwidth}
    \includegraphics[width = 1\columnwidth]{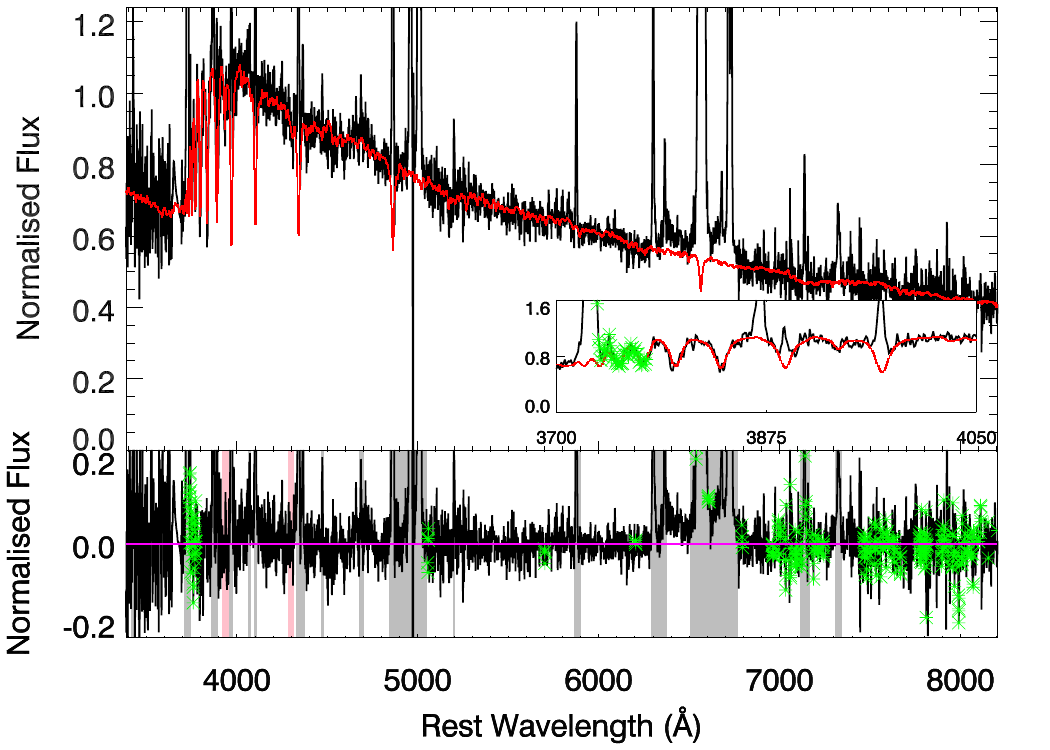}
	\end{subfigure}
    \begin{subfigure}{\columnwidth}
    \includegraphics[width = 1\columnwidth]{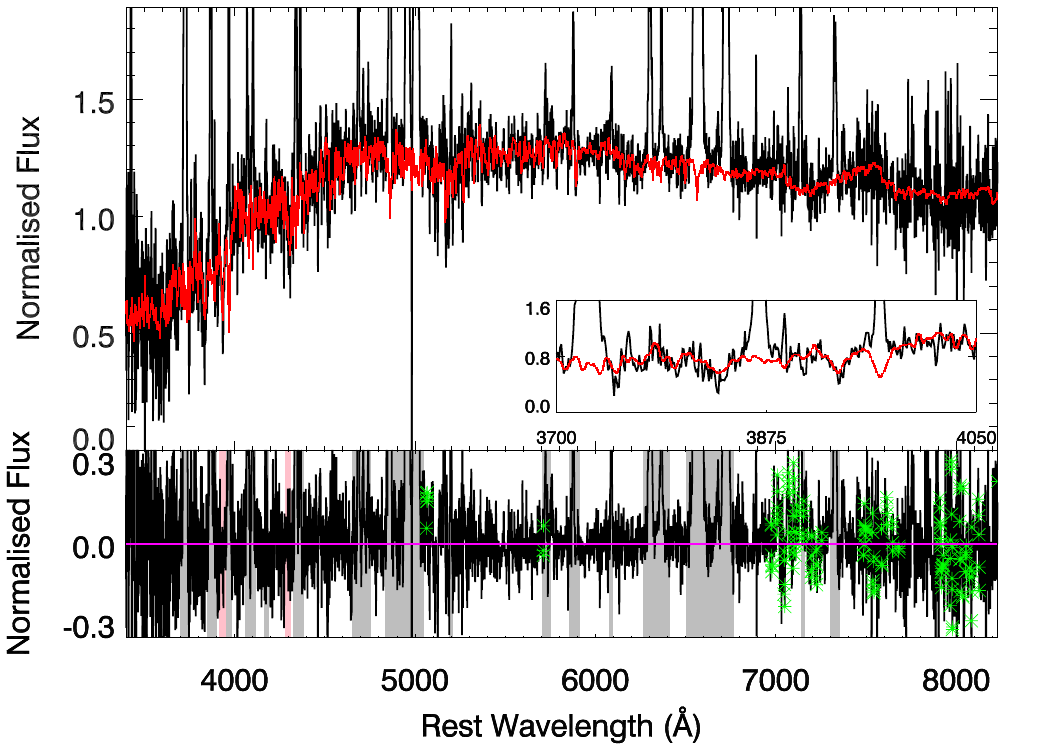}
	\end{subfigure}   
    
    \begin{subfigure}{\columnwidth}
    \includegraphics[width = 1\columnwidth]{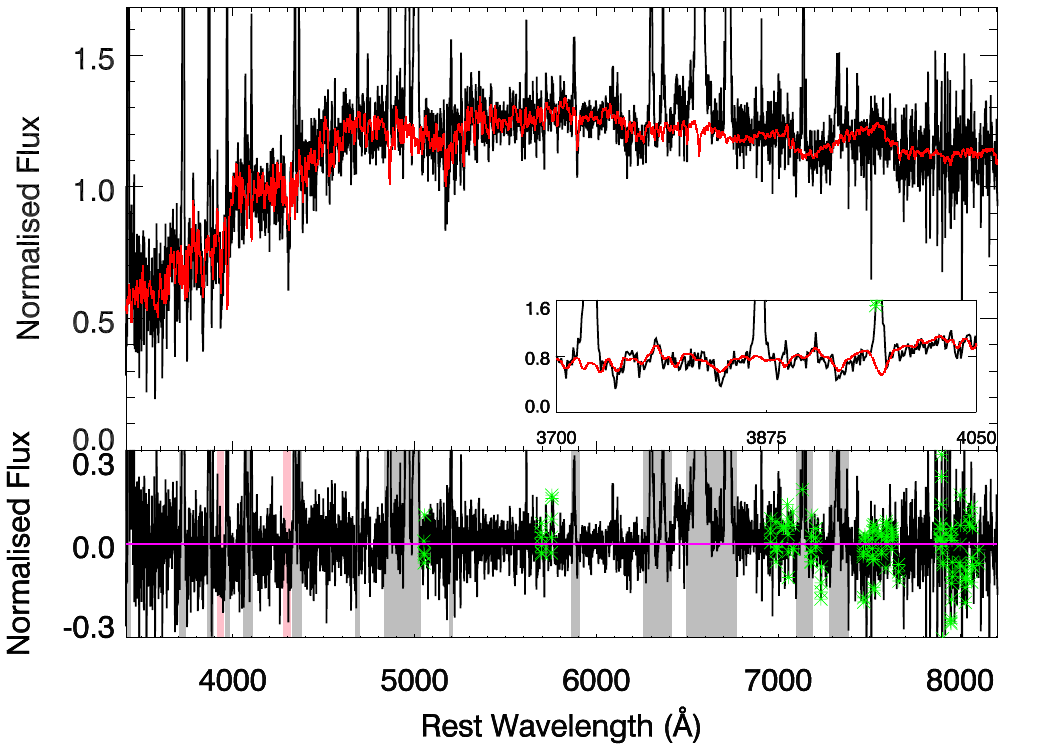}
	\end{subfigure}
    \begin{subfigure}{\columnwidth}
    \includegraphics[width = 1\columnwidth]{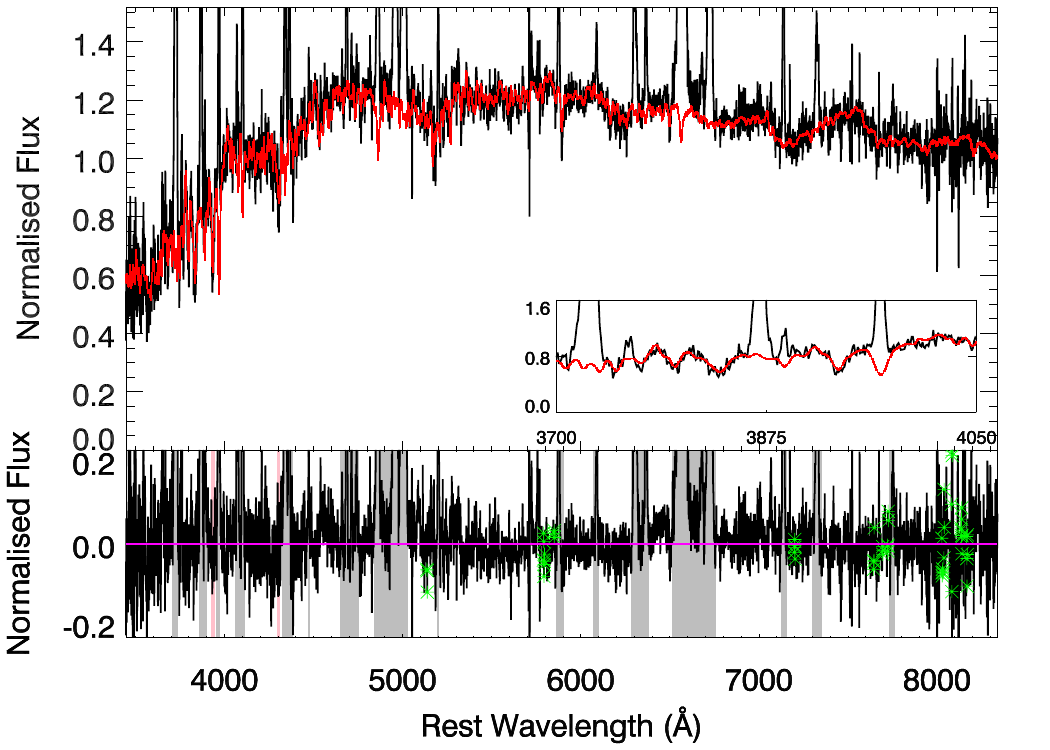}
	\end{subfigure}

    \begin{subfigure}{\columnwidth}
    \includegraphics[width = 1\columnwidth]{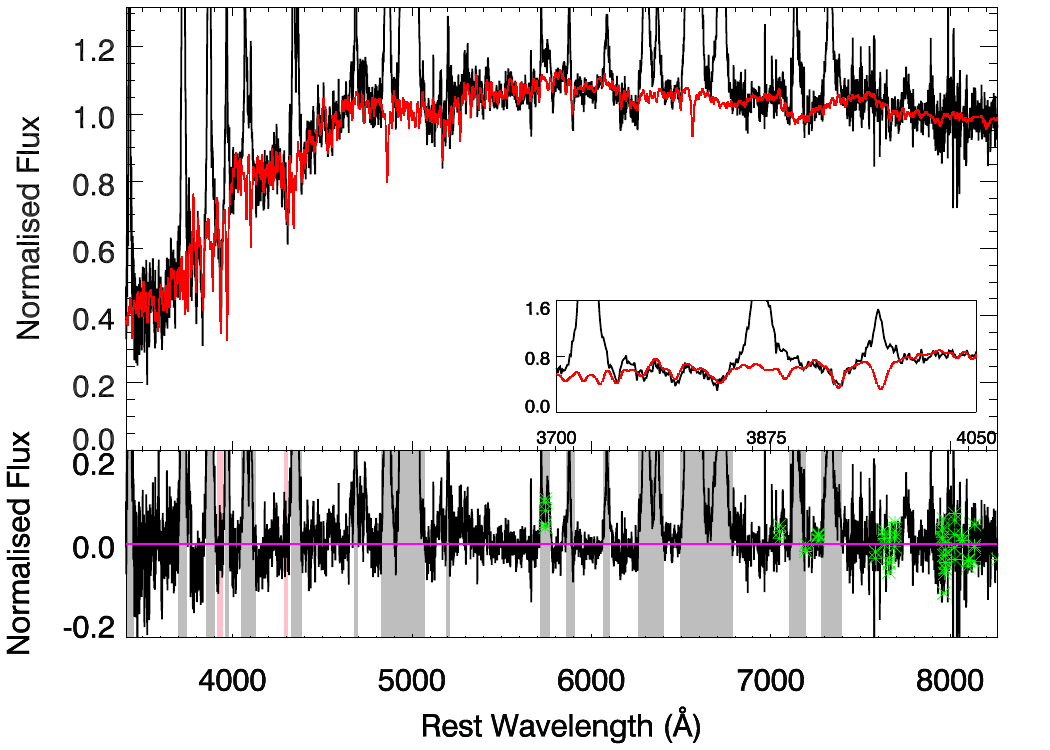}
	\end{subfigure}
    \begin{subfigure}{\columnwidth}
    \includegraphics[width = 1\columnwidth]{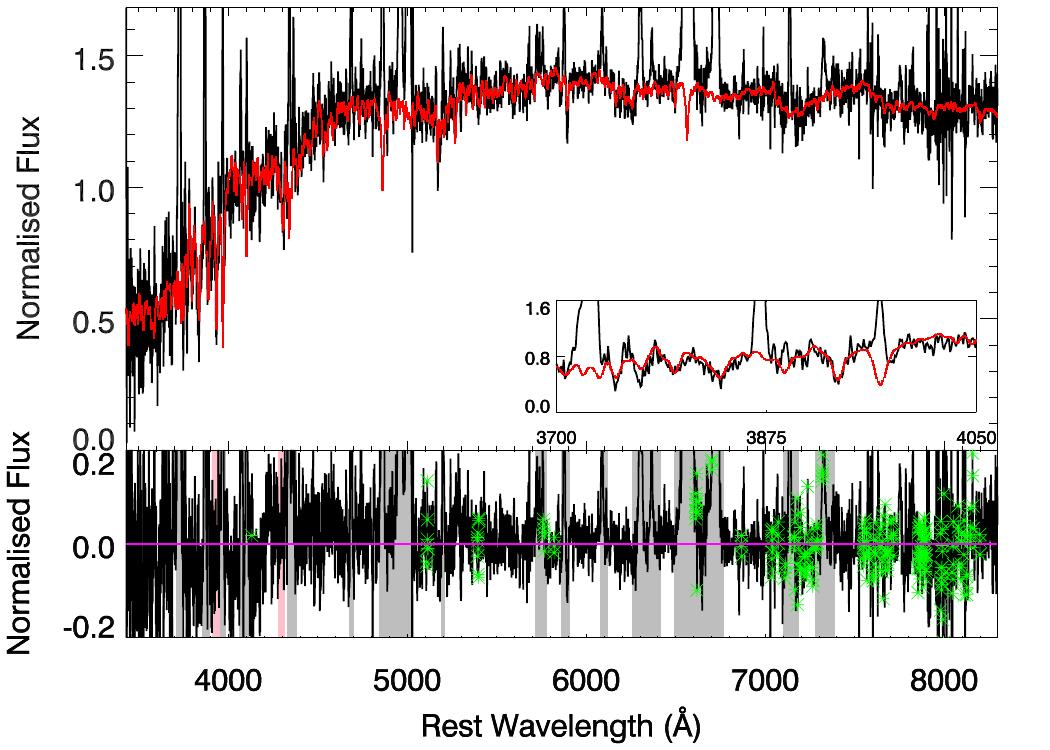}
	\end{subfigure}
    \caption{Stellar population fits of J1548-01 and J1558+35 (top row), J1624+33 and J1653+23 (middle row), and J1713+57 and J2154+11 (bottom row).}
    \label{fig:sl_pg8}
\end{figure*}

\clearpage

\section{Emission line fitting and non-parametric measurements}
\label{app:emlines}

\begin{figure*}
\centering
\begin{subfigure}{0.49\textwidth}
 \includegraphics[width = 0.49\linewidth]{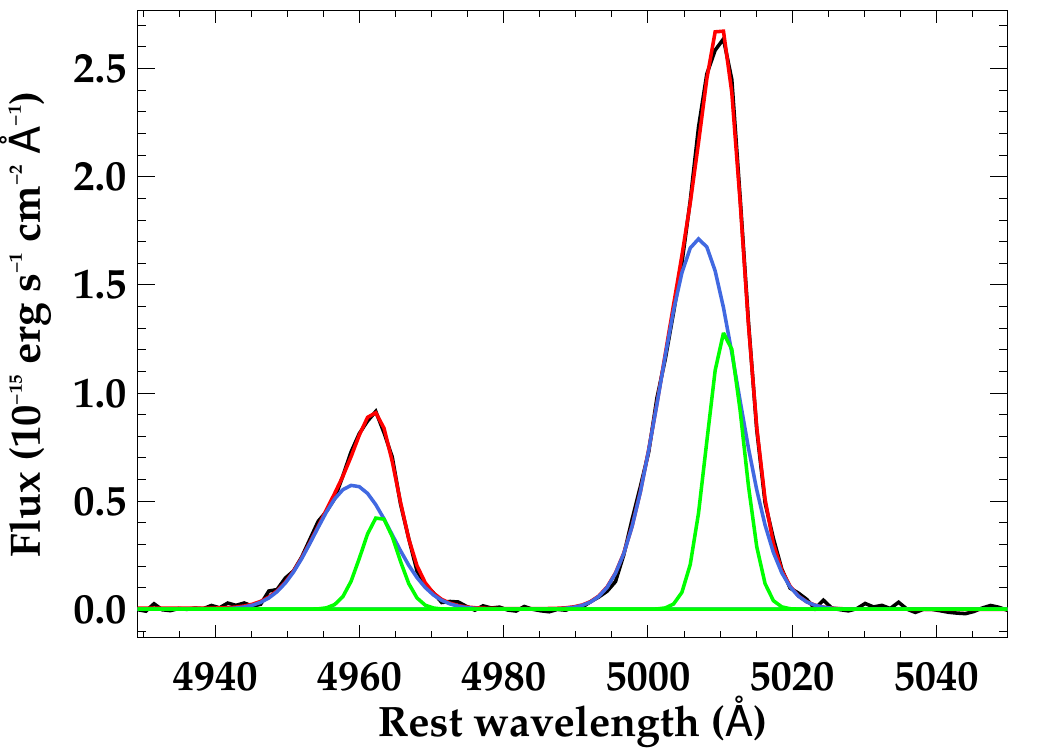}
 \includegraphics[width = 0.49\linewidth]{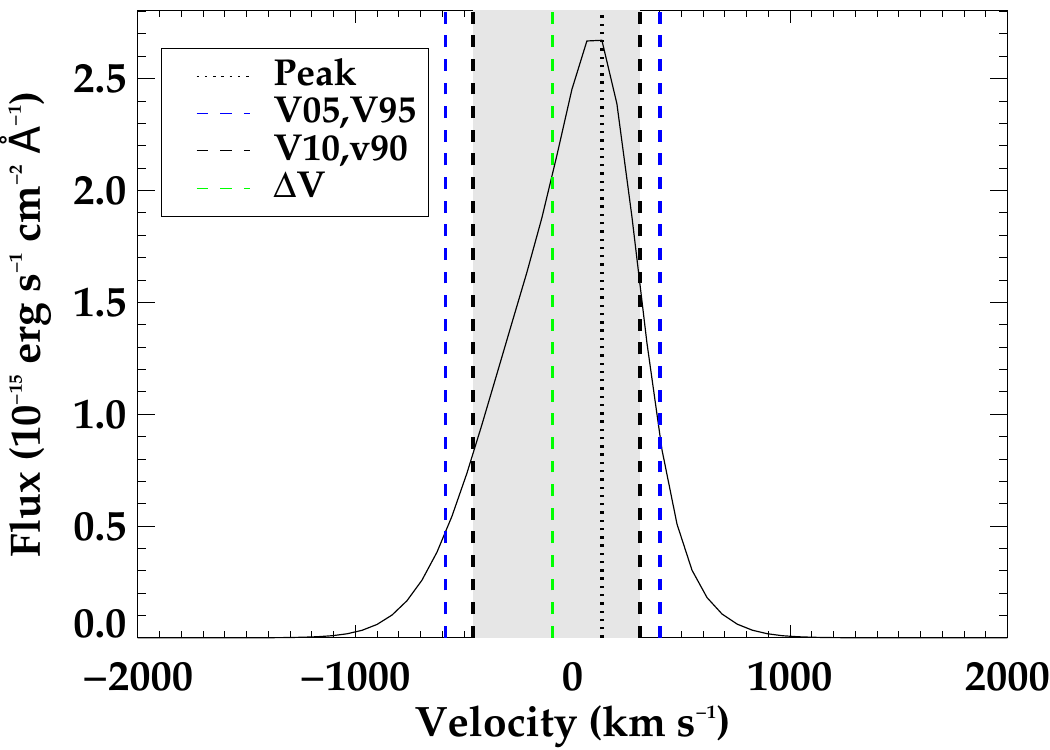}
\caption{J0052-01}
\end{subfigure}
\begin{subfigure}{0.49\textwidth}
 \includegraphics[width = 0.49\linewidth]{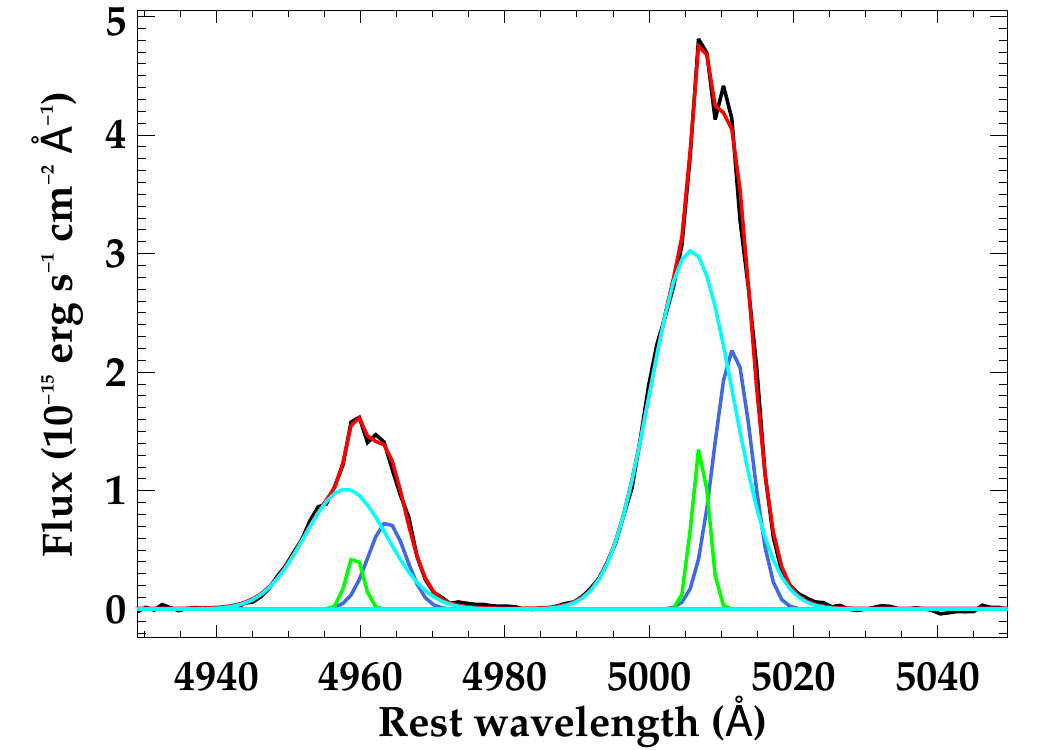}
 \includegraphics[width = 0.49\linewidth]{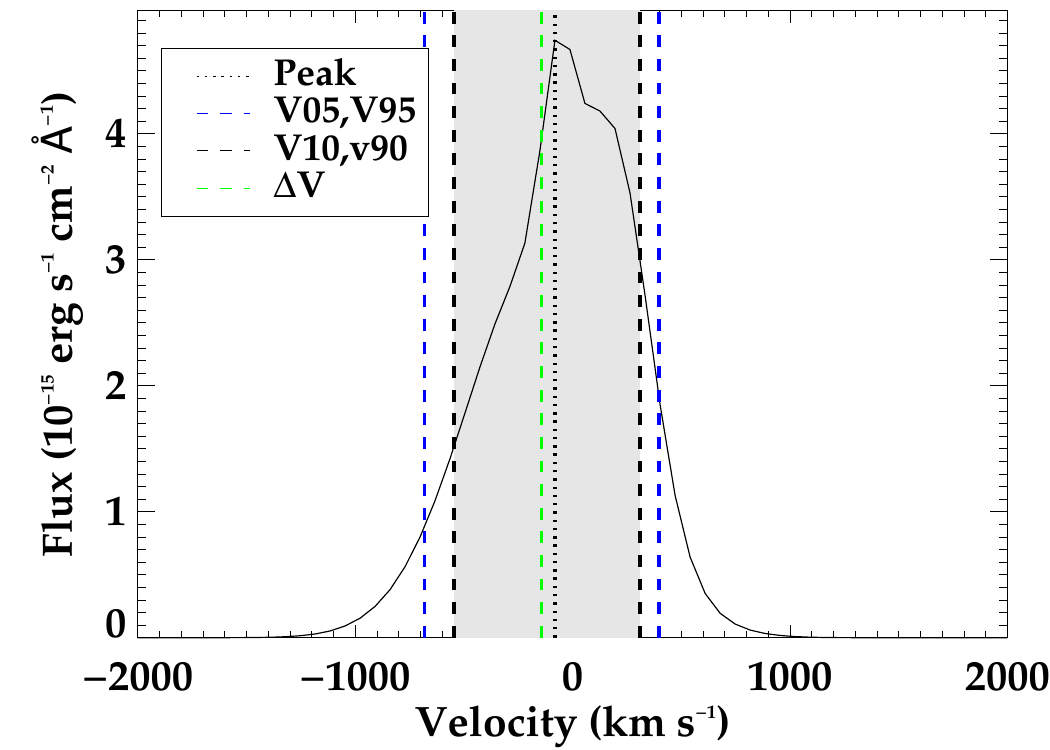}
\caption{J0232-08}
\end{subfigure}

\begin{subfigure}{0.49\textwidth}
 \includegraphics[width = 0.49\linewidth]{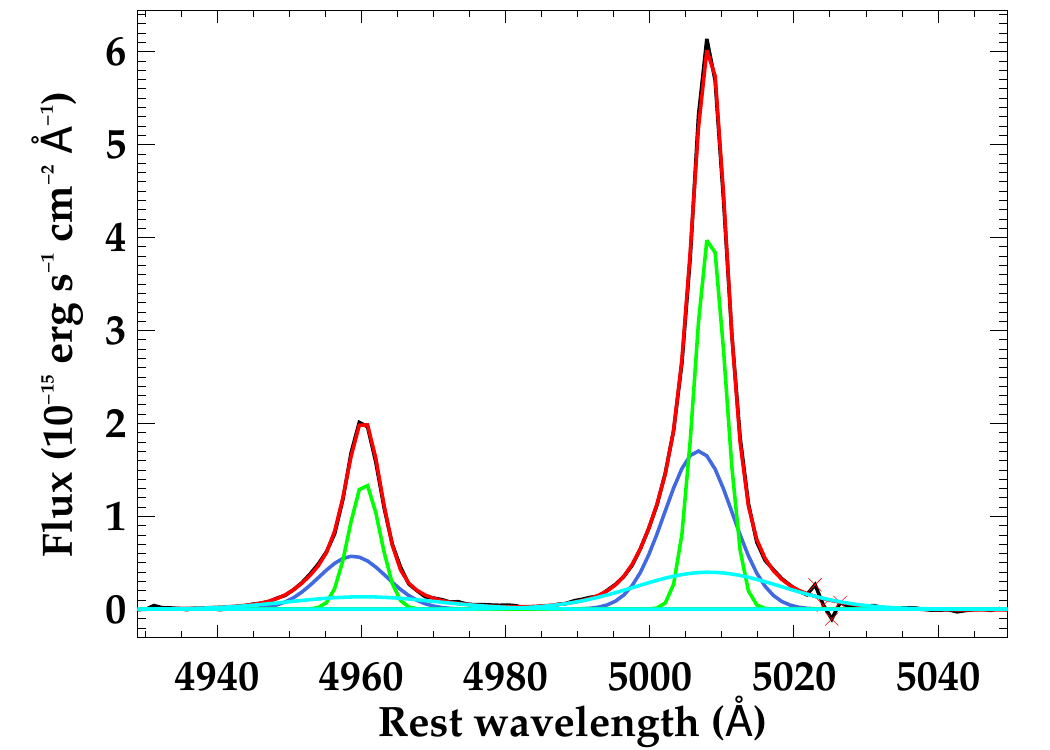}
 \includegraphics[width = 0.49\linewidth]{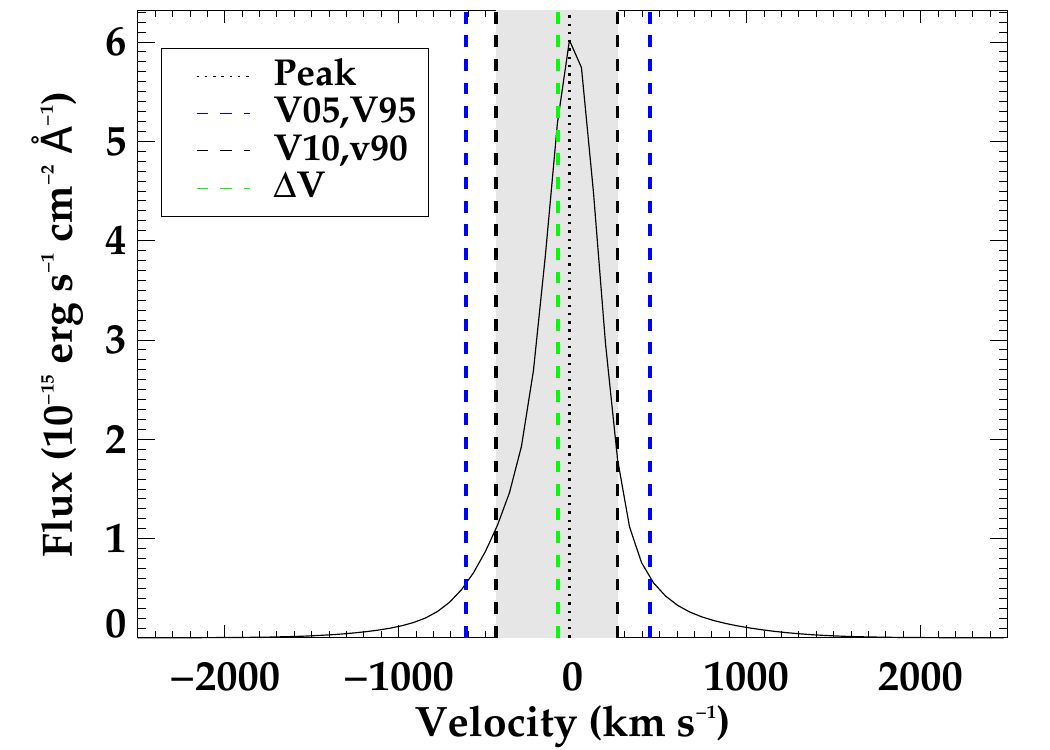}
\caption{J0731+39}
\end{subfigure}
\begin{subfigure}{0.49\textwidth}
 \includegraphics[width = 0.49\linewidth]{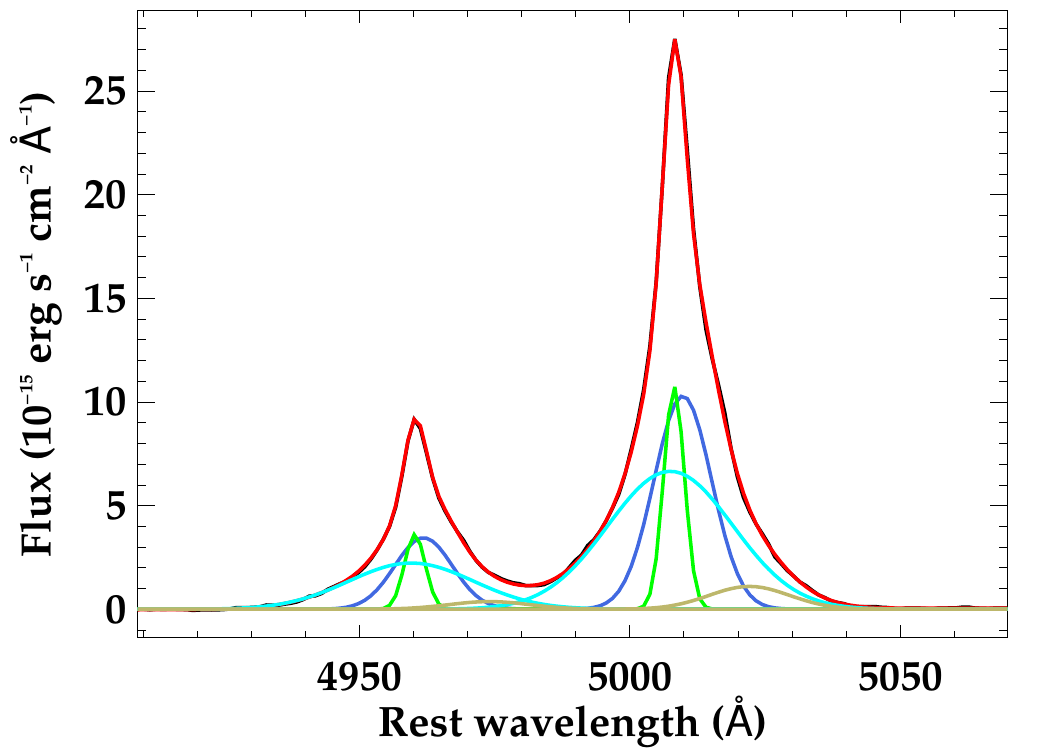}
 \includegraphics[width = 0.49\linewidth]{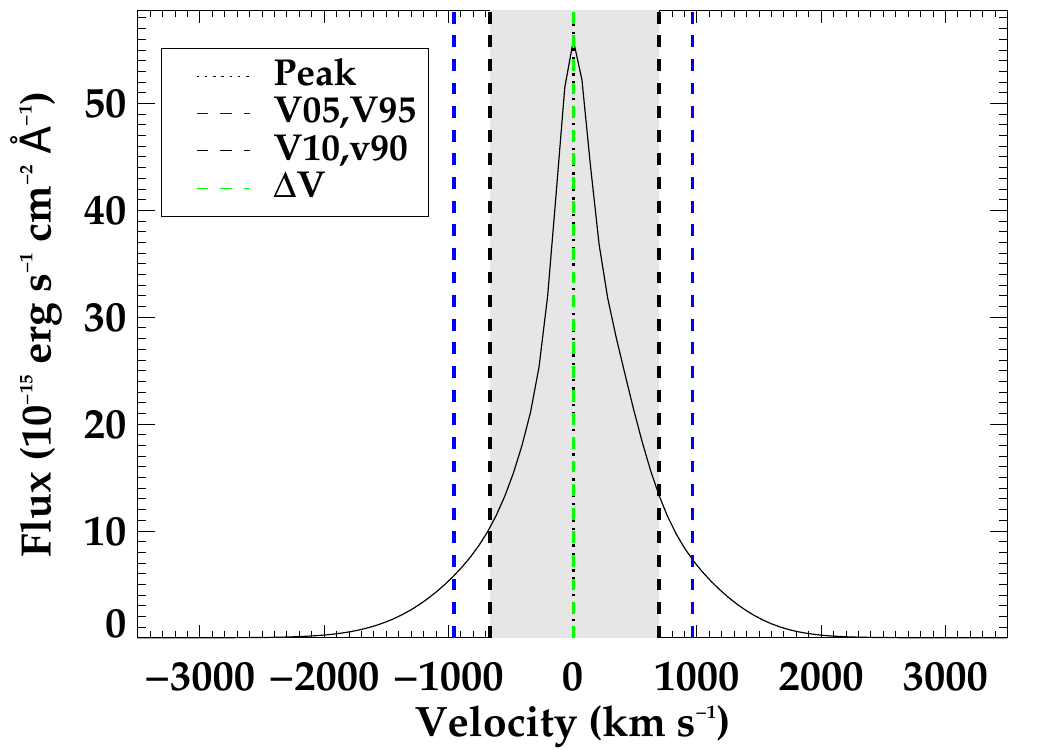}
\caption{J0759+50}
\end{subfigure}

\begin{subfigure}{0.49\textwidth}
 \includegraphics[width = 0.49\linewidth]{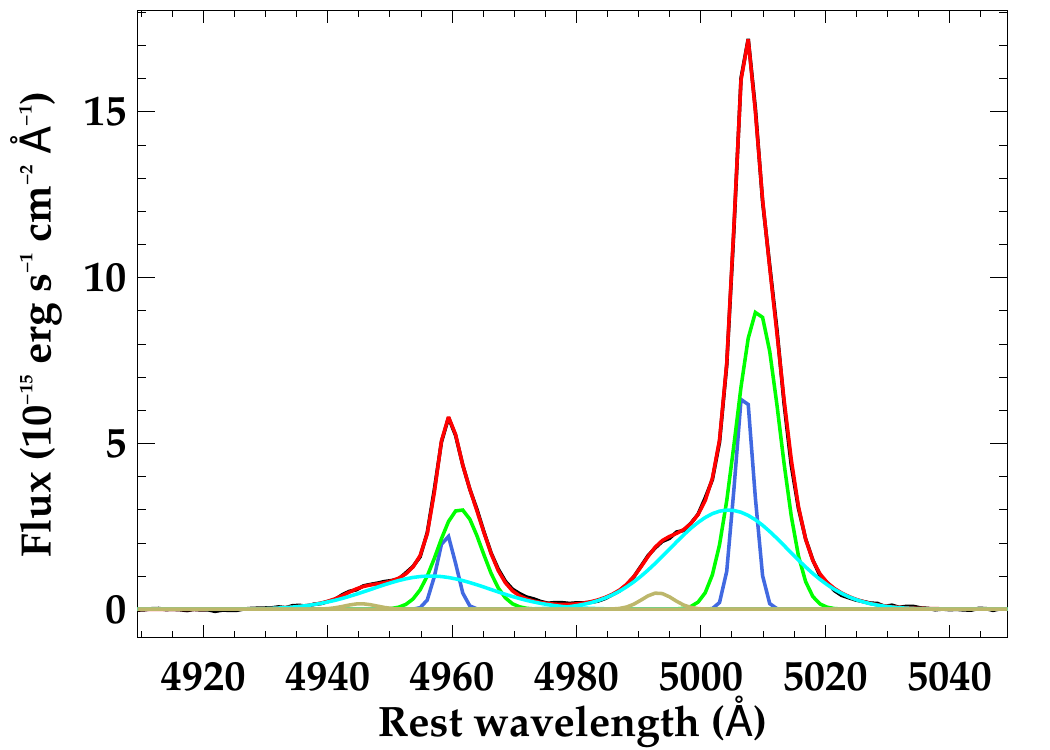}
 \includegraphics[width = 0.49\linewidth]{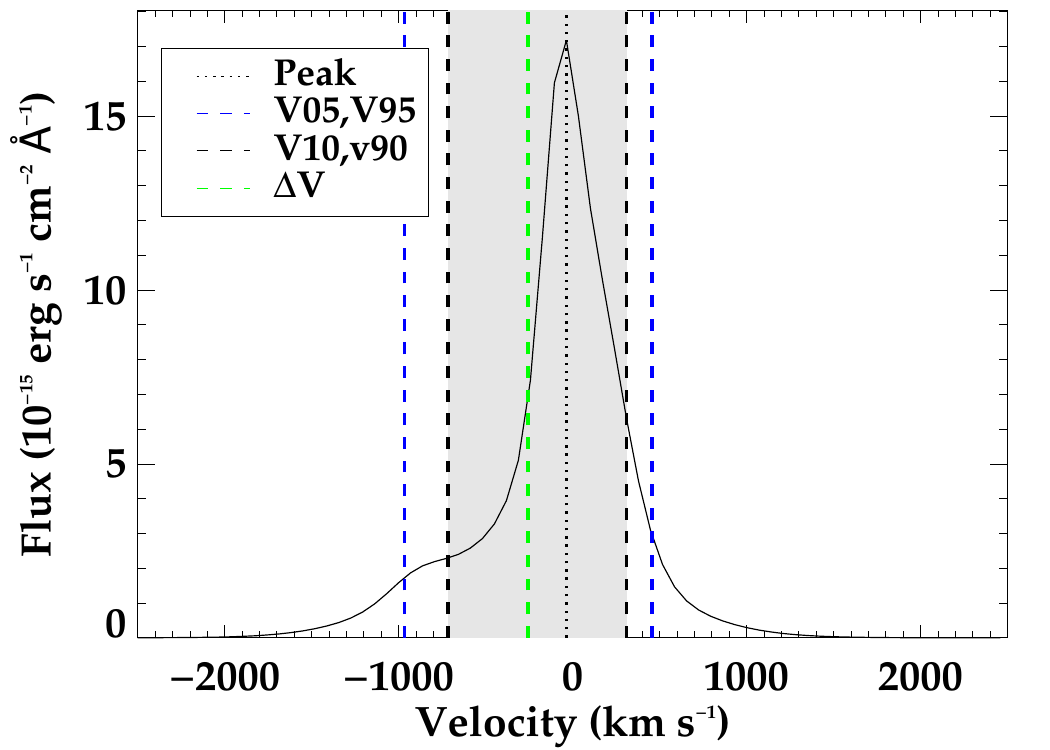}
\caption{J0802+25}
\end{subfigure}
\begin{subfigure}{0.49\textwidth}
 \includegraphics[width = 0.49\linewidth]{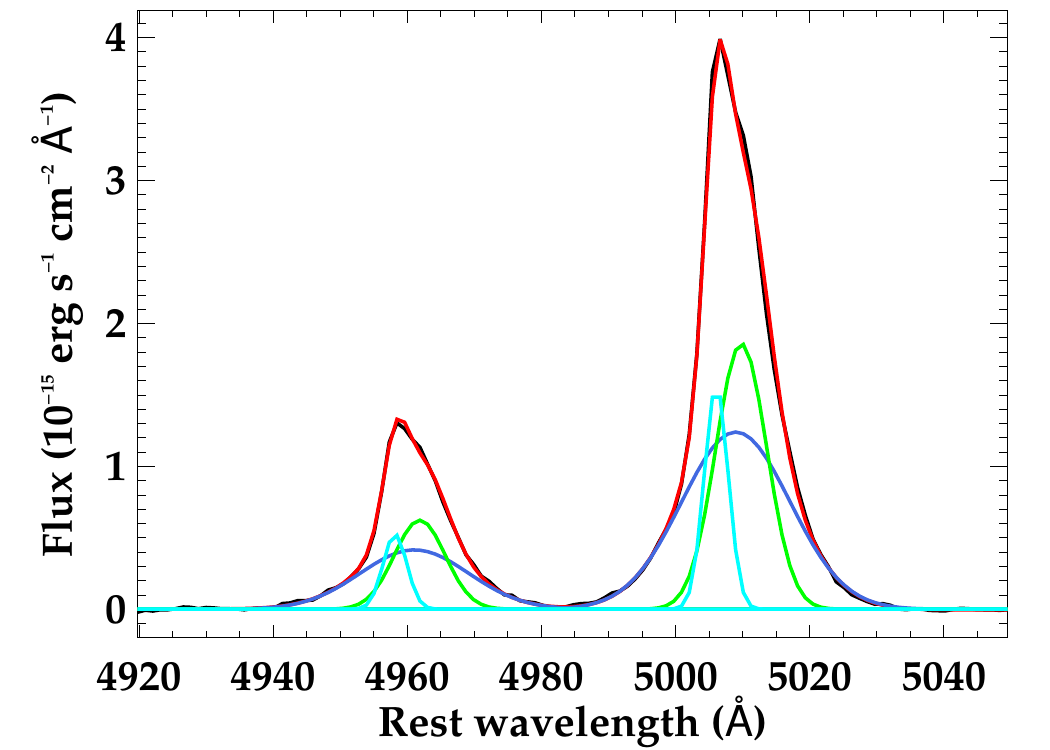}
 \includegraphics[width = 0.49\linewidth]{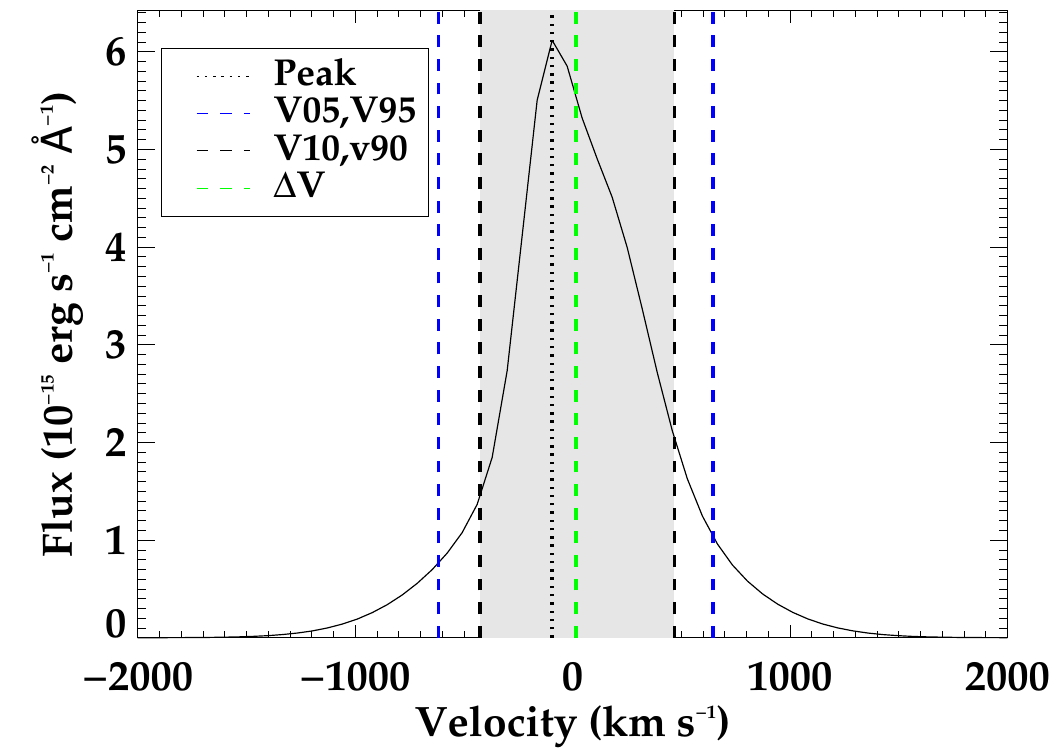}
\caption{J0802+46}
\end{subfigure}

\begin{subfigure}{0.49\textwidth}
 \includegraphics[width = 0.49\linewidth]{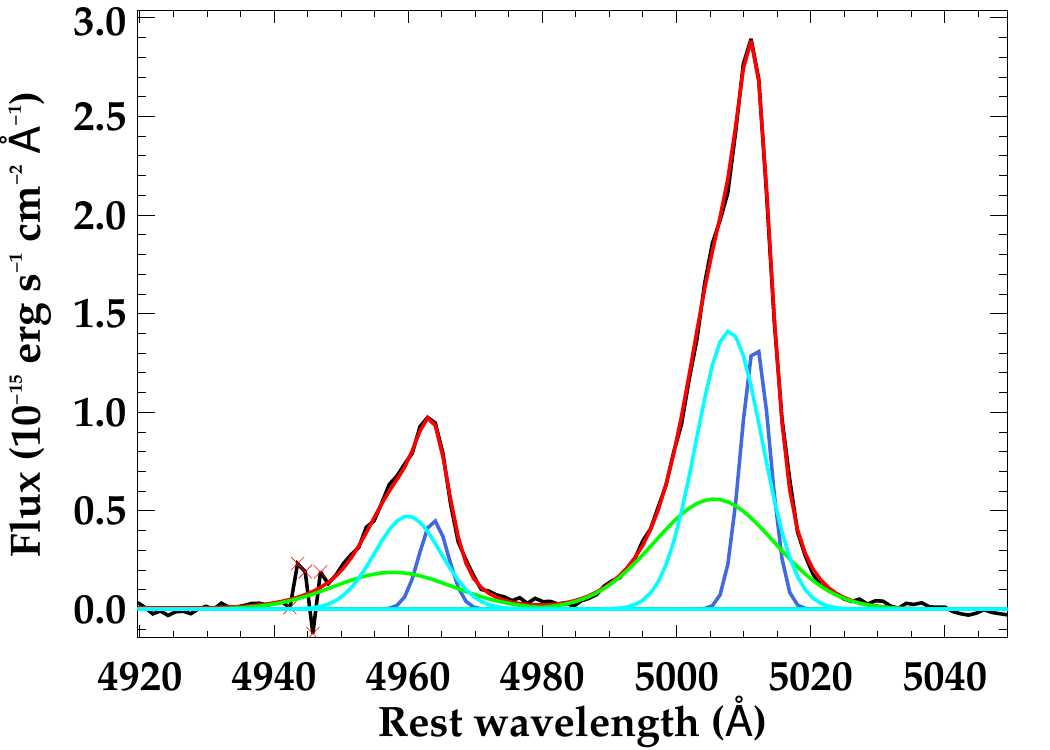}
 \includegraphics[width = 0.49\linewidth]{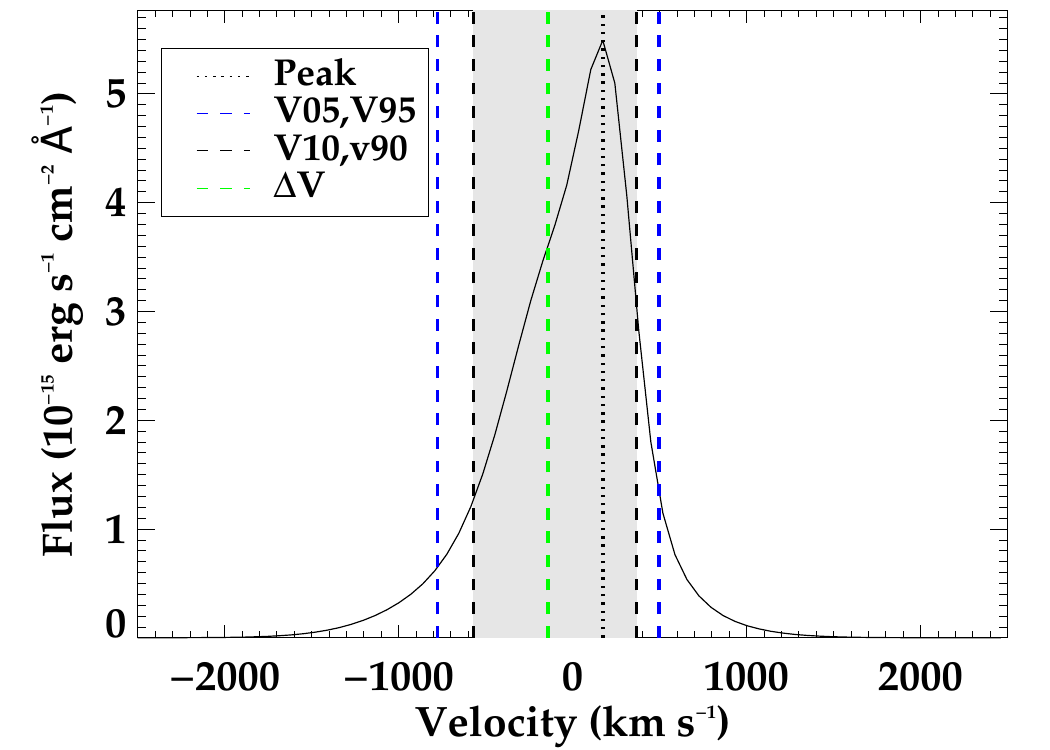}
\caption{J0805+28}
\end{subfigure}
\begin{subfigure}{0.49\textwidth}
 \includegraphics[width = 0.49\linewidth]{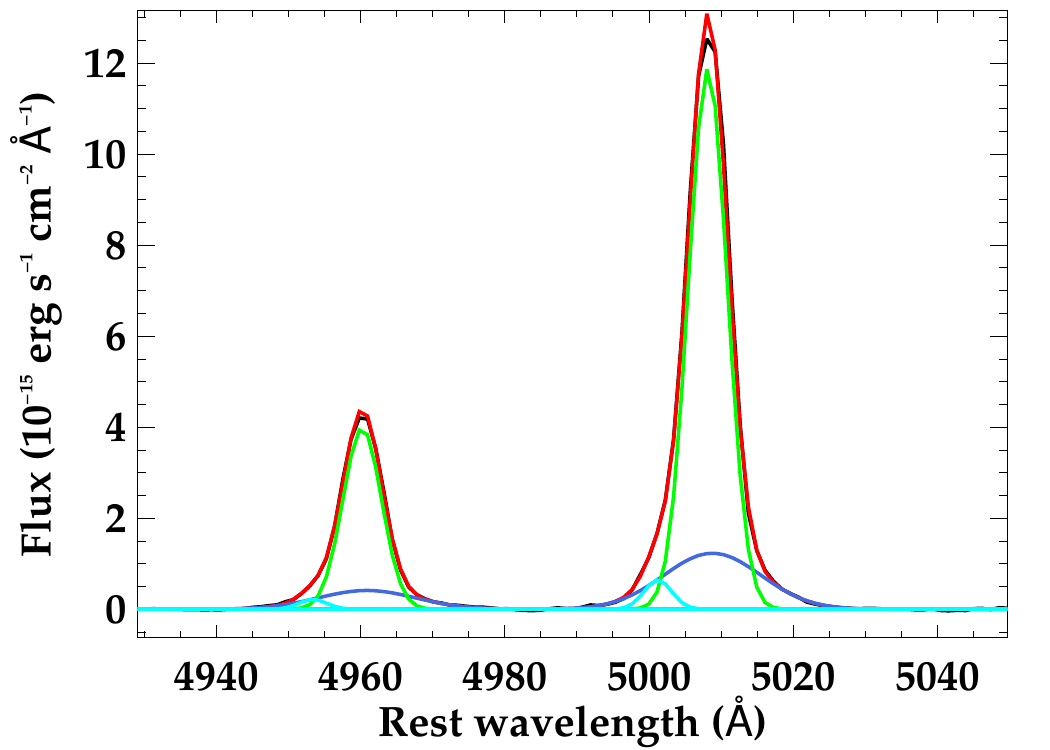}
 \includegraphics[width = 0.49\linewidth]{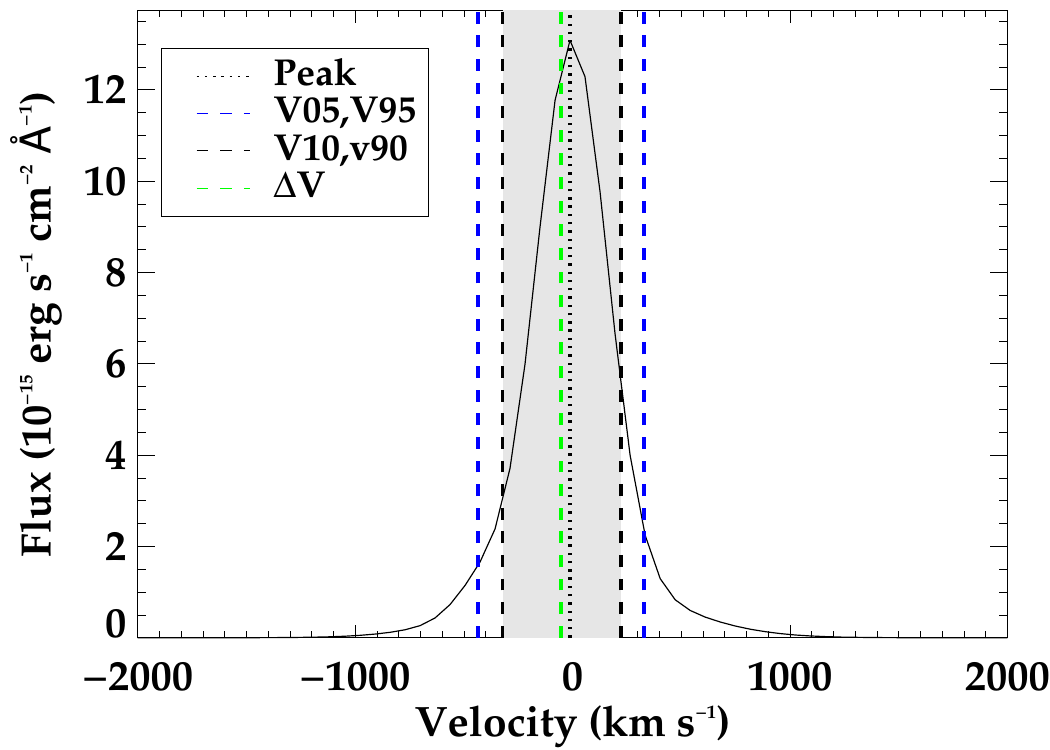}
\caption{J0818+36}
\end{subfigure}

\begin{subfigure}{0.49\textwidth}
 \includegraphics[width = 0.49\linewidth]{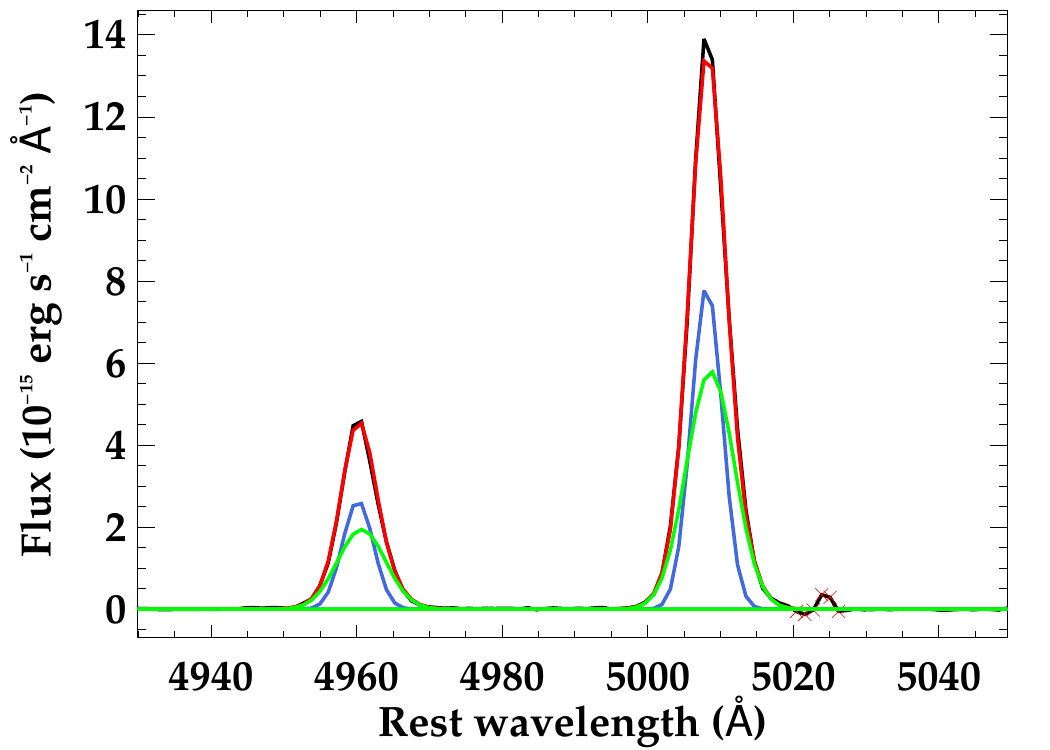}
 \includegraphics[width = 0.49\linewidth]{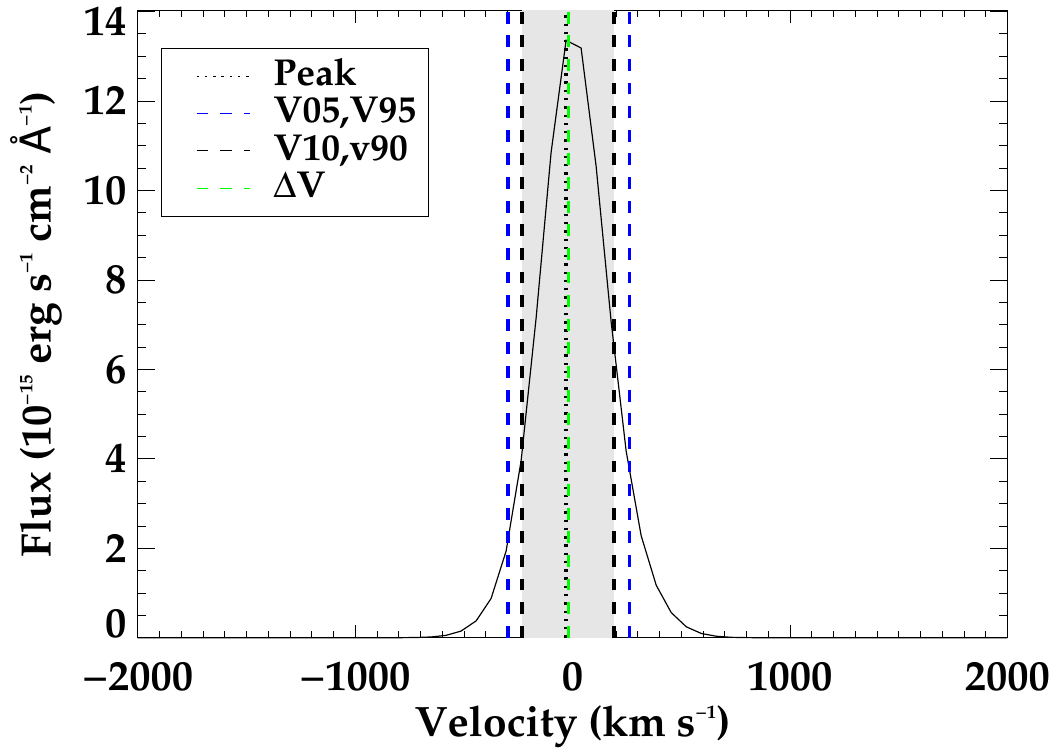}
\caption{J0841+01}
\end{subfigure}
\begin{subfigure}{0.49\textwidth}
 \includegraphics[width = 0.49\linewidth]{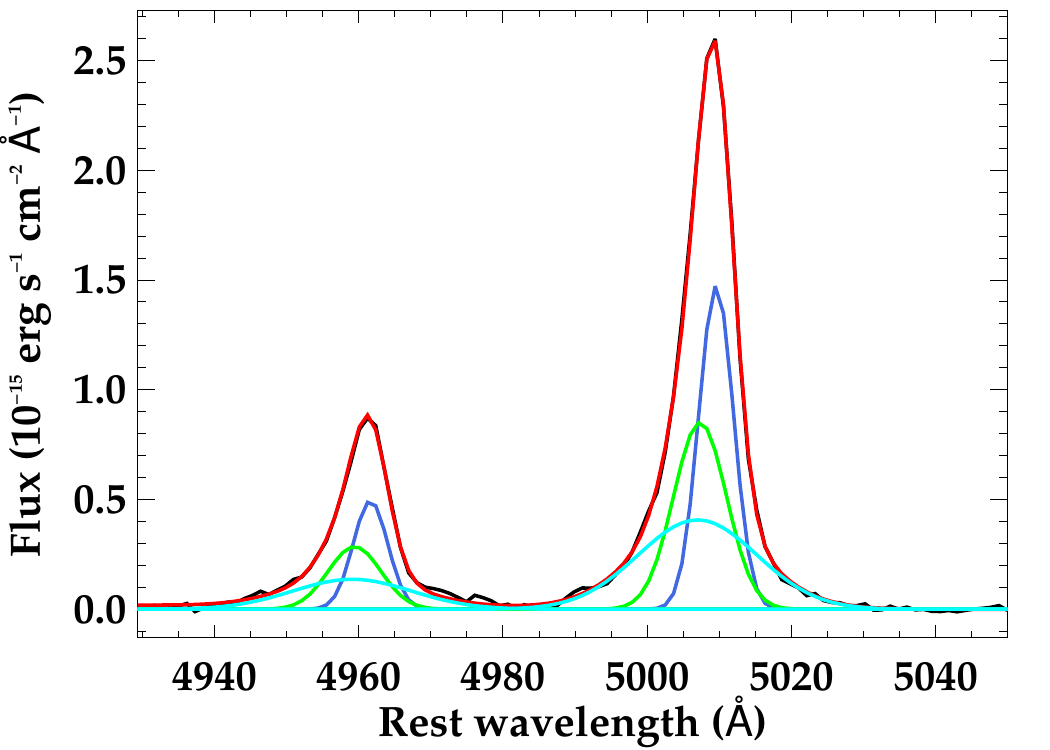}
 \includegraphics[width = 0.49\linewidth]{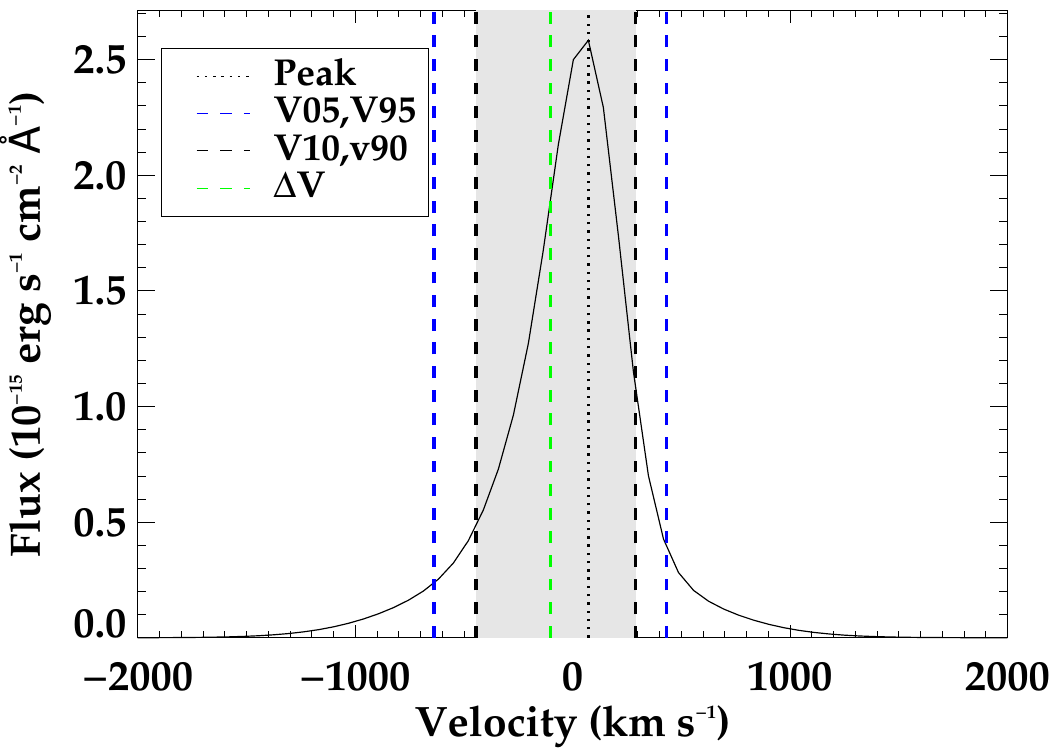}
\caption{J0858+31}
\end{subfigure}

\begin{subfigure}{0.49\textwidth}
 \includegraphics[width = 0.49\linewidth]{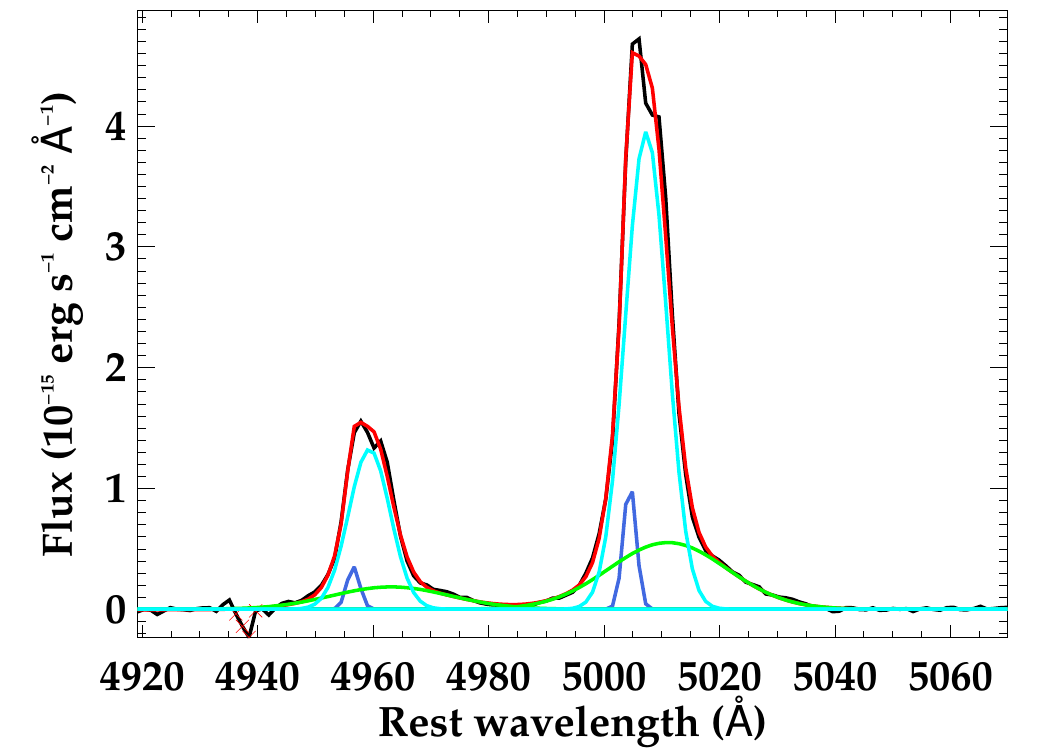}
 \includegraphics[width = 0.49\linewidth]{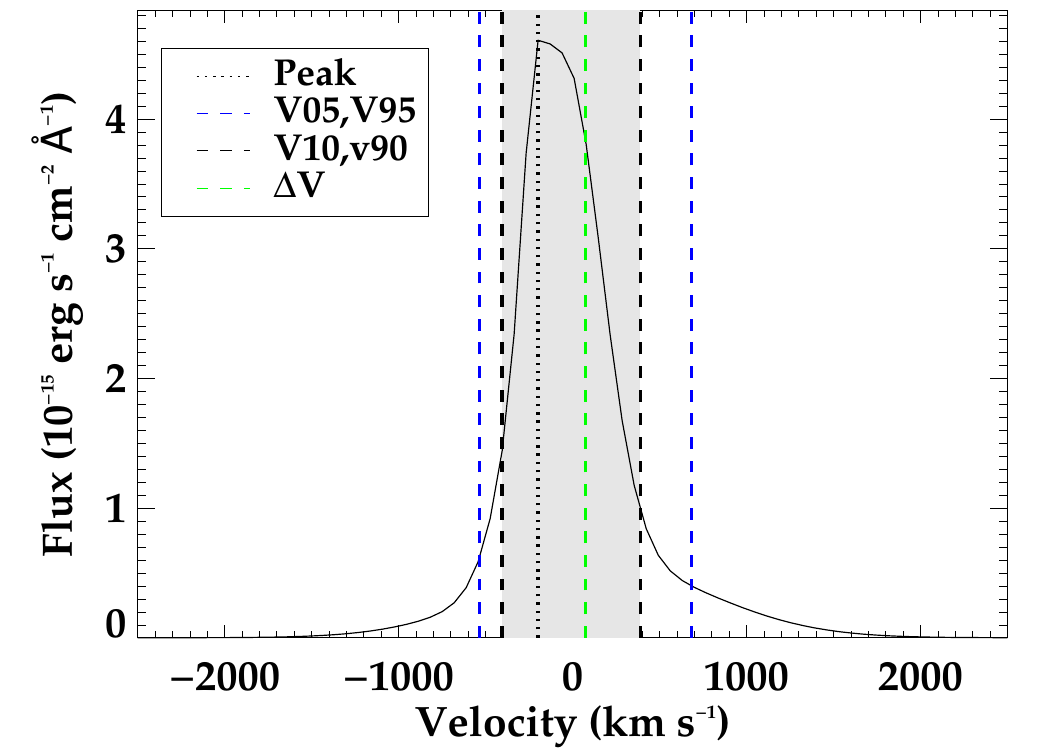}
\caption{J0915+30}
\end{subfigure}
\begin{subfigure}{0.49\textwidth}
 \includegraphics[width = 0.49\linewidth]{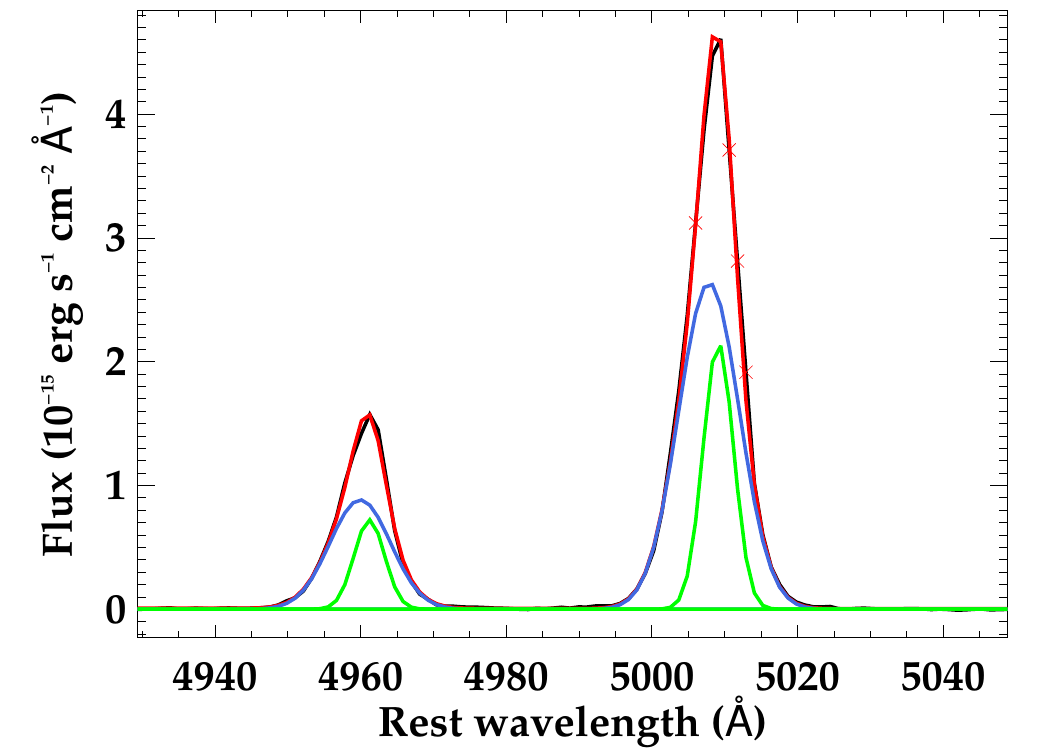}
 \includegraphics[width = 0.49\linewidth]{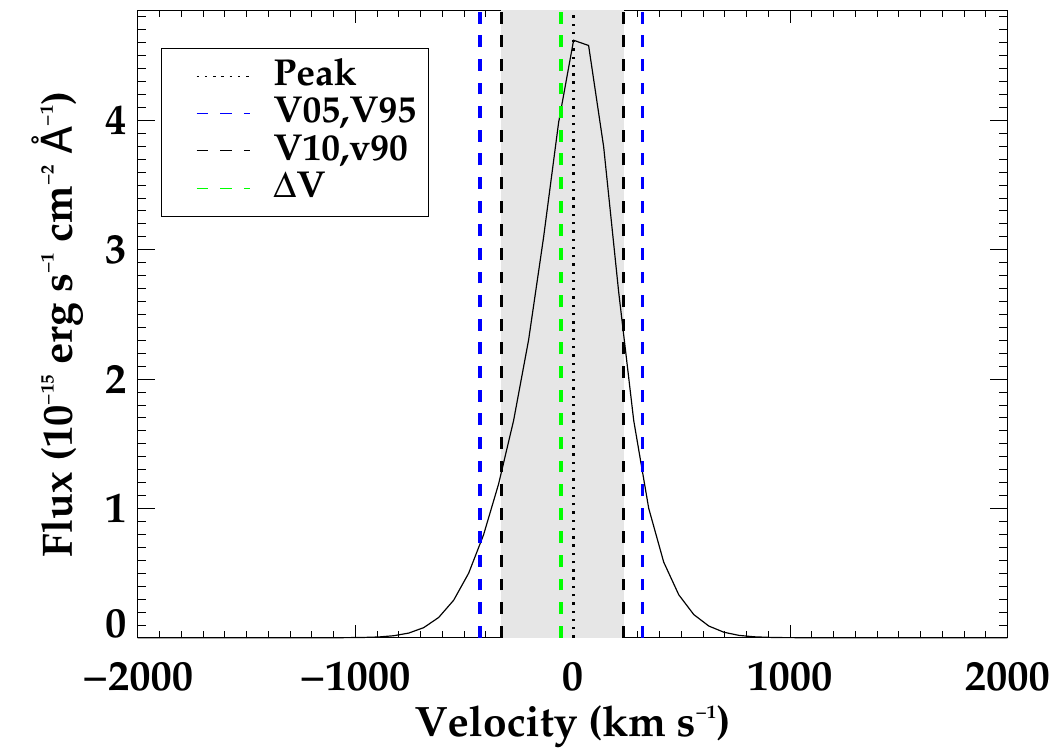}
\caption{J0939+35}
\end{subfigure}
\caption{Gaussian fits to [OIII]$\lambda\lambda 4959,5007$ and non-parametric fits to [OIII]$\lambda 5007$ emission lines. For each object the Gaussian fit is shown in the left panel, where he black line shows the data and and the red line shows the sum of the Gaussian components which are denoted in shades of blue and green. The red crosses show pixels that were masked out of the fit. The right panels show the corresponding non-parametric values derived from the emission line models.}
\label{fig:linefit_pg1}
\end{figure*}

\begin{figure*}
\centering
\begin{subfigure}{0.49\textwidth}
 \includegraphics[width = 0.49\linewidth]{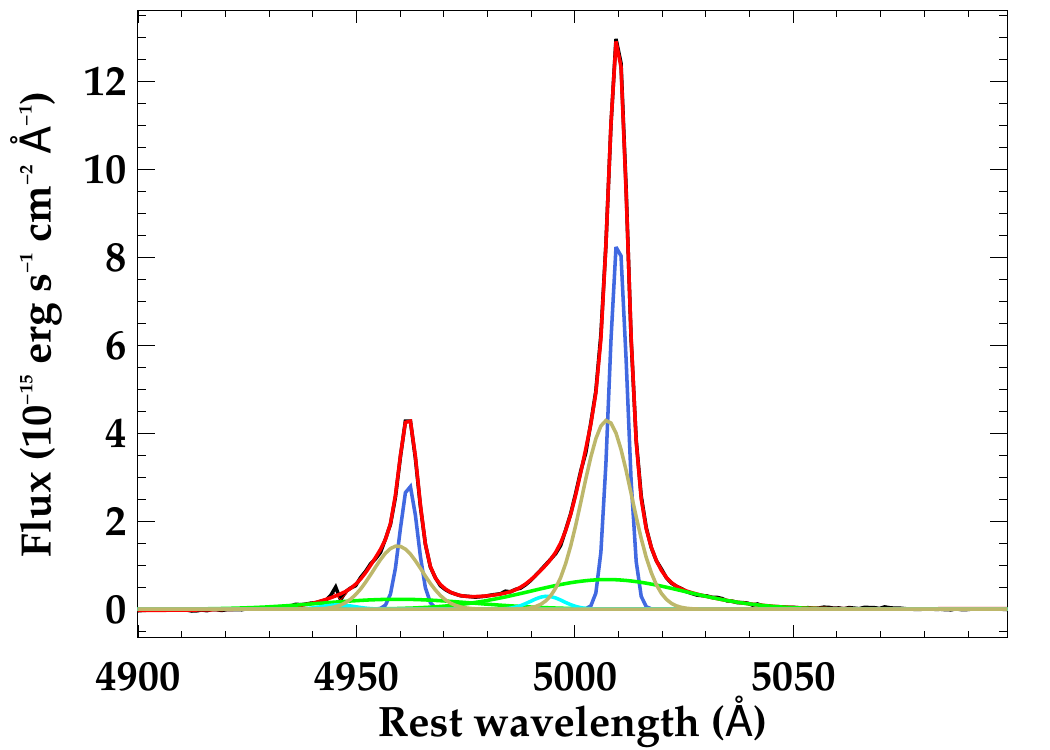}
 \includegraphics[width = 0.49\linewidth]{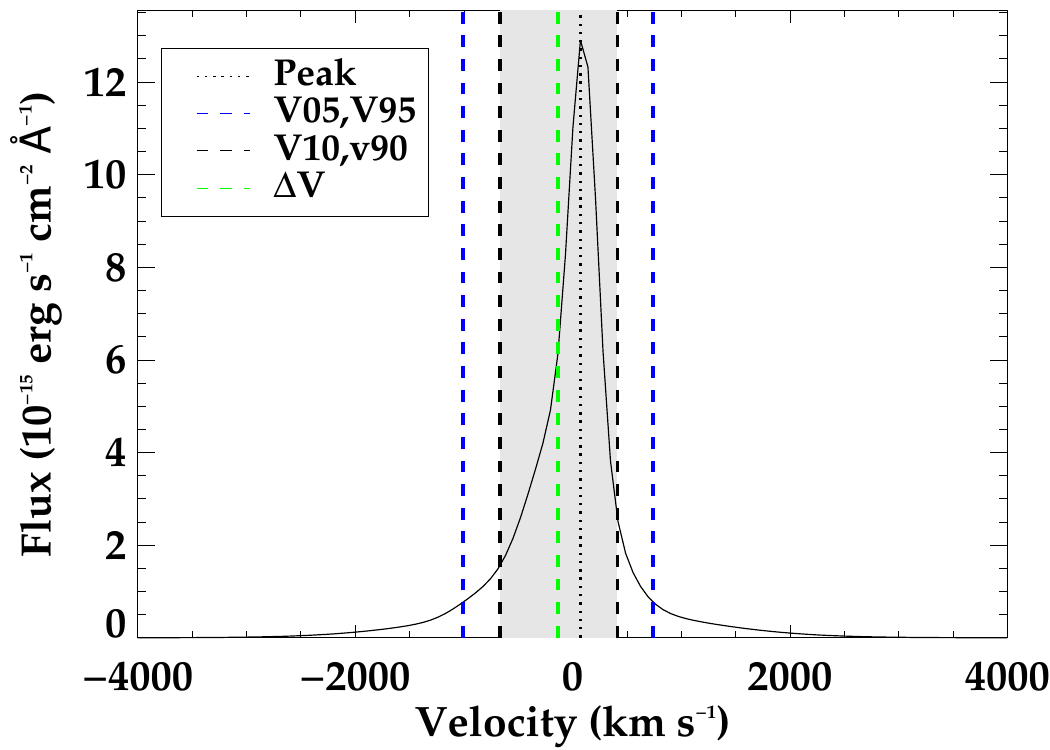}
\caption{J0945+17}
\end{subfigure}
\begin{subfigure}{0.49\textwidth}
 \includegraphics[width = 0.49\linewidth]{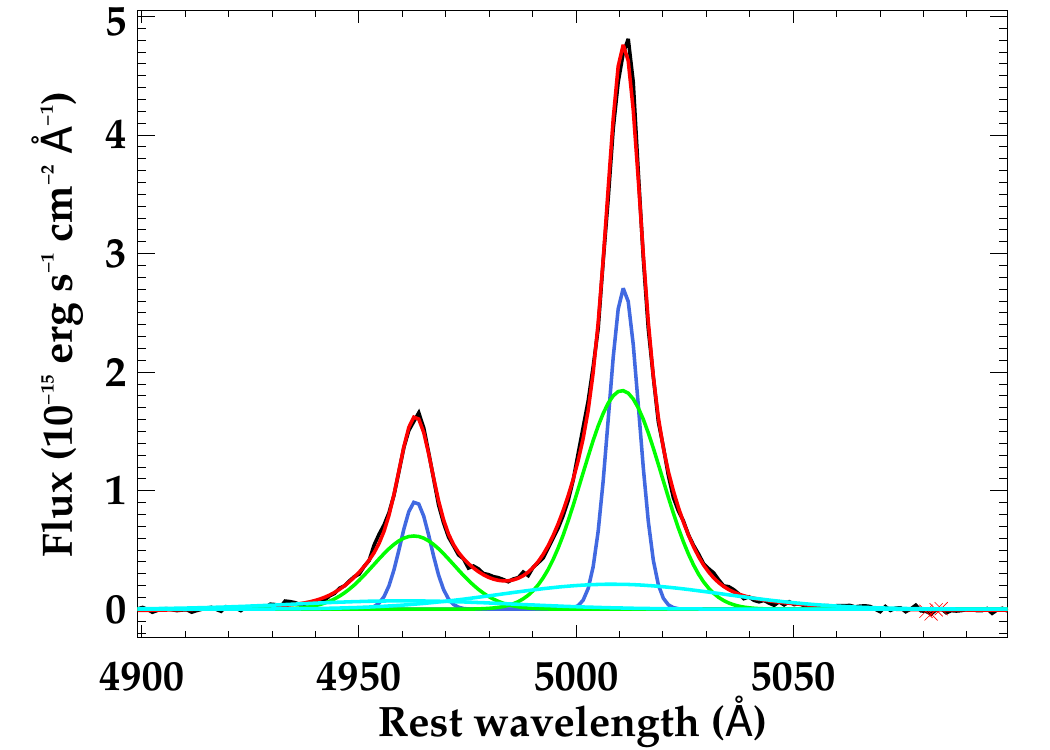}
 \includegraphics[width = 0.49\linewidth]{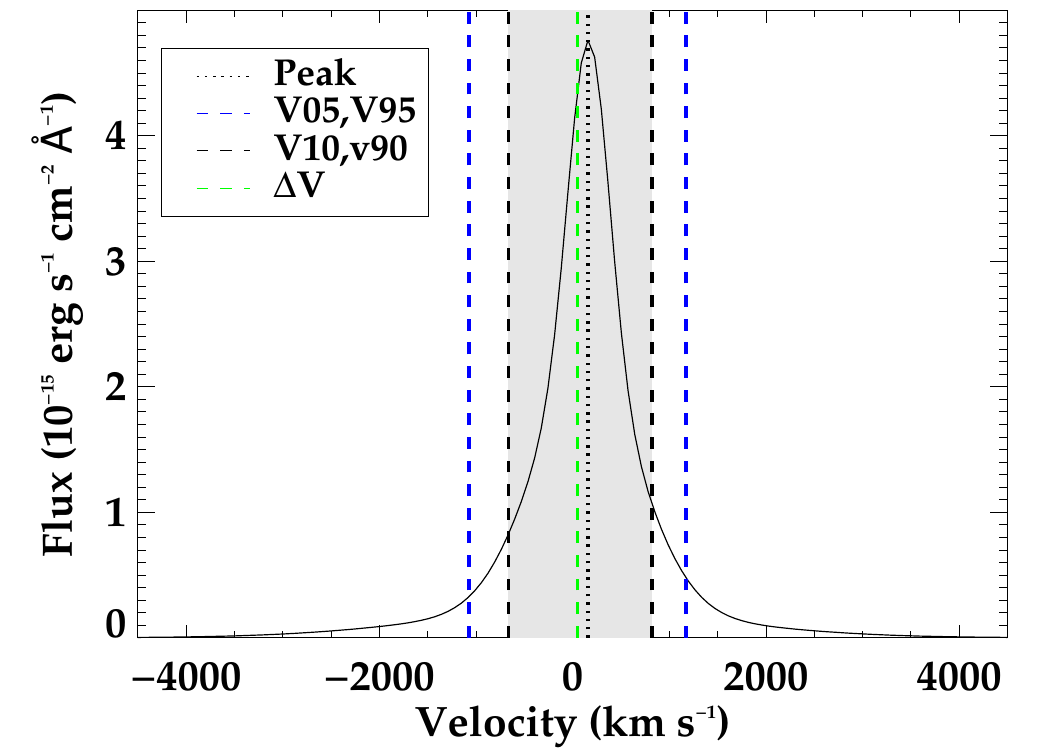}
\caption{J1010+06}
\end{subfigure}

\begin{subfigure}{0.49\textwidth}
 \includegraphics[width = 0.49\linewidth]{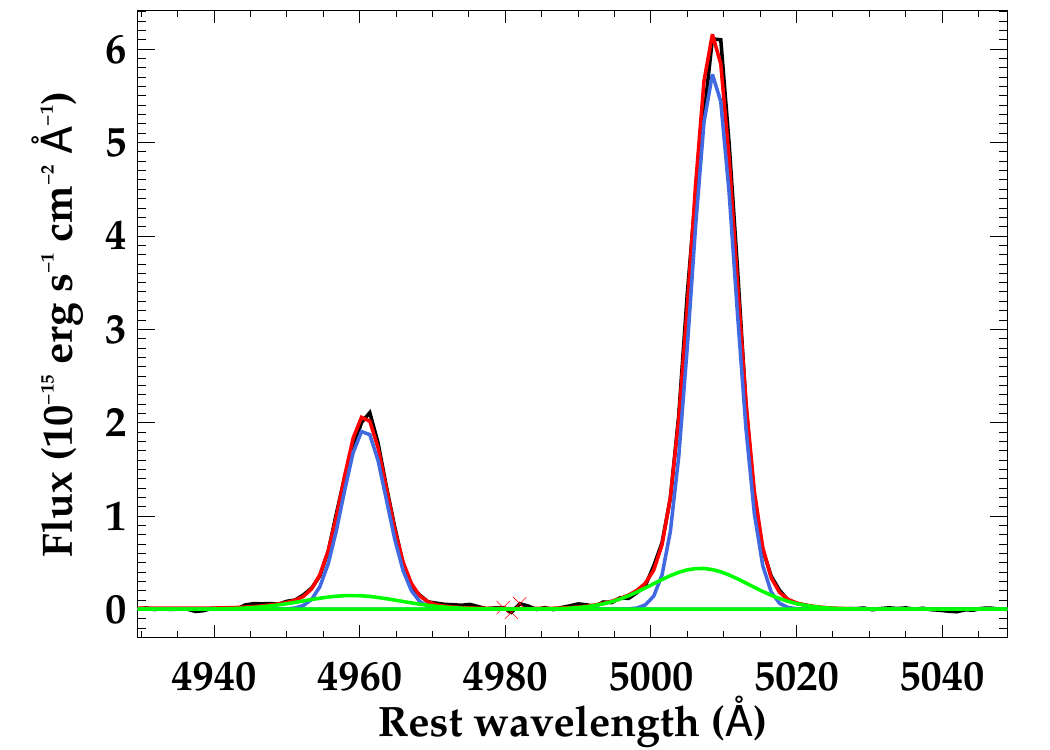}
 \includegraphics[width = 0.49\linewidth]{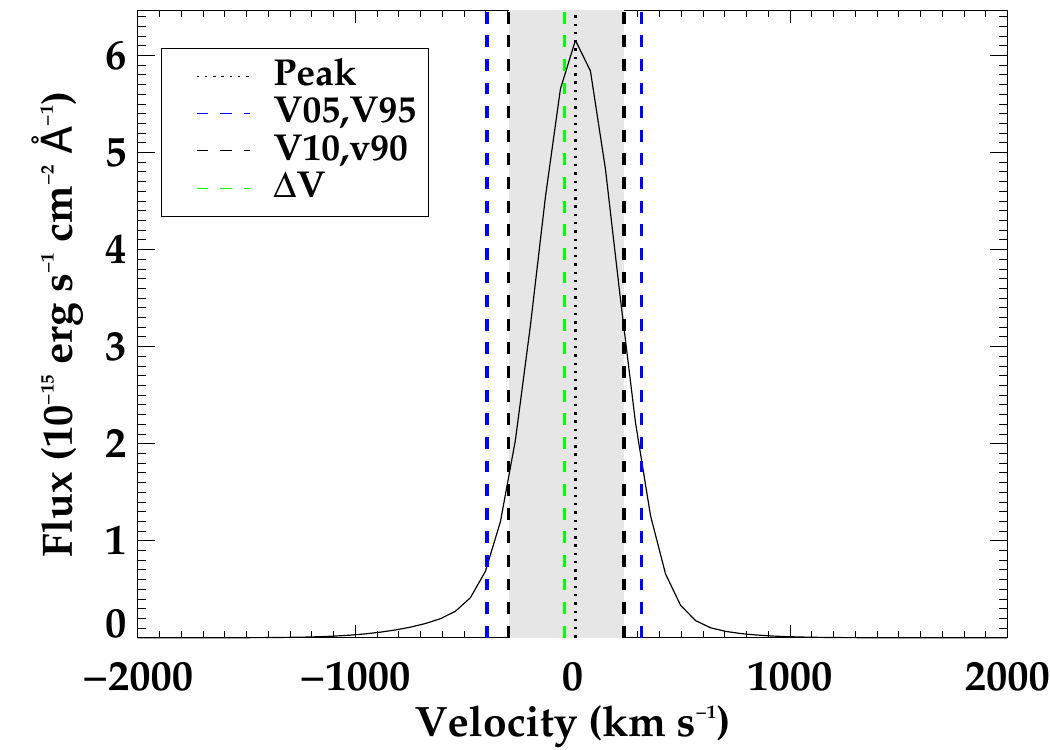}
\caption{J1015+00}
\end{subfigure}
\begin{subfigure}{0.49\textwidth}
 \includegraphics[width = 0.49\linewidth]{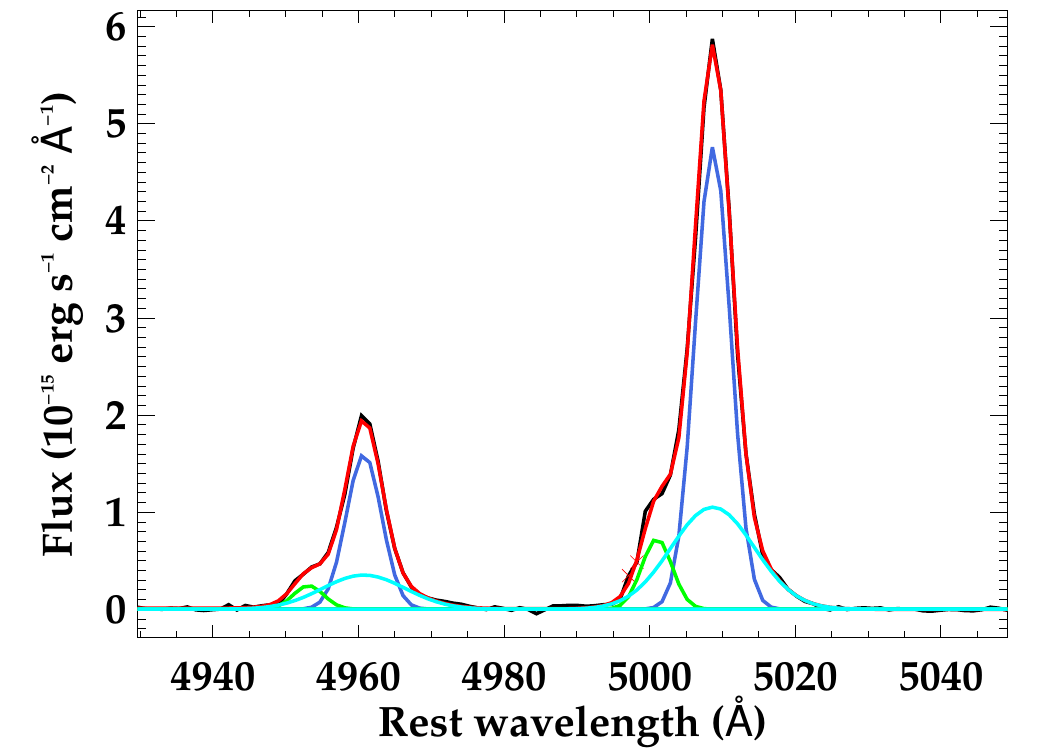}
 \includegraphics[width = 0.49\linewidth]{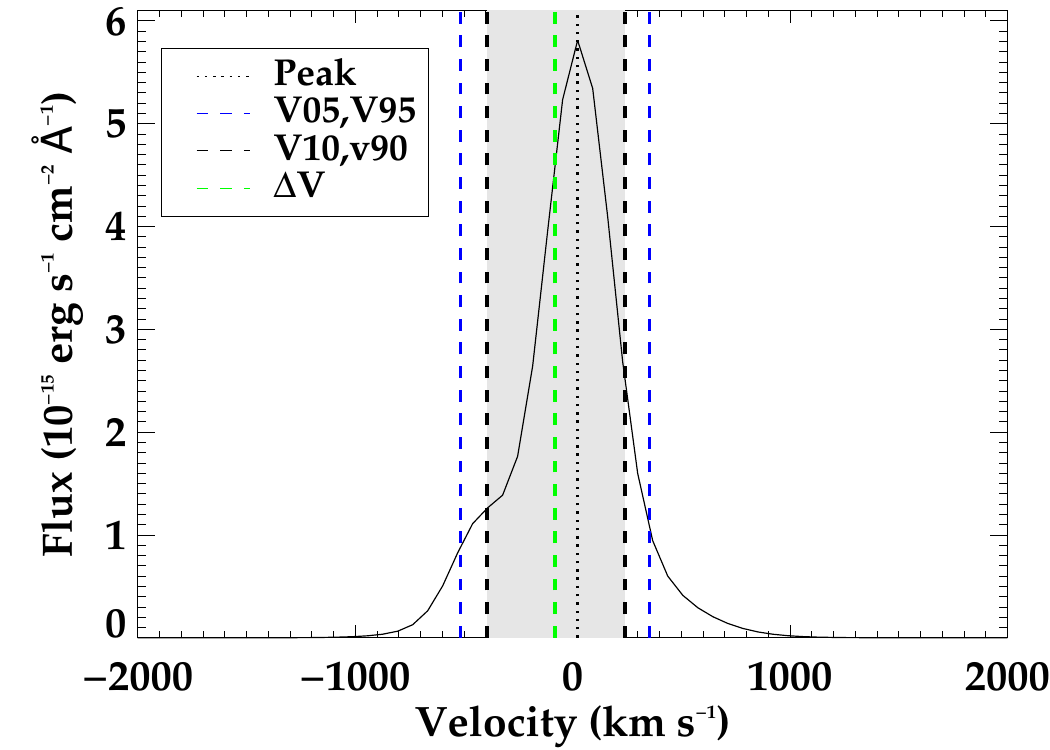}
\caption{J1016+00}
\end{subfigure}

\begin{subfigure}{0.49\textwidth}
 \includegraphics[width = 0.49\linewidth]{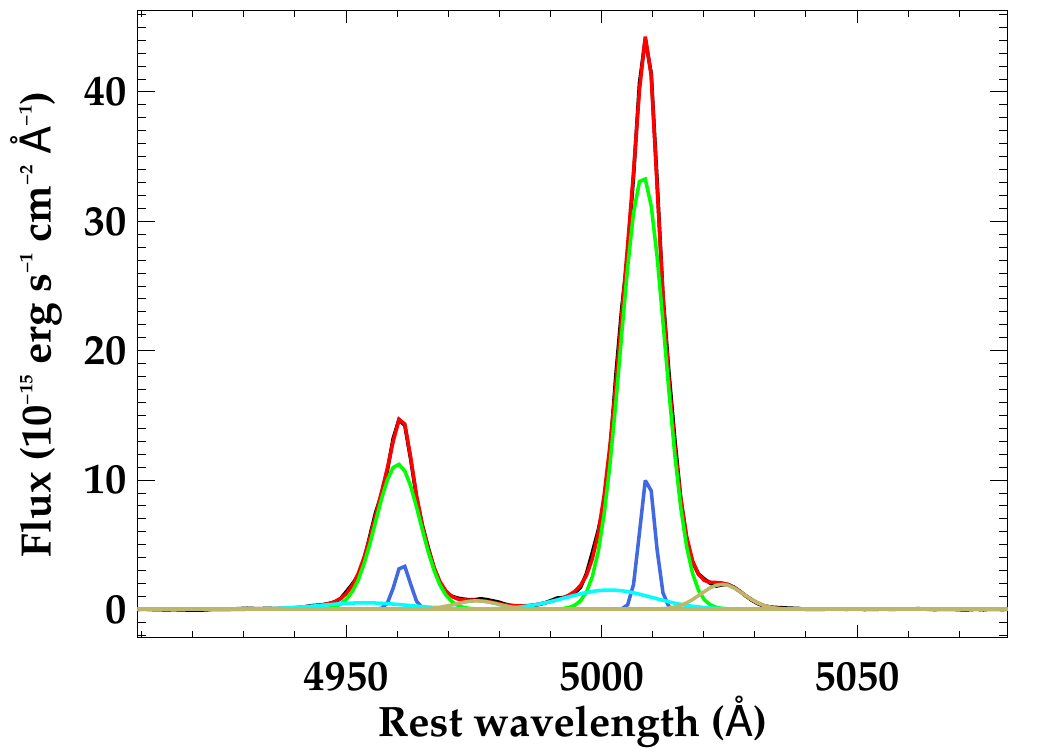}
 \includegraphics[width = 0.49\linewidth]{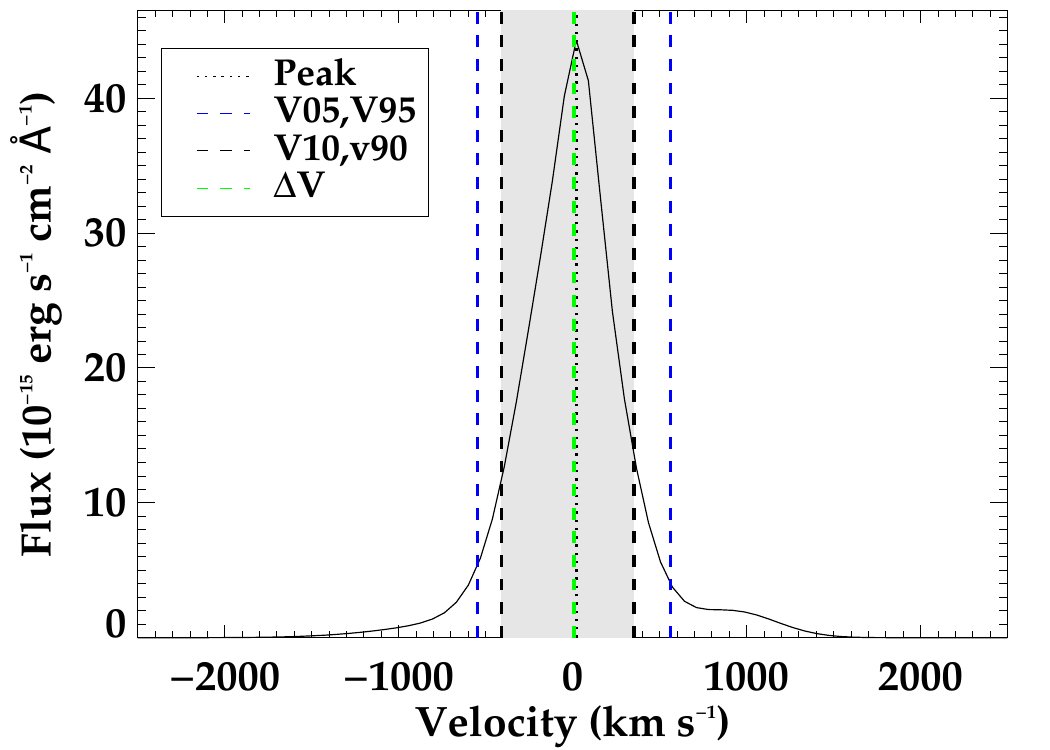}
\caption{J1034+60}
\end{subfigure}
\begin{subfigure}{0.49\textwidth}
 \includegraphics[width = 0.49\linewidth]{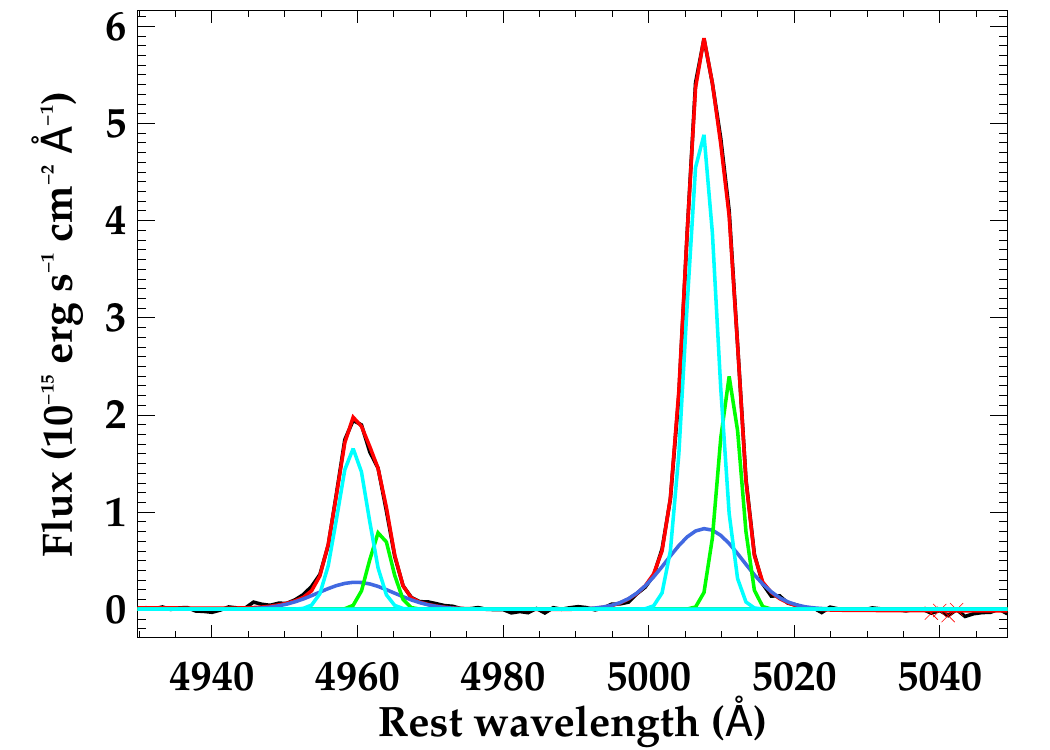}
 \includegraphics[width = 0.49\linewidth]{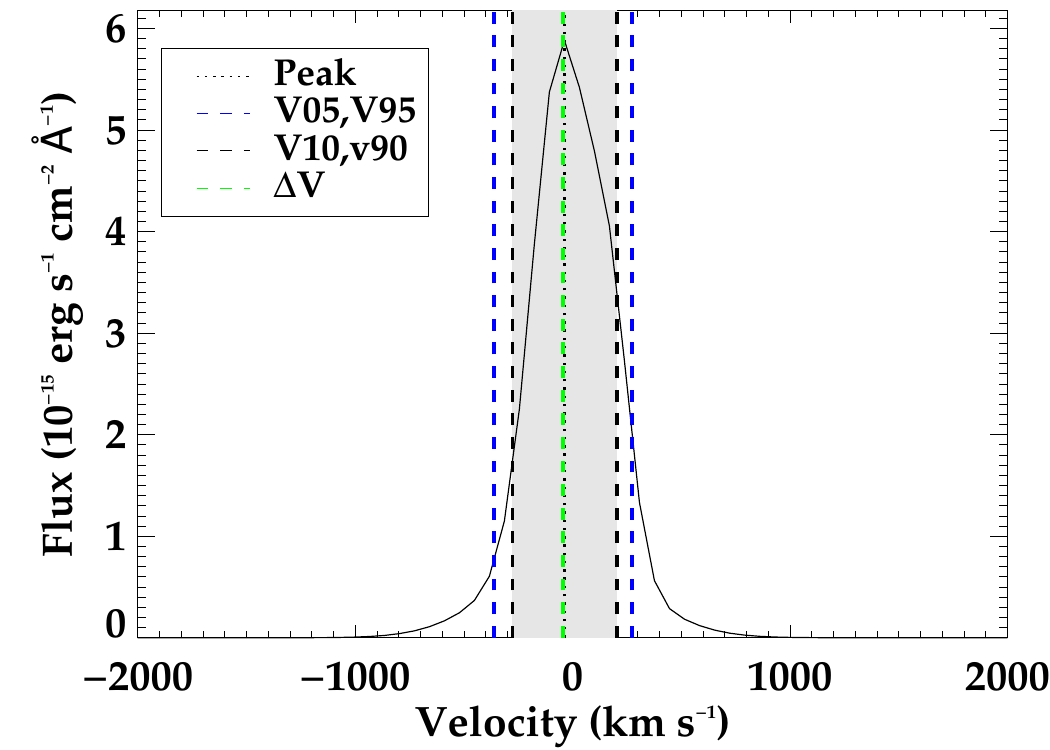}
\caption{J1036+01}
\end{subfigure}

\begin{subfigure}{0.49\textwidth}
 \includegraphics[width = 0.49\linewidth]{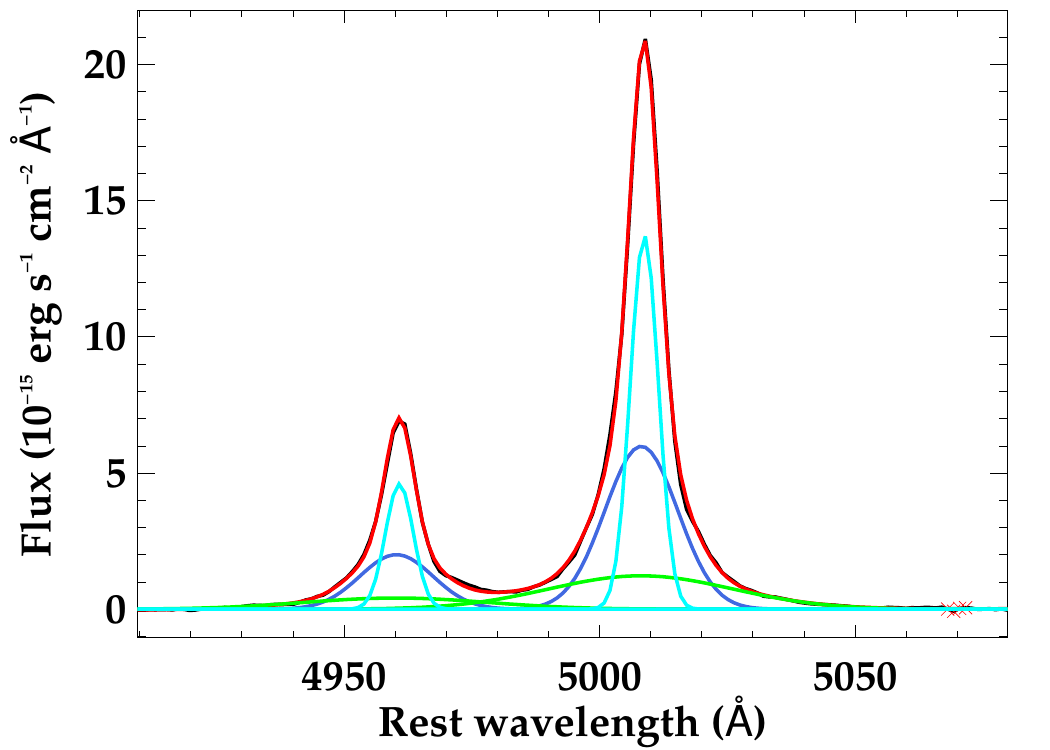}
 \includegraphics[width = 0.49\linewidth]{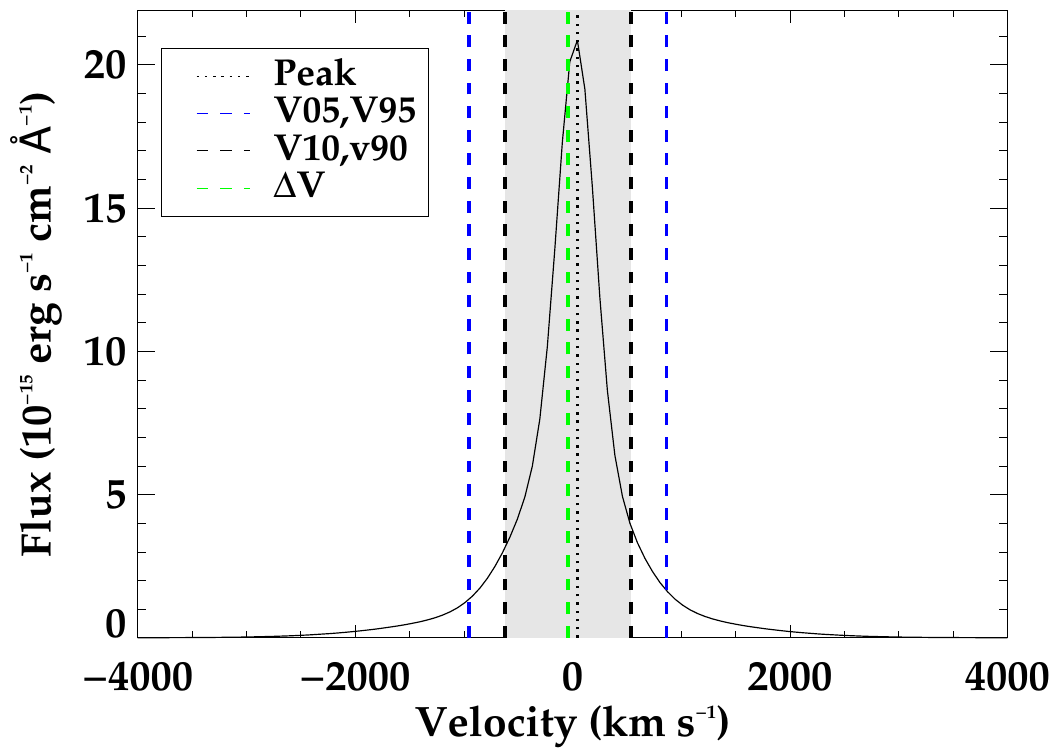}
\caption{J1100+08}
\end{subfigure}
\begin{subfigure}{0.49\textwidth}
 \includegraphics[width = 0.49\linewidth]{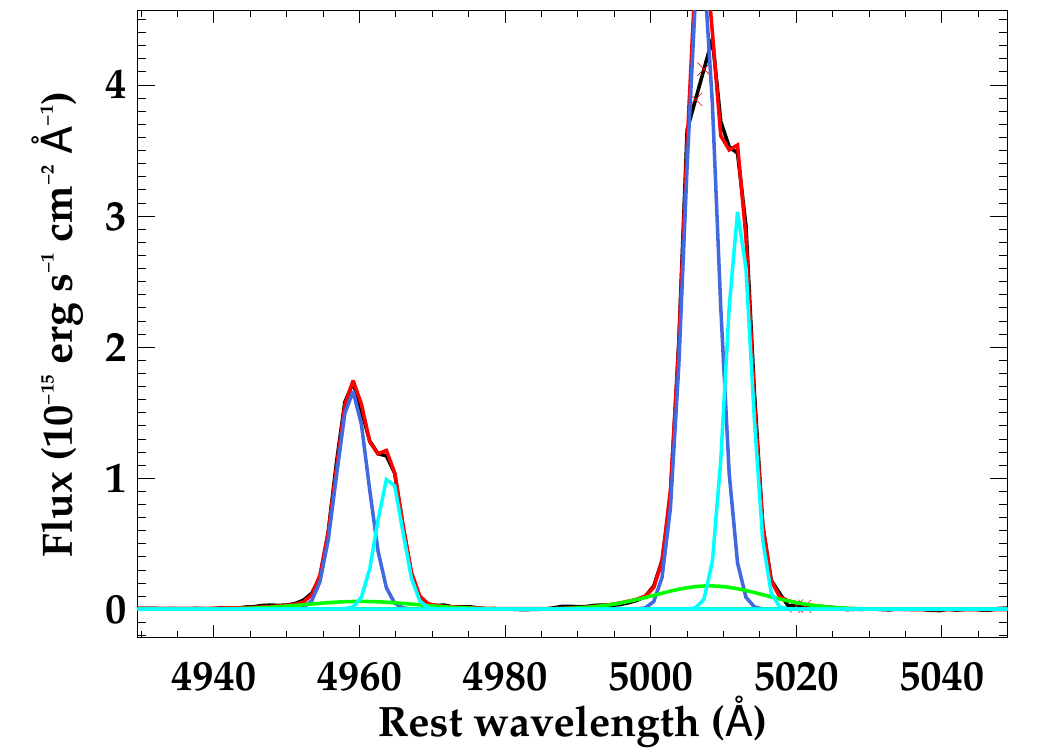}
 \includegraphics[width = 0.49\linewidth]{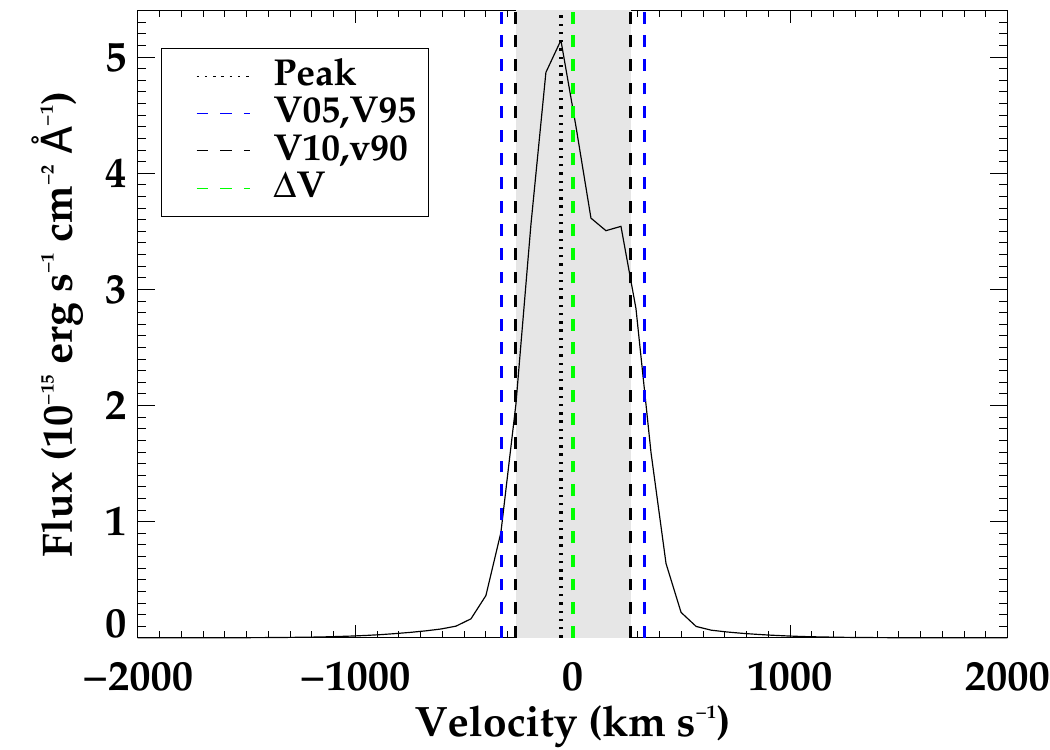}
\caption{J1137+61}
\end{subfigure}

\begin{subfigure}{0.49\textwidth}
 \includegraphics[width = 0.49\linewidth]{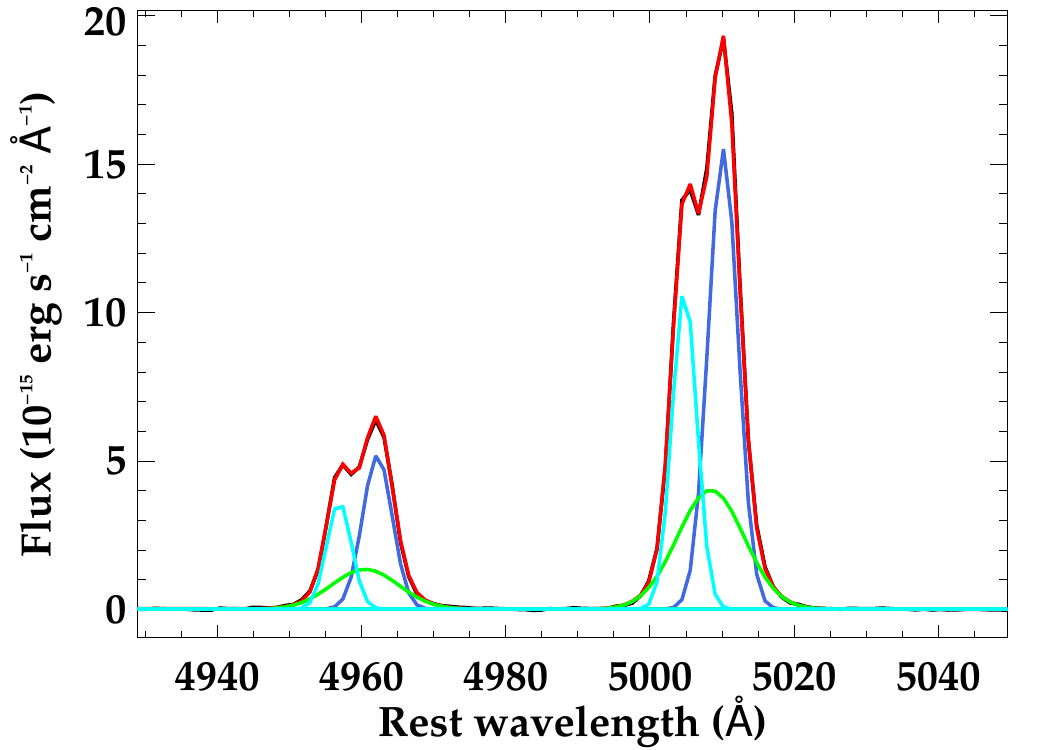}
 \includegraphics[width = 0.49\linewidth]{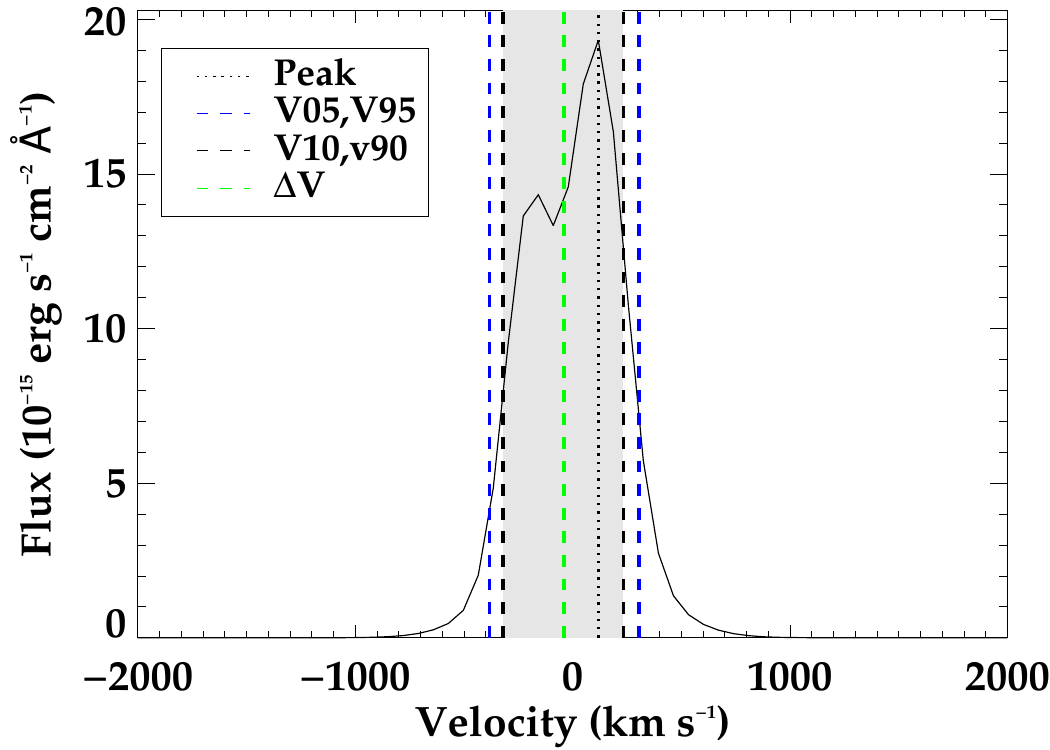}
\caption{J1152+10}
\end{subfigure}
\begin{subfigure}{0.49\textwidth}
 \includegraphics[width = 0.49\linewidth]{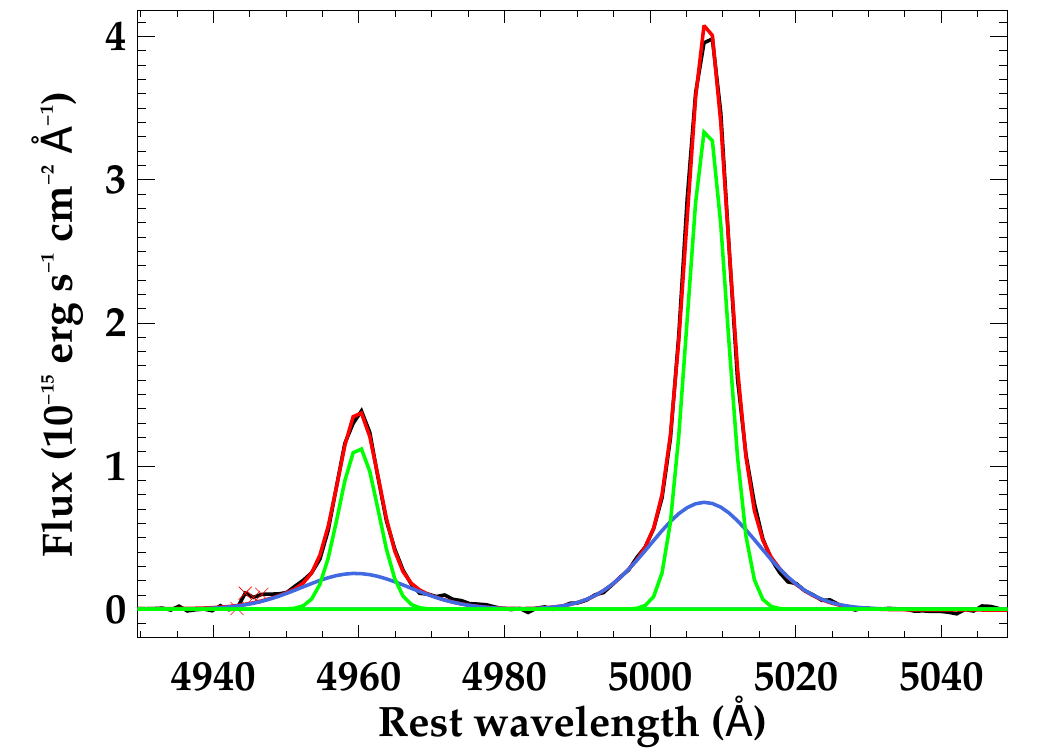}
 \includegraphics[width = 0.49\linewidth]{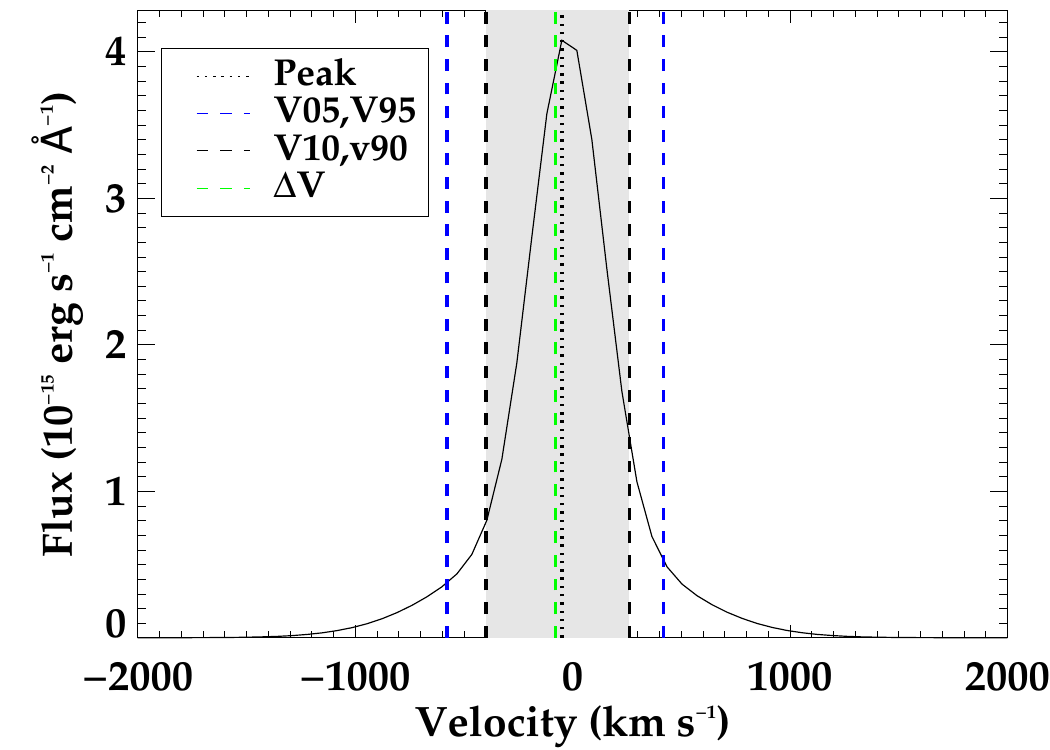}
\caption{J1157+37}
\end{subfigure}

\begin{subfigure}{0.49\textwidth}
 \includegraphics[width = 0.49\linewidth]{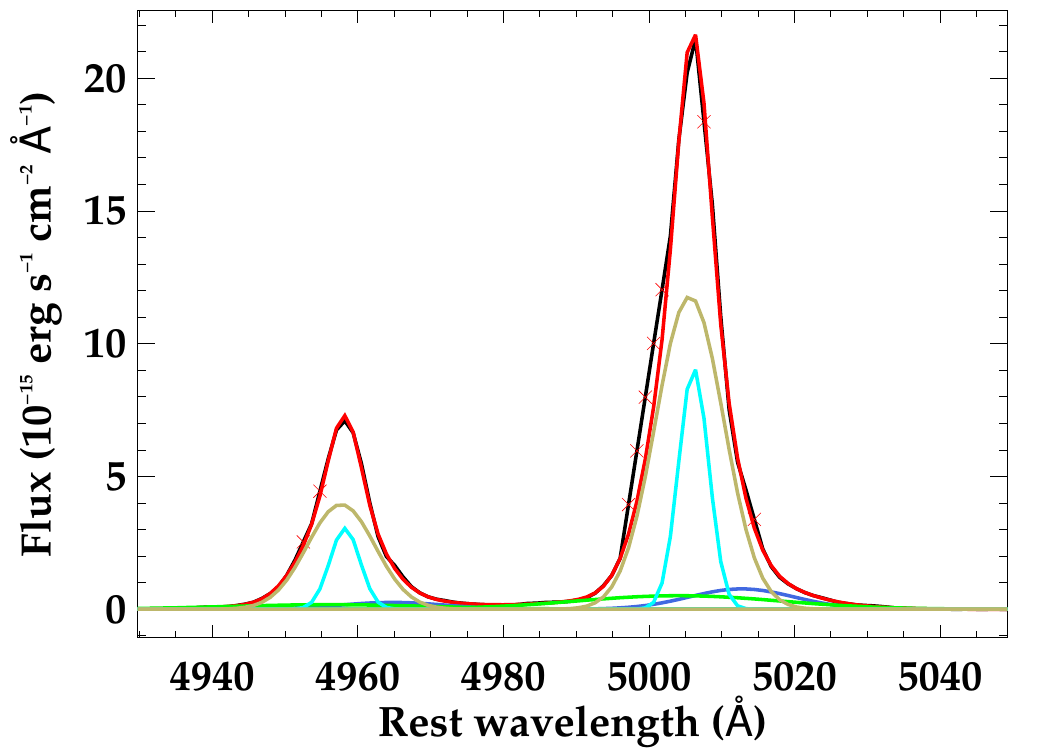}
 \includegraphics[width = 0.49\linewidth]{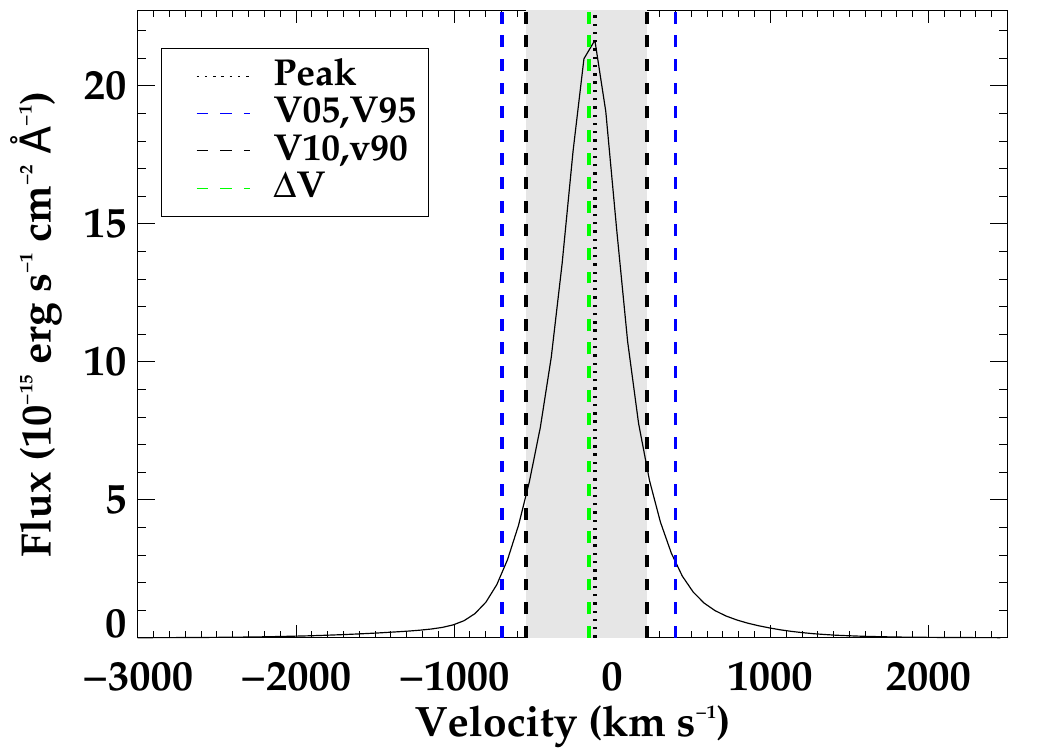}
\caption{J1200+31}
\end{subfigure}
\begin{subfigure}{0.49\textwidth}
 \includegraphics[width = 0.49\linewidth]{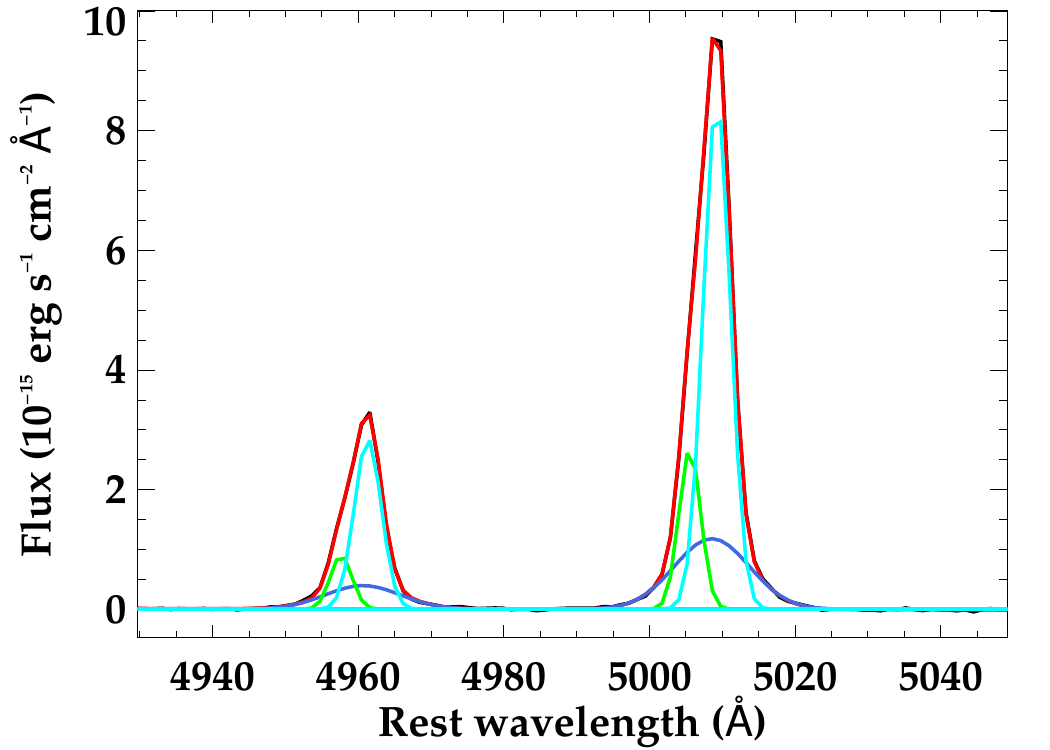}
 \includegraphics[width = 0.49\linewidth]{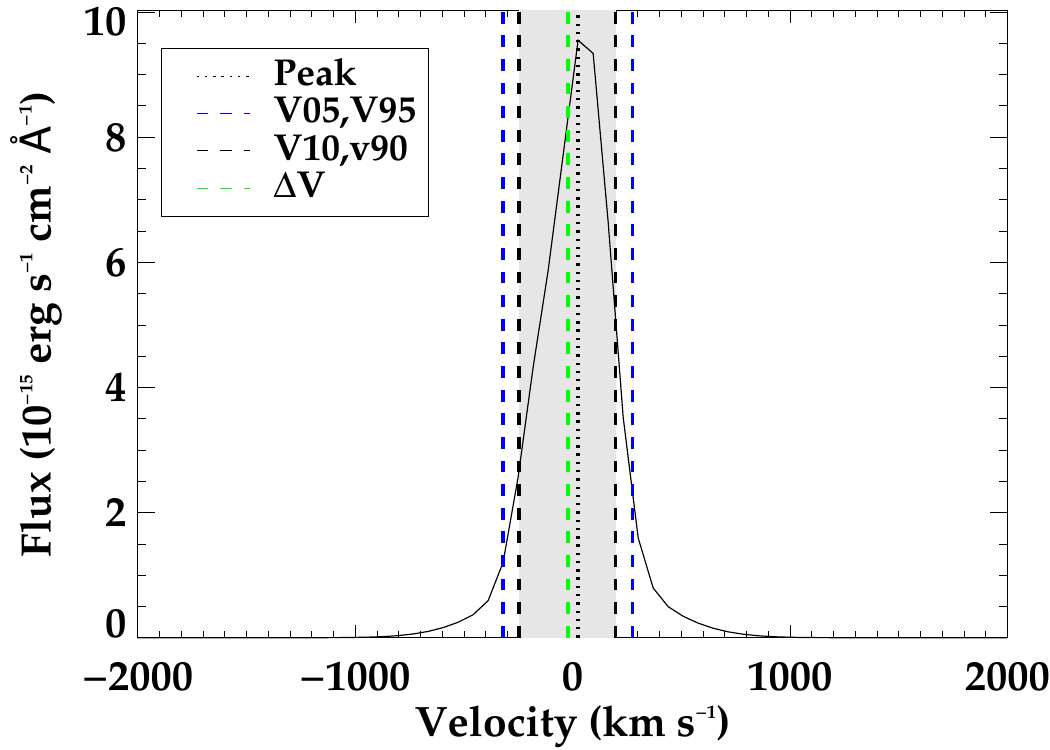}
\caption{J1218+47}
\end{subfigure}
\caption{Same as Figure \ref{fig:linefit_pg1}}
\label{fig:linefit_pg2}
\end{figure*}

\begin{figure*}
\centering
\begin{subfigure}{0.49\textwidth}
 \includegraphics[width = 0.49\linewidth]{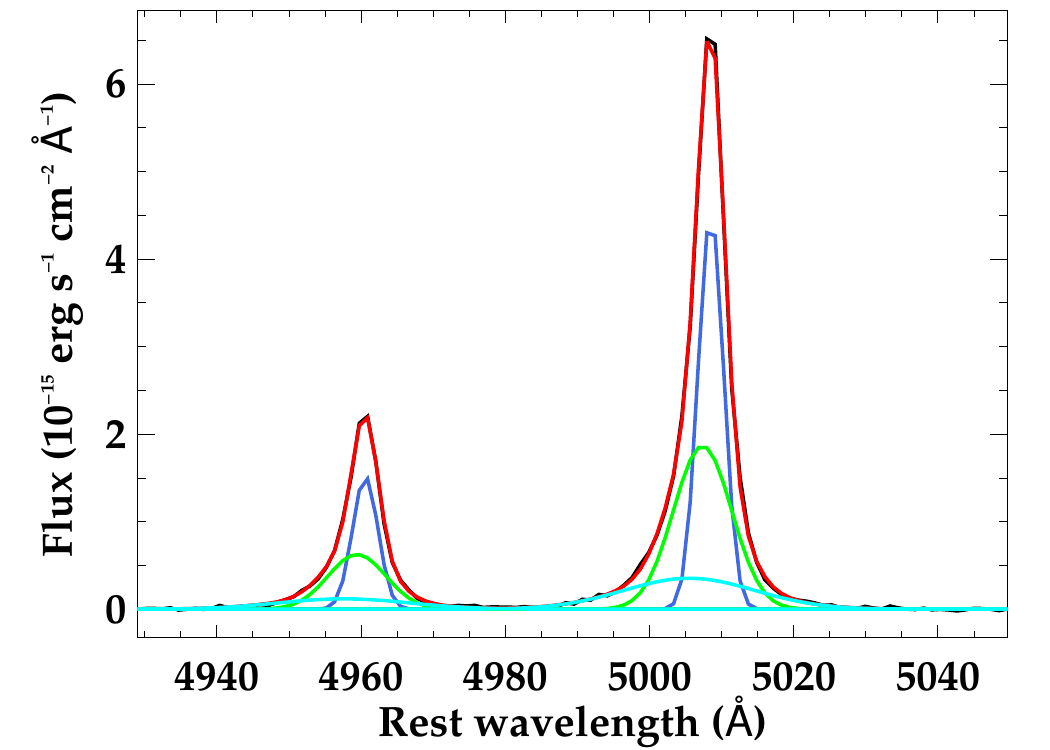}
 \includegraphics[width = 0.49\linewidth]{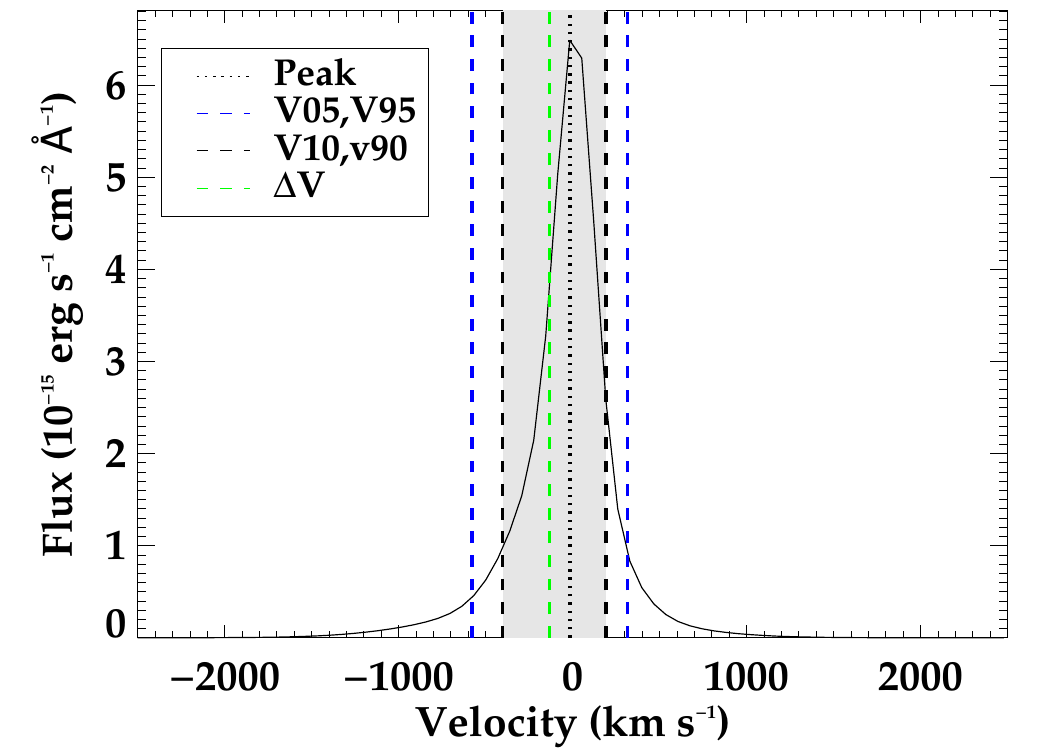}
\caption{J1223+08}
\end{subfigure}
\begin{subfigure}{0.49\textwidth}
 \includegraphics[width = 0.49\linewidth]{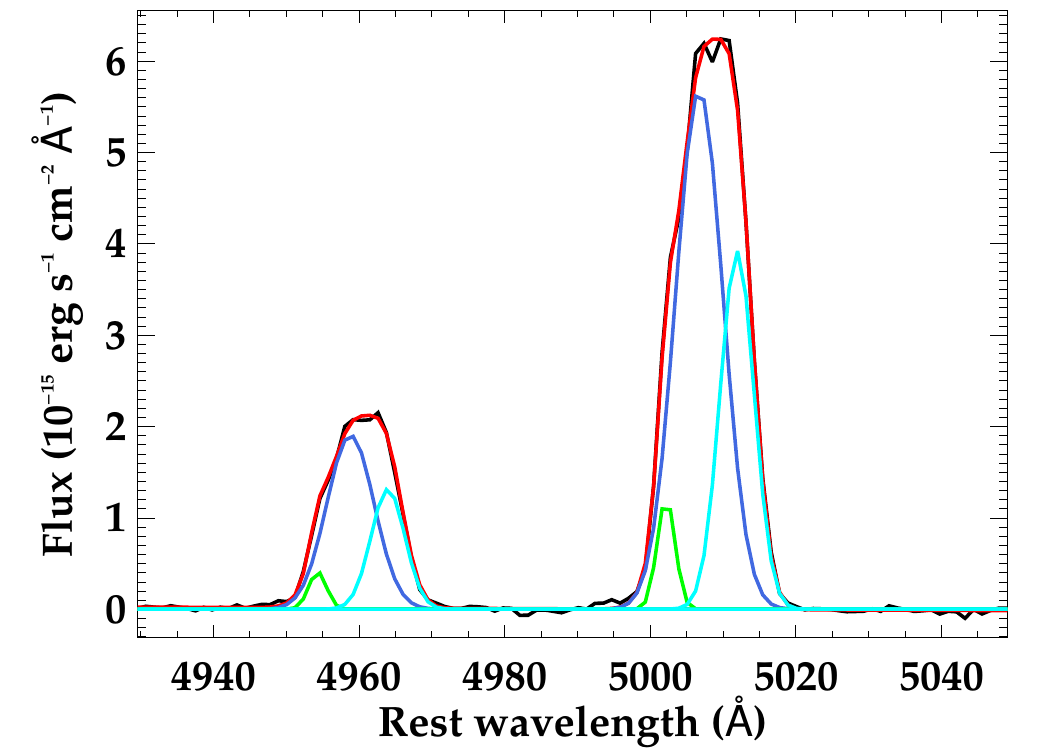}
 \includegraphics[width = 0.49\linewidth]{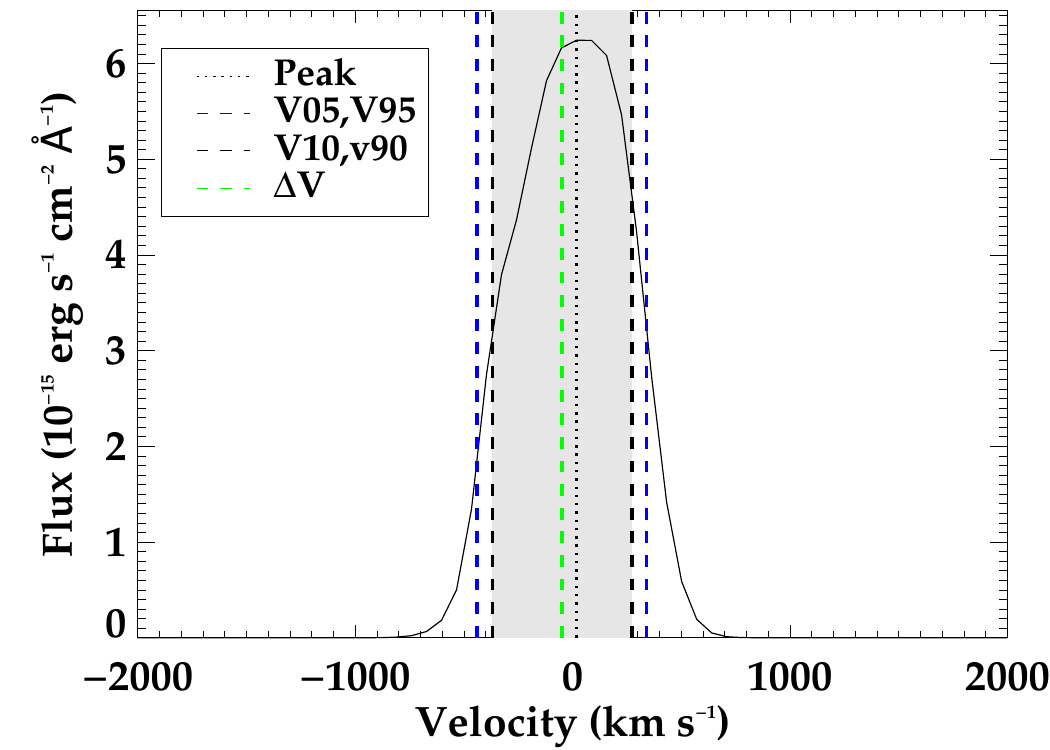}
\caption{J1238+09}
\end{subfigure}

\begin{subfigure}{0.49\textwidth}
 \includegraphics[width = 0.49\linewidth]{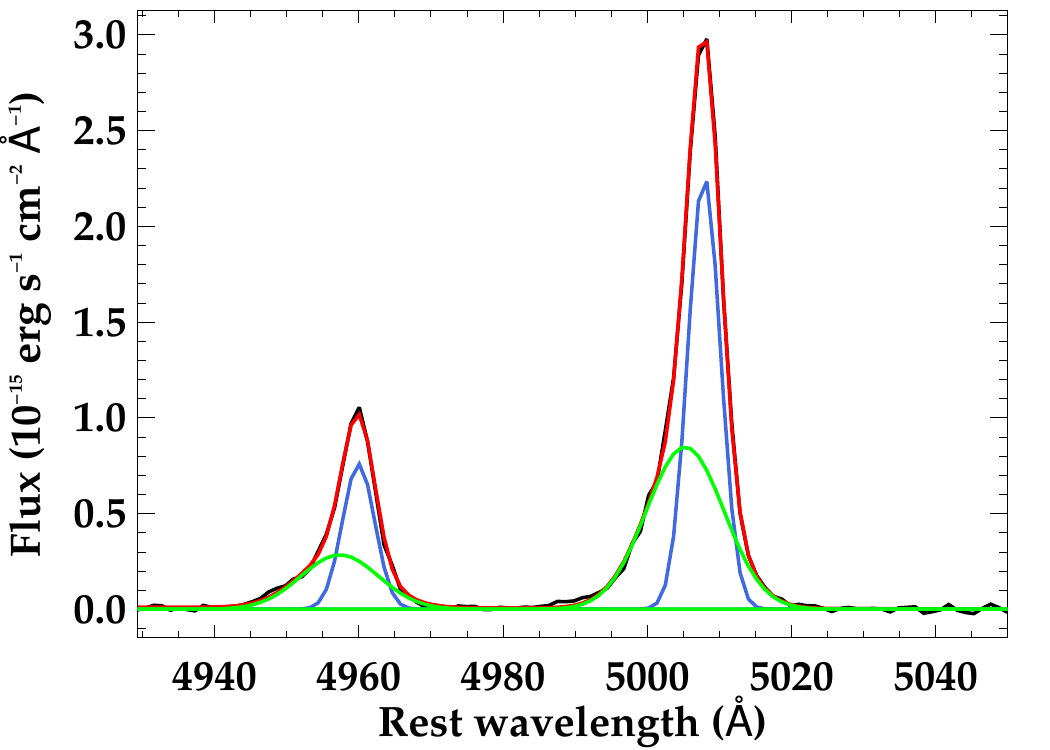}
 \includegraphics[width = 0.49\linewidth]{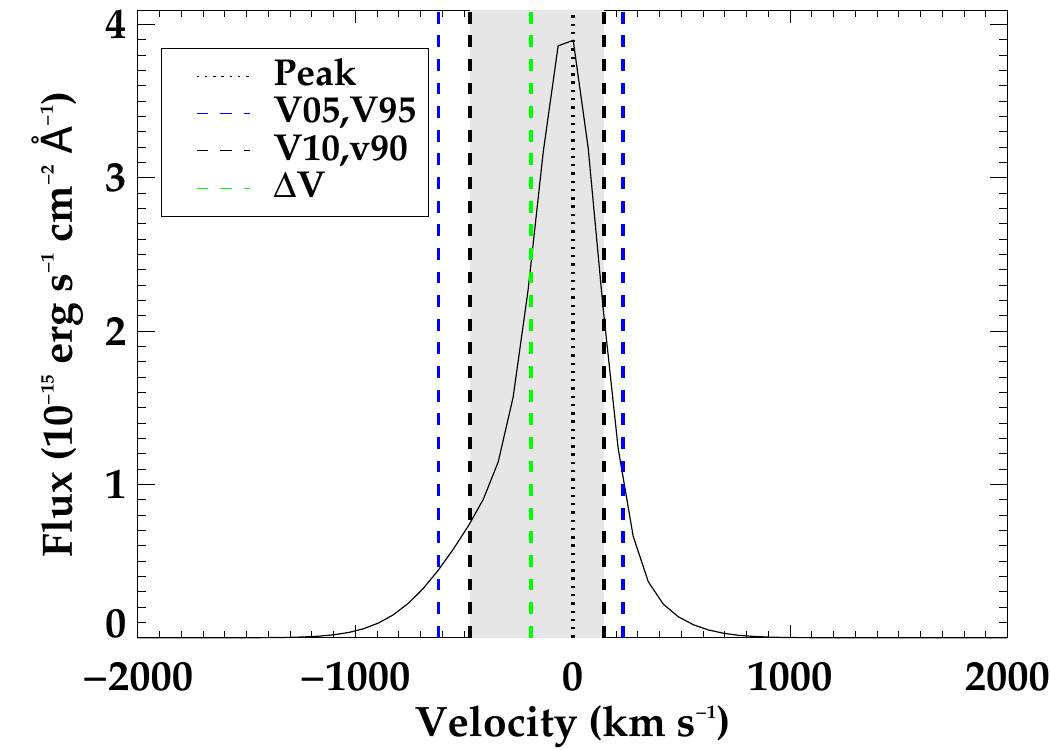}
\caption{J1241+61}
\end{subfigure}
\begin{subfigure}{0.49\textwidth}
 \includegraphics[width = 0.49\linewidth]{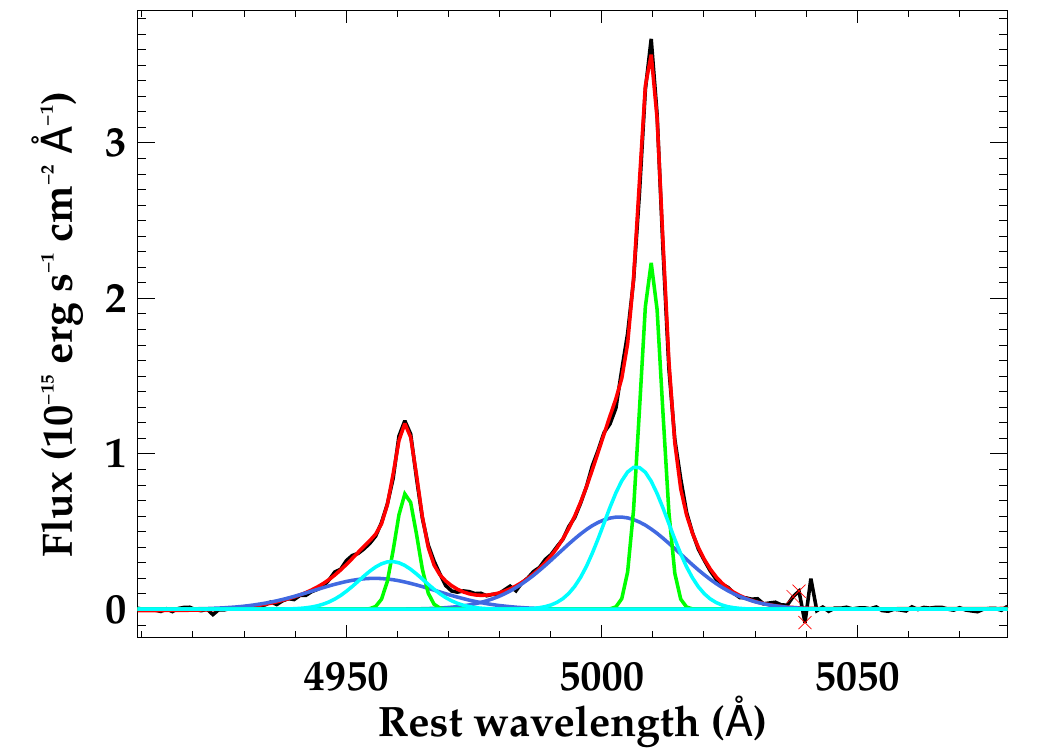}
 \includegraphics[width = 0.49\linewidth]{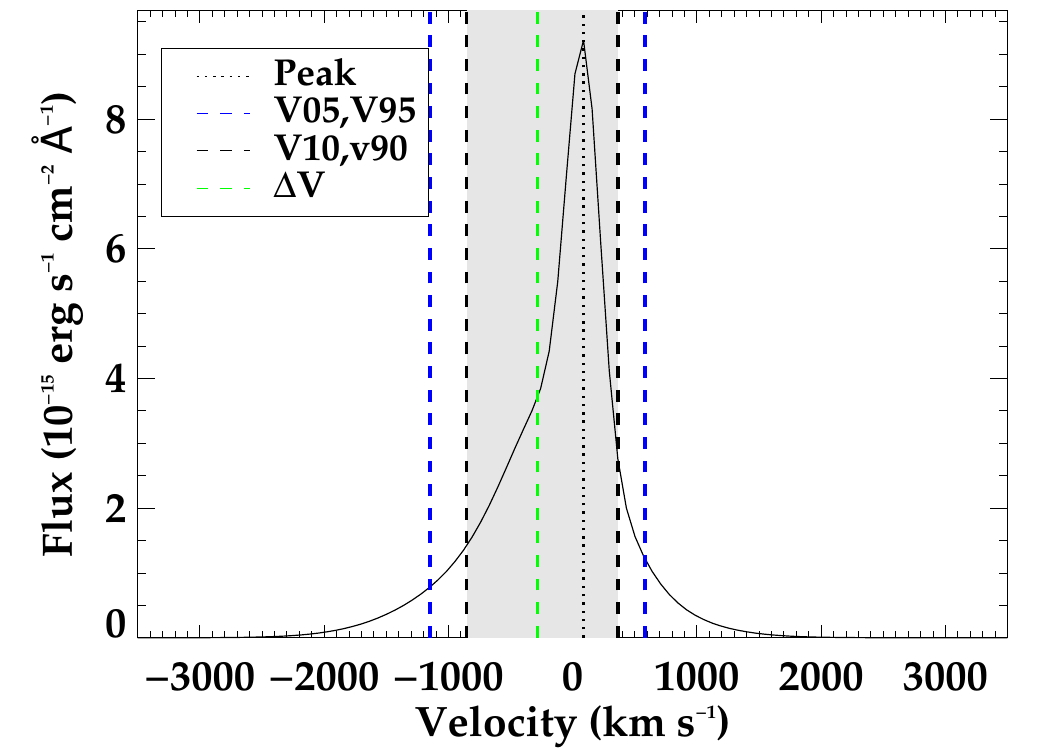}
\caption{J1244+65}
\end{subfigure}

\begin{subfigure}{0.49\textwidth}
 \includegraphics[width = 0.49\linewidth]{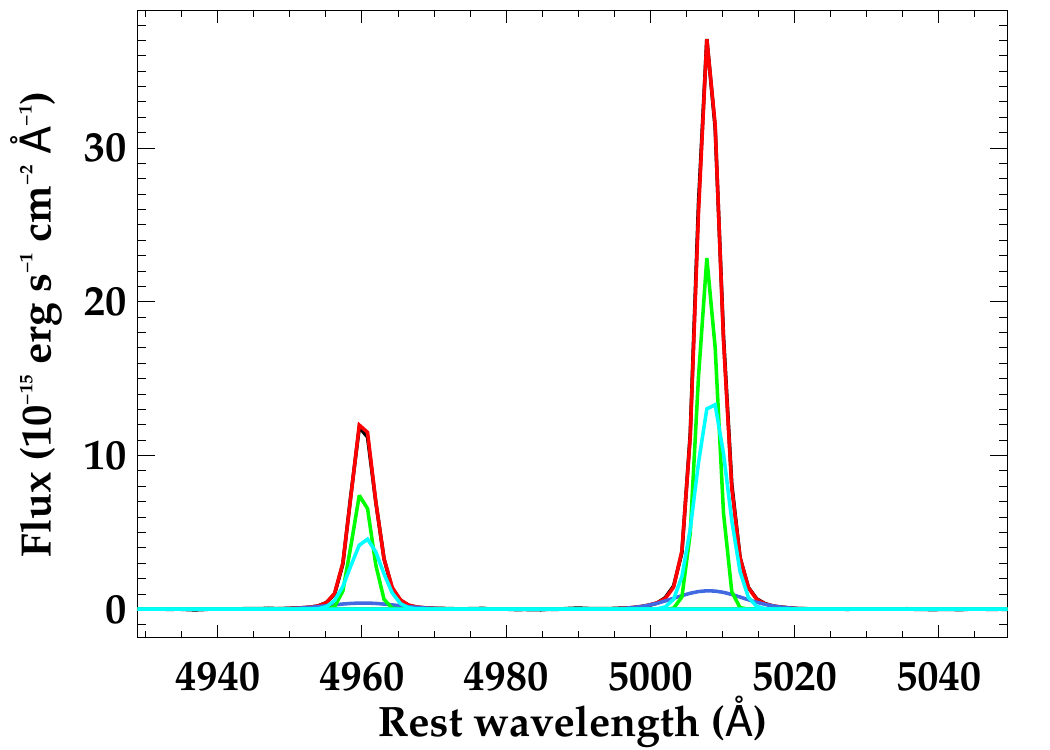}
 \includegraphics[width = 0.49\linewidth]{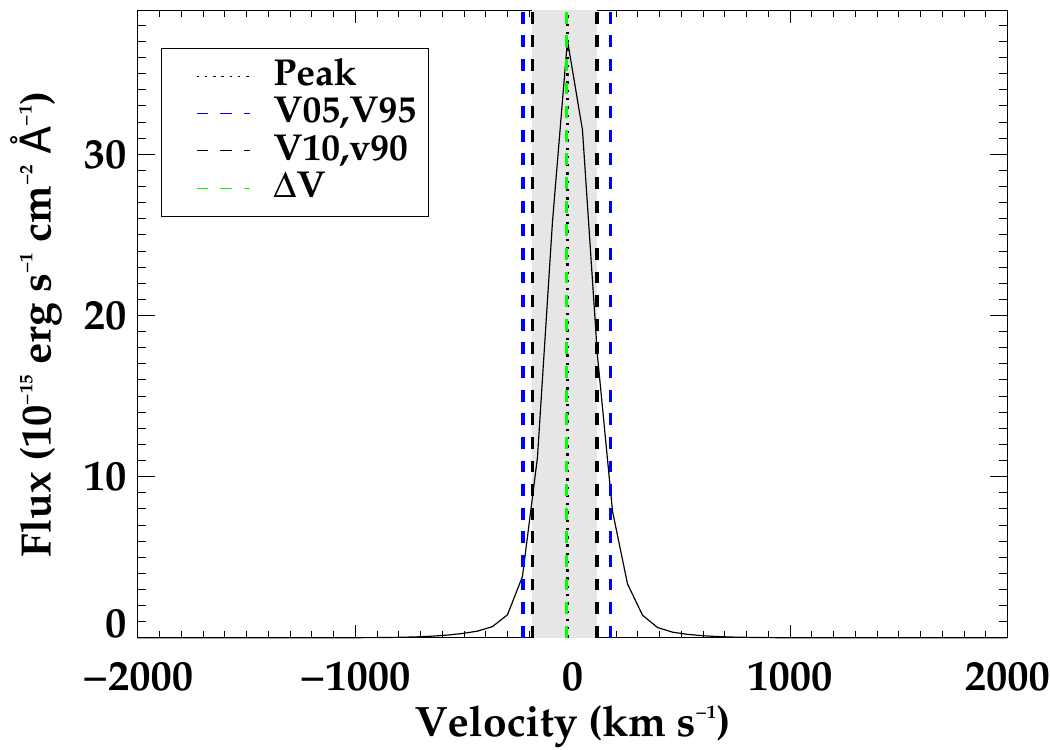}
\caption{J1300+54}
\end{subfigure}
\begin{subfigure}{0.49\textwidth}
 \includegraphics[width = 0.49\linewidth]{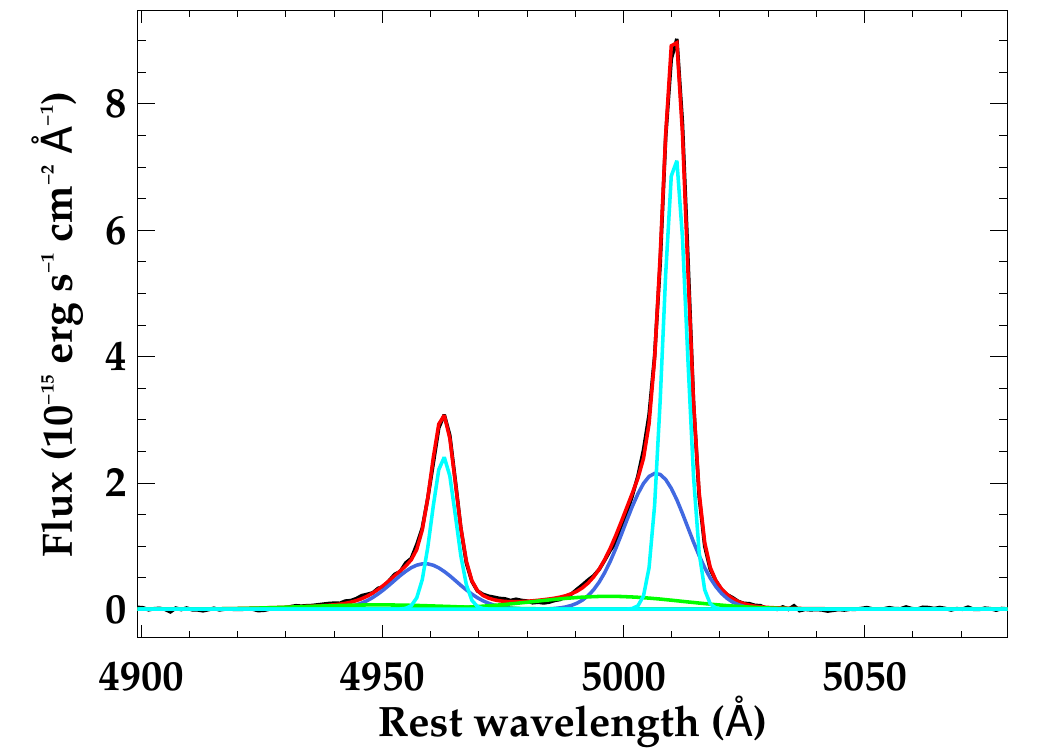}
 \includegraphics[width = 0.49\linewidth]{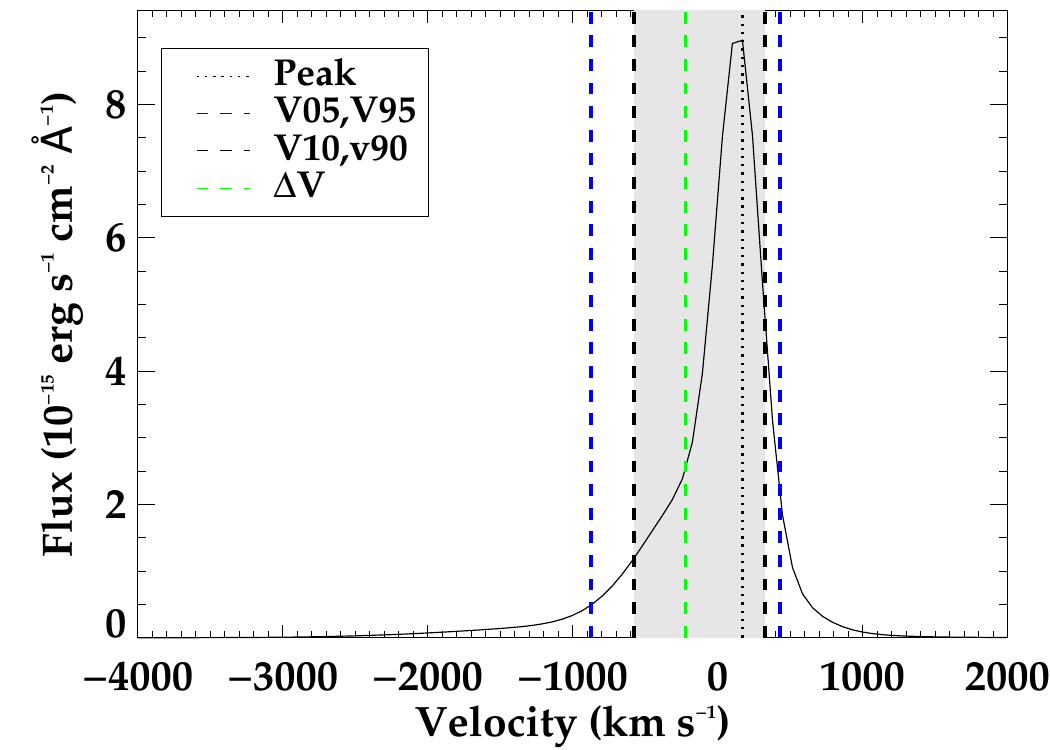}
\caption{J1316+44}
\end{subfigure}

\begin{subfigure}{0.49\textwidth}
 \includegraphics[width = 0.49\linewidth]{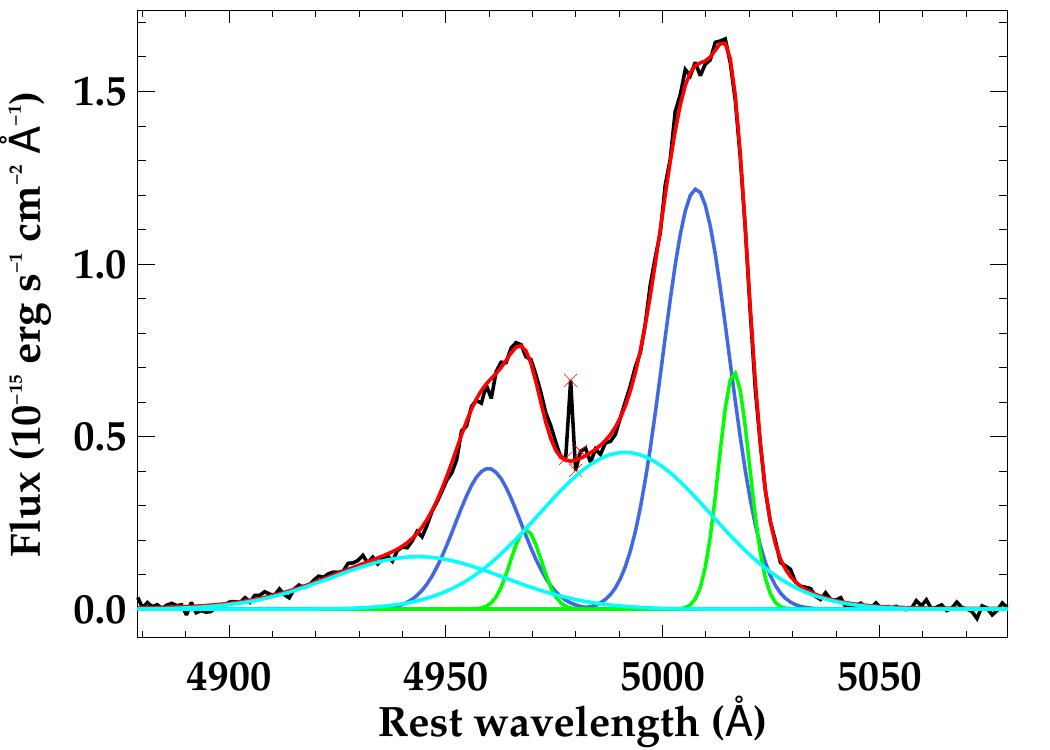}
 \includegraphics[width = 0.49\linewidth]{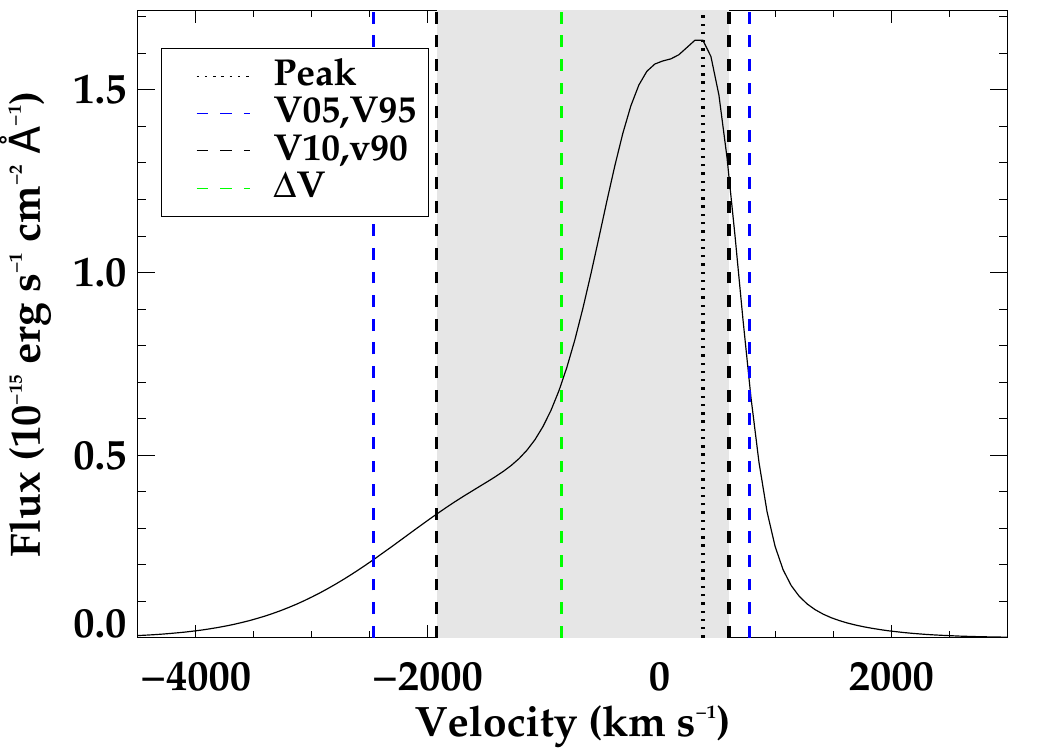}
\caption{J1347+12}
\end{subfigure}
\begin{subfigure}{0.49\textwidth}
 \includegraphics[width = 0.49\linewidth]{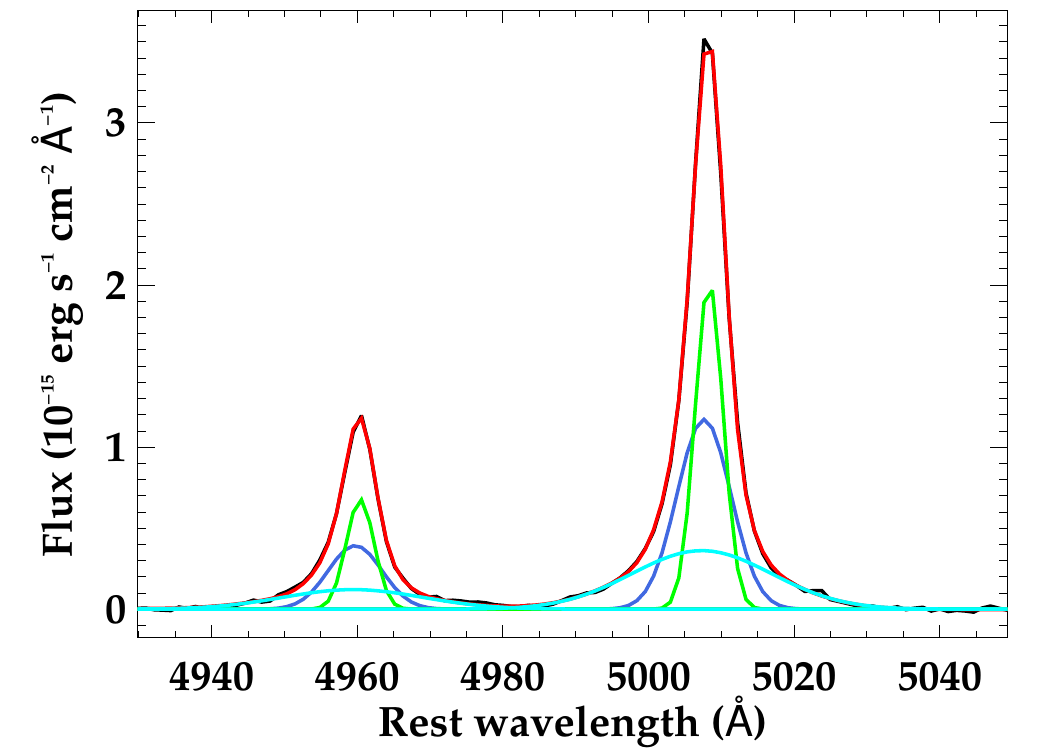}
 \includegraphics[width = 0.49\linewidth]{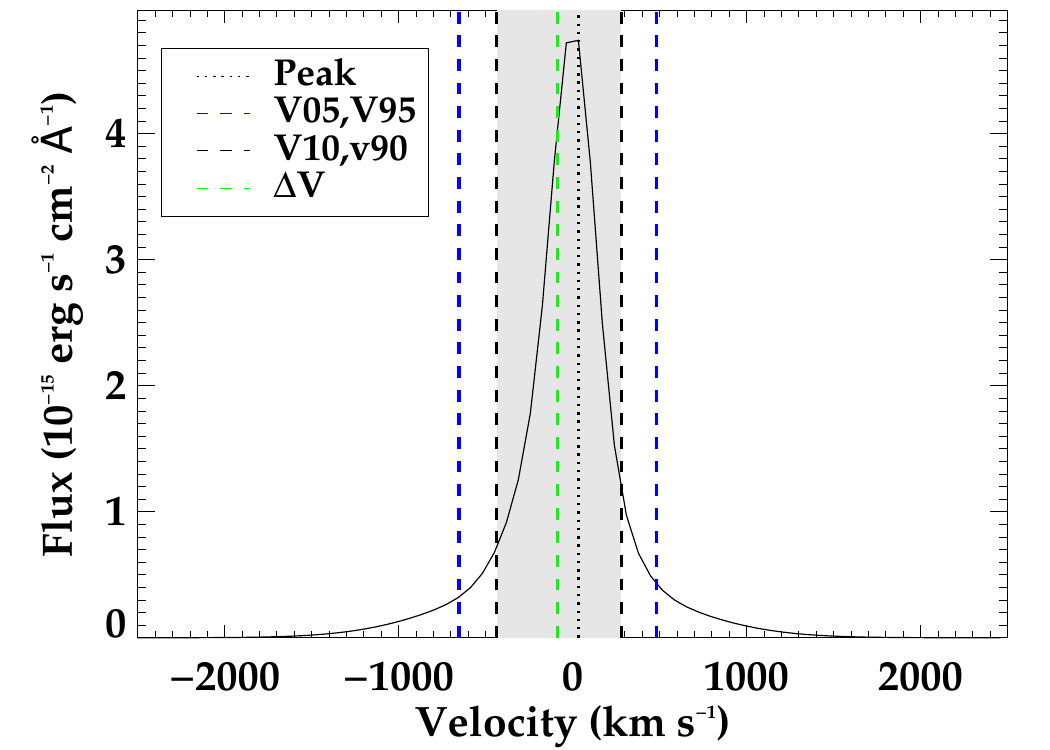}
\caption{J1356-02}
\end{subfigure}

\begin{subfigure}{0.49\textwidth}
 \includegraphics[width = 0.49\linewidth]{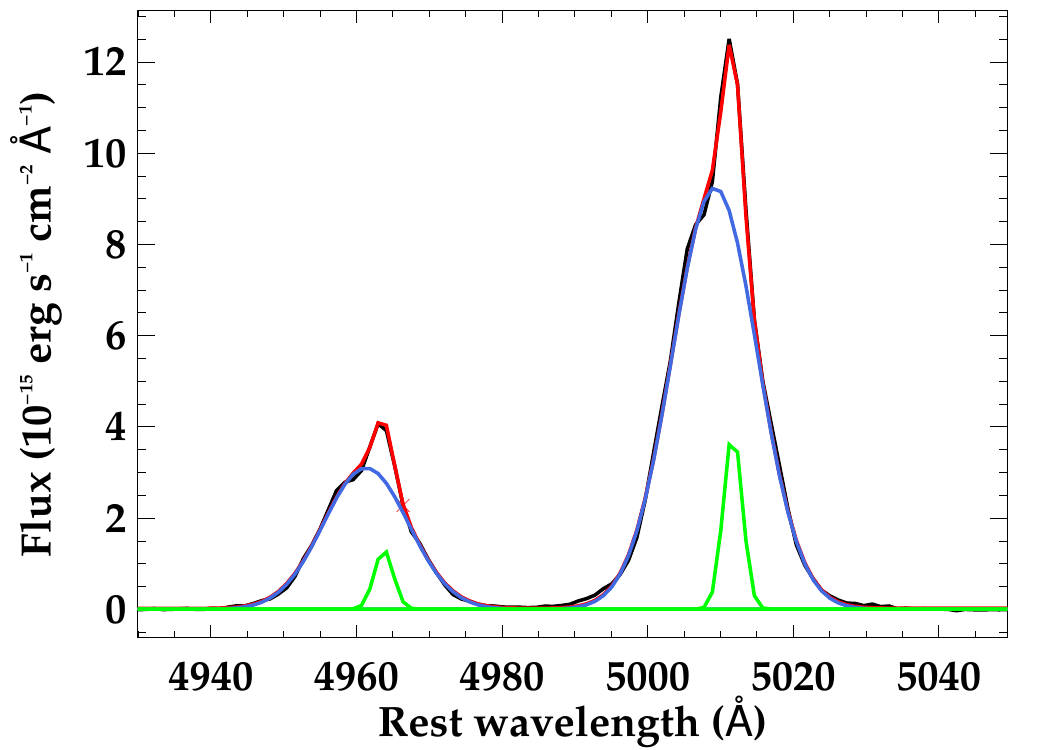}
 \includegraphics[width = 0.49\linewidth]{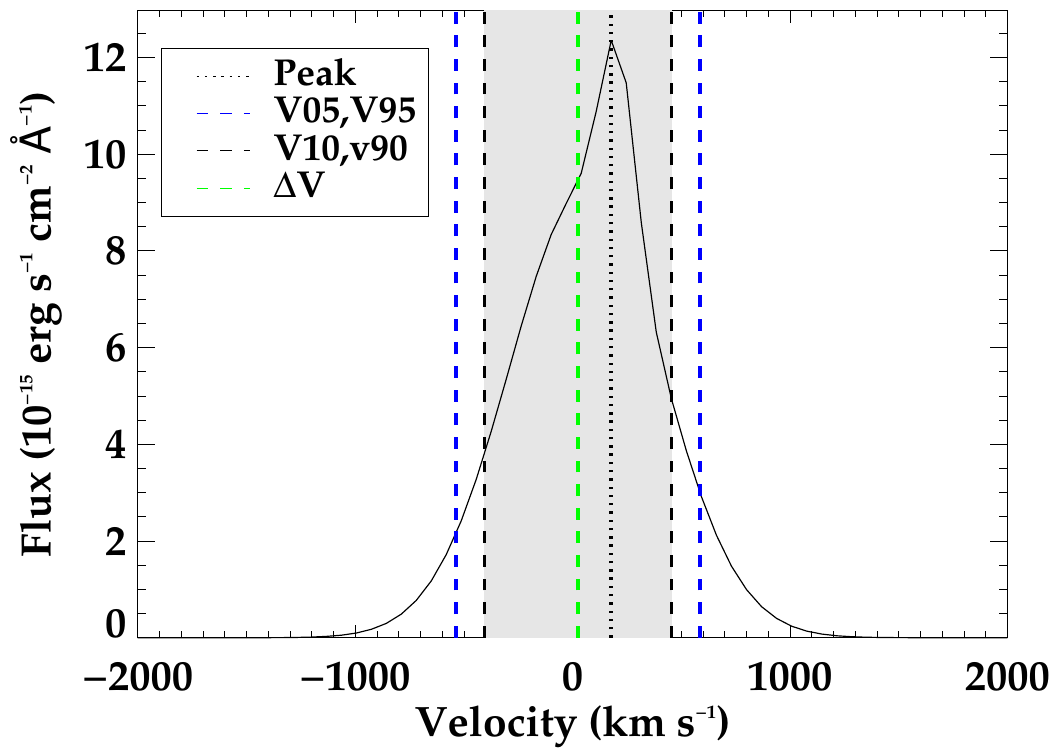}
\caption{J1356+10}
\end{subfigure}
\begin{subfigure}{0.49\textwidth}
 \includegraphics[width = 0.49\linewidth]{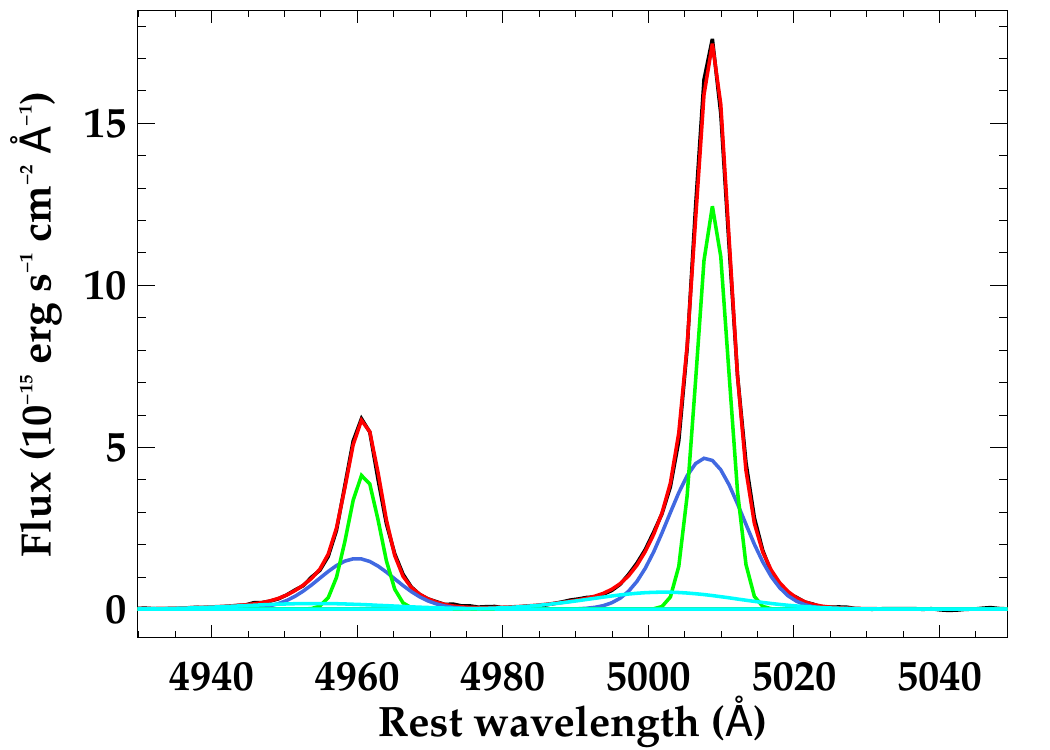}
 \includegraphics[width = 0.49\linewidth]{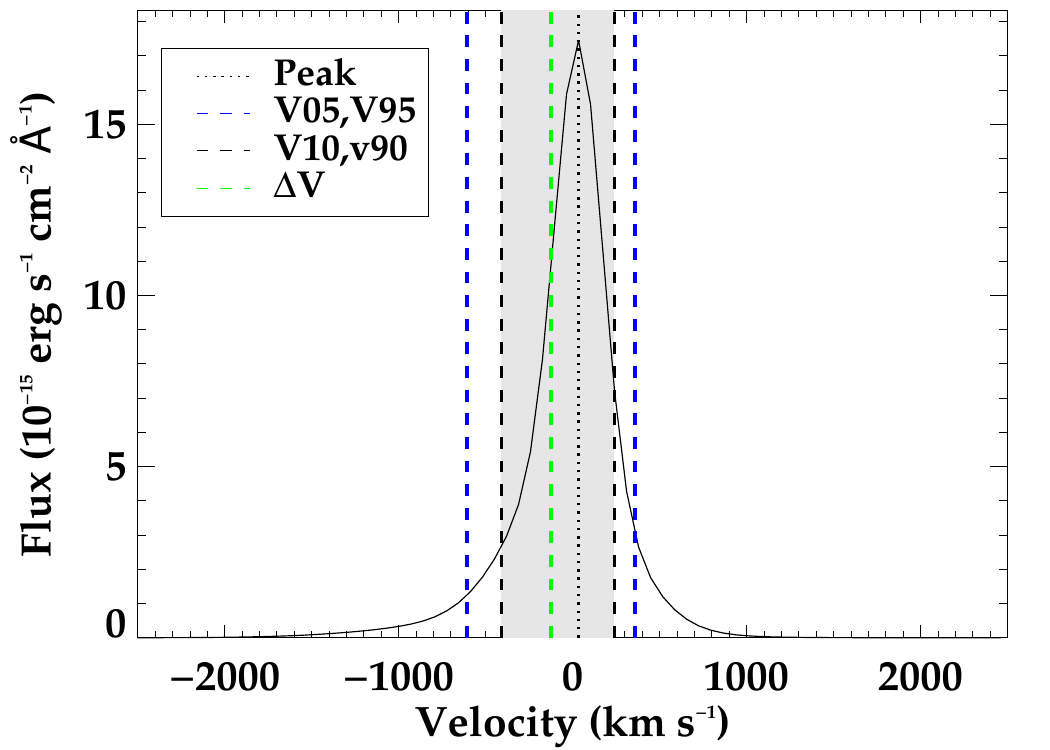}
\caption{J1405+40}
\end{subfigure}

\begin{subfigure}{0.49\textwidth}
 \includegraphics[width = 0.49\linewidth]{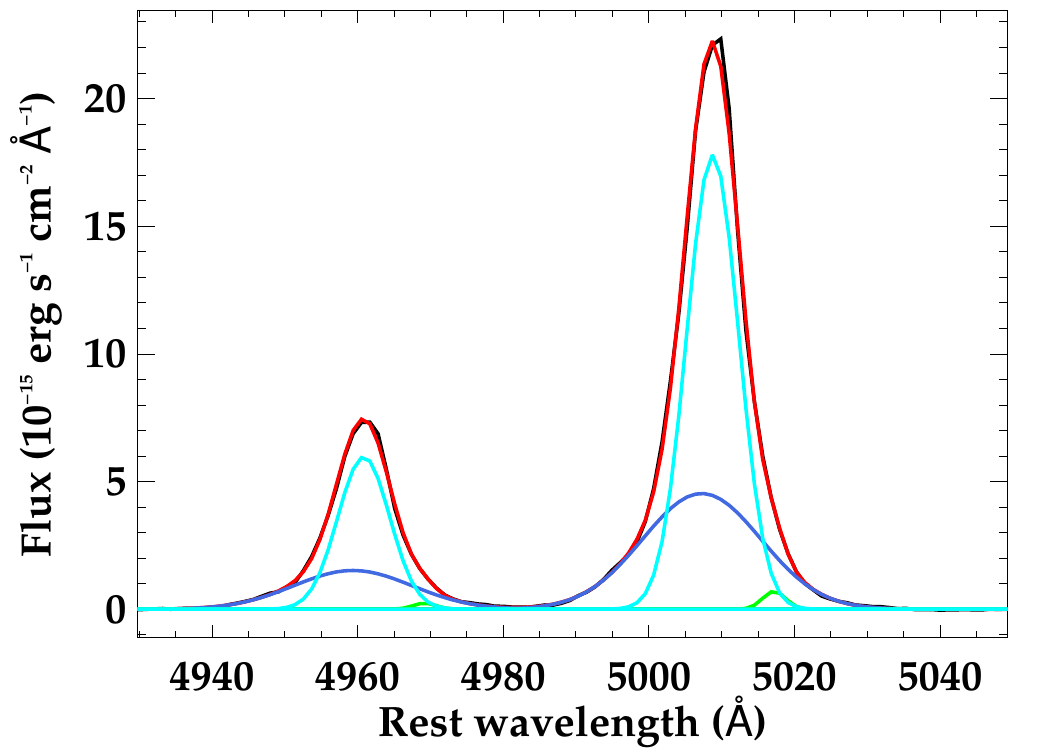}
 \includegraphics[width = 0.49\linewidth]{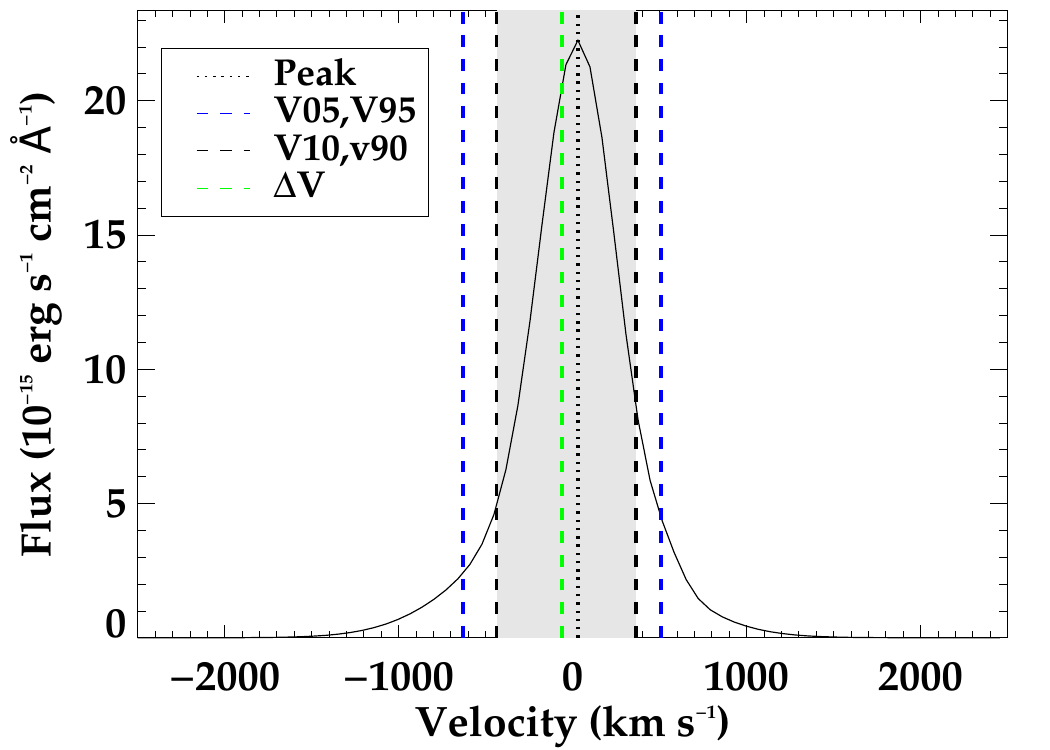}
\caption{J1430+13}
\end{subfigure}
\begin{subfigure}{0.49\textwidth}
 \includegraphics[width = 0.49\linewidth]{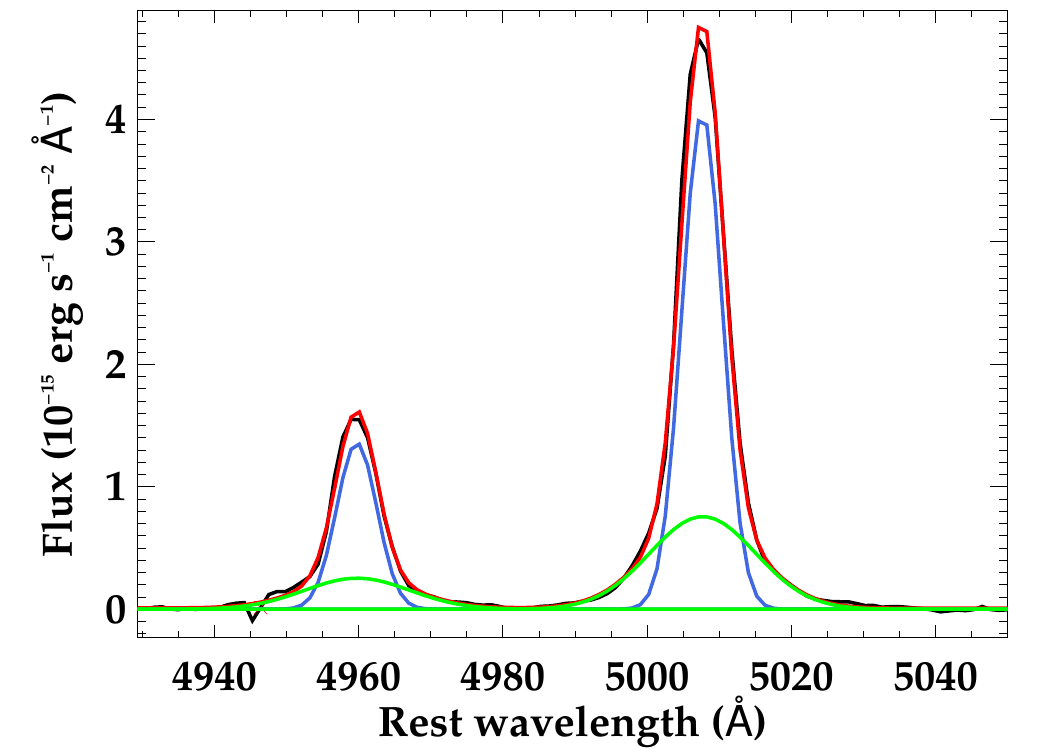}
 \includegraphics[width = 0.49\linewidth]{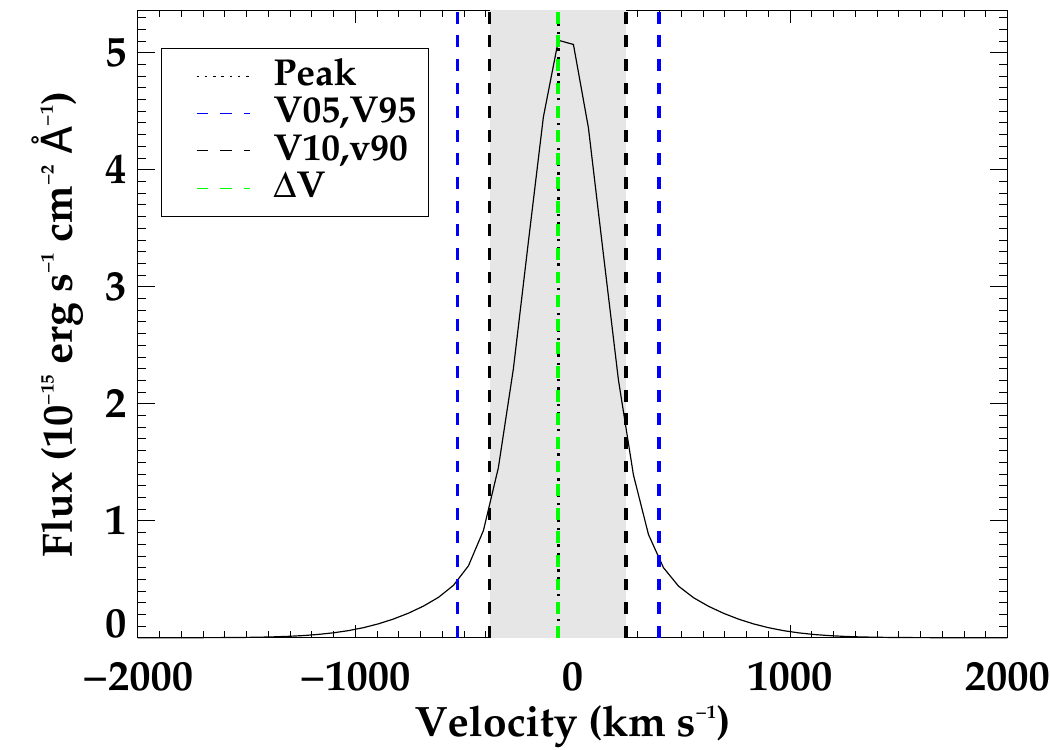}
\caption{J1436+13}
\end{subfigure}
\caption{Same as Figure \ref{fig:linefit_pg1}}
\label{fig:linefit_pg3}
\end{figure*}

\begin{figure*}
\centering
\begin{subfigure}{0.49\textwidth}
 \includegraphics[width = 0.49\linewidth]{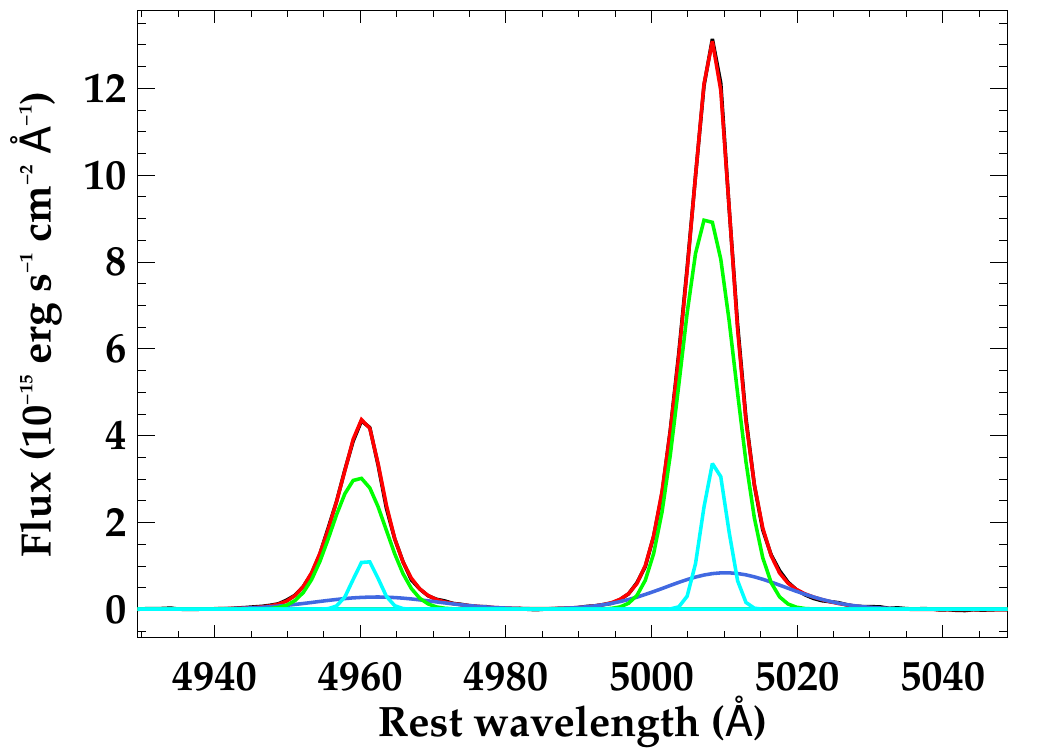}
 \includegraphics[width = 0.49\linewidth]{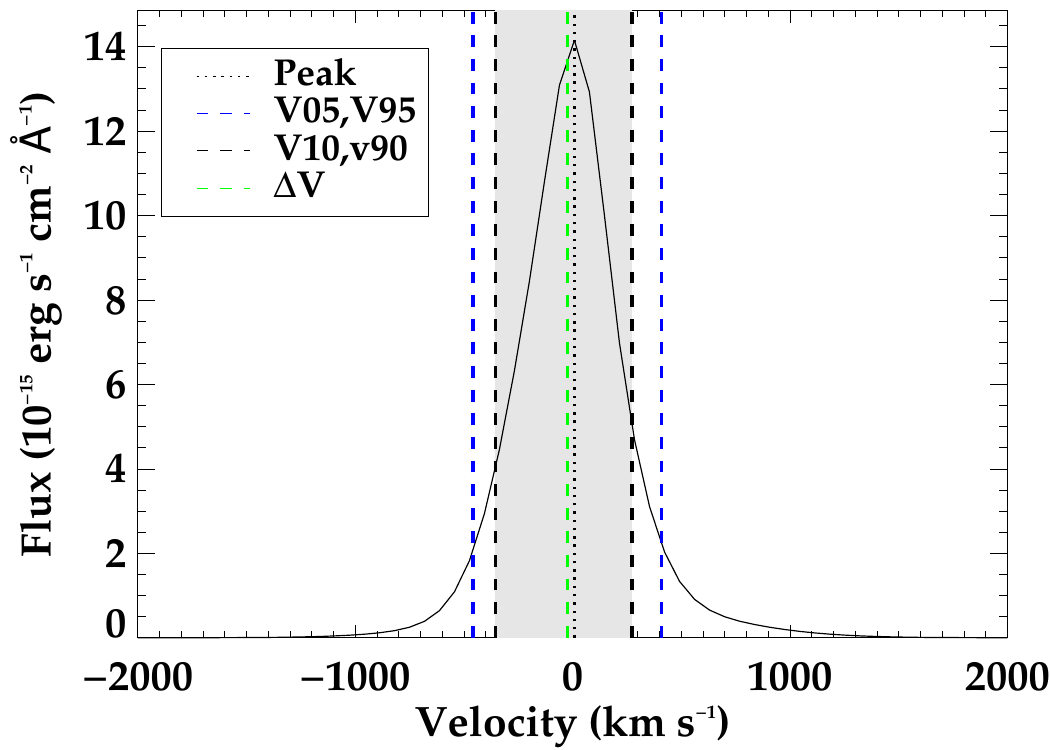}
\caption{J1437+30}
\end{subfigure}
\begin{subfigure}{0.49\textwidth}
 \includegraphics[width = 0.49\linewidth]{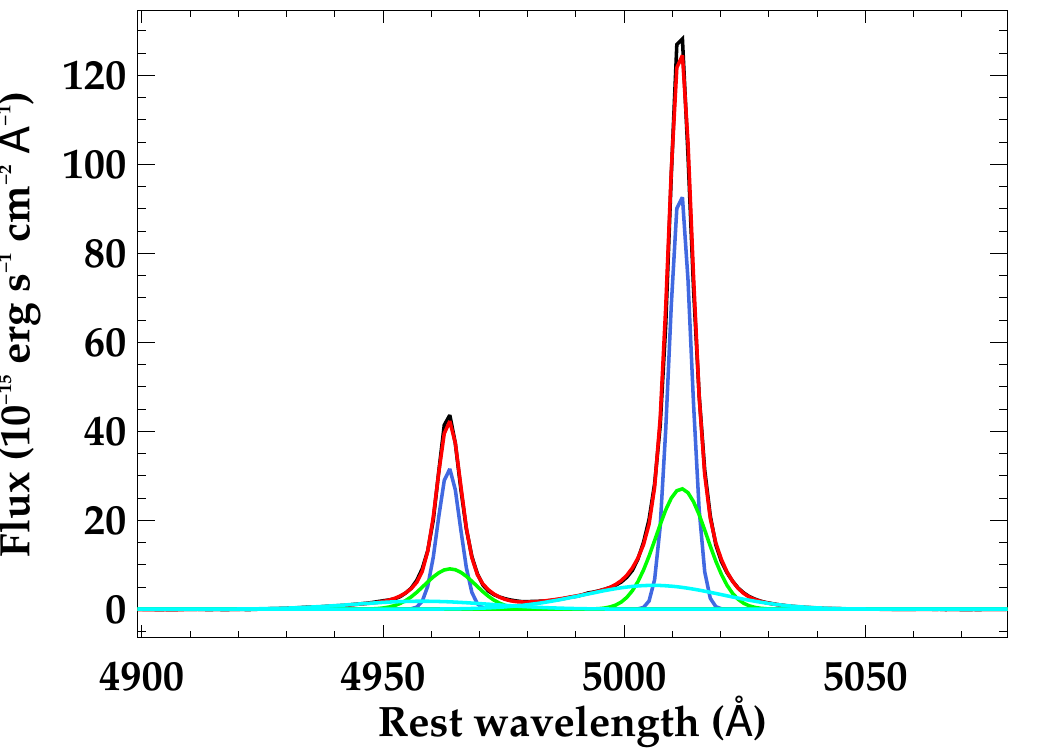}
 \includegraphics[width = 0.49\linewidth]{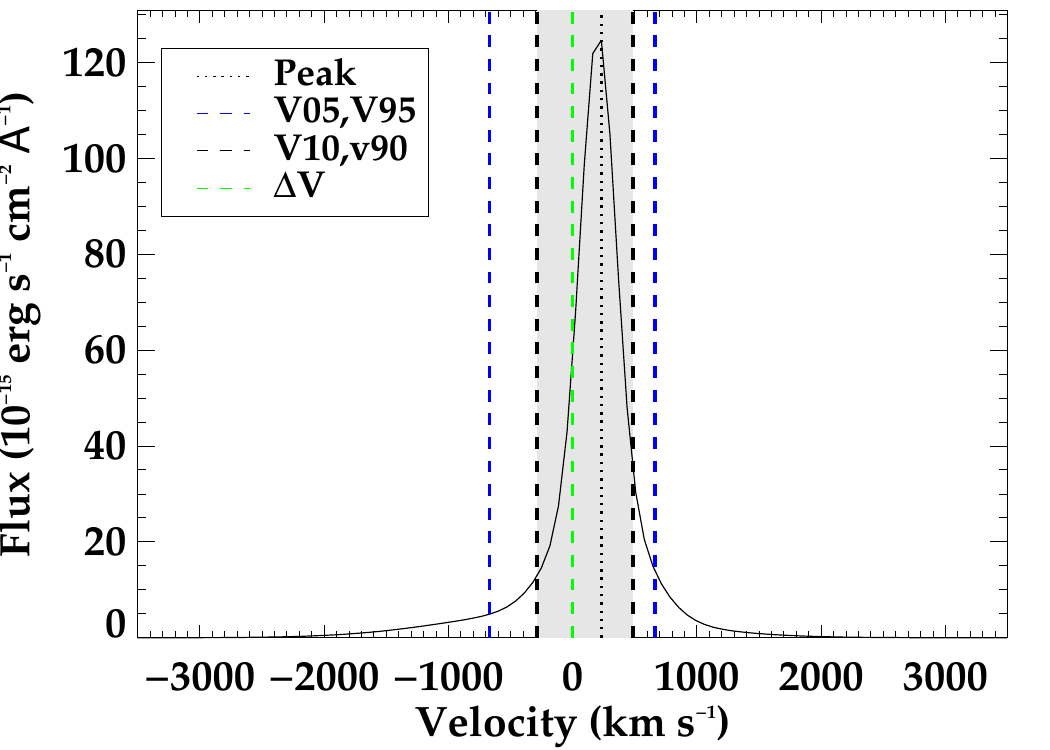}
\caption{J1440+53}
\end{subfigure}

\begin{subfigure}{0.49\textwidth}
 \includegraphics[width = 0.49\linewidth]{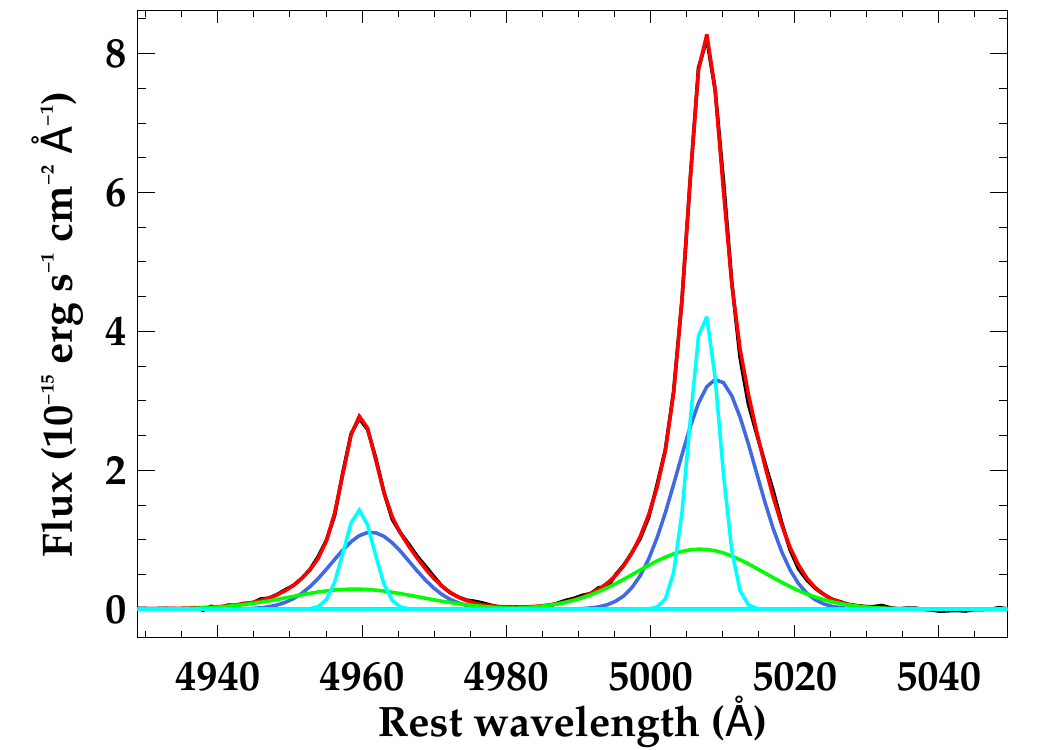}
 \includegraphics[width = 0.49\linewidth]{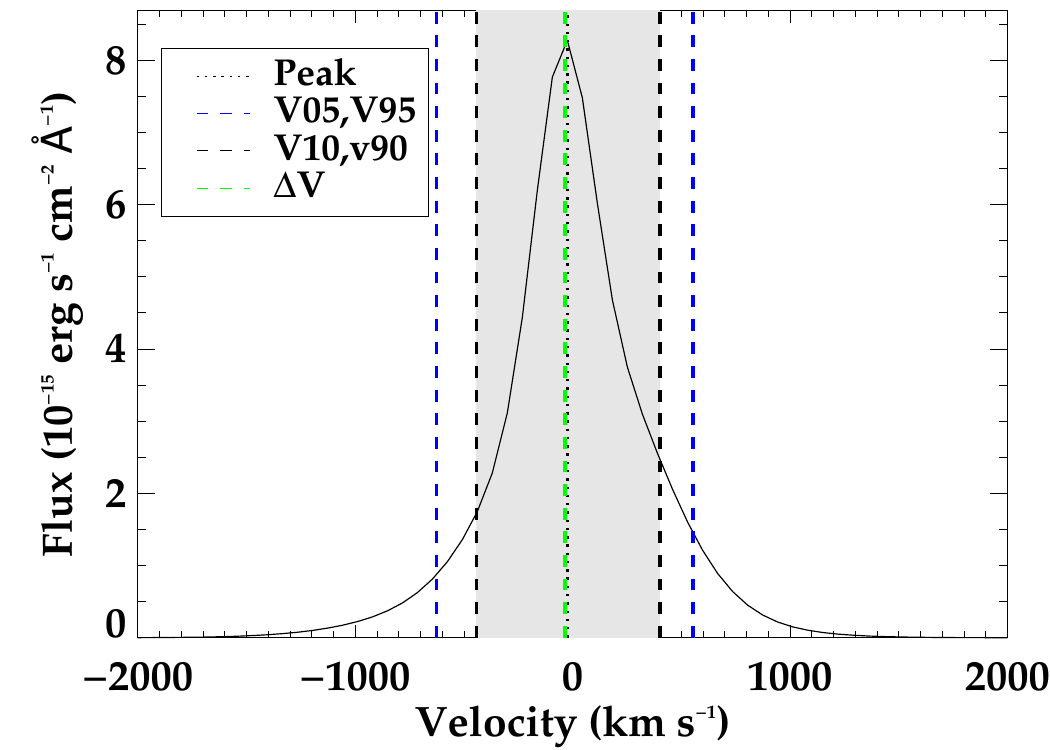}
\caption{J1455+32}
\end{subfigure}
\begin{subfigure}{0.49\textwidth}
 \includegraphics[width = 0.49\linewidth]{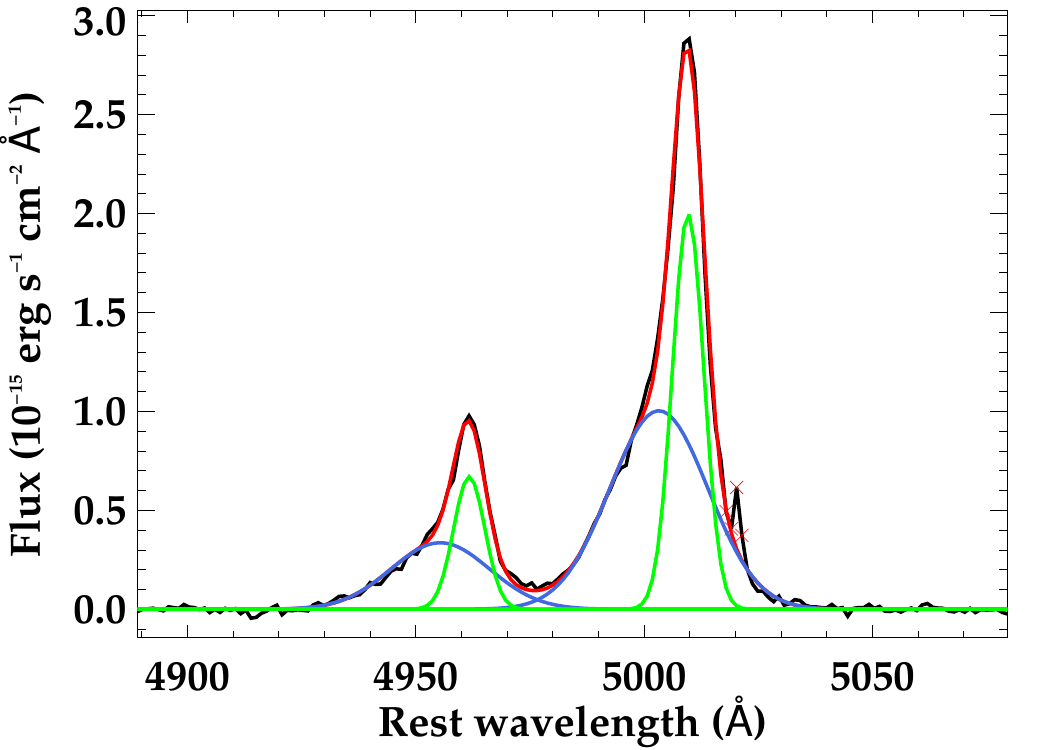}
 \includegraphics[width = 0.49\linewidth]{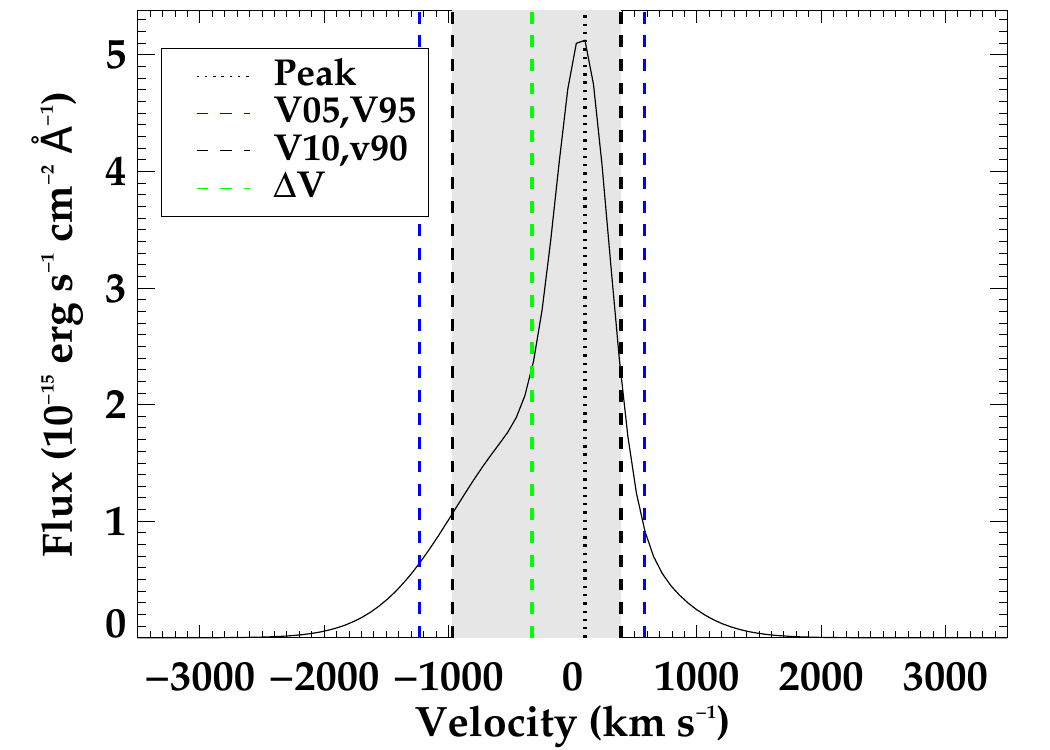}
\caption{J1509+04}
\end{subfigure}

\begin{subfigure}{0.49\textwidth}
 \includegraphics[width = 0.49\linewidth]{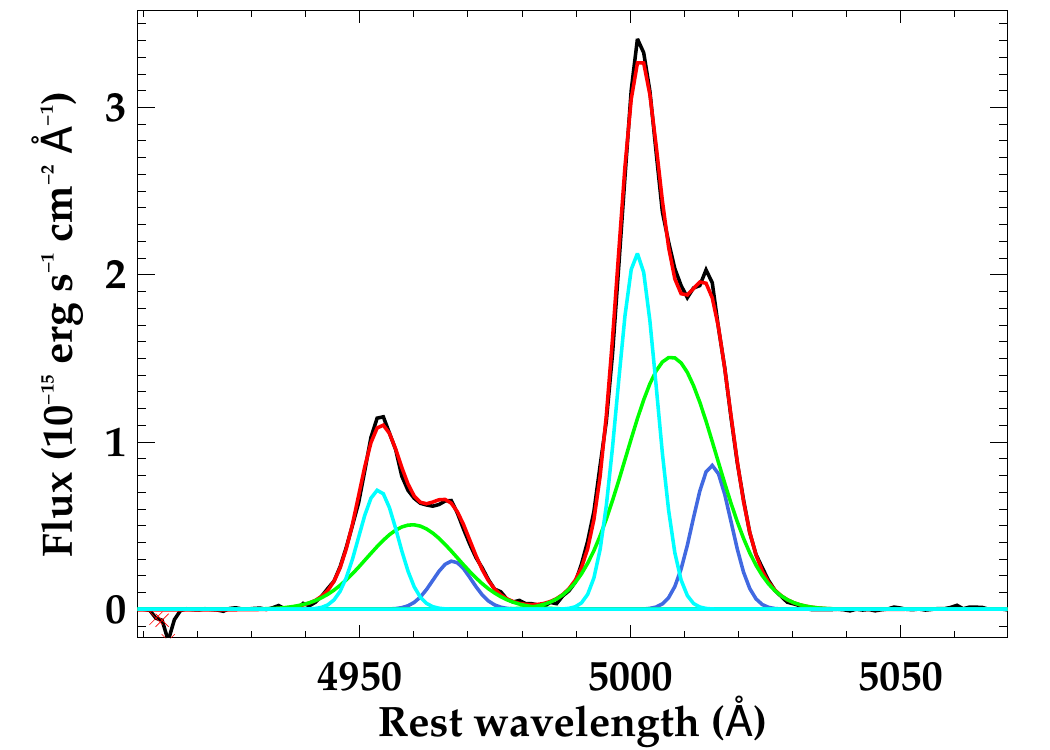}
 \includegraphics[width = 0.49\linewidth]{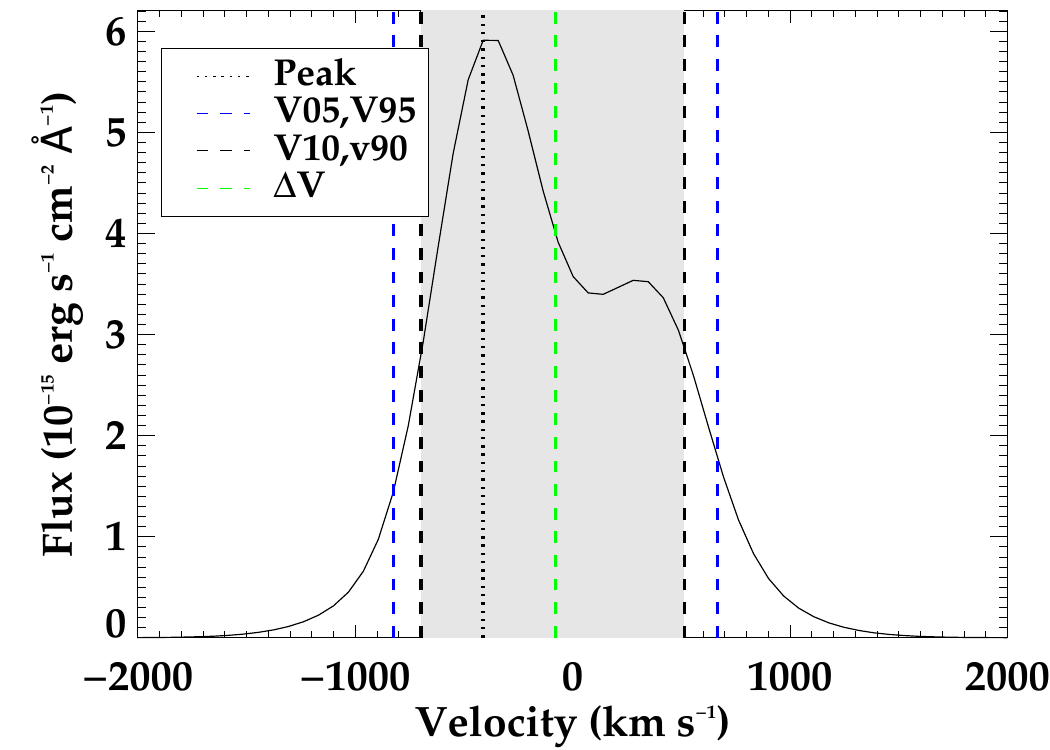}
\caption{J1517+33}
\end{subfigure}
\begin{subfigure}{0.49\textwidth}
 \includegraphics[width = 0.49\linewidth]{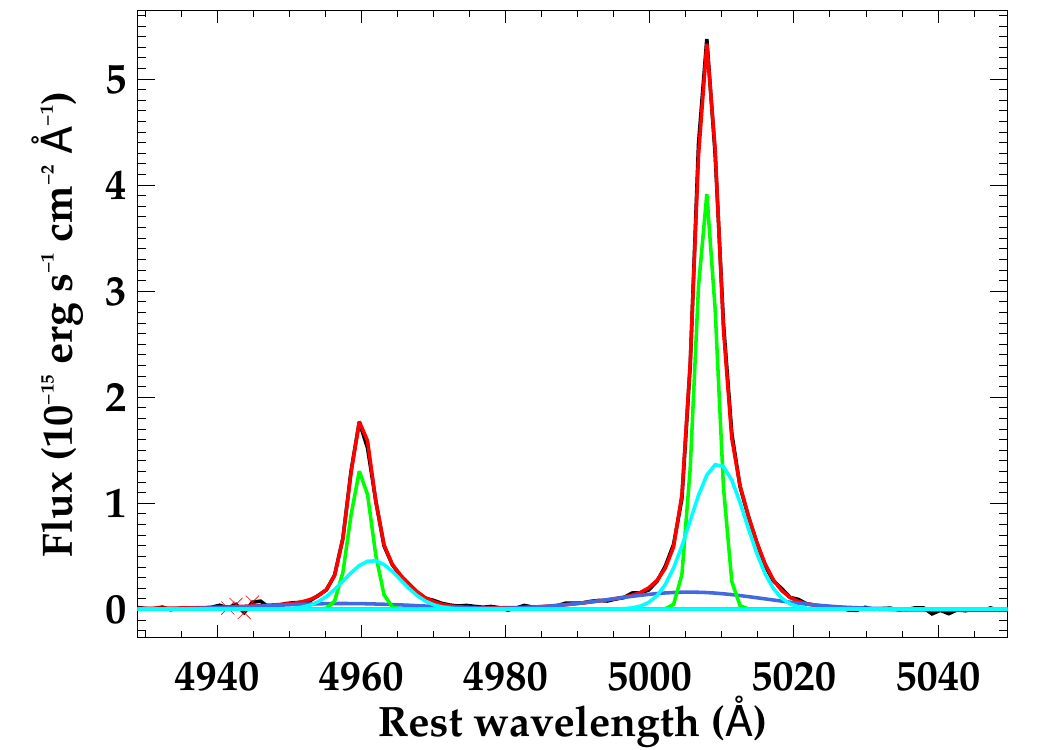}
 \includegraphics[width = 0.49\linewidth]{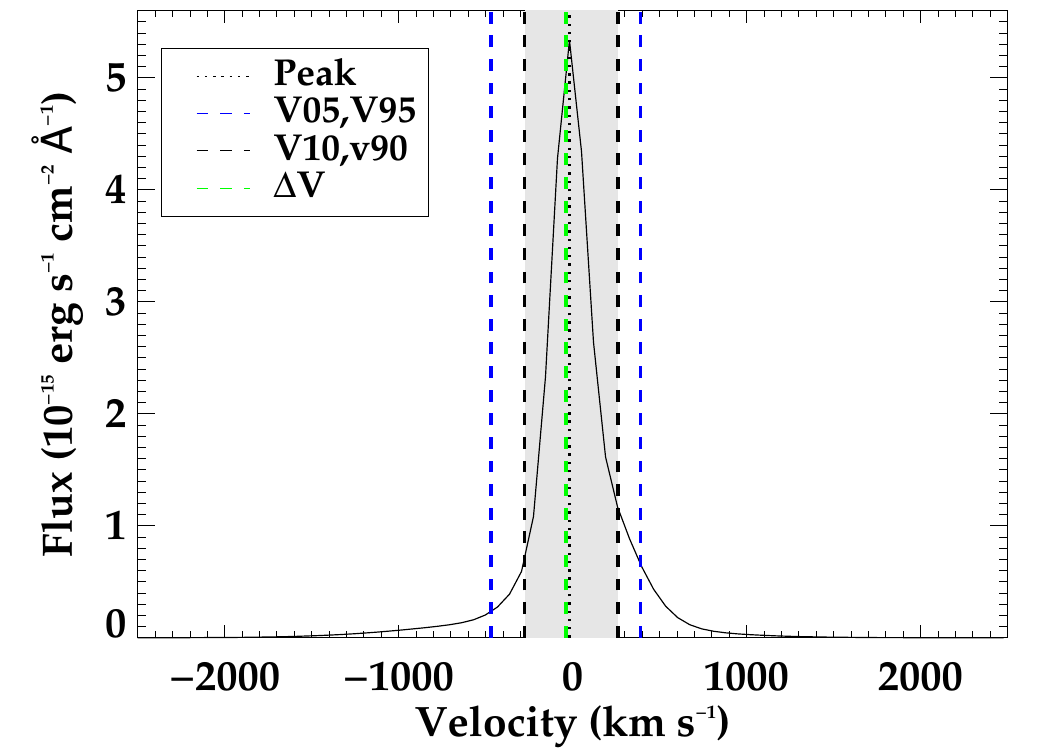}
\caption{J1533+35}
\end{subfigure}

\begin{subfigure}{0.49\textwidth}
 \includegraphics[width = 0.49\linewidth]{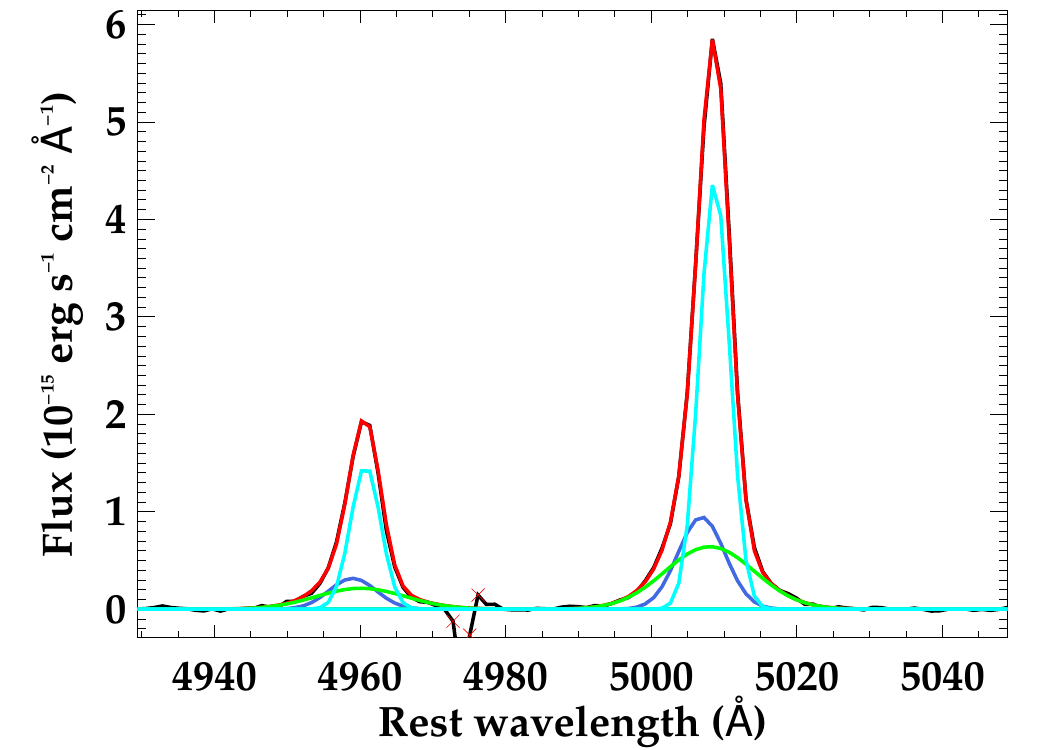}
 \includegraphics[width = 0.49\linewidth]{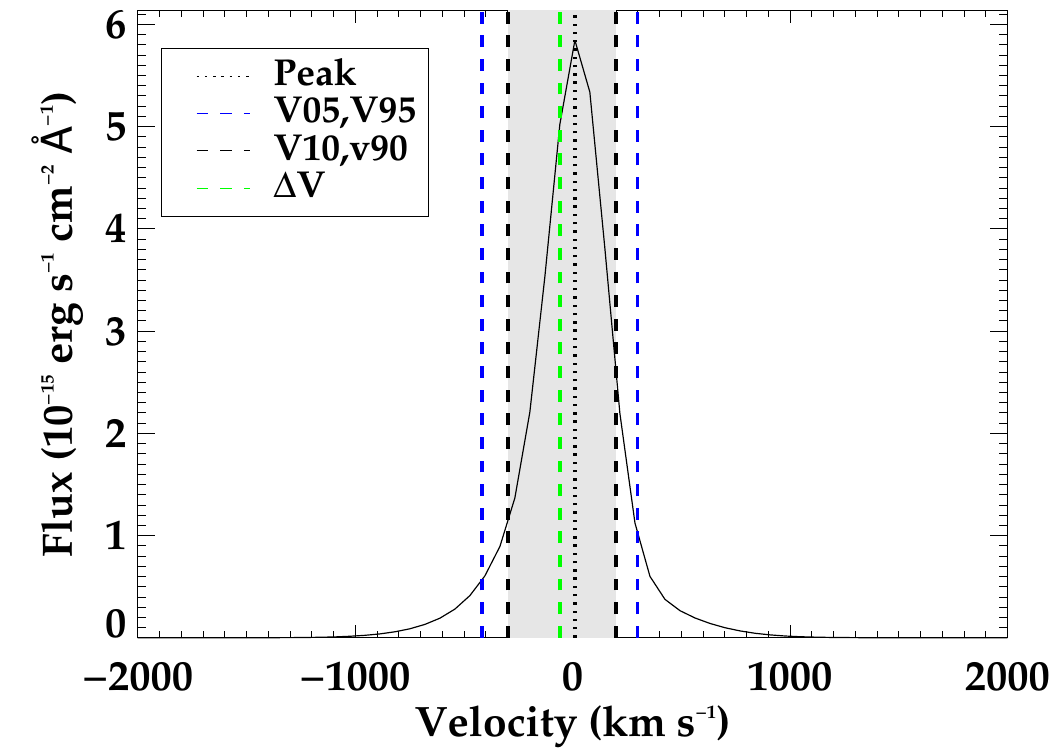}
\caption{J1548-01}
\end{subfigure}
\begin{subfigure}{0.49\textwidth}
 \includegraphics[width = 0.49\linewidth]{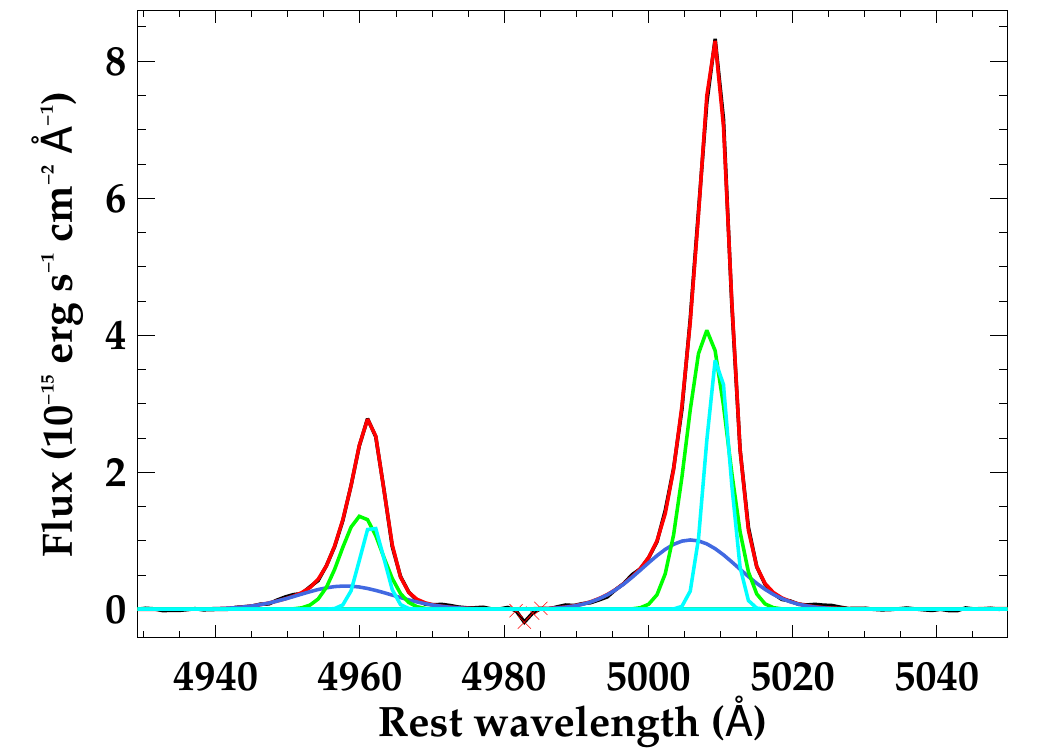}
 \includegraphics[width = 0.49\linewidth]{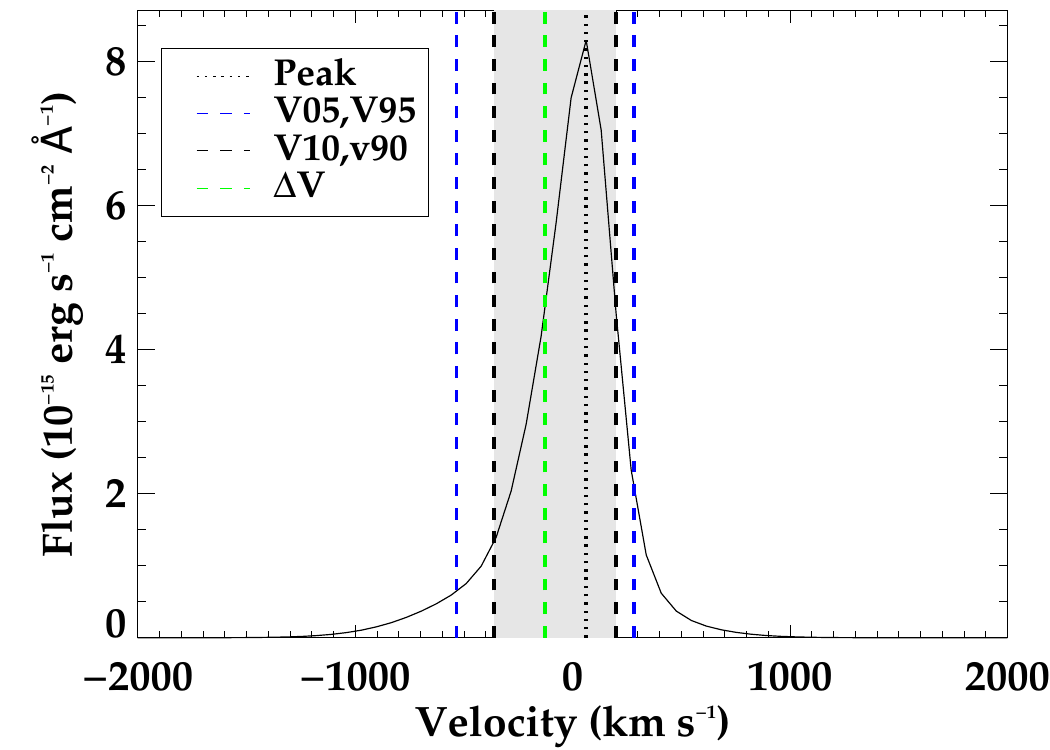}
\caption{J1558+35}
\end{subfigure}

\begin{subfigure}{0.49\textwidth}
 \includegraphics[width = 0.49\linewidth]{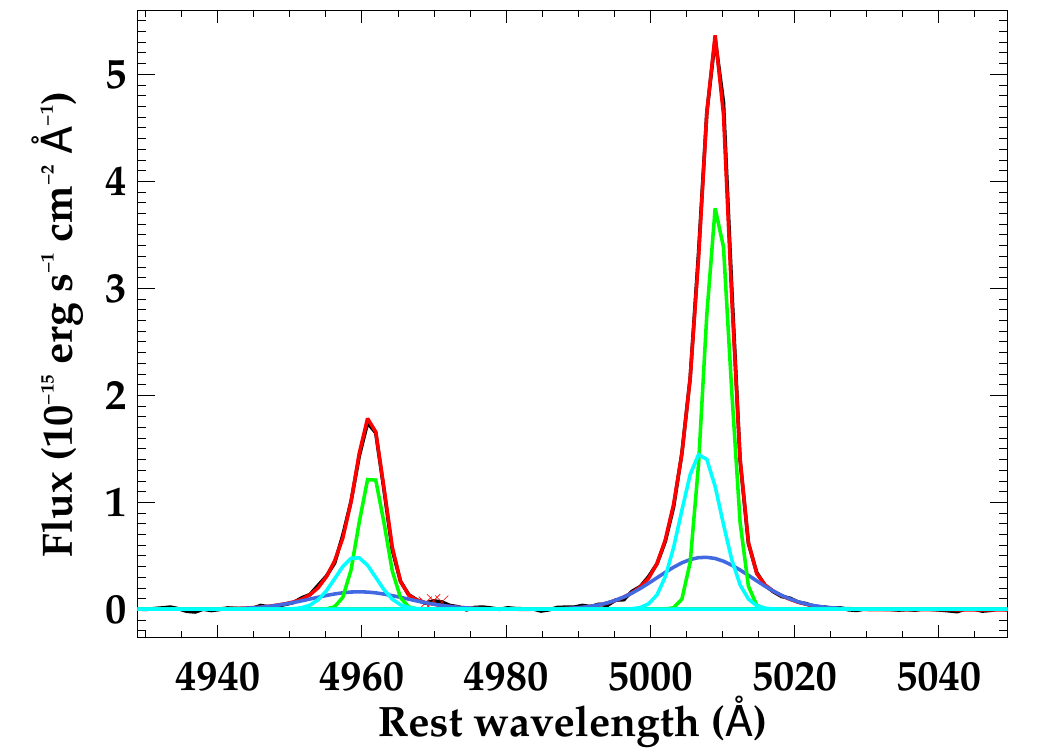}
 \includegraphics[width = 0.49\linewidth]{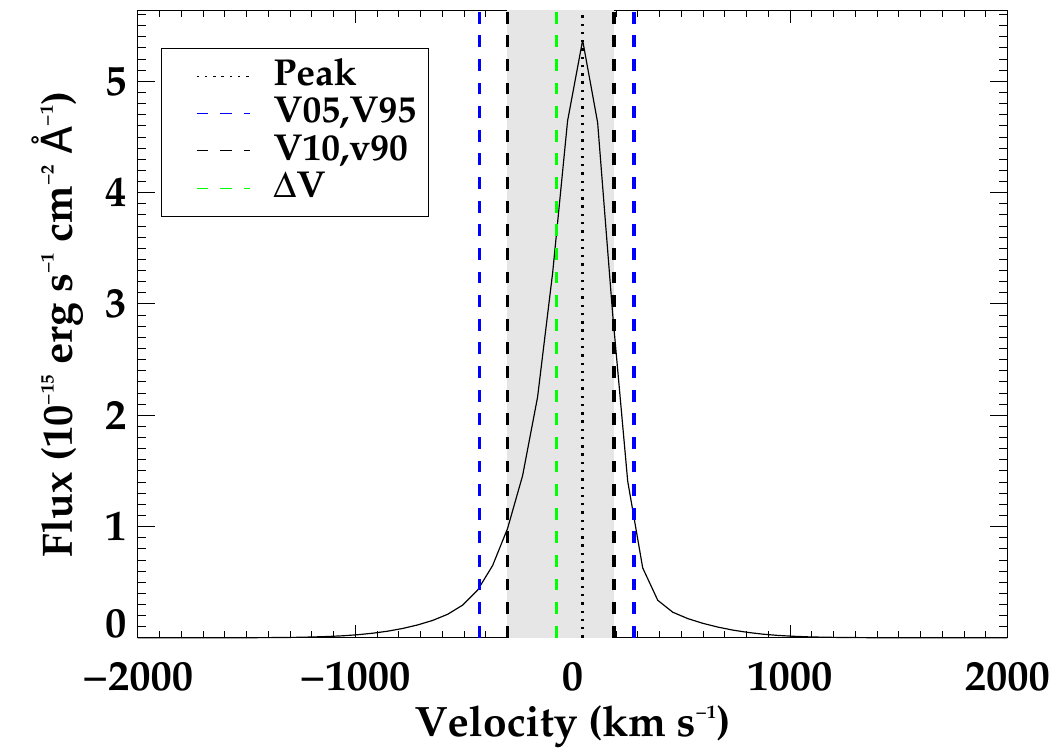}
\caption{J1624+33}
\end{subfigure}
\begin{subfigure}{0.49\textwidth}
 \includegraphics[width = 0.49\linewidth]{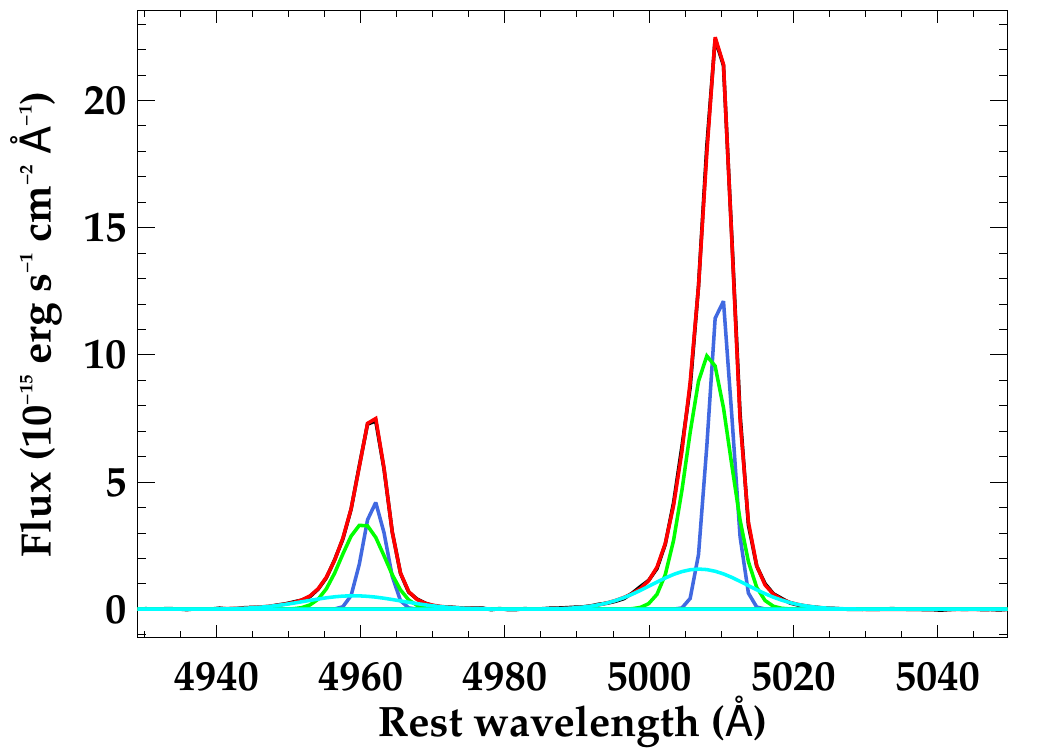}
 \includegraphics[width = 0.49\linewidth]{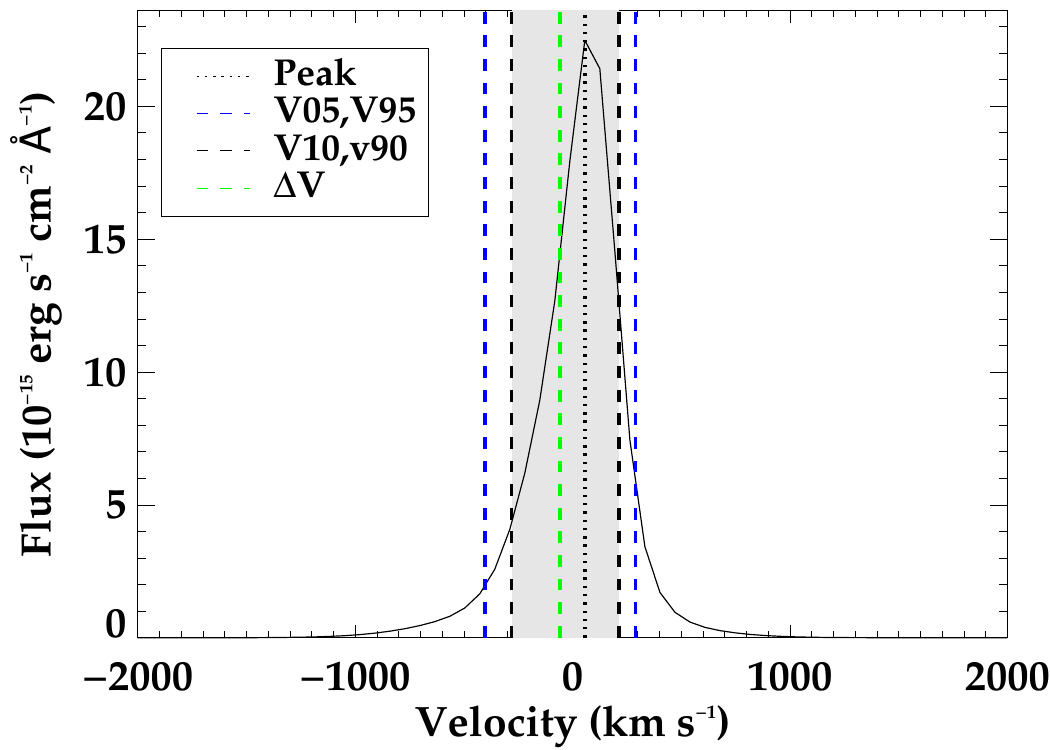}
\caption{J1653}
\end{subfigure}

\begin{subfigure}{0.49\textwidth}
 \includegraphics[width = 0.49\linewidth]{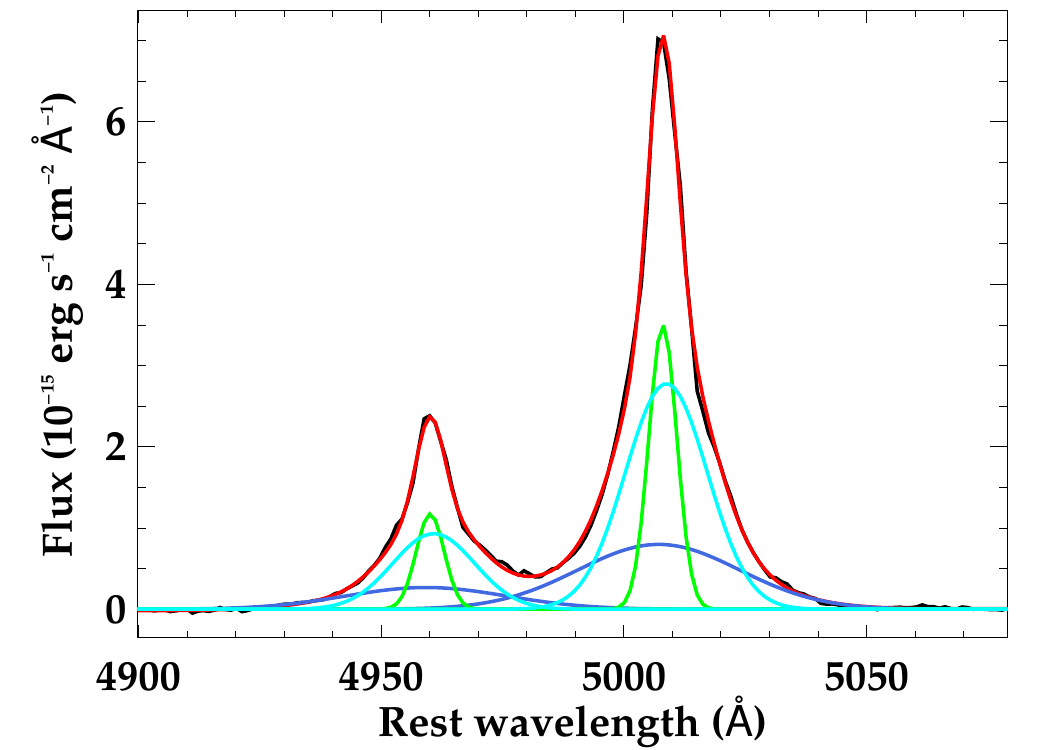}
 \includegraphics[width = 0.49\linewidth]{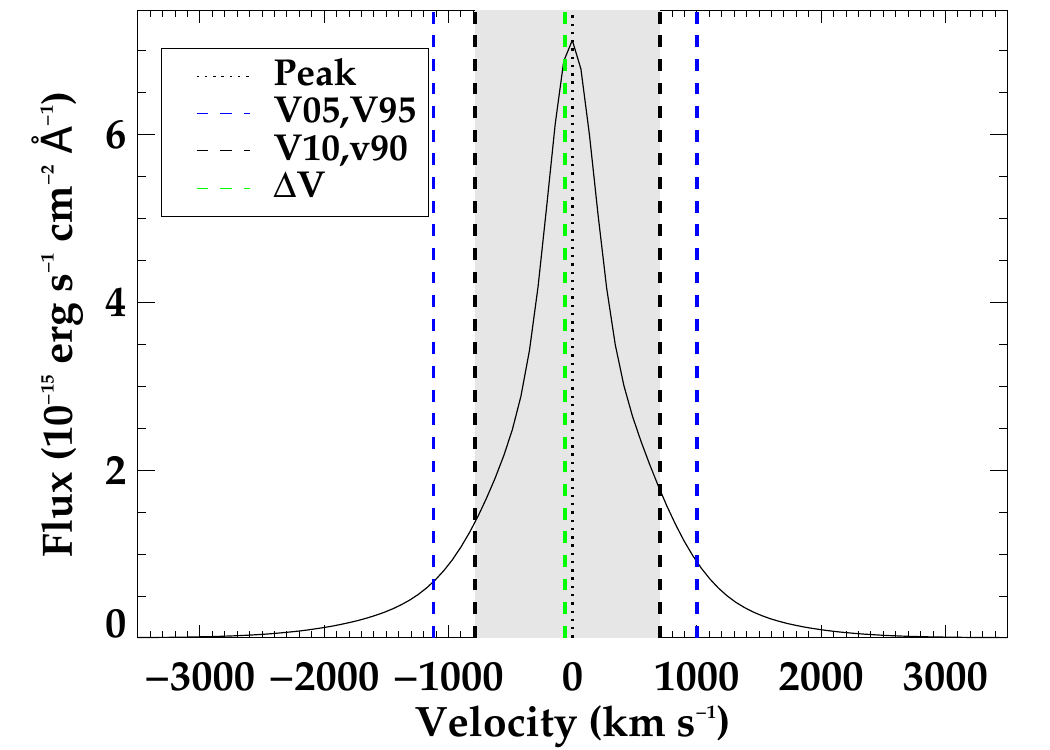}
\caption{J1713+57}
\end{subfigure}
\begin{subfigure}{0.49\textwidth}
 \includegraphics[width = 0.49\linewidth]{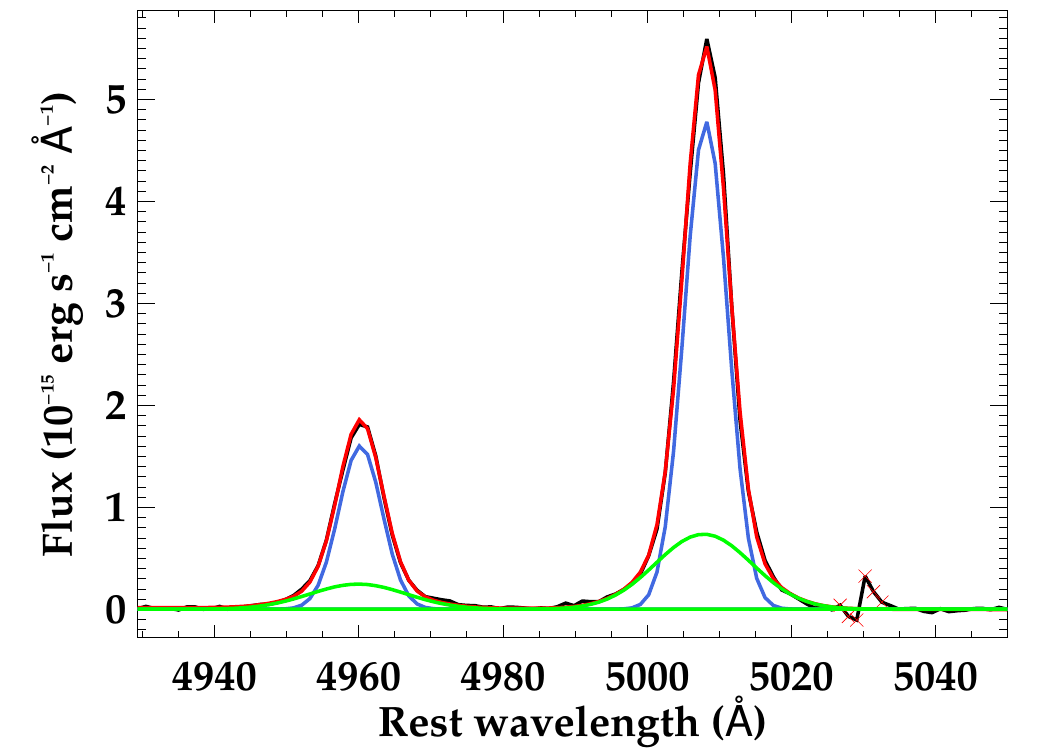}
 \includegraphics[width = 0.49\linewidth]{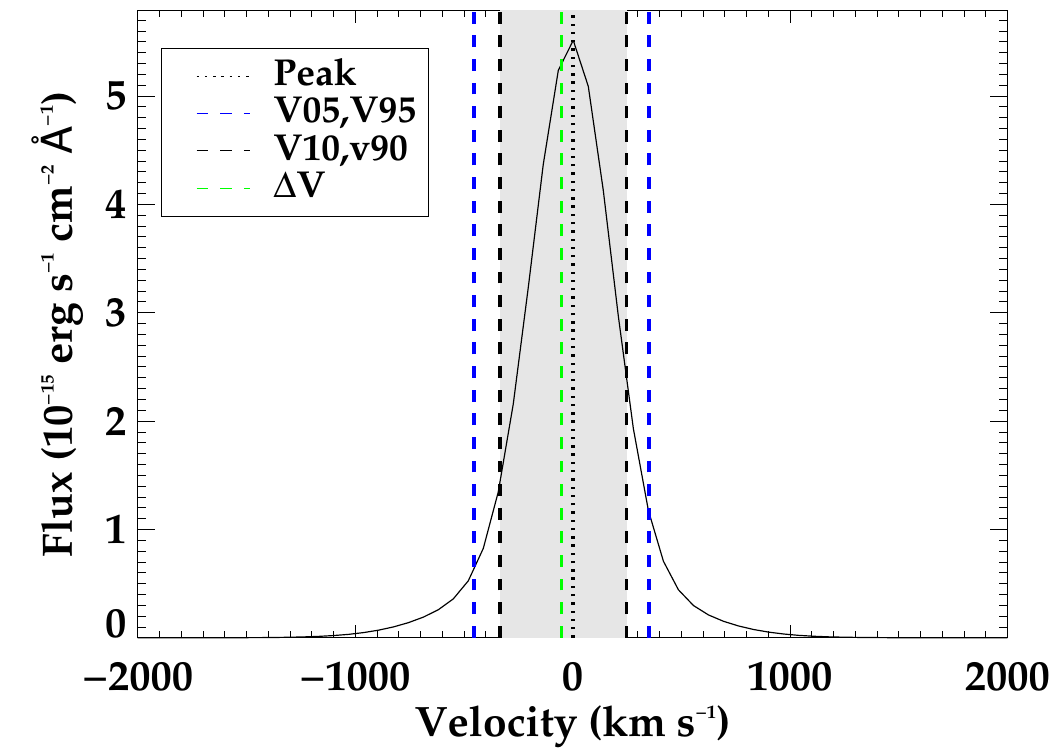}
\caption{J2154+11}
\end{subfigure}
\caption{Same as Figure \ref{fig:linefit_pg1}}
\label{fig:linefit_pg4}
\end{figure*}

\clearpage

\section{Tables of results}
\label{tabs}


\begin{table*}
\small
\centering
 \caption{The main properties of the \qfd\ sample: Columns 1 and 2 give the SDSS identifier and abbreviated identifier that will be used throughout this paper. Objects for which the BOSS spectra have been utilised are maked with an asterik in column 1. Columns 3, 4, and 5 list the SDSS spectroscopic redshift, luminosity distance, and cosmology corrected physical scale adopted in this paper. {\color {black} Column 6 lists the observed and extinction-corrected values of $\log L_{\rm [OIII]}$ taken from \citet{reyes08} and from \citet{kong18} respectively, the latter between parenthesis. Column 7 gives the values of $\log L_{\rm bol}$ derived from the extinction-corrected luminosities} using a bolometric correction factor of 454 \citep{lamastra09}. Column 8 lists the $\log L_{\rm 1.4 GHz}$ radio luminosity based on flux densities taken from the FIRST radio survey \citep{becker95} where available and otherwise from the NVSS survey \citep{condon98}. Colums 9 and 10 list black hole masses and Eddington ratios taken from \citet{kong18} and the final column gives the stellar mass of the host galaxy from \citet{pierce23}.}
   \label{tab:sample} 
\begin{tabular}{lcccccccccc}
    \hline
	\hline
SDSS ID & Short & z & $D_L$ & Scale & log \lo & log \lbol & $\log L_{\rm 1.4 GHz}$ & log $M_{\rm BH}$ & $\log \frac{L_{\rm bol}}{L_{\rm Edd}}$ & $\log \mbox{M}_*$\\ 
    & name & SDSS &  (Mpc)   & (kpc/ \arcsec) &   (\lsun) & (\ergs) & (W Hz$^{-1}$) & (\msun)& & (\msun) \\
\hline
J005230.59-011548.4	&	J0052-01	&	0.1348	&	635	&	2.390	&	8.58	(	8.66	)	&	44.90	&	22.98	&	$	7.67	\pm	0.4	$	&	$	-0.75	\pm	0.5	$	&	10.8	\\
J023224.24-081140.2	&	J0232-08	&	0.1001	&	461	&	1.846	&	8.60	(	8.87	)	&	45.11	&	22.96	&	$	7.46	\pm	0.3	$	&	$	-0.33	\pm	0.4	$	&	10.8	\\
J073142.37+392623.7	&	J0731+39	&	0.1103	&	511	&	2.010	&	8.59	(	9.20	)	&	45.44	&	23.14	&	$	7.47	\pm	0.7	$	&	$	-0.02	\pm	0.8	$	&	11.0	\\
J075940.95+505023.9	&	J0759+50	&	0.0544	&	243	&	1.058	&	8.83	(	9.34	)	&	45.58	&	23.52	&	$	8.24	\pm	0.7	$	&	$	-0.64	\pm	0.7	$	&	10.6	\\
J080224.34+464300.7*	&	J0802+46	&	0.1206	&	563	&	2.173	&	8.58	(	9.06	)	&	45.30	&	23.56	&	$	7.77	\pm	0.4	$	&	$	-0.46	\pm	0.4	$	&	11.1	\\
J080252.92+255255.5	&	J0802+25	&	0.0811	&	369	&	1.529	&	8.86	(	9.26	)	&	45.50	&	23.69	&	$	8.21	\pm	0.3	$	&	$	-0.69	\pm	0.3	$	&	11.3	\\
J080523.29+281815.7	&	J0805+28	&	0.1284	&	602	&	2.293	&	8.62	(	9.19	)	&	45.43	&	23.36	&	$	7.37	\pm	0.4	$	&	$	0.08	\pm	0.4	$	&	11.4	\\
J081842.35+360409.6	&	J0818+36	&	0.0758	&	343	&	1.438	&	8.53	(	9.02	)	&	45.26	&	22.50	&	$	7.88	\pm	0.4	$	&	$	-0.60	\pm	0.4	$	&	10.6	\\
J084135.09+010156.3	&	J0841+01	&	0.1106	&	513	&	2.015	&	8.87	(	9.44	)	&	45.68	&	23.08	&	$	8.26	\pm	0.4	$	&	$	-0.56	\pm	0.5	$	&	11.1	\\
J085810.63+312136.2	&	J0858+31	&	0.1387	&	655	&	2.448	&	8.53	(	8.88	)	&	45.12	&	23.00	&	$	7.25	\pm	0.4	$	&	$	-0.11	\pm	0.4	$	&	11.1	\\
J091544.18+300922.0	&	J0915+30	&	0.1298	&	609	&	2.314	&	8.78	(	9.06	)	&	45.30	&	23.24	&	$	8.47	\pm	0.3	$	&	$	-1.15	\pm	0.3	$	&	11.2	\\
J093952.75+355358.9*	&	J0939+35	&	0.1366	&	644	&	2.417	&	8.77	(	9.29	)	&	45.53	&	25.77	&	$	8.26	\pm	0.3	$	&	$	-0.72	\pm	0.4	$	&	11.0	\\
J094521.33+173753.2	&	J0945+17	&	0.1280	&	600	&	2.287	&	9.05	(	9.74	)	&	45.98	&	24.27	&	$	8.27	\pm	0.8	$	&	$	-0.27	\pm	0.8	$	&	10.9	\\
J101043.36+061201.4	&	J1010+06	&	0.0977	&	449	&	1.807	&	8.68	(	9.30	)	&	45.54	&	24.37	&	$	8.36	\pm	0.8	$	&	$	-0.80	\pm	0.8	$	&	11.4	\\
J101536.21+005459.4	&	J1015+00	&	0.1202	&	561	&	2.166	&	8.69	(	8.95	)	&	45.19	&	22.96	&	$	7.28	\pm	0.4	$	&	$	-0.07	\pm	0.4	$	&	10.9	\\
J101653.82+002857.2	&	J1016+00	&	0.1163	&	541	&	2.105	&	8.63	(	8.93	)	&	45.17	&	23.52	&	$	7.99	\pm	0.4	$	&	$	-0.80	\pm	0.4	$	&	11.0	\\
J103408.59+600152.2	&	J1034+60	&	0.0511	&	227	&	0.998	&	8.85	(	9.16	)	&	45.40	&	23.07	&	$	7.83	\pm	0.3	$	&	$	-0.41	\pm	0.3	$	&	11.1	\\
J103600.37+013653.5	&	J1036+01	&	0.1068	&	494	&	1.954	&	8.53	(	9.09	)	&	45.33	&	<22.45	&	$	8.10	\pm	0.3	$	&	$	-0.75	\pm	0.3	$	&	11.3	\\
J110012.39+084616.3	&	J1100+08	&	0.1004	&	462	&	1.851	&	9.20	(	9.60	)	&	45.84	&	24.18	&	$	7.82	\pm	0.4	$	&	$	0.04	\pm	0.5	$	&	11.4	\\
J113721.36+612001.1*	&	J1137+61	&	0.1112	&	516	&	2.025	&	8.64	(	8.77	)	&	45.01	&	25.16	&	$	8.37	\pm	0.4	$	&	$	-1.34	\pm	0.4	$	&	10.9	\\
J115245.66+101623.8	&	J1152+10	&	0.0699	&	315	&	1.335	&	8.72	(	8.94	)	&	45.18	&	22.67	&	$	7.91	\pm	0.3	$	&	$	-0.72	\pm	0.4	$	&	10.8	\\
J115759.50+370738.2	&	J1157+37	&	0.1282	&	601	&	2.290	&	8.62	(	9.27	)	&	45.51	&	23.30	&	$	8.26	\pm	0.3	$	&	$	-0.73	\pm	0.4	$	&	11.2	\\
J120041.39+314746.2*	&	J1200+31	&	0.1156	&	538	&	2.094	&	9.36	(	9.51	)	&	45.75	&	23.37	&	$	7.27	\pm	0.5	$	&	$	0.50	\pm	0.5	$	&	11.0	\\
J121839.40+470627.7	&	J1218+47	&	0.0939	&	430	&	1.744	&	8.58	(	8.95	)	&	45.19	&	22.71	&	$	7.56	\pm	0.4	$	&	$	-0.35	\pm	0.5	$	&	10.6	\\
J122341.47+080651.3	&	J1223+08	&	0.1393	&	658	&	2.457	&	8.81	(	9.24	)	&	45.48	&	<22.70	&	$	6.84	\pm	0.7	$	&	$	0.66	\pm	0.8	$	&	11.0	\\
J123843.44+092736.6	&	J1238+09	&	0.0829	&	377	&	1.559	&	8.51	(	8.66	)	&	44.90	&	22.30	&	$	8.76	\pm	0.3	$	&	$	-1.84	\pm	0.4	$	&	11.3	\\
J124136.22+614043.4	&	J1241+61	&	0.1353	&	637	&	2.397	&	8.51	(	9.29	)	&	45.53	&	23.41	&	$	7.06	\pm	0.7	$	&	$	0.49	\pm	0.8	$	&	11.4	\\
J124406.61+652925.2	&	J1244+65	&	0.1071	&	495	&	1.959	&	8.52	(	9.36	)	&	45.60	&	23.36	&	$	8.18	\pm	0.8	$	&	$	-0.56	\pm	0.8	$	&	11.3	\\
J130038.09+545436.8	&	J1300+54	&	0.0883	&	403	&	1.651	&	8.94	(	9.14	)	&	45.38	&	22.67	&	$	6.94	\pm	0.4	$	&	$	0.46	\pm	0.4	$	&	10.8	\\
J131639.74+445235.0	&	J1316+44	&	0.0906	&	414	&	1.689	&	8.65	(	9.13	)	&	45.37	&	22.96	&	$	7.59	\pm	0.8	$	&	$	-0.20	\pm	0.8	$	&	11.7	\\
J134733.36+121724.3	&	J1347+12	&	0.1204	&	562	&	2.169	&	8.70	(	9.28	)	&	45.52	&	26.25	&	$	-	$	&	$	-	$	&	11.7	\\				
J135617.79-023101.5	&	J1356-02	&	0.1344	&	633	&	2.384	&	8.53	(	9.32	)	&	45.56	&	22.92	&	$	7.48	\pm	0.7	$	&	$	0.10	\pm	0.8	$	&	11.1	\\
J135646.10+102609.0	&	J1356+10	&	0.1232	&	576	&	2.213	&	9.21	(	9.29	)	&	45.53	&	24.36	&	$	8.58	\pm	0.3	$	&	$	-1.03	\pm	0.4	$	&	11.3	\\
J140541.21+402632.6	&	J1405+40	&	0.0806	&	366	&	1.520	&	8.78	(	9.20	)	&	45.44	&	23.44	&	$	7.13	\pm	0.7	$	&	$	0.33	\pm	0.8	$	&	10.8	\\
J143029.88+133912.0	&	J1430+13	&	0.0851	&	388	&	1.597	&	9.08	(	9.58	)	&	45.82	&	23.67	&	$	8.19	\pm	0.4	$	&	$	-0.35	\pm	0.4	$	&	11.1	\\
J143607.21+492858.6*	&	J1436+13	&	0.1280	&	600	&	2.287	&	8.61	(	8.88	)	&	45.12	&	23.46	&	$	8.30	\pm	0.3	$	&	$	-1.17	\pm	0.4	$	&	11.0	\\
J143737.85+301101.1	&	J1437+30	&	0.0922	&	422	&	1.716	&	8.82	(	9.20	)	&	45.44	&	24.14	&	$	8.40	\pm	0.3	$	&	$	-0.94	\pm	0.3	$	&	11.2	\\
J144038.10+533015.9	&	J1440+53	&	0.0370	&	163	&	0.735	&	8.94	(	9.31	)	&	45.55	&	23.27	&	$	7.16	\pm	0.8	$	&	$	0.41	\pm	0.8	$	&	10.6	\\
J145519.41+322601.8	&	J1455+32	&	0.0873	&	398	&	1.634	&	8.64	(	8.88	)	&	45.12	&	22.78	&	$	7.72	\pm	0.3	$	&	$	-0.58	\pm	0.4	$	&	10.6	\\
J150904.22+043441.8	&	J1509+04	&	0.1114	&	517	&	2.028	&	8.56	(	9.79	)	&	46.03	&	23.81	&	$	8.27	\pm	0.8	$	&	$	-0.22	\pm	0.8	$	&	10.9	\\
J151709.20+335324.7	&	J1517+33	&	0.1353	&	637	&	2.397	&	8.91	(	9.51	)	&	45.75	&	24.75	&	$	8.64	\pm	0.4	$	&	$	-0.87	\pm	0.4	$	&	11.5	\\
J153338.03+355708.1	&	J1533+35	&	0.1286	&	603	&	2.296	&	8.56	(	8.71	)	&	44.95	&	<22.62	&	$	7.64	\pm	0.5	$	&	$	-0.67	\pm	0.5	$	&	10.9	\\
J154832.37-010811.8	&	J1548-01	&	0.1215	&	567	&	2.187	&	8.52	(	9.01	)	&	45.25	&	<22.57	&	$	7.61	\pm	0.8	$	&	$	-0.34	\pm	0.8	$	&	10.8	\\
J155829.36+351328.6	&	J1558+35	&	0.1195	&	557	&	2.155	&	8.77	(	9.03	)	&	45.27	&	23.17	&	$	7.60	\pm	0.7	$	&	$	-0.31	\pm	0.7	$	&	10.9	\\
J162436.40+334406.7	&	J1624+33	&	0.1224	&	572	&	2.200	&	8.56	(	8.87	)	&	45.11	&	22.94	&	$	7.66	\pm	0.4	$	&	$	-0.53	\pm	0.4	$	&	11.0	\\
J165315.05+234942.9	&	J1653+23	&	0.1034	&	477	&	1.900	&	9.00	(	9.34	)	&	45.58	&	23.29	&	$	7.85	\pm	0.4	$	&	$	-0.25	\pm	0.4	$	&	11.0	\\
J171350.32+572954.9	&	J1713+57	&	0.1128	&	524	&	2.050	&	8.99	(	9.41	)	&	45.65	&	23.37	&	$	7.37	\pm	0.4	$	&	$	0.29	\pm	0.4	$	&	11.1	\\
J215425.74+113129.4	&	J2154+11	&	0.1092	&	506	&	1.993	&	8.54	(	9.12	)	&	45.36	&	23.32	&	$	7.88	\pm	0.4	$	&	$	-0.51	\pm	0.4	$	&	10.9	\\
\hline
\end{tabular}
\end{table*}

\clearpage

\begin{table*}
\tiny
\centering
\caption{The results of the {\sc starlight} spectral fitting. Column 1 gives the abbreviated name, column 2 denotes whether the higher order Balmer absorption lines were detected before nebular subtraction, and column 3 gives the percentage of the total flux attributed to the nebular component below 3646 \AA. Where possible, the flux was measured in the wavelength bin 3540 -- 3640 \AA, otherwise it was measured between the shortest wavelength and 3640\AA. Columns 4, 5, and 6 give the percentage of the total flux, before reddening is applied, associated with the YSP, ISP, and OSP respectively, measured in the normalising bin (4190 -- 4210 \AA). Columns 8 and 9 give the A$_{\rm V}$ values used by {\sc starlight} for the fitting and the additional reddening that was applied to templates with ages <7 Myr (where applicable). The final column gives the metallicity allocated to the OSP, ISP, and YSP respectively: H=super-solar, S=Solar, and L=sub-solar.}
\label{tab:starlight}
\begin{tabular}{lccccccccccc}
\hline
\hline
Name	&	Balmer	&	Nebular	&	YSP	&	ISP	&	OSP	&	log(SFR)	&	$\Delta V_{SL}$	&	$\sigma_{SL}$	&		A$_{\rm V}$	&	YSP A$_{\rm V}$	&	Metallicity	\\																															
	&	Lines?	&	(per cent)	&	(per cent)	&	(per cent)	&	(per cent)	&	(\msunyr)	&	(\kms)	&	(\kms)	&		(mag)	&	(mag)	&		\\		
\hline
J0052-01	&	N	&	38	&	$	7	\pm	7	$	&	$	48	\pm	10	$	&	$	35	\pm	15	$	&	$	0.48	\pm	3.9	$	&	$	-8	\pm	21	$	&	$	181	\pm	25	$	&	$	0.465	\pm	0.14	$	&	$	3.312	\pm	1.72	$	&	S,L,H	\\
J0232-08	&	Y	&	23	&	$	29	\pm	2	$	&	$	24	\pm	3	$	&	$	37	\pm	5	$	&	$	1.52	\pm	0.1	$	&	$	41	\pm	8	$	&	$	164	\pm	8	$	&	$	0.025	\pm	0.00	$	&	$	4.057	\pm	0.06	$	&	S,S,S	\\
J0731+39	&	Y	&	19	&	$	42	\pm	4	$	&	$	11	\pm	3	$	&	$	42	\pm	4	$	&	$	0.68	\pm	0.2	$	&	$	30	\pm	10	$	&	$	142	\pm	9	$	&	$	0.621	\pm	0.03	$	&	$	2.299	\pm	0.40	$	&	L,S,H	\\
J0759+50	&	Y	&	89	&	$	18	\pm	2	$	&	$	15	\pm	2	$	&	$	57	\pm	3	$	&	$	0.94	\pm	0.0	$	&	$	-49	\pm	7	$	&	$	146	\pm	10	$	&	$	0.397	\pm	0.03	$	&	$	3.182	\pm	0.12	$	&	L,S,S	\\
J0802+25	&	Y	&	22	&	$	41	\pm	1	$	&	$	16	\pm	1	$	&	$	34	\pm	3	$	&	$	1.65	\pm	0.1	$	&	$	-16	\pm	11	$	&	$	198	\pm	5	$	&	$	0.257	\pm	0.04	$	&	$	3.148	\pm	0.13	$	&	S,L,L	\\
J0802+46	&	Y	&	40	&	$	11	\pm	5	$	&	$	13	\pm	3	$	&	$	66	\pm	5	$	&	$	0.86	\pm	3.5	$	&	$	-248	\pm	10	$	&	$	171	\pm	8	$	&	$	0.903	\pm	0.02	$	&	$	2.922	\pm	0.70	$	&	L,H,H	\\
J0805+28	&	Y	&	24	&	$	37	\pm	3	$	&	$	28	\pm	3	$	&	$	28	\pm	2	$	&	$	0.68	\pm	0.1	$	&	$	84	\pm	9	$	&	$	158	\pm	9	$	&	$	0.669	\pm	0.02	$	&	$			0.00	$	&	L,L,S	\\
J0818+36	&	N	&	35	&	$	7	\pm	4	$	&	$	33	\pm	4	$	&	$	51	\pm	3	$	&	$	-0.06	\pm	1.9	$	&	$	-10	\pm	13	$	&	$	181	\pm	13	$	&	$	0.710	\pm	0.03	$	&	$	1.970	\pm	0.46	$	&	L,S,S	\\
J0841+01	&	N	&	30	&	$	31	\pm	3	$	&	$	20	\pm	2	$	&	$	38	\pm	3	$	&	$	0.25	\pm	0.2	$	&	$	36	\pm	28	$	&	$	202	\pm	16	$	&	$	0.865	\pm	0.03	$	&	$	0.227	\pm	0.15	$	&	L,S,L	\\
J0858+31	&	Y	&	18	&	$	0	\pm	0	$	&	$	7	\pm	2	$	&	$	83	\pm	3	$	&	$	\dots			$	&	$	59	\pm	4	$	&	$	158	\pm	4	$	&	$	0.325	\pm	0.02	$	&	$		 \dots	$	&	L,S,S	\\
J0915+30	&	Y	&	23	&	$	18	\pm	2	$	&	$	33	\pm	3	$	&	$	39	\pm	6	$	&	$	1.18	\pm	0.1	$	&	$	-147	\pm	14	$	&	$	197	\pm	11	$	&	$	0.418	\pm	0.08	$	&	$	3.302	\pm	0.29	$	&	S,L,H	\\
J0939+35	&	N	&	36	&	$	23	\pm	2	$	&	$	19	\pm	2	$	&	$	48	\pm	5	$	&	$	1.26	\pm	0.1	$	&	$	125	\pm	16	$	&	$	235	\pm	24	$	&	$	0.379	\pm	0.11	$	&	$	3.747	\pm	0.31	$	&	S,L,S	\\
J0945+17	&	N	&	57	&	$	76	\pm	2	$	&	$	4	\pm	3	$	&	$	14	\pm	4	$	&	$	1.15	\pm	0.1	$	&	$	133	\pm	25	$	&	$	172	\pm	20	$	&	$	0.003	\pm	0.00	$	&	$	2.204	\pm	0.18	$	&	S,S,L	\\
J1010+06	&	N	&	39	&	$	65	\pm	4	$	&	$	13	\pm	2	$	&	$	23	\pm	3	$	&	$	1.53	\pm	0.1	$	&	$	156	\pm	27	$	&	$	218	\pm	16	$	&	$	1.231	\pm	0.03	$	&	$	2.395	\pm	0.24	$	&	L,S,S	\\
J1015+00	&	Y	&	25	&	$	41	\pm	2	$	&	$	0	\pm	0	$	&	$	49	\pm	3	$	&	$	1.66	\pm	0.0	$	&	$	-22	\pm	17	$	&	$	168	\pm	10	$	&	$	0.012	\pm	0.00	$	&	$	3.915	\pm	0.03	$	&	S,S,S	\\
J1016+00	&	Y	&	22	&	$	35	\pm	2	$	&	$	15	\pm	3	$	&	$	40	\pm	4	$	&	$	1.42	\pm	0.1	$	&	$	-14	\pm	12	$	&	$	192	\pm	8	$	&	$	0.192	\pm	0.04	$	&	$	3.760	\pm	0.11	$	&	S,L,H	\\
J1034+60	&	N	&	43	&	$	36	\pm	1	$	&	$	9	\pm	2	$	&	$	45	\pm	3	$	&	$	1.11	\pm	0.1	$	&	$	-17	\pm	9	$	&	$	192	\pm	6	$	&	$	0.179	\pm	0.04	$	&	$	3.146	\pm	0.21	$	&	S,L,L	\\
J1036+01	&	Y	&	13	&	$	31	\pm	1	$	&	$	24	\pm	2	$	&	$	34	\pm	4	$	&	$	1.55	\pm	0.1	$	&	$	24	\pm	13	$	&	$	188	\pm	8	$	&	$	0.443	\pm	0.06	$	&	$	3.438	\pm	0.16	$	&	S,L,H	\\
J1100+08	&	Y	&	30	&	$	44	\pm	2	$	&	$	0	\pm	0	$	&	$	54	\pm	2	$	&	$	1.13	\pm	0.1	$	&	$	-59	\pm	17	$	&	$	165	\pm	9	$	&	$	0.386	\pm	0.03	$	&	$	2.087	\pm	0.16	$	&	L,L,S	\\
J1137+61	&	N	&	33	&	$	45	\pm	2	$	&	$	0	\pm	0	$	&	$	45	\pm	3	$	&	$	1.35	\pm	0.0	$	&	$	23	\pm	13	$	&	$	237	\pm	10	$	&	$	0.141	\pm	0.06	$	&	$	3.858	\pm	0.11	$	&	H,L,S	\\
J1152+10	&	N	&	43	&	$	25	\pm	2	$	&	$	14	\pm	1	$	&	$	51	\pm	4	$	&	$	0.89	\pm	0.1	$	&	$	-57	\pm	9	$	&	$	184	\pm	7	$	&	$	0.373	\pm	0.08	$	&	$	3.536	\pm	0.24	$	&	S,L,H	\\
J1157+37	&	Y	&	16	&	$	29	\pm	1	$	&	$	34	\pm	3	$	&	$	27	\pm	4	$	&	$	1.46	\pm	0.1	$	&	$	-59	\pm	14	$	&	$	188	\pm	9	$	&	$	0.309	\pm	0.04	$	&	$	3.299	\pm	0.10	$	&	S,L,S	\\
J1200+31	&	N	&	60	&	$	73	\pm	2	$	&	$	0	\pm	0	$	&	$	18	\pm	4	$	&	$	1.61	\pm	0.1	$	&	$	-104	\pm	21	$	&	$	157	\pm	11	$	&	$	0.386	\pm	0.08	$	&	$	3.465	\pm	0.21	$	&	H,H,S	\\
J1218+47	&	Y	&	26	&	$	55	\pm	3	$	&	$	0	\pm	1	$	&	$	36	\pm	5	$	&	$	1.38	\pm	0.1	$	&	$	46	\pm	10	$	&	$	141	\pm	15	$	&	$	0.150	\pm	0.05	$	&	$	3.611	\pm	0.23	$	&	S,L,S	\\
J1223+08	&	Y	&	23	&	$	41	\pm	2	$	&	$	28	\pm	3	$	&	$	21	\pm	4	$	&	$	1.22	\pm	0.0	$	&	$	-70	\pm	16	$	&	$	162	\pm	12	$	&	$	0.000	\pm	0.00	$	&	$	3.411	\pm	0.05	$	&	S,S,H	\\
J1238+09	&	Y	&	18	&	$	37	\pm	2	$	&	$	11	\pm	1	$	&	$	42	\pm	4	$	&	$	1.34	\pm	0.1	$	&	$	23	\pm	17	$	&	$	226	\pm	15	$	&	$	0.582	\pm	0.08	$	&	$	3.185	\pm	0.33	$	&	S,L,S	\\
J1241+61	&	Y	&	14	&	$	100	\pm	3	$	&	$	0	\pm	2	$	&	$	0	\pm	1	$	&	$	1.96	\pm	0.0	$	&	$	-3	\pm	11	$	&	$	140	\pm	13	$	&	$	0.491	\pm	0.07	$	&	$	3.626	\pm	0.06	$	&	S,S,S	\\
J1244+65	&	Y	&	86	&	$	48	\pm	4	$	&	$	26	\pm	6	$	&	$	19	\pm	4	$	&	$	1.78	\pm	0.4	$	&	$	142	\pm	23	$	&	$	143	\pm	16	$	&	$	1.004	\pm	0.31	$	&	$	3.511	\pm	1.13	$	&	S,L,S	\\
J1300+54	&	N	&	37	&	$	56	\pm	2	$	&	$	11	\pm	3	$	&	$	26	\pm	3	$	&	$	1.06	\pm	0.1	$	&	$	-15	\pm	8	$	&	$	147	\pm	6	$	&	$	0.020	\pm	0.00	$	&	$	2.725	\pm	0.11	$	&	H,S,L	\\
J1316+44	&	N	&	18	&	$	72	\pm	2	$	&	$	1	\pm	2	$	&	$	24	\pm	3	$	&	$	1.10	\pm	0.1	$	&	$	115	\pm	16	$	&	$	148	\pm	12	$	&	$	0.714	\pm	0.12	$	&	$	1.492	\pm	0.29	$	&	L,S,S	\\
J1347+12	&	Y	&	40	&	$	58	\pm	2	$	&	$	11	\pm	1	$	&	$	26	\pm	3	$	&	$	1.77	\pm	0.0	$	&	$	322	\pm	25	$	&	$	212	\pm	15	$	&	$	0.666	\pm	0.07	$	&	$	3.833	\pm	0.13	$	&	S,S,H	\\
J1356-02	&	Y	&	28	&	$	30	\pm	4	$	&	$	17	\pm	4	$	&	$	48	\pm	4	$	&	$	0.40	\pm	0.2	$	&	$	-40	\pm	14	$	&	$	133	\pm	11	$	&	$	1.083	\pm	0.06	$	&	$	0.718	\pm	0.61	$	&	L,H,H	\\
J1356+10	&	N	&	37	&	$	27	\pm	1	$	&	$	24	\pm	1	$	&	$	43	\pm	2	$	&	$	0.13	\pm	0.0	$	&	$	100	\pm	21	$	&	$	208	\pm	13	$	&	$	0.778	\pm	0.01	$	&	$			0.00	$	&	L,S,L	\\
J1405+40	&	Y	&	21	&	$	26	\pm	3	$	&	$	22	\pm	3	$	&	$	44	\pm	4	$	&	$	0.77	\pm	0.2	$	&	$	-10	\pm	8	$	&	$	121	\pm	7	$	&	$	0.311	\pm	0.07	$	&	$	2.166	\pm	0.51	$	&	L,S,S	\\
J1430+13	&	Y	&	27	&	$	31	\pm	1	$	&	$	46	\pm	2	$	&	$	19	\pm	2	$	&	$	0.50	\pm	0.0	$	&	$	-16	\pm	11	$	&	$	197	\pm	10	$	&	$	0.328	\pm	0.02	$	&	$			0.00	$	&	L,L,S	\\
J1436+13	&	N	&	38	&	$	25	\pm	2	$	&	$	21	\pm	3	$	&	$	43	\pm	4	$	&	$	1.40	\pm	0.1	$	&	$	18	\pm	14	$	&	$	192	\pm	5	$	&	$	0.645	\pm	0.08	$	&	$	3.226	\pm	0.18	$	&	L,H,H	\\
J1437+30	&	N	&	34	&	$	50	\pm	2	$	&	$	0	\pm	1	$	&	$	41	\pm	4	$	&	$	1.37	\pm	0.2	$	&	$	62	\pm	12	$	&	$	199	\pm	9	$	&	$	0.533	\pm	0.15	$	&	$	2.919	\pm	0.56	$	&	S,S,S	\\
J1440+53	&	N	&	8	&	$	92	\pm	2	$	&	$	4	\pm	2	$	&	$	3	\pm	2	$	&	$	1.40	\pm	0.1	$	&	$	101	\pm	17	$	&	$	143	\pm	18	$	&	$	0.462	\pm	0.26	$	&	$	3.922	\pm	0.56	$	&	H,L,L	\\
J1455+32	&	Y	&	28	&	$	32	\pm	1	$	&	$	29	\pm	3	$	&	$	29	\pm	4	$	&	$	1.10	\pm	0.0	$	&	$	-14	\pm	7	$	&	$	153	\pm	7	$	&	$	0.333	\pm	0.04	$	&	$	3.423	\pm	0.10	$	&	S,L,H	\\
J1509+04	&	Y	&	29	&	$	33	\pm	3	$	&	$	10	\pm	4	$	&	$	49	\pm	4	$	&	$	0.43	\pm	0.2	$	&	$	134	\pm	11	$	&	$	144	\pm	14	$	&	$	0.890	\pm	0.04	$	&	$			0.00	$	&	L,S,S	\\
J1517+33	&	N	&	149	&	$	22	\pm	6	$	&	$	28	\pm	6	$	&	$	41	\pm	5	$	&	$	1.63	\pm	1.0	$	&	$	62	\pm	21	$	&	$	228	\pm	15	$	&	$	1.169	\pm	0.06	$	&	$	4.112	\pm	0.23	$	&	S,L,H	\\
J1533+35	&	N	&	23	&	$	30	\pm	5	$	&	$	30	\pm	6	$	&	$	30	\pm	7	$	&	$	0.72	\pm	0.4	$	&	$	-16	\pm	17	$	&	$	149	\pm	13	$	&	$	0.618	\pm	0.17	$	&	$	2.343	\pm	0.78	$	&	S,L,H	\\
J1548-01	&	N	&	24	&	$	88	\pm	3	$	&	$	6	\pm	4	$	&	$	1	\pm	1	$	&	$	1.16	\pm	0.0	$	&	$	39	\pm	17	$	&	$	213	\pm	21	$	&	$	0.265	\pm	0.08	$	&	$	1.327	\pm	0.47	$	&	H,S,L	\\
J1558+35	&	N	&	38	&	$	44	\pm	3	$	&	$	12	\pm	3	$	&	$	35	\pm	4	$	&	$	1.48	\pm	0.0	$	&	$	-9	\pm	11	$	&	$	118	\pm	12	$	&	$	0.075	\pm	0.02	$	&	$	3.816	\pm	0.07	$	&	H,S,S	\\
J1624+33	&	N	&	20	&	$	50	\pm	2	$	&	$	8	\pm	4	$	&	$	32	\pm	5	$	&	$	1.32	\pm	0.0	$	&	$	53	\pm	13	$	&	$	153	\pm	10	$	&	$	0.081	\pm	0.01	$	&	$	3.617	\pm	0.07	$	&	S,S,H	\\
J1653+23	&	N	&	36	&	$	43	\pm	1	$	&	$	15	\pm	2	$	&	$	32	\pm	3	$	&	$	1.48	\pm	0.0	$	&	$	14	\pm	12	$	&	$	198	\pm	10	$	&	$	0.089	\pm	0.01	$	&	$	3.503	\pm	0.05	$	&	S,H,S	\\
J1713+57	&	Y	&	31	&	$	27	\pm	2	$	&	$	21	\pm	2	$	&	$	44	\pm	2	$	&	$	1.26	\pm	0.1	$	&	$	32	\pm	11	$	&	$	163	\pm	10	$	&	$	0.604	\pm	0.05	$	&	$	3.036	\pm	0.23	$	&	L,S,H	\\
J2154+11	&	Y	&	25	&	$	12	\pm	6	$	&	$	19	\pm	4	$	&	$	62	\pm	6	$	&	$	0.71	\pm	0.9	$	&	$	-13	\pm	14	$	&	$	194	\pm	13	$	&	$	0.656	\pm	0.04	$	&	$	2.591	\pm	0.43	$	&	L,S,H	\\
\hline
\end{tabular}
\end{table*}

\clearpage

\begin{table*}
\centering

     \caption{Results of the non-parametric analysis of the gas kinematics using the [OIII] emission line. Column 1 is the abbreviated name of the QSO2, columns 2 and 3 give the derived quantities W80 and $\Delta$V whilst columns 4 and 5 give V05 and V95. All quantities are given in \kms\ . Column 6 lists the electron densities measured using either the transauroral technique or the [SII] ratios (between parenthesis). For the two objects where it was not possible to measure the densities, the assumed value of log n$_e$=3.66 is given and denoted by an \emph{a}. Columns 7, 8 and 9 show the total outflow mass, mass outflow rate and the kinetic energy, assuming an outflow radius of 0.62 kpc. The errors given for mass outflow rates are obtained from deriving the outflow rates at the minimum (0.15 kpc) and maximum (1.89 kpc) outflow radii found by \citet{fischer18}, whilst 0.62 kpc is the mean of the values presented there. Errors for the values of $\log \dot{E}_{of}$ are $+0.6,-0.5$. }
    \label{tab:results-sfr-gas}
	\begin{tabular}{l c c c c c c c c}
	\hline																																								
\hline		
																																								
Name			&	W80	&	$\Delta$V	&	V05	&	V95	&	log $n_e$	& $\log M_{of}$ 	&$\dot{M}_{of}$	&	$\log \dot{E}_{of}$\\																									
			&	(\kms)	&	(\kms)	&	(\kms)	&	(\kms)	&($\mbox{cm}^{-3}$)	&(\msun)	&(\msunyr)	&	(\ergs)	\\																										
																																									
\hline																																									
J0052-01	&	$	767	\pm	3	$	&	$	-91	\pm	3	$	&	$	-584	\pm	5	$	&	$	401	\pm	3	$	&	$	(<2.94)	$	&	$	5.530	\pm	0.002	$	&	$	0.7	\substack{	+	2.3	\\	-	0.5	}	$	&	$	40.8	$	\\
J0232-08	&	$	854	\pm	2	$	&	$	-142	\pm	1	$	&	$	-681	\pm	2	$	&	$	398	\pm	1	$	&	$	3.43\substack{+0.03\\-0.12}	$	&	$	5.109	\pm	0.001	$	&	$	0.3	\substack{	+	0.9	\\	-	0.2	}	$	&	$	40.5	$	\\
J0731+39	&	$	699	\pm	6	$	&	$	-82	\pm	4	$	&	$	-611	\pm	6	$	&	$	447	\pm	9	$	&	$	3.97\substack{+0.06\\-0.05}	$	&	$	4.38	\pm	0.01	$	&	$	0.1	\substack{	+	0.2	\\	-	0.0	}	$	&	$	39.9	$	\\
J0759+50	&	$	1361	\pm	4	$	&	$	6	\pm	6	$	&	$	-953	\pm	7	$	&	$	965	\pm	6	$	&	$	4.09\substack{+0.03\\-0.03}	$	&	$	4.52	\pm	0.02	$	&	$	0.2	\substack{	+	0.6	\\	-	0.1	}	$	&	$	40.9	$	\\
J0802+25	&	$	1028	\pm	6	$	&	$	-255	\pm	3	$	&	$	-967	\pm	5	$	&	$	456	\pm	5	$	&	$	3.56\substack{+0.12\\-0.14}	$	&	$	5.04	\pm	0.03	$	&	$	0.4	\substack{	+	1.3	\\	-	0.3	}	$	&	$	41.0	$	\\
J0802+46	&	$	893	\pm	4	$	&	$	15	\pm	3	$	&	$	-615	\pm	4	$	&	$	645	\pm	3	$	&	$	3.88\substack{+0.04\\-0.05}	$	&	$	4.575	\pm	0.001	$	&	$	0.1	\substack{	+	0.4	\\	-	0.1	}	$	&	$	40.3	$	\\
J0805+28	&	$	938	\pm	180	$	&	$	-140	\pm	241	$	&	$	-776	\pm	121	$	&	$	496	\pm	362	$	&	$	3.59\substack{+0.08\\-0.16}	$	&	$	4.84	\pm	0.71	$	&	$	0.2	\substack{	+	0.7	\\	-	0.1	}	$	&	$	40.6	$	\\
J0818+36	&	$	545	\pm	2	$	&	$	-53	\pm	2	$	&	$	-436	\pm	3	$	&	$	329	\pm	1	$	&	$	3.67\substack{+0.06\\-0.1}	$	&	$	4.749	\pm	0.001	$	&	$	0.1	\substack{	+	0.3	\\	-	0.1	}	$	&	$	39.7	$	\\
J0841+01	&	$	423	\pm	1	$	&	$	-18	\pm	1	$	&	$	-297	\pm	1	$	&	$	262	\pm	1	$	&	$	2.32\substack{+0.24\\-0.13}	$	&	$	…			$	&	$	…								$	&	$	…	$	\\
J0858+31	&	$	735	\pm	8	$	&	$	-102	\pm	6	$	&	$	-636	\pm	11	$	&	$	433	\pm	10	$	&	$	(2.71\substack{+0.1\\-0.12})	$	&	$	5.54	\pm	0.03	$	&	$	1.1	\substack{	+	3.3	\\	-	0.7	}	$	&	$	41.2	$	\\
J0915+30	&	$	795	\pm	8	$	&	$	74	\pm	8	$	&	$	-535	\pm	5	$	&	$	683	\pm	15	$	&	$	3.41\substack{+0.05\\-0.06}	$	&	$	5.10	\pm	0.02	$	&	$	0.5	\substack{	+	1.5	\\	-	0.3	}	$	&	$	40.9	$	\\
J0939+35	&	$	563	\pm	2	$	&	$	-53	\pm	1	$	&	$	-426	\pm	2	$	&	$	321	\pm	2	$	&	$	3.16\substack{+0.05\\-0.11}	$	&	$	...			$	&	$	…								$	&	$	…	$	\\
J0945+17	&	$	1079	\pm	6	$	&	$	-132	\pm	8	$	&	$	-1005	\pm	9	$	&	$	742	\pm	14	$	&	$	3.39\substack{+0.08\\-0.1}	$	&	$	5.46	\pm	0.01	$	&	$	1.5	\substack{	+	4.9	\\	-	1.0	}	$	&	$	41.8	$	\\
J1010+06	&	$	1490	\pm	53	$	&	$	51	\pm	31	$	&	$	-1073	\pm	87	$	&	$	1176	\pm	37	$	&	$	4.58\substack{+0.03\\-0.03}	$	&	$	3.75	\pm	0.04	$	&	$	0.04	\substack{	+	0.12	\\	-	0.03	}	$	&	$	40.4	$	\\
J1015+00	&	$	531	\pm	3	$	&	$	-38	\pm	2	$	&	$	-393	\pm	3	$	&	$	318	\pm	3	$	&	$	3.16\substack{+0.06\\-0.18}	$	&	$	5.396	\pm	0.003	$	&	$	0.4	\substack{	+	1.3	\\	-	0.3	}	$	&	$	40.2	$	\\
J1016+00	&	$	636	\pm	6	$	&	$	-81	\pm	3	$	&	$	-515	\pm	5	$	&	$	354	\pm	4	$	&	$	(<2.67)	$	&	$	5.813	\pm	0.004	$	&	$	1.3	\substack{	+	4.1	\\	-	0.9	}	$	&	$	40.9	$	\\
J1034+60	&	$	763	\pm	2	$	&	$	8	\pm	2	$	&	$	-546	\pm	3	$	&	$	562	\pm	3	$	&	$	2.99\substack{+0.19\\-0.2}	$	&	$	5.586	\pm	0.001	$	&	$	1.3	\substack{	+	4.2	\\	-	0.9	}	$	&	$	41.3	$	\\
J1036+01	&	$	482	\pm	3	$	&	$	-44	\pm	2	$	&	$	-362	\pm	4	$	&	$	273	\pm	3	$	&	$	3.28\substack{+0.06\\-0.18}	$	&	$	…			$	&	$	…								$	&	$	…	$	\\
J1100+08	&	$	1162	\pm	7	$	&	$	-43	\pm	7	$	&	$	-952	\pm	13	$	&	$	865	\pm	8	$	&	$	3.99\substack{+0.07\\-0.08}	$	&	$	4.92	\pm	0.01	$	&	$	0.5	\substack{	+	1.5	\\	-	0.3	}	$	&	$	41.3	$	\\
J1137+61	&	$	529	\pm	64	$	&	$	2	\pm	10	$	&	$	-327	\pm	31	$	&	$	331	\pm	51	$	&	$	3.26\substack{+0.07\\-{\color{black}0}}	$	&	$	…			$	&	$	…								$	&	$	…	$	\\
J1152+10	&	$	553	\pm	1	$	&	$	-39	\pm	1	$	&	$	-382	\pm	1	$	&	$	305	\pm	1	$	&	$	3.39\substack{+0.02\\-0.13}	$	&	$	5.244	\pm	0.001	$	&	$	0.2	\substack{	+	0.8	\\	-	0.2	}	$	&	$	39.9	$	\\
J1157+37	&	$	660	\pm	6	$	&	$	-79	\pm	4	$	&	$	-577	\pm	8	$	&	$	419	\pm	7	$	&	$	3.67\substack{+0.06\\-0.07}	$	&	$	4.72	\pm	0.01	$	&	$	0.1	\substack{	+	0.5	\\	-	0.1	}	$	&	$	40.2	$	\\
J1200+31	&	$	766	\pm	4	$	&	$	-146	\pm	2	$	&	$	-695	\pm	3	$	&	$	402	\pm	4	$	&	$	3.42\substack{+0.08\\-0.06}	$	&	$	5.564	\pm	0.001	$	&	$	1.1	\substack{	+	3.4	\\	-	0.7	}	$	&	$	41.2	$	\\
J1218+47	&	$	444	\pm	2	$	&	$	-21	\pm	1	$	&	$	-319	\pm	2	$	&	$	277	\pm	2	$	&	$	3.66^a	$	&	$	4.83	\pm	0.03	$	&	$	0.1	\substack{	+	0.3	\\	-	0.1	}	$	&	$	39.3	$	\\
J1223+08	&	$	594	\pm	7	$	&	$	-132	\pm	4	$	&	$	-580	\pm	10	$	&	$	315	\pm	6	$	&	$	3.7\substack{+0.08\\-0.13}	$	&	$	4.90	\pm	0.02	$	&	$	0.2	\substack{	+	0.5	\\	-	0.1	}	$	&	$	40.2	$	\\
J1238+09	&	$	641	\pm	2	$	&	$	-49	\pm	1	$	&	$	-438	\pm	2	$	&	$	341	\pm	2	$	&	$	(2.68\substack{+0.09\\-0.09})	$	&	$	5.763	\pm	0.001	$	&	$	1.0	\substack{	+	3.0	\\	-	0.6	}	$	&	$	40.6	$	\\
J1241+61	&	$	616	\pm	7	$	&	$	-192	\pm	5	$	&	$	-615	\pm	9	$	&	$	231	\pm	6	$	&	$	2.95\substack{+0.12\\-0.46}	$	&	$	5.38	\pm	0.02	$	&	$	0.4	\substack{	+	1.4	\\	-	0.3	}	$	&	$	40.6	$	\\
J1244+65	&	$	1218	\pm	6	$	&	$	-282	\pm	6	$	&	$	-1149	\pm	9	$	&	$	585	\pm	9	$	&	$	3.31\substack{+0.05\\-0.22}	$	&	$	4.94	\pm	0.02	$	&	$	0.4	\substack{	+	1.3	\\	-	0.3	}	$	&	$	41.2	$	\\
J1300+54	&	$	298	\pm	1	$	&	$	-27	\pm	0	$	&	$	-229	\pm	1	$	&	$	175	\pm	1	$	&	$	3.33\substack{+0.1\\-0.4}	$	&	$	…			$	&	$	…								$	&	$	…	$	\\
J1316+44	&	$	905	\pm	15	$	&	$	-220	\pm	18	$	&	$	-872	\pm	37	$	&	$	432	\pm	3	$	&	$	3.66^a	$	&	$	4.84	\pm	0.01	$	&	$	0.2	\substack{	+	0.6	\\	-	0.1	}	$	&	$	40.6	$	\\
J1347+12	&	$	2519	\pm	150	$	&	$	-845	\pm	337	$	&	$	-2466	\pm	483	$	&	$	775	\pm	192	$	&	$	4.27\substack{+0.08\\-0.05}	$	&	$	4.21	\pm	0.24	$	&	$	0.1	\substack{	+	0.4	\\	-	0.1	}	$	&	$	41.4	$	\\
J1356-02	&	$	717	\pm	8	$	&	$	-85	\pm	7	$	&	$	-654	\pm	11	$	&	$	483	\pm	13	$	&	$	3.66\substack{{+0.09}\\-0.20}	$	&	$	4.64	\pm	0.02	$	&	$	0.1	\substack{	+	0.5	\\	-	0.1	}	$	&	$	40.4	$	\\
J1356+10	&	$	861	\pm	2	$	&	$	26	\pm	1	$	&	$	-535	\pm	2	$	&	$	586	\pm	2	$	&	$	3.21\substack{+0.0\\-0.15}	$	&	$	5.794	\pm	0.002	$	&	$	1.9	\substack{	+	5.9	\\	-	1.2	}	$	&	$	41.3	$	\\
J1405+40	&	$	650	\pm	5	$	&	$	-124	\pm	4	$	&	$	-606	\pm	8	$	&	$	358	\pm	2	$	&	$	4.1\substack{+0.15\\-0.19}	$	&	$	4.476	\pm	0.003	$	&	$	0.1	\substack{	+	0.2	\\	-	0.0	}	$	&	$	39.8	$	\\
J1430+13	&	$	803	\pm	3	$	&	$	-61	\pm	2	$	&	$	-630	\pm	4	$	&	$	509	\pm	2	$	&	$	3.24\substack{+0.05\\-0.3}	$	&	$	5.646	\pm	0.004	$	&	$	1.3	\substack{	+	4.1	\\	-	0.9	}	$	&	$	41.2	$	\\
J1436+13	&	$	627	\pm	4	$	&	$	-66	\pm	3	$	&	$	-530	\pm	4	$	&	$	397	\pm	5	$	&	$	3.4\substack{+0.08\\-0.17}	$	&	$	5.068	\pm	0.001	$	&	$	0.3	\substack{	+	0.9	\\	-	0.2	}	$	&	$	40.4	$	\\
J1437+30	&	$	628	\pm	2	$	&	$	-24	\pm	1	$	&	$	-457	\pm	2	$	&	$	409	\pm	3	$	&	$	3.3\substack{+0.01\\-0.13}	$	&	$	5.301	\pm	0.001	$	&	$	0.4	\substack{	+	1.4	\\	-	0.3	}	$	&	$	40.5	$	\\
J1440+53	&	$	775	\pm	3	$	&	$	-2	\pm	3	$	&	$	-667	\pm	5	$	&	$	664	\pm	2	$	&	$	3.92\substack{+0.09\\-0.08}	$	&	$	4.71	\pm	0.01	$	&	$	0.2	\substack{	+	0.7	\\	-	0.1	}	$	&	$	40.7	$	\\
J1455+32	&	$	845	\pm	4	$	&	$	-36	\pm	3	$	&	$	-626	\pm	4	$	&	$	555	\pm	4	$	&	$	3.88\substack{+0.05\\-0.06}	$	&	$	4.507	\pm	0.002	$	&	$	0.1	\substack{	+	0.3	\\	-	0.1	}	$	&	$	40.2	$	\\
J1509+04	&	$	1356	\pm	9	$	&	$	-327	\pm	7	$	&	$	-1233	\pm	10	$	&	$	579	\pm	9	$	&	$	3.41\substack{+0.11\\-0.21}	$	&	$	4.94	\pm	0.03	$	&	$	0.4	\substack{	+	1.3	\\	-	0.3	}	$	&	$	41.3	$	\\
J1517+33	&	$	1210	\pm	4	$	&	$	-79	\pm	2	$	&	$	-823	\pm	2	$	&	$	666	\pm	3	$	&	$	2.98\substack{+0.23\\-0.31}	$	&	$	5.67	\pm	0.02	$	&	$	1.8	\substack{	+	5.6	\\	-	1.2	}	$	&	$	41.5	$	\\
J1533+35	&	$	538	\pm	6	$	&	$	-38	\pm	7	$	&	$	-467	\pm	15	$	&	$	391	\pm	6	$	&	$	(2.44\substack{+0.11\\-0.13})	$	&	$	5.88	\pm	0.04	$	&	$	1.6	\substack{	+	5.1	\\	-	1.1	}	$	&	$	41.0	$	\\
J1548-01	&	$	497	\pm	2	$	&	$	-59	\pm	2	$	&	$	-415	\pm	4	$	&	$	298	\pm	5	$	&	$	(2.74\substack{+0.04\\-0.05})	$	&	$	…			$	&	$	…								$	&	$	…	$	\\
J1558+35	&	$	561	\pm	3	$	&	$	-125	\pm	2	$	&	$	-533	\pm	5	$	&	$	282	\pm	3	$	&	$	3.23\substack{+0.12\\-0.26}	$	&	$	5.358	\pm	0.002	$	&	$	0.5	\substack{	+	1.4	\\	-	0.3	}	$	&	$	40.6	$	\\
J1624+33	&	$	490	\pm	4	$	&	$	-73	\pm	2	$	&	$	-428	\pm	4	$	&	$	282	\pm	4	$	&	$	3.63\substack{+0.12\\-0.09}	$	&	$	4.77	\pm	0.01	$	&	$	0.1	\substack{	+	0.3	\\	-	0.1	}	$	&	$	39.7	$	\\
J1653+23	&	$	494	\pm	2	$	&	$	-57	\pm	1	$	&	$	-403	\pm	2	$	&	$	289	\pm	1	$	&	$	3.26\substack{+0.14\\-0.04}	$	&	$	…			$	&	$	…								$	&	$	…	$	\\
J1713+57	&	$	1491	\pm	10	$	&	$	-60	\pm	10	$	&	$	-1121	\pm	18	$	&	$	1001	\pm	7	$	&	$	4.06\substack{+0.05\\-0.06}	$	&	$	4.64	\pm	0.01	$	&	$	0.3	\substack{	+	0.9	\\	-	0.2	}	$	&	$	41.2	$	\\
J2154+11	&	$	582	\pm	206	$	&	$	-50	\pm	135	$	&	$	-452	\pm	252	$	&	$	353	\pm	17	$	&	$	3.27\substack{+0.16\\-0.05}	$	&	$	5.06	\pm	0.54	$	&	$	0.3	\substack{	+	0.8	\\	-	0.2	}	$	&	$	40.2	$	\\

	\hline
    
	\end{tabular}
\end{table*}

\clearpage

\begin{table*}
\centering

     \caption{Results of calculations of the outflow masses, outflow mass rates, and kinetic energy rates assuming that all gas with velocities greater than 1.5 times the FWHM of the stellar velocity dispersion derived from the {\sc starlight} modelling. As previously discussed, the outflow radius is unknown, so we have calculated $\dot{M}_{of}$ assuming outflow radii of 0.15, 0.62, and 1.89 kpc. Column 2 gives the total mass of the outflow whilst columns 3 to 8 show $\dot{M}_{of}$ and $\log \dot{E}_{of}$ for each assumed radius. Only objects which have an outflow according to our definition are included in the table.}
    \label{tab:test_vd}
	\begin{tabular}{l c |c c |c c |c c}
	\hline																																								
\hline		
				&					&	\multicolumn{2}{c}{0.15 kpc}										&	\multicolumn{2}{c}{0.62 kpc}										&	\multicolumn{2}{c}{1.89 kpc}									\\
				&					&					&						&					&						&					&					\\
Name			&	$\log$(Mass)	&	$\dot{M}_{of}$	&	$\log \dot{E}_{of}$	&	$\dot{M}_{of}$	&	$\log \dot{E}_{of}$	&	$\dot{M}_{of}$	&	$\log \dot{E}_{of}$\\																									
				&	(\msun)		&	(\msunyr)		&	(\msunyr)			&	(\ergs)			&(\msunyr)				&	(\ergs)			&	(\ergs)	\\																										
																																									
\hline																																									
J0052-01	&	$	5.87	\pm	0.08	$	&	$	5.65	\pm	1.04	$	&	$	41.43	\pm	0.10	$	&	$	1.36	\pm	0.25	$	&	$	40.81	\pm	0.10	$	&	$	0.45	\pm	0.08	$	&	$	40.33	\pm	0.10	$	\\
J0232-08	&	$	5.46	\pm	0.03	$	&	$	2.39	\pm	0.20	$	&	$	41.14	\pm	0.06	$	&	$	0.58	\pm	0.05	$	&	$	40.52	\pm	0.06	$	&	$	0.19	\pm	0.02	$	&	$	40.04	\pm	0.06	$	\\
J0731+39	&	$	4.90	\pm	0.04	$	&	$	0.51	\pm	0.05	$	&	$	40.28	\pm	0.06	$	&	$	0.12	\pm	0.01	$	&	$	39.66	\pm	0.06	$	&	$	0.041	\pm	0.004	$	&	$	39.18	\pm	0.06	$	\\
J0759+50	&	$	5.23	\pm	0.02	$	&	$	1.64	\pm	0.08	$	&	$	41.06	\pm	0.04	$	&	$	0.39	\pm	0.02	$	&	$	40.45	\pm	0.04	$	&	$	0.13	\pm	0.01	$	&	$	39.96	\pm	0.04	$	\\
J0802+25	&	$	5.51	\pm	0.00	$	&	$	3.54	\pm	0.01	$	&	$	41.577	\pm	0.002	$	&	$	0.851	\pm	0.002	$	&	$	40.958	\pm	0.002	$	&	$	0.281	\pm	0.001	$	&	$	40.477	\pm	0.002	$	\\
J0802+46	&	$	5.09	\pm	0.03	$	&	$	1.03	\pm	0.09	$	&	$	40.76	\pm	0.06	$	&	$	0.25	\pm	0.02	$	&	$	40.14	\pm	0.06	$	&	$	0.08	\pm	0.01	$	&	$	39.66	\pm	0.06	$	\\
J0805+28	&	$	5.41	\pm	0.01	$	&	$	1.99	\pm	0.07	$	&	$	41.01	\pm	0.02	$	&	$	0.48	\pm	0.02	$	&	$	40.39	\pm	0.02	$	&	$	0.16	\pm	0.01	$	&	$	39.91	\pm	0.02	$	\\
J0818+36	&	$	4.86	\pm	0.06	$	&	$	0.51	\pm	0.07	$	&	$	40.31	\pm	0.08	$	&	$	0.12	\pm	0.02	$	&	$	39.69	\pm	0.08	$	&	$	0.04	\pm	0.01	$	&	$	39.21	\pm	0.08	$	\\
J0858+31	&	$	6.06	\pm	0.05	$	&	$	7.95	\pm	0.98	$	&	$	41.55	\pm	0.07	$	&	$	1.91	\pm	0.24	$	&	$	40.93	\pm	0.07	$	&	$	0.63	\pm	0.08	$	&	$	40.45	\pm	0.07	$	\\
J0915+30	&	$	5.46	\pm	0.05	$	&	$	2.75	\pm	0.36	$	&	$	41.27	\pm	0.09	$	&	$	0.66	\pm	0.09	$	&	$	40.65	\pm	0.09	$	&	$	0.22	\pm	0.03	$	&	$	40.17	\pm	0.09	$	\\
J0945+17	&	$	6.06	\pm	0.05	$	&	$	11.39	\pm	1.46	$	&	$	41.96	\pm	0.07	$	&	$	2.74	\pm	0.35	$	&	$	41.34	\pm	0.07	$	&	$	0.90	\pm	0.12	$	&	$	40.86	\pm	0.07	$	\\
J1010+06	&	$	4.46	\pm	0.03	$	&	$	0.36	\pm	0.03	$	&	$	40.61	\pm	0.05	$	&	$	0.09	\pm	0.01	$	&	$	39.99	\pm	0.05	$	&	$	0.028	\pm	0.002	$	&	$	39.51	\pm	0.05	$	\\
J1015+00	&	$	5.47	\pm	0.09	$	&	$	1.96	\pm	0.40	$	&	$	40.83	\pm	0.09	$	&	$	0.47	\pm	0.10	$	&	$	40.22	\pm	0.09	$	&	$	0.16	\pm	0.03	$	&	$	39.73	\pm	0.09	$	\\
J1016+00	&	$	5.95	\pm	0.03	$	&	$	7.42	\pm	0.50	$	&	$	41.61	\pm	0.04	$	&	$	1.79	\pm	0.12	$	&	$	41.00	\pm	0.04	$	&	$	0.59	\pm	0.04	$	&	$	40.51	\pm	0.04	$	\\
J1034+60	&	$	5.98	\pm	0.02	$	&	$	8.27	\pm	0.41	$	&	$	41.67	\pm	0.03	$	&	$	1.99	\pm	0.10	$	&	$	41.05	\pm	0.03	$	&	$	0.66	\pm	0.03	$	&	$	40.57	\pm	0.03	$	\\
J1100+08	&	$	5.58	\pm	0.02	$	&	$	3.50	\pm	0.21	$	&	$	41.36	\pm	0.05	$	&	$	0.84	\pm	0.05	$	&	$	40.74	\pm	0.05	$	&	$	0.28	\pm	0.02	$	&	$	40.26	\pm	0.05	$	\\
J1152+10	&	$	5.32	\pm	0.08	$	&	$	1.23	\pm	0.22	$	&	$	40.54	\pm	0.08	$	&	$	0.30	\pm	0.05	$	&	$	39.92	\pm	0.08	$	&	$	0.10	\pm	0.02	$	&	$	39.44	\pm	0.08	$	\\
J1157+37	&	$	4.95	\pm	0.04	$	&	$	0.79	\pm	0.09	$	&	$	40.69	\pm	0.07	$	&	$	0.19	\pm	0.02	$	&	$	40.07	\pm	0.07	$	&	$	0.06	\pm	0.01	$	&	$	39.59	\pm	0.07	$	\\
J1200+31	&	$	6.13	\pm	0.03	$	&	$	10.64	\pm	0.84	$	&	$	41.70	\pm	0.05	$	&	$	2.56	\pm	0.20	$	&	$	41.09	\pm	0.05	$	&	$	0.84	\pm	0.07	$	&	$	40.60	\pm	0.05	$	\\
J1218+47	&	$	4.93	\pm	0.12	$	&	$	0.44	\pm	0.12	$	&	$	39.94	\pm	0.14	$	&	$	0.10	\pm	0.03	$	&	$	39.32	\pm	0.14	$	&	$	0.03	\pm	0.01	$	&	$	38.84	\pm	0.14	$	\\
J1223+08	&	$	5.16	\pm	0.05	$	&	$	1.04	\pm	0.14	$	&	$	40.63	\pm	0.09	$	&	$	0.25	\pm	0.03	$	&	$	40.01	\pm	0.09	$	&	$	0.08	\pm	0.01	$	&	$	39.53	\pm	0.09	$	\\
J1238+09	&	$	5.57	\pm	0.12	$	&	$	2.94	\pm	0.88	$	&	$	41.17	\pm	0.15	$	&	$	0.71	\pm	0.21	$	&	$	40.55	\pm	0.15	$	&	$	0.23	\pm	0.07	$	&	$	40.07	\pm	0.15	$	\\
J1241+61	&	$	5.72	\pm	0.04	$	&	$	3.46	\pm	0.41	$	&	$	41.11	\pm	0.08	$	&	$	0.83	\pm	0.10	$	&	$	40.49	\pm	0.08	$	&	$	0.27	\pm	0.03	$	&	$	40.01	\pm	0.08	$	\\
J1244+65	&	$	5.63	\pm	0.03	$	&	$	4.07	\pm	0.34	$	&	$	41.51	\pm	0.06	$	&	$	0.98	\pm	0.08	$	&	$	40.89	\pm	0.06	$	&	$	0.32	\pm	0.03	$	&	$	40.41	\pm	0.06	$	\\
J1316+44	&	$	5.34	\pm	0.05	$	&	$	1.63	\pm	0.16	$	&	$	40.95	\pm	0.05	$	&	$	0.39	\pm	0.04	$	&	$	40.33	\pm	0.05	$	&	$	0.13	\pm	0.01	$	&	$	39.85	\pm	0.05	$	\\
J1347+12	&	$	4.92	\pm	0.02	$	&	$	1.47	\pm	0.07	$	&	$	41.63	\pm	0.04	$	&	$	0.35	\pm	0.02	$	&	$	41.01	\pm	0.04	$	&	$	0.12	\pm	0.01	$	&	$	40.53	\pm	0.04	$	\\
J1356-02	&	$	5.21	\pm	0.04	$	&	$	0.96	\pm	0.14	$	&	$	40.44	\pm	0.11	$	&	$	0.23	\pm	0.03	$	&	$	39.82	\pm	0.11	$	&	$	0.08	\pm	0.01	$	&	$	39.34	\pm	0.11	$	\\
J1356+10	&	$	6.20	\pm	0.04	$	&	$	13.74	\pm	1.39	$	&	$	41.90	\pm	0.06	$	&	$	3.31	\pm	0.34	$	&	$	41.28	\pm	0.06	$	&	$	1.09	\pm	0.11	$	&	$	40.80	\pm	0.06	$	\\
J1405+40	&	$	4.97	\pm	0.01	$	&	$	0.51	\pm	0.02	$	&	$	40.13	\pm	0.02	$	&	$	0.122	\pm	0.004	$	&	$	39.51	\pm	0.02	$	&	$	0.040	\pm	0.001	$	&	$	39.03	\pm	0.02	$	\\
J1430+13	&	$	5.98	\pm	0.02	$	&	$	8.46	\pm	0.47	$	&	$	41.71	\pm	0.03	$	&	$	2.04	\pm	0.11	$	&	$	41.09	\pm	0.03	$	&	$	0.67	\pm	0.04	$	&	$	40.61	\pm	0.03	$	\\
J1436+13	&	$	5.33	\pm	0.04	$	&	$	1.66	\pm	0.18	$	&	$	40.89	\pm	0.08	$	&	$	0.40	\pm	0.04	$	&	$	40.27	\pm	0.08	$	&	$	0.13	\pm	0.01	$	&	$	39.79	\pm	0.08	$	\\
J1437+30	&	$	5.48	\pm	0.06	$	&	$	2.34	\pm	0.34	$	&	$	41.04	\pm	0.08	$	&	$	0.56	\pm	0.08	$	&	$	40.43	\pm	0.08	$	&	$	0.19	\pm	0.03	$	&	$	39.94	\pm	0.08	$	\\
J1440+53	&	$	5.47	\pm	0.05	$	&	$	2.06	\pm	0.27	$	&	$	40.97	\pm	0.07	$	&	$	0.50	\pm	0.07	$	&	$	40.36	\pm	0.07	$	&	$	0.16	\pm	0.02	$	&	$	39.87	\pm	0.07	$	\\
J1455+32	&	$	5.02	\pm	0.02	$	&	$	0.82	\pm	0.04	$	&	$	40.59	\pm	0.03	$	&	$	0.20	\pm	0.01	$	&	$	39.97	\pm	0.03	$	&	$	0.066	\pm	0.003	$	&	$	39.49	\pm	0.03	$	\\
J1509+04	&	$	5.58	\pm	0.03	$	&	$	4.11	\pm	0.30	$	&	$	41.65	\pm	0.05	$	&	$	0.99	\pm	0.07	$	&	$	41.04	\pm	0.05	$	&	$	0.33	\pm	0.02	$	&	$	40.55	\pm	0.05	$	\\
J1517+33	&	$	6.29	\pm	0.03	$	&	$	20.67	\pm	1.63	$	&	$	42.26	\pm	0.04	$	&	$	4.97	\pm	0.39	$	&	$	41.64	\pm	0.04	$	&	$	1.64	\pm	0.13	$	&	$	41.16	\pm	0.04	$	\\
J1533+35	&	$	6.11	\pm	0.06	$	&	$	8.75	\pm	1.41	$	&	$	41.49	\pm	0.09	$	&	$	2.11	\pm	0.34	$	&	$	40.87	\pm	0.09	$	&	$	0.69	\pm	0.11	$	&	$	40.39	\pm	0.09	$	\\
J1558+35	&	$	5.79	\pm	0.08	$	&	$	3.14	\pm	0.59	$	&	$	40.83	\pm	0.11	$	&	$	0.76	\pm	0.14	$	&	$	40.21	\pm	0.11	$	&	$	0.25	\pm	0.05	$	&	$	39.73	\pm	0.11	$	\\
J1624+33	&	$	4.90	\pm	0.08	$	&	$	0.51	\pm	0.09	$	&	$	40.22	\pm	0.07	$	&	$	0.12	\pm	0.02	$	&	$	39.60	\pm	0.07	$	&	$	0.04	\pm	0.01	$	&	$	39.12	\pm	0.07	$	\\
J1713+57	&	$	5.34	\pm	0.02	$	&	$	2.53	\pm	0.17	$	&	$	41.42	\pm	0.04	$	&	$	0.61	\pm	0.04	$	&	$	40.80	\pm	0.04	$	&	$	0.20	\pm	0.01	$	&	$	40.32	\pm	0.04	$	\\
J2154+11	&	$	5.25	\pm	0.07	$	&	$	1.34	\pm	0.23	$	&	$	40.80	\pm	0.10	$	&	$	0.32	\pm	0.06	$	&	$	40.18	\pm	0.10	$	&	$	0.11	\pm	0.02	$	&	$	39.70	\pm	0.10	$	\\

\hline
    
	\end{tabular}
\end{table*}

\end{appendix}

\end{document}